%% file: pubVBFW.tex
\newcommand*{\ATLASLATEXPATH}{latex/}
\documentclass[cernpreprint,texlive=2016,txfonts,UKenglish]{latex/atlasdoc} 
\pdfoutput=1

\usepackage[subfigure]{\ATLASLATEXPATH atlaspackage}
\usepackage{\ATLASLATEXPATH atlasbiblatex}

\usepackage{\ATLASLATEXPATH atlascontribute}

\usepackage{\ATLASLATEXPATH atlasphysics}

\graphicspath{{logos/}{figures/}}

\usepackage{pubVBFW-defs}

\usepackage{multirow}
\usepackage{pdflscape}

\input{pubVBFW-metadata}

\hypersetup{pdftitle={ATLAS draft},pdfauthor={The ATLAS Collaboration}}

\begin{document}

\maketitle

\tableofcontents

\clearpage
\section{Introduction} %
\label{sec:intro}
\input{intro}

\section{ATLAS detector and data reconstruction} 
\label{sec:detector}
\input{detector}

\section{Event selection}
\label{sec:evtsel}
\input{atlas-evtsel}

\section{Modelling of signal and background processes}
\label{sec:theory}
\input{modelintro}

\subsection{Monte Carlo simulation}
\label{sec:mc}
\input{montecarlo}

\subsection{Multijet background}
\label{sec:multijet}
\input{multijet}

\subsection{Distributions and yields}
\label{sec:plots}
\input{plots}

\section{Fiducial and total electroweak \wjets cross sections}
\label{sec:xsec}
\input{fidxsec}

\subsection{Control-region constraint}
\input{cr}

\subsection{Uncertainties in $\muew$}
\input{fidxsuncertainties}

\subsection{Electroweak \wjets cross-section results}
\label{sec:fidxsresults}
\input{fidxsresults}

\section{Differential cross sections}
\label{sec:diffxsec}
\input{diffxsecintro}

\subsection{Unfolding and uncertainties}
\label{sec:diff:unf}
\input{unfolding}

\subsection{Fiducial regions and integrated cross sections}
\input{diffregions}

\subsection{Observables distinguishing QCD \wjets and EW \wjets}
\label{sec:diff:obsdist}
\input{diffqcdvsew}

\subsubsection{Dijet observables}
\input{diffdijet}

\subsubsection{Object topology relative to the rapidity gap}
\input{difftopology}

\subsection{Observables sensitive to anomalous gauge couplings}
\input{diffatgc}

\section{Anomalous triple-gauge-boson couplings}
\label{sec:aTGC}
\input{atlas-aTGC}

\section{Summary}
\label{sec:summary}
\input{summary}


\section*{Acknowledgements}

\input{Acknowledgements}

\clearpage
\appendix
\section{Appendix}
\label{sec:appendix}

\input{appendix}

\clearpage

\printbibliography


\newpage \input{atlas_authlist}

\end{document}

%% file: pubVBFW-metadata.tex
\AtlasTitle{Measurements of electroweak $\wjets$ production and constraints on anomalous gauge couplings with the ATLAS detector}

\author{The ATLAS Collaboration}

\AtlasRefCode{STDM-2014-11}

\PreprintIdNumber{CERN-EP-2017-008}

\AtlasDate{\today}

\arXivId{1703.04362}

 \AtlasJournalRef{Eur. Phys. J. C77 (2017) 474}
 \AtlasDOI{10.1140/epjc/s10052-017-5007-2}

\AtlasAbstract{%
Measurements of the electroweak production of a $W$ boson in association with two jets at high dijet invariant 
mass are performed using $\sqrt{s} = 7$~and~$8$~\TeV~proton--proton collision data produced by the Large Hadron 
Collider, corresponding respectively to 4.7~and~20.2~fb$^{-1}$ of integrated luminosity collected by the ATLAS 
detector.  The measurements are sensitive to the production of a $W$ boson via a triple-gauge-boson vertex and 
include both the fiducial and differential cross sections of the electroweak process.
}

%% file: intro.tex
The non-Abelian nature of the Standard Model (SM) electroweak theory predicts the self-interactions of the weak 
gauge bosons.  These triple and quartic gauge-boson couplings provide a unique means to test for new fundamental 
interactions.  The fusion of electroweak (EW) bosons is a particularly important process for measuring particle 
properties, such as the couplings of the Higgs boson, and for searching for new particles beyond the 
Standard Model\,\cite{vbfsearchinvatlas,Khachatryan:2016whc,Chatrchyan:2014tja,Aad:2015nfa,
Khachatryan:2016mbu,Khachatryan:2015kxa,vbfsearchdm,vbfsearchsusy,vbfsearchsusy2,vbfsearchdchiggs,vbfsearchdchiggs2}.  
In proton--proton ($pp$) collisions, a characteristic signature of these 
processes is the production of two high-momentum jets of hadrons at small angles with respect to the incoming 
proton beams\,\cite{vbftopology}.  Measurements of this vector-boson-fusion (VBF) topology have been performed 
in $W$\,\cite{cmsvbfw}, $Z$\,\cite{Aad:2014dta,cmsvbfz} and Higgs\,\cite{higgscoupling} boson production, though 
the observation of purely electroweak processes in this topology has only been achieved in individual 
measurements of $Z$-boson production.  This paper presents a precise measurement of electroweak $W$-boson 
production in the VBF topology, with a significance well above the standard for claiming observation, as well 
as differential cross section measurements and constraints on anomalous triple-gauge-boson couplings (aTGCs).

The production of a $W$~boson in association with two or more jets (\wjets) is dominated by processes involving 
strong interactions (strong \wjets or QCD \wjets).  These processes have been extensively studied by experiments 
at the Large Hadron Collider (LHC)\,\cite{ATLASWjets, CMSWjets} and the Tevatron collider\,\cite{CDFWjets, D0Wjets}, 
motivating the development of precise perturbative predictions\,\cite{mcfm2jets,mcfm2jetslhc,mcfm3jets,
blackhat3jets,blackhat4jets,blackhat5jets,mepsatnlo, madgraph, powhegwjj,Andersen:2009nu,Andersen:2009he,
Andersen:2011hs,Andersen:2012gk}.  The large cross section for $W$-boson production provides greater sensitivity 
to the VBF topology and to the electroweak production of \wjets (electroweak \wjets or EW \wjets) than 
corresponding measurements of $Z$- or Higgs-boson production.  

\begin{figure}[htbp]
  \centering
  \subfigure[Vector boson fusion]{
      \includegraphics[width=0.28\textwidth]{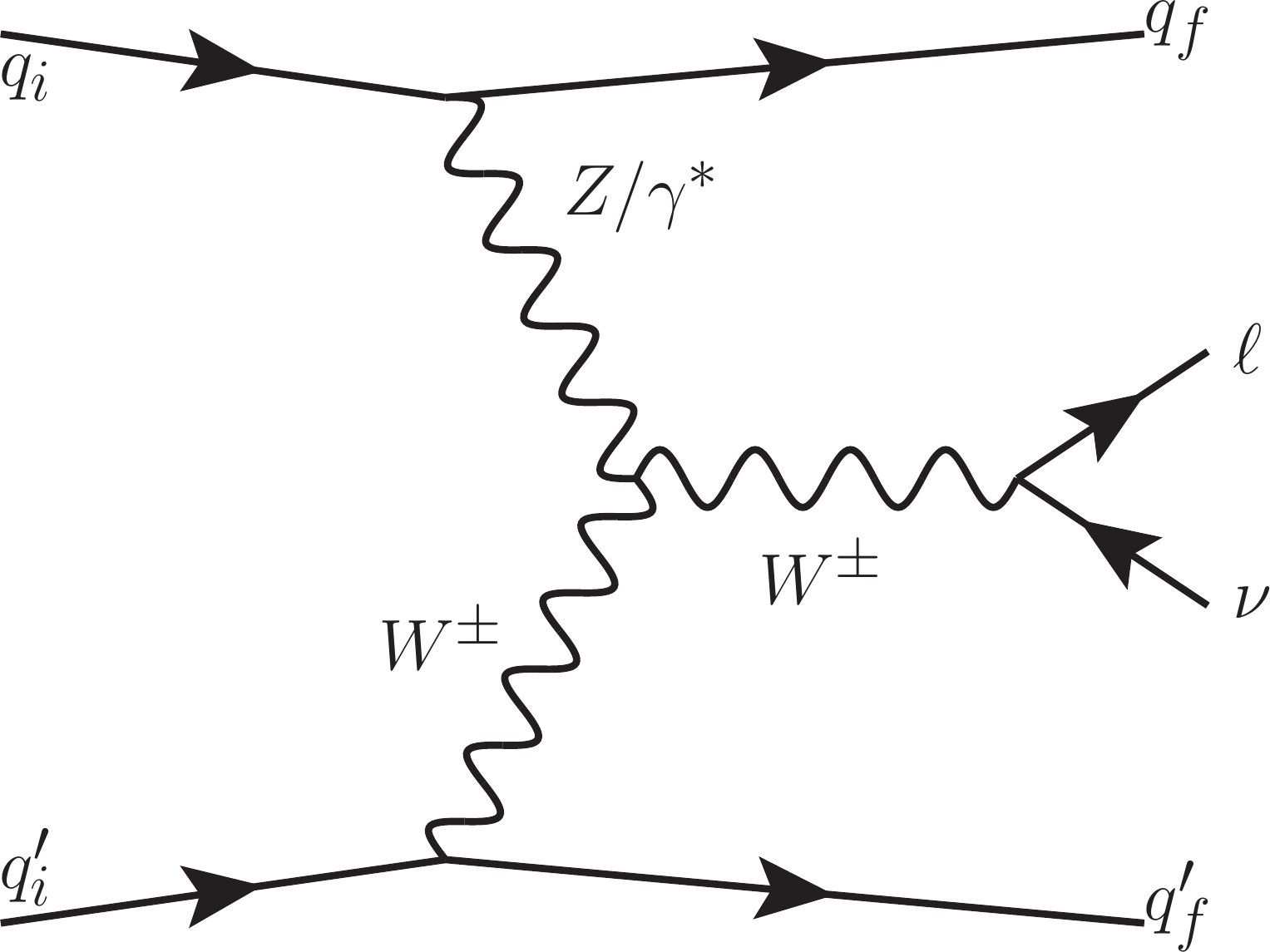}
  }     
  \qquad
  \subfigure[$W$~bremsstrahlung]{
    \includegraphics[width=0.28\textwidth]{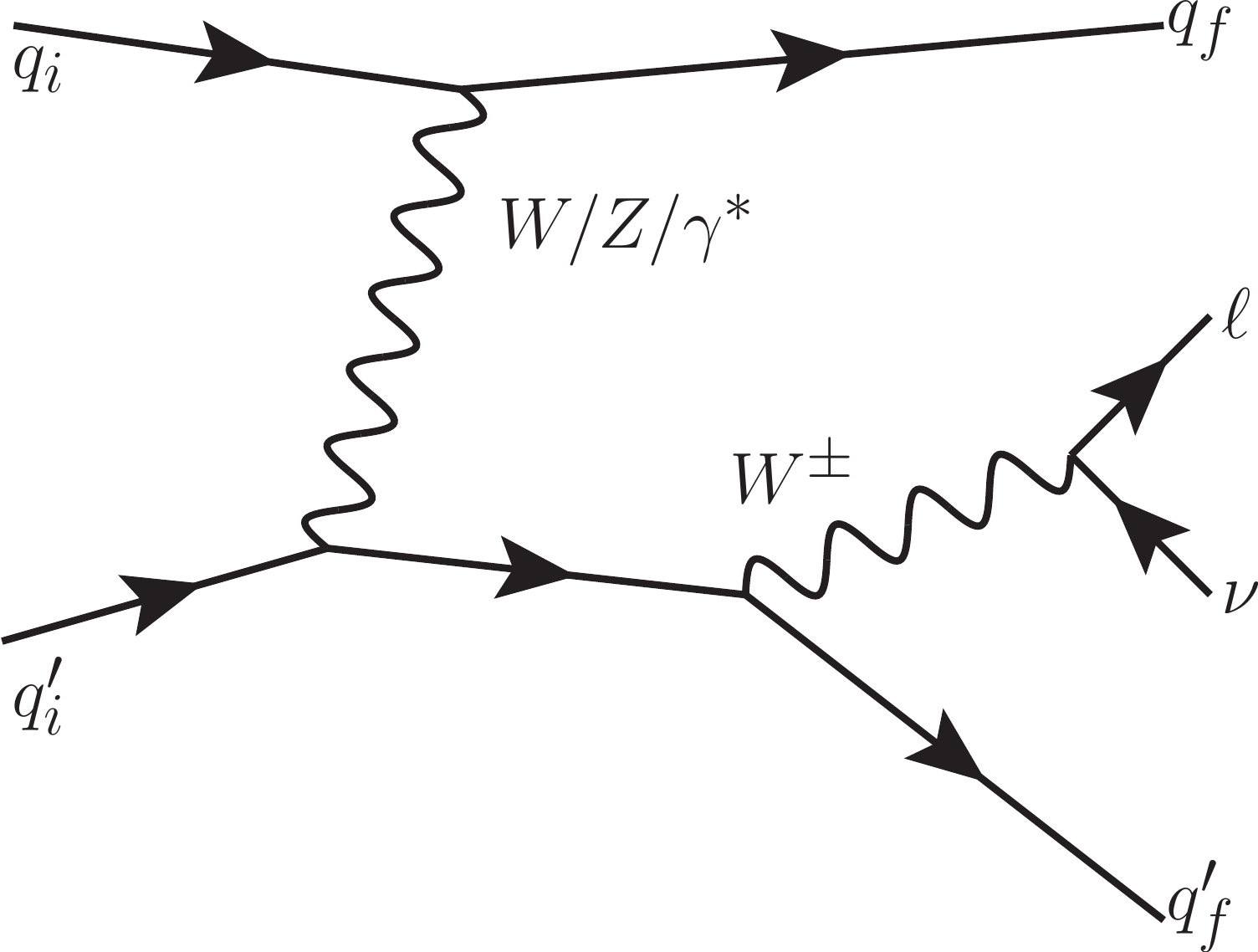}
  }
  \qquad
  \subfigure[Non-resonant]{  
    \includegraphics[width=0.28\textwidth]{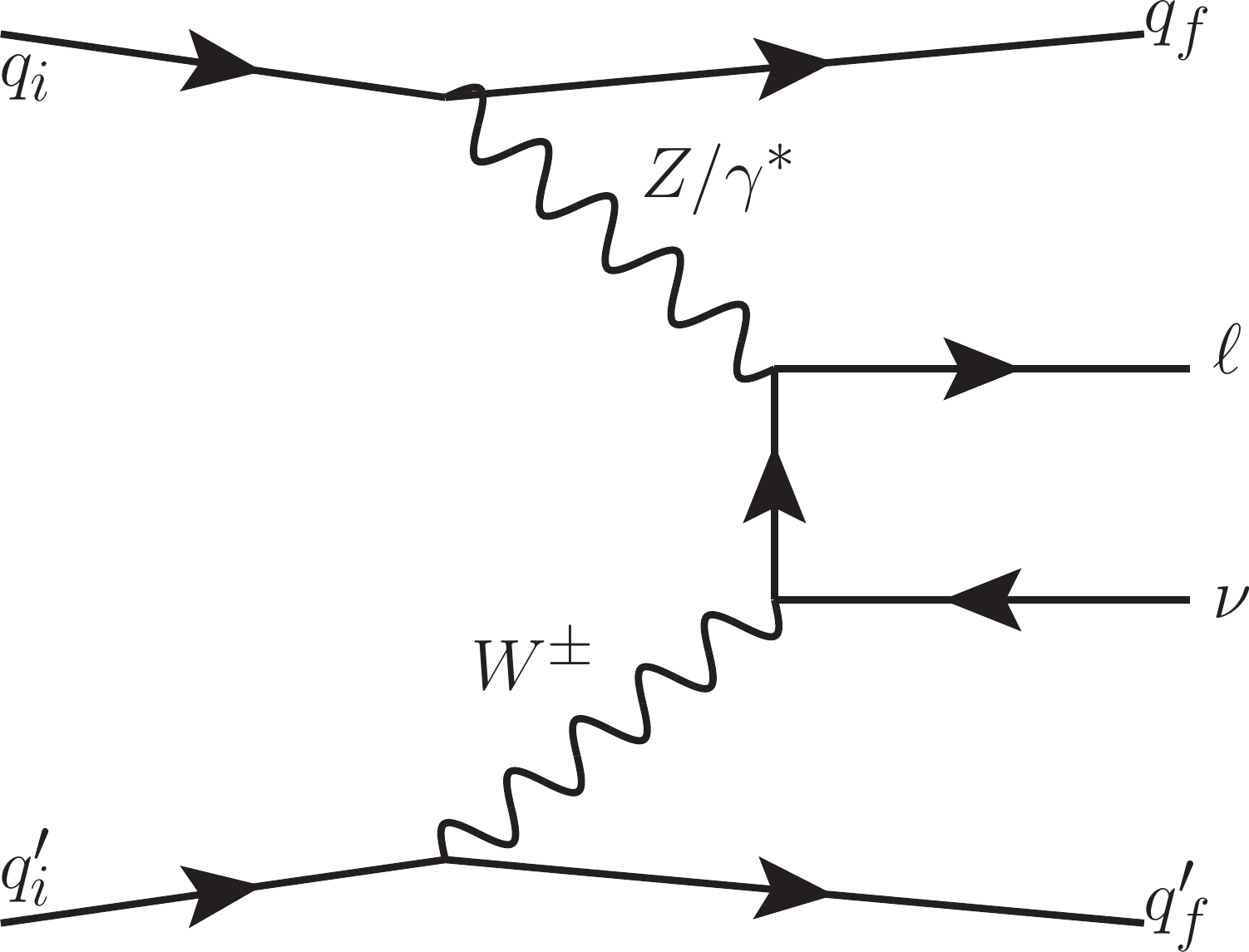}
  }
  \caption{Representative leading-order diagrams for electroweak $Wjj$ production at the LHC. In addition to 
(a) the vector boson fusion process, there are four (b) $W$~bremsstrahlung diagrams, corresponding to $W^\pm$ boson 
radiation by any incoming or outgoing quark, and two (c) non-resonant diagrams, corresponding to $W^\pm$ boson 
radiation by either incoming quark.}
  \label{intro:Fig:EWK2Jets}
\end{figure}

\begin{figure}[htbp]
\centering
    \includegraphics[width=0.28\textwidth]{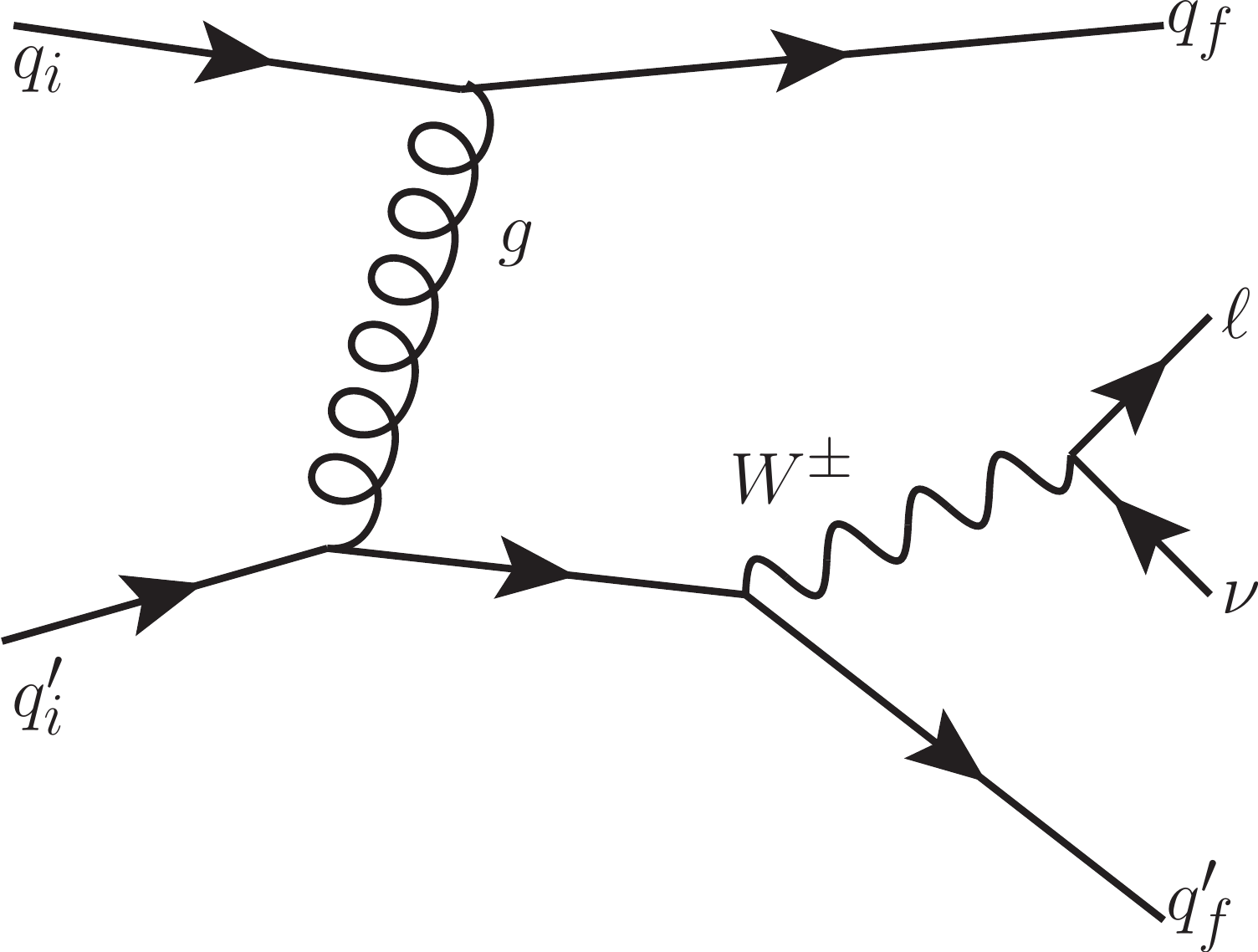}
  \qquad
    \includegraphics[width=0.28\textwidth]{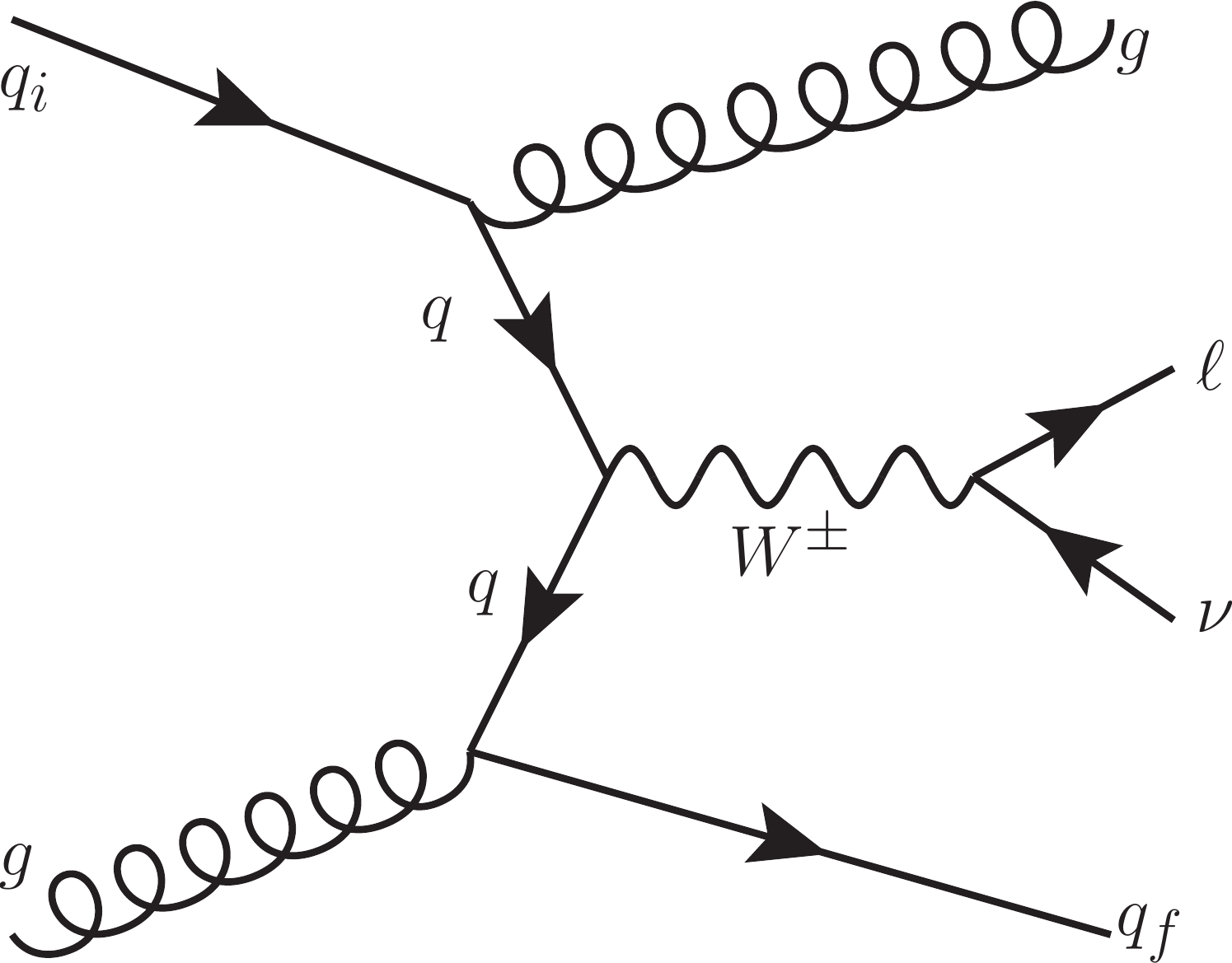}
  \caption{Examples of leading-order diagrams for strong $Wjj$ production at the LHC.  The left-hand diagram 
interferes with the electroweak diagrams of Figure~\ref{intro:Fig:EWK2Jets} when the final-state quarks have the 
same colours as the initial-state quarks.}
  \label{intro:Fig:QCD2Jets}
\end{figure}

The VBF process is inseparable from other electroweak \wjets~processes, so it is not measured directly; sensitivity 
to the VBF production mechanism is quantified by determining constraints on operator coefficients in an effective 
Lagrangian approach~\cite{atgc}.  The classes of electroweak diagrams constituting the signal are shown in 
Figure~\ref{intro:Fig:EWK2Jets}~\cite{EWvbfQCDcorr} and contain at least three vertices where an electroweak gauge 
boson connects to a pair of fermions.  Diboson production, where the final-state quarks result from the decay of 
an $s$-channel gauge boson, is not shown and is considered as a background; it is small for the VBF topology 
defined in the analysis.  The large background from a $W$ boson associated with strongly produced jets is shown 
in Figure~\ref{intro:Fig:QCD2Jets} and has only two electroweak vertices.  This background has ${\cal{O}}$(10) 
times the yield of the signal process, and can interfere with the signal.  This interference is suppressed because 
only a small subset of the background diagrams have the same initial and final state as the signal. 

The analysis signature consists of a neutrino and either an electron or a muon, two jets with a high dijet invariant 
mass, and no additional jets at a wide angle from the beam.  This signature discriminates signal events from the 
copious background events consisting of strongly produced jets associated with a $W$ (or $Z$) boson, top-quark 
production, or multijet production.  The purity of electroweak \wjets~production increases with increasing dijet 
invariant mass, increasing the sensitivity to anomalous triple-gauge-boson couplings.

Measurements of the inclusive and fiducial cross sections of electroweak \wjets~production in proton--proton 
collisions at centre-of-mass energies $\sqrt{s}=7$~and~8~\TeV~are performed in a fiducial region with a 
signal-to-background ratio of approximately 1:8.  The electroweak signal is extracted with a binned likelihood 
fit to the dijet invariant mass distribution.  The fit determines the ratio $\muew$ of the measured signal 
cross section to that of a Standard Model calculation~\cite{powhegvbfw}; this 
ratio is then multiplied by the prediction to provide the measured cross section.
To reduce the uncertainties in the modelling of the strong \wjets~events, data are used to constrain their 
dijet mass distribution, resulting in a precise measurement of the electroweak \wjets fiducial cross 
section.  The quantum-mechanical interference between electroweak and strong \wjets processes is not modelled 
and its impact on the measurement is estimated using a Monte Carlo simulation and taken as an uncertainty.  

In order to explore the kinematics of the \wjets~topology, and the interplay between strong and electroweak 
production, the 8~\TeV~data are unfolded differentially to particle level in many variables and 
phase-space regions, and compared to theoretical predictions.  Electroweak \wjets~production is measured 
in regions where the signal purity is relatively high ($\gtrsim 10\%$); combined strong and electroweak 
\wjets~production is measured in the other regions.  These measurements are then integrated to obtain 
fiducial cross sections in the different phase-space regions, albeit with larger uncertainties than the 
measurement with the constrained background.

Sensitivity to the VBF diagram is determined by modifying the triple-gauge-boson couplings.  Anomalous 
couplings arising from new processes at a high energy scale would cause increasing deviations from the 
SM prediction for increasing momentum transfer between the incoming partons.  Hence, a region of high 
momentum transfer is defined, and constraints on anomalous gauge couplings are set in the context of an 
effective field theory (EFT), including limits on interactions that violate charge-parity (CP) conservation.

The paper is organized as follows.  The ATLAS detector and reconstruction of the final-state particles 
are described in Section~\ref{sec:detector}.  The definitions of the measurement phase-space regions and 
the event selection are given in Section~\ref{sec:evtsel}.  The modelling of signal and background 
processes is discussed in Section~\ref{sec:theory}.  Section~\ref{sec:xsec} is dedicated to the precise 
extraction of the inclusive and fiducial cross sections, while Section~\ref{sec:diffxsec} 
presents differential cross sections unfolded for detector effects.  Section~\ref{sec:aTGC} describes limits 
on aTGCs and parameters of an effective field theory.   Section~\ref{sec:summary} summarizes the results and the 
Appendix provides a comprehensive set of differential cross-section measurements.

%% file: detector.tex
The data set corresponds to LHC $pp$ collisions at $\sqrt{s}=7$~\TeV~in 2011 and at 
$\sqrt{s}=8$~\TeV~in 2012, with final-state particles measured by the ATLAS detector.  
This section describes the detector and the reconstruction of the data to produce 
the final-state physics objects used in the measurements.

\subsection{ATLAS detector}

ATLAS is a multi-purpose detector used to measure LHC particle collisions.  A detailed description of 
the detector can be found in Ref.~\cite{PERF-2007-01}.  A tracking system comprises the inner 
detector (ID) surrounding the collision point, with silicon pixel and microstrip detectors most 
centrally located, followed by a transition radiation tracker at higher 
radii~\cite{IDtracking,ATLAS-CONF-2012-042}.  These tracking detectors are used to 
measure the trajectories and momenta of charged particles up to pseudorapidities of 
$|\eta|$ = 2.5.\footnote{ATLAS uses a right-handed coordinate system with its origin at the nominal 
interaction point in the centre of the detector and the $z$-axis along the beam pipe. The $x$-axis 
points from the interaction point to the centre of the LHC ring, and the $y$-axis points upward. 
Cylindrical coordinates $(r,\phi)$ are used in the transverse plane, $\phi$ being the azimuthal angle 
around the $z$-axis. The pseudorapidity is defined in terms of the polar angle $\theta$ as $\eta=-\ln\tan(\theta/2)$. 
The rapidity is defined as $y=0.5\ln [(E+p_z)/(E-p_z)]$, where $E$ and $p_z$ are the energy and 
longitudinal momentum, respectively.  Momentum in the transverse plane is denoted by $\pt$. }  
The ID is surrounded by a superconducting solenoid, providing a 2~T magnetic field for the 
tracking detectors.   

A calorimeter system surrounds the solenoid magnet and consists of electromagnetic and hadronic sections.  
The electromagnetic section is segmented along the $z$-axis into a barrel region covering $|\eta| < 1.475$, 
two end-cap components spanning $1.375 < |\eta| < 3.2$, and two forward components ($3.1 < |\eta| < 4.9$).  
Similarly, the hadronic section comprises a barrel region ($|\eta| < 1.7$), two end-cap regions 
($1.5 < |\eta| < 3.2$), and two forward regions ($3.1 < |\eta| < 4.9$).  The barrel region of the hadronic 
section uses scintillator tiles as the active medium, while the remaining regions use liquid argon.  

A muon spectrometer surrounds the calorimeter system and contains superconducting coils, drift 
tubes and cathode strip chambers to provide precise measurements of muon momenta within $|\eta| < 2.7$.   
The spectrometer also includes resistive-plate and thin-gap chambers to trigger on muons in the 
region $|\eta| < 2.4$. 

The ATLAS trigger system uses three consecutive stages to select events for permanent storage.  
The first level uses custom electronics and the second level uses fast software algorithms to inspect 
regions of interest flagged by the first trigger level.  At the third level, the full event is 
reconstructed using software algorithms similar to those used offline.

\subsection{Object reconstruction}

Electrons, muons, and hadronic jets are reconstructed in the ATLAS detector.   Each type of object has a 
distinctive signature and is identified using the criteria described below.  The object identification 
includes track and vertex positions relative to the primary event vertex, defined as the reconstructed 
vertex with the highest summed $\pt^2$ of all associated tracks.  Each object is calibrated and modelled 
in Monte Carlo simulation, corrected to match data measurements of the trigger, reconstruction, and 
identification efficiencies, and of the energy and momentum scales and 
resolutions~\cite{electronCalib, electronID, electronCP, MuonCP, JetEnergy}.

\textbf{Electrons}

Electron candidates are reconstructed from energy clusters in the electromagnetic section of the calorimeter 
which are matched to tracks reconstructed in the ID.  Candidates for signal events are required to satisfy 
`tight' selection criteria\,\cite{electronID, electronCP}, which include requirements on calorimeter shower 
shape, track hit multiplicity, the ratio of reconstructed energy to track momentum, $E/p$, and the matching of 
the energy clusters to the track.  In order to build templates to model the multijet background (see 
Section~\ref{sec:multijet}), a set of criteria is employed based on `loose' or `medium' selection, which drops 
the $E/p$ requirement and uses less restrictive selection criteria for the other discriminating variables.  

Electron candidates are required to be isolated to reject possible misidentified jets or heavy-flavour hadron decays.  
Isolation is calculated as the ratio of energy in an isolation cone around the primary track or calorimeter deposit 
to the energy of the candidate.  Different isolation requirements are made in the 7~\TeV~and 8~\TeV~data sets, due 
to the different LHC and detector operating conditions.  For 7~\TeV~data taking, the requirements on track and 
calorimeter isolation variables associated with the electron candidate achieve a constant identification efficiency as 
a function of the candidate transverse energy ($\et$) and pseudorapidity.  The 8~\TeV~trigger includes a requirement 
on track isolation, so the selection is more restrictive and requires the summed $\pt$ of surrounding tracks 
to be $<5\%$ of the electron candidate $\et$, excluding the electron track and using a cone of size 
$R \equiv \sqrt{(\Delta\phi)^2 + (\Delta\eta)^2} = 0.2$ around the shower centroid. 

\textbf{Muons}

Muon candidates are identified as reconstructed tracks in the muon spectrometer which are matched to and 
combined with ID tracks to form a `combined' muon candidate\,\cite{MuonCP}.  Quality 
requirements on the ID track include a minimum number of hits in each subdetector to ensure good 
track reconstruction.  Candidates in 7~\TeV~data are selected using a track-based fractional isolation 
requiring the scalar sum of the $\pt$ values of tracks within a cone of size $R = 0.2$ of the muon track to 
be less than 10\% of the candidate $\pt$.  For 8~\TeV~data taking, requirements are applied to track and calorimeter 
fractional isolation using a cone of size $R = 0.3$.  The upper bound on each type of isolation increases 
with increasing muon $\pt$, and is 15\% for $\pt>30$~\GeV.  

Additional transverse ($d_0$) and longitudinal ($z_0$) impact parameter requirements of $|d_0/\sigma_{d_0}|<3$ 
(where $\sigma_{d_0}$ is the $d_0$ uncertainty) and $|z_0\sin\theta|<0.5$~mm are imposed on all muon and 
electron candidates to suppress contributions from hadron decays to leptons.

\textbf{Jets} 

Jets are reconstructed using the anti-$k_t$ algorithm\,\cite{antikt} with a jet-radius parameter of 0.4, from 
three-dimensional clustered energy deposits in the calorimeters\,\cite{Topocluster}.  Jets are required to have 
$\pt>30$~\GeV~and $|\eta|<4.4$, and must be separated from the lepton in $\eta$--$\phi$ space, 
$\Delta R (\ell, j) \geq 0.3$.  Quality requirements are imposed to remove events where jets are associated with 
noisy calorimeter cells.  Jet energies 
are corrected for the presence of low-energy contributions from additional in-time or out-of-time collisions 
(pile-up), the non-compensating response of the calorimeter, detector material variations, and energy losses in 
uninstrumented regions.  This calibration is performed in bins of $\pt$ and $\eta$, using correction factors 
determined using a combination of Monte Carlo simulations and in-situ calibrations with 
data\,\cite{JetEnergy,MCJESinsituComb}.  The systematic uncertainties in these correction factors are determined 
from the same control samples in data.  A significant source of uncertainty in this analysis arises from the 
modelling of the $\eta$ dependence of the jet energy response.

To suppress the contribution of jets from additional coincident $pp$ collisions, the jet vertex fraction 
(JVF)\,\cite{Pileup} is used to reject central jets ($|\eta|<2.4$) that are not compatible with originating from 
the primary vertex.  The JVF is defined as the scalar sum of the $\pt$ values of tracks associated with both the 
primary vertex and the jet, divided by the summed $\pt$ of all tracks associated with the jet.  For the 7~\TeV~data 
taking, the requirement is $|\mathrm{JVF}|\geq 0.75$; this requirement is loosened in 8~\TeV~data taking to 
$|\mathrm{JVF}|\geq 0.5$ if the jet has $\pt<50$~\GeV.  The relaxed requirement in 8~\TeV~data is due to the larger 
pile-up rate causing signal events to be rejected when using the 7~\TeV~selection, and the requirement of 
$|\eta|<2.4$ is to ensure the jets are within the ID tracking acceptance.

Jets that are consistent with originating from heavy-flavour quarks are identified using a neural network algorithm 
trained on input variables related to the impact parameter significance of tracks in the jet and the secondary 
vertices reconstructed from these tracks~\cite{bjetid}.  Jets are identified as $b$-jets with a selection on 
the output of the neural network corresponding to an identification efficiency of 80\%.  

\textbf{Missing transverse momentum}

In events with a leptonically decaying $W$ boson, one expects large missing momentum in the transverse plane due 
to the escaping neutrino.  The magnitude of this missing transverse momentum (\met) is constructed from the vector 
sum of muon momenta and three-dimensional energy clusters in the calorimeter~\cite{Aad:2012re,MET8TeV}.  The 
clusters are corrected to account for the different response to hadrons compared to electrons or photons, as well 
as dead material and out-of-cluster energy losses.  Additional tracking information is used to extrapolate 
low-momentum particles to the primary vertex to reduce the contribution from pile-up.

%% file: atlas-evtsel.tex
The proton--proton collision data samples correspond to a total integrated luminosity of 4.7~fb${}^{-1}$ 
for the 7~\TeV~data and 20.2~fb${}^{-1}$ for the 8~\TeV~data with uncertainties of 
1.8\%\,\cite{Aad:2013ucp} and 1.9\%\,\cite{Lumi8TeV}, respectively.

The measurements use data collected with single-electron and single-muon triggers.  The triggers identify 
candidate muons by combining an ID track with a muon-spectrometer track, and candidate electrons by matching an 
inner detector track to an energy cluster in the calorimeter consistent with an electromagnetic shower.  The 
triggers in the 7~\TeV~data require $\pt > 18$~\GeV~for muons and either $\et > 20$~\GeV~or $\et > 22$~\GeV~for 
electrons, depending on the data-taking period.  The 8~\TeV~data events are selected by two triggers in each 
channel.  The electron-channel triggers have \et thresholds of 24~\GeV~and 60~\GeV, where 
the lower-threshold trigger includes a calorimeter isolation criterion: the measured \et within a cone 
of radius $R=0.2$ around the electron candidate, excluding the electron candidate's \et, must be less than 
10\% of the \et of the electron.  
The muon-channel triggers have \pt thresholds of 24~\GeV~and 36~\GeV.  The lower-threshold 
trigger has a track-isolation requirement, where the scalar summed $\pt$ of tracks within a cone of radius 
$R = 0.2$ around the muon is required to be less than 12\% of the $\pt$ of the muon.  

The analysis defines many measurement regions varying in electroweak \wjets purity.  
Table~\ref{tab:selection} shows the regions at the generated particle level based on the variables defined below.  
Particle-level objects are reconstructed as follows: jets are reconstructed using the anti-$k_t$ algorithm with a 
radius parameter of $0.4$ using final-state particles with a proper lifetime longer than 10~ps; and leptons are 
reconstructed by combining the final-state lepton with photons within a cone of $R=0.1$ around the lepton.  The 
requirements in Table~\ref{tab:selection} are also used to select data events, except for the following 
differences: (1) electrons must have $|\eta|<2.47$ and cannot be in the crack region of the calorimeter 
($1.37 < |\eta| < 1.52$); (2) muons must have $|\eta|<2.4$; and (3) jets are selected using pseudorapidity 
($|\eta|<4.4$) rather than rapidity.  Also, a $b$-jet veto is applied to the validation region in data when 
performing the measurement of the fiducial electroweak \wjets cross section described in Section~\ref{sec:xsec}. 

\begin{table*}[tb!]
\caption{
Phase-space definitions at the generated particle level.  Each phase-space region includes the 
preselection and the additional requirements listed for that region.  The variables are defined 
in Sections~\ref{sec:presel}~and~\ref{sec:sel}.
}
\label{tab:selection}
\begin{center}
\begin{tabular*}{0.95\textwidth}{ll}
\toprule
Region name & Requirements
\\
\midrule
Preselection               & Lepton $\pt > 25$~\GeV \\
                           & Lepton $|\eta|<2.5$ \\
                           & $\met > 25$~\GeV \\
                           & $\mT > 40$~\GeV \\
                           & $\ptjlead > 80$~\GeV \\
                           & $\ptjsub > 60$~\GeV \\
                           & Jet $|y|<4.4$ \\ 
                           & $\mjj > 500$~\GeV \\
                           & $\dyjj>2$ \\    
                           & $\Delta R(j,\ell)>0.3$ \\    
\hline
Fiducial and differential measurements &                      \\
~~Signal region  & $\ncenlep = 1, \ncenjet = 0$ \\
~~Forward-lepton control region  & $\ncenlep = 0, \ncenjet = 0$ \\
~~Central-jet validation region  & $\ncenlep = 1, \ncenjet \geq 1$ \\
\hline
Differential measurements only & \\
~~Inclusive regions & $\mjj > 0.5$~\TeV, 1~\TeV, 1.5~\TeV, or 2~\TeV \\
~~Forward-lepton/central-jet region & $\ncenlep = 0, \ncenjet \geq 1$ \\
~~High-mass signal region & $\mjj > 1$~\TeV, $\ncenlep = 1, \ncenjet = 0$ \\
\hline
Anomalous coupling measurements only & \\
~~High-$q^2$ region & $\mjj > 1$~\TeV, $\ncenlep = 1, \ncenjet = 0$, $\ptjlead > 600$~\GeV \\
\bottomrule
\end{tabular*}
\end{center}
\end{table*}

\subsection{Event preselection}
\label{sec:presel}
Signal candidate events are initially defined by the presence of missing transverse momentum ($\met>20$~\GeV), 
exactly one charged lepton (electron or muon) candidate with $\pt > 25$~\GeV, and at least two jets.  The 
highest-$\pt$ jet is required to have $\ptjlead>80$~\GeV~and the second jet must have $\ptjsub>60$~\GeV.
To isolate events with a $W$ boson, a veto is imposed on events with a second same-flavour lepton with 
$\pt > 20$~\GeV; these leptons are identified in data using relaxed isolation and impact parameter criteria.  
A minimum cut on the transverse mass, $\mT>40$~\GeV, of the $W$-boson candidate is additionally imposed, 
where $\mT$ is defined by:
\begin{equation}
\mT = \sqrt{\,2\pt \cdot\met\left[1-\cos\Delta\phi(\ell, \met)\right]\, }.
\nonumber
\end{equation}

Jets are selected in data if they have $|\eta|<4.4$ and $\Delta R(j,\ell)>0.3$.  A VBF topology is selected 
by requiring the invariant mass of the dijet system defined by the two highest-$\pt$ jets to satisfy 
$\mjj>500$~\GeV, and the absolute value of the rapidity separation of the jets to satisfy $\dyjj > 2$.

\subsection{Definitions of the measurement regions}
\label{sec:sel}
\begin{figure}[htbp]
\centering
\includegraphics[width=0.7\textwidth]{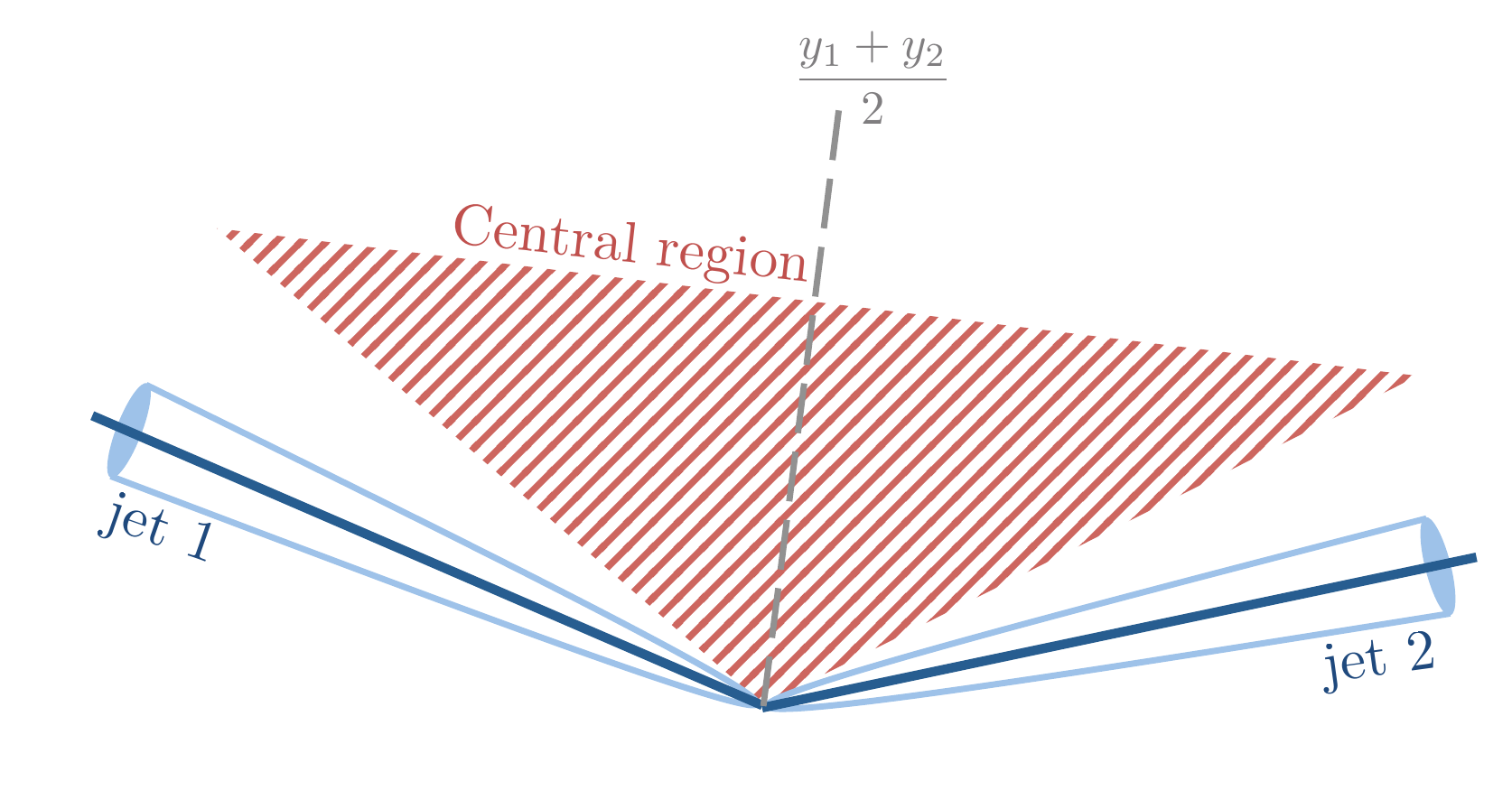}
\caption{Illustration of the central region used to count leptons and jets in the definition of the signal, 
control, and validation regions.  The rapidity range of the region corresponds to $C_{\textrm{max}}=0.4$ in 
Eq.~(\ref{evtsel:rapinterval}).  An object in the direction of the dashed line has $C = 0$. }
\label{evtsel:vetoIllustration}
\end{figure}

The above preselection defines an {\em inclusive} fiducial region, which is then split into four orthogonal fiducial 
regions defined by the presence or absence of the lepton or an additional jet in a ``central'' rapidity range between 
the two highest-$\pt$ jets.  The signal EW \wjets process is characterized by a lepton and no jets in the central rapidity 
range.  This range is determined by the centrality variable $C_{\ell}$ or $C_j$ for the lepton or jets respectively:
                  \begin{equation}
                   C_{\ell~(j)} \equiv \left|\frac{y_{\ell~(j)} - \frac{y_1+y_2}{2}}{y_1-y_2}\right|,
                   \label{eqn:lepcentrality}
                   \end{equation}
where $y_{\ell~(j)}$ is the rapidity of the candidate lepton (jet), and $y_1$ and $y_2$ are the rapidities of the 
highest-\pt (leading) and next-highest-\pt (subleading) jets.  Requiring the centrality to be below a value 
$C_{\textrm{max}}$ defines the selection of a rapidity range centred on the mean rapidity of the leading jets, i.e.,
\begin{equation}
\left[\frac{y_1+y_2}{2}-C_{\textrm{max}}\times |y_1-y_2|,\quad \frac{y_1+y_2}{2}+C_{\textrm{max}}\times |y_1-y_2|\right],
\label{evtsel:rapinterval}
\end{equation}
as illustrated in Figure~\ref{evtsel:vetoIllustration}.  For $C_{\textrm{max}}=0.5$, the interval spans the entire 
rapidity region between the two jets; the number of jets within this interval is denoted \ngapjet.  In defining the 
electroweak \wjets signal region, $C_{\textrm{max}}=0.4$ is used to count the number of leptons (\ncenlep) or jets 
(\ncenjet) within the range.  A value of $C_{\textrm{max}}=0.4$ permits an event with the emission of an additional 
jet close to one of the two highest-\pt jets to be retained as a candidate signal event.  

\begin{figure}[htbp]
\centering
\includegraphics[width=0.65\textwidth]{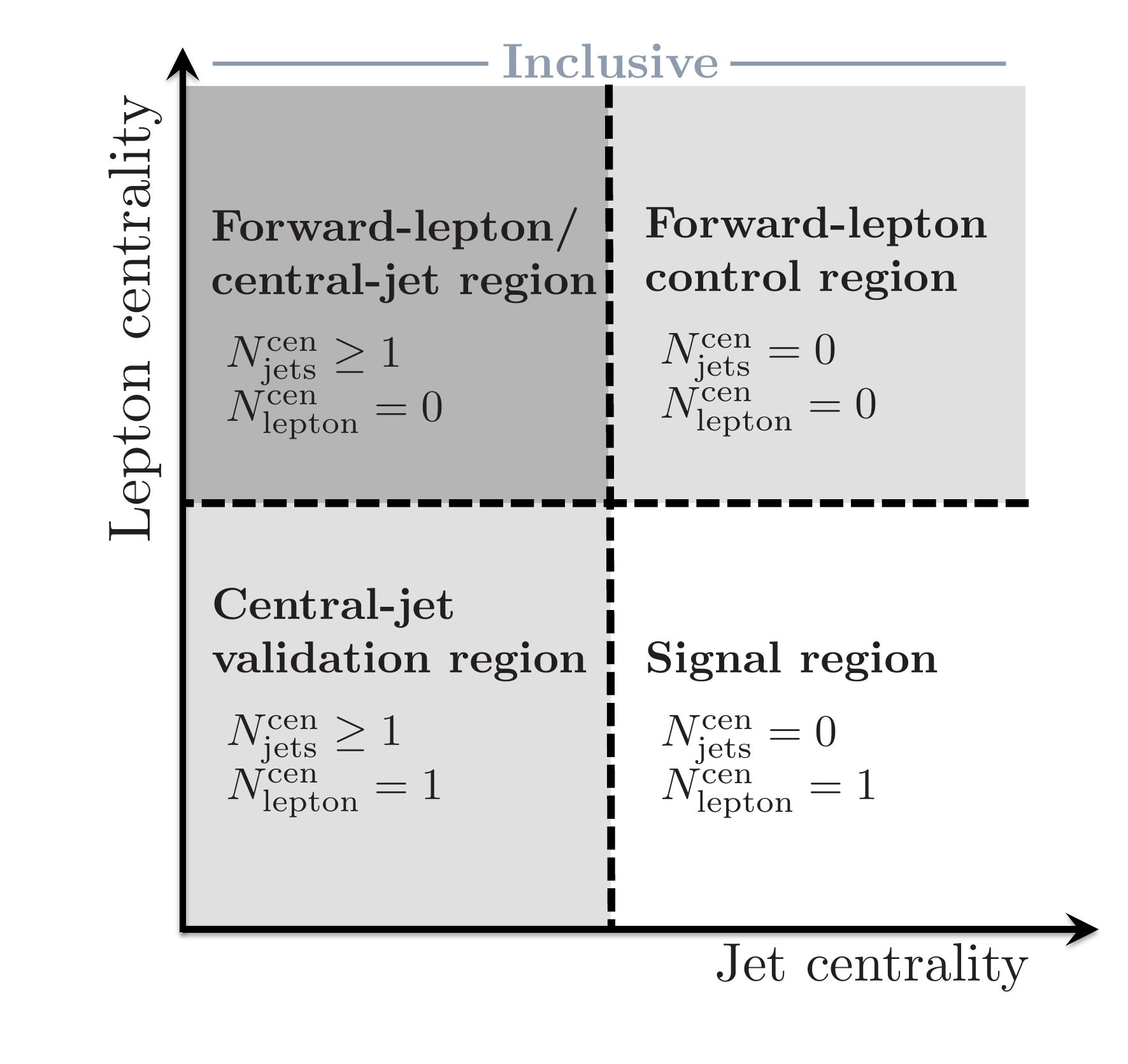}
\caption{Illustration of the relationship between the signal, control, and validation fiducial regions.  The signal 
region is defined by both a veto on additional jets (beyond the two highest-$\pt$ jets) and the presence of a lepton 
in the rapidity region defined in Eq.~(\ref{evtsel:rapinterval}).  The signal region is studied with either 
$\mjj>0.5$~\TeV~or $1$~\TeV.  A forward-lepton/central-jet fiducial region is also defined, for which the centrality 
requirements on the jets and the lepton are inverted with respect to the signal region.  The inclusive region 
corresponds to the union of all four regions, and is studied with $\mjj>0.5,~1.0,~1.5,$ or 2.0~TeV.
The quantities $\ncenjet$ and $\ncenlep$ refer to the number of reconstructed leptons and additional jets 
reconstructed in the rapidity interval defined by Eq.~(\ref{evtsel:rapinterval}) and illustrated in 
Figure~\ref{evtsel:vetoIllustration}, with $C_{\textrm{max}} = 0.4$.
}
\label{evtsel:phasespaceIllustration}
\end{figure}

The fiducial regions are illustrated in Figure~\ref{evtsel:phasespaceIllustration}.  The signal process is characterized 
by a $W$ boson in the rapidity range spanned by the two jets (Figure~\ref{intro:Fig:EWK2Jets}), with no jets in this 
range due to the absence of colour flow between the interacting partons.  An event is therefore defined as being in 
the electroweak-enhanced {\em signal} region if the identified lepton is reconstructed in the rapidity region 
defined by Eq.~(\ref{evtsel:rapinterval}) and no additional jets are reconstructed in this interval.
A QCD-enhanced {\em forward-lepton control} fiducial region is defined by the requirement that neither the identified 
lepton nor any additional jets be present in the central rapidity interval.
A second QCD-enhanced {\em central-jet validation} region is defined by events having both the identified lepton and 
at least one additional jet reconstructed in the central rapidity interval.  
These three orthogonal fiducial regions are used in Section~\ref{sec:xsec} 
to extract the EW~\wjets production cross section, constrain the modelling of QCD~\wjets 
production from data, and validate the QCD~\wjets  modelling, respectively.

For the determination of unfolded differential cross sections presented in Section~\ref{sec:diffxsec}, four additional 
fiducial regions are studied: the inclusive region for the progressively more restrictive dijet invariant mass 
thresholds of 1.0~\TeV, 1.5~\TeV, and 2.0~\TeV, and an orthogonal {\em forward-lepton/central-jet} region
defined by events with the lepton outside the central region, but at least one additional jet reconstructed in the 
interval.  For the study of EW \wjets differential cross sections, the signal fiducial region with an increased dijet 
invariant mass requirement of $\mjj>1$~\TeV~({\em high-mass signal region}) is also analyzed; a further requirement 
that the leading-jet $\pt$ be greater than 600~\GeV~defines a {\em high-$q^2$} region used for constraints on aTGCs 
(discussed in Section~\ref{sec:aTGC}).

%% file: modelintro.tex
Simulated Monte Carlo (MC) samples are used to model \wjets production, with small data-derived corrections 
applied to reduce systematic uncertainties.  Other processes producing a prompt charged lepton are also 
modelled with MC samples.  The multijet background, where a photon or hadronic jet is misreconstructed as a 
prompt lepton, or where a lepton is produced in a hadron decay, is modelled using data.

%% file: montecarlo.tex
The measurements described in this paper focus on the electroweak production of~\wjets.
This process has different kinematic properties to strong \wjets production, but there is 
nonetheless some small interference between the processes.  The other significant background 
processes are top-quark, $Z$-boson, and diboson production, which are modelled with MC 
simulation.  All MC samples used to model the data are passed through a detector 
simulation\,\cite{SIMREF} based on \geant\,\cite{geant}.  Pile-up interactions are 
modelled with \pythia (v. 8.165)\,\cite{pythia8}.  Table~\ref{tbl:mc} lists the MC 
samples and the cross sections used in the MC normalization.

\begin{table}[tb!]
\caption{
 Monte Carlo samples used to model the signal and background processes.  The cross 
 sections times branching fractions, $\sigma \cdot \mathcal{B}$, are quoted for 
 $\sqrt{s} = 7$~and~8~\TeV.  The branching fraction corresponds to the decay to a 
 single lepton flavour, and here $\ell$ refers to $e$, $\mu$, or $\tau$.  The neutral 
 current $Z/\gamma^*$ process is denoted by $Z$.  To remove overlap between 
 $W(\rightarrow \tau\nu)$ + 2 jets and $WW$/$WZ$ in 7~\TeV~samples, events with 
 a generated $\tau$ lepton are removed from the 7~\TeV~$WW$/$WZ$ samples.  Jets 
 refer to a quark or gluon in the final state of the matrix-element calculation.
}
\label{tbl:mc}
\begin{center}
\begin{tabular}{llllrr}
\toprule
\multicolumn{3}{l}{Process}
& \multicolumn{1}{l}{MC generator }
& \multicolumn{2}{c}{$\sigma \cdot \mathcal{B}$~[pb]}
\\
& & & 
& \multicolumn{1}{r}{7~\TeV} & \multicolumn{1}{r}{8~\TeV}
\\
\midrule
\multicolumn{2}{l}{$W(\rightarrow e\nu, \mu\nu)$ + 2 jets                                       } & & \\
\multicolumn{3}{l}{\quad 2 EW vertices                                                          } & \powheg + \pythia  & 4670 & 5340 \\
\multicolumn{3}{l}{\quad 4 EW vertices (no dibosons)                                            } & \powheg + \pythia  & 2.7 & 3.4 \\
\\
\multicolumn{2}{l}{$W(\rightarrow \tau\nu)$ inclusive                                           } & & \\
\multicolumn{3}{l}{\quad 2 EW vertices                                                          } & \sherpa  & 10100 & 11900 \\
\\
\multicolumn{2}{l}{$W(\rightarrow \tau\nu)$ + 2 jets                                            } & & \\
\multicolumn{3}{l}{\quad 4 EW vertices (with dibosons)                                          } & \sherpa  & 8.4 &  \\
\multicolumn{3}{l}{\quad 4 EW vertices (no dibosons)                                            } & \sherpa  & & 4.2  \\
\\
\multicolumn{3}{l}{Top quarks                                                                   } & & \\
\multicolumn{3}{l}{\quad $\ttbar(\rightarrow\ell\nu b\bar{q}q\bar{b}, \ell\nu b\ell\nu\bar{b})$ } & \mcatnlo + \herwig   & 90.0 &  \\
\multicolumn{3}{l}{                                                                             } & \powheg + \pythiasix &      & 114 \\
\multicolumn{3}{l}{\quad $tW$                                                                   } & \acermc + \pythiasix & 15.3 &  \\
\multicolumn{3}{l}{                                                                             } & \mcatnlo + \herwig   &      & 20.7 \\
\multicolumn{3}{l}{\quad $t\bar{b}q \rightarrow \ell\nu b\bar{b}q$                              } & \acermc + \pythiasix & 23.5 & 25.8 \\
\multicolumn{3}{l}{\quad $t\bar{b} \rightarrow \ell\nu b\bar{b}$                                } & \acermc + \pythiasix & 1.0  &  \\
\multicolumn{3}{l}{                                                                             } & \mcatnlo + \herwig   &      & 1.7 \\
\\
\multicolumn{3}{l}{$Z(\rightarrow \ell\ell)$ inclusive, $m_{\ell\ell}>40$ \GeV                  } & & \\
\multicolumn{3}{l}{\quad 2 EW vertices                                                          } & \sherpa            & 3140 & 3620 \\
\\
\multicolumn{3}{l}{$Z(\rightarrow ee,\mu\mu)$ + 2 jets, $m_{ee,\mu\mu} >40$ \GeV                 } & & \\
\multicolumn{3}{l}{\quad 4 EW vertices (no dibosons)                                            } & \sherpa            & 0.7 & 0.9 \\
\\
\multicolumn{3}{l}{Dibosons} & & \\
\multicolumn{3}{l}{\quad $WW$                                                                   } & \herwigpp          & 45.9 & 56.8 \\
\multicolumn{3}{l}{\quad $WZ$                                                                   } & \herwigpp          & 18.4 & 22.5 \\
\multicolumn{3}{l}{\quad $ZZ$                                                                   } & \herwigpp          & 6.0  & 7.2 \\
\bottomrule
\end{tabular}
\end{center}
\end{table}

\textbf{\wjets}

The primary model of the signal and background \wjets processes in the analysis is the 
next-to-leading-order (NLO) \powheg Monte Carlo generator\,\cite{powheg1,powheg2,powhegvbfw,powhegwjj}, 
interfaced with \pythia using the AU2 parameter values\,\cite{AU2tune} for the simulation of parton 
showering, underlying event, and hadronization.  Two final-state partons with $\pt >20$~\GeV~are required 
for the signal.  A generator-level suppression is applied in the background generation to enhance events 
with one parton with $\pt > 80$~\GeV~and a second parton with $\pt > 60$~\GeV, and the mass of the pair 
larger than 500~\GeV.  Parton momentum distributions are modelled using the CT10\,\cite{ct10} set of parton 
distribution functions (PDFs).  The QCD factorization and renormalization scales are set to the $W$-boson 
mass for the sample with jets produced via the electroweak interaction.  For the sample with strongly 
produced jets, the hard-process scale is also the $W$-boson mass while the QCD emission 
scales are set with the multiscale-improved NLO (MiNLO) procedure\,\cite{Hamilton:2012np} to 
improve the modelling and reduce the scale dependence.  Uncertainties due to missing higher-order 
contributions are estimated by doubling and halving the factorization and renormalization 
scales independently, but keeping their ratio within the range $0.5$--$2.0$.  Uncertainties due 
to parton distribution functions are estimated using CT10 eigenvector variations rescaled to 
68\% confidence level, and an uncertainty due to the parton shower and hadronization 
model is taken from the difference between predictions using the~\pythia 
and~\herwigpp\,\cite{herwigpp,jimmy} generators.  

Measured particle-level differential distributions are also compared to the \sherpa (v. 1.4)\,\cite{sherpa} 
generation of QCD+EW \wjets production at leading-order accuracy, including interference.  An uncertainty 
due to the neglect of interference in the EW \wjets measurement is estimated using this sample and 
individual \sherpa QCD and EW \wjets samples.  The individual samples are also used to model the small 
contribution from $W\rightarrow \tau\nu$ decays.  Measured distributions of QCD+EW \wjets production are 
compared to the combined QCD+EW and to the QCD \wjets samples, the latter to demonstrate the effect 
of the EW \wjets process.  The QCD \wjets sample is a $W+(n)$-parton prediction with $n\leq 4$ partons 
with $\pt > 15$~\GeV~produced via QCD interactions.  The EW \wjets sample has two partons produced via 
electroweak vertices, and up to one additional parton produced by QCD interactions.  The CKKW matching 
scheme\,\cite{Catani:2001cc} is used to remove the overlap between different parton multiplicities at the 
matrix-element level.  The predictions use the CT10 PDFs and the default parameter values for simulating 
the underlying event.  Renormalization and factorization scales are set using the standard dynamical scale 
scheme in \sherpa.  The interference uncertainty is cross-checked with the \madgraph\,\cite{madgraph} 
generator interfaced to \pythia.

For unfolded distributions with a low purity of electroweak \wjets production, an additional 
comparison is made to the all-order resummation calculation of \textsc{hej} (High Energy 
Jets)\,\cite{Andersen:2012gk} for strong \wjets production.  The calculation improves the 
accuracy of predictions in wide-angle or high-invariant-mass dijet configurations, where 
logarithmic corrections are significant.  To allow a comparison to unfolded data and to other 
generators, the small electroweak \wjets contribution is added using \powheg interfaced to 
\pythia and the sum is labelled~\textsc{hej (qcd) + pow+py (ew)}.

Both the \powheg and \sherpa predictions for electroweak \wjets production omit the small 
contribution from diboson production processes, assuming negligible interference with these 
processes.  Higher-order electroweak corrections to the background \wjets process are studied 
with OpenLoops\,\cite{openloops1,openloops2} and found to affect the measured fiducial cross 
section by $<1\%$.

\textbf{Other processes}

Background contributions from top-quark, \zjets, and diboson processes are estimated using MC 
simulation. 

The top-quark background consists of pair-production and single-production processes, with the latter 
including $s$-channel production and production in association with a $b$ quark or $W$ boson.  
Top-quark pair production is normalized using the cross section calculated at next-to-next-to-leading 
order (NNLO) in $\alphas$, with resummation to next-to-next-to-leading logarithm (NNLL) using 
TOP++2.0\,\cite{toppp}.  Kinematic distributions are modelled at NLO using the \mcatnlo\,\cite{mcatnlo} 
generator and the \herwig\,\cite{herwig,jimmy} parton shower model for 7~\TeV~data, and with \powheg and 
\pythiasix (v. 6.427)\,\cite{pythia6} for 8~\TeV~data; both use the CT10 PDF set.  An uncertainty due to 
the parton shower model, and its interface to the matrix-element generator, is estimated by comparing 
the \powheg sample to an \mcatnlo sample interfaced to \herwig.  Single-top-quark production 
in the $t$-channel, $t\bar{b}q \rightarrow \ell\nu b\bar{b}q$, is modelled using the leading-order 
generator \acermc (v. 3.8)\,\cite{acermc} interfaced with \pythiasix and the CTEQ6L1\,\cite{cteq6} PDF 
set, and the sample is normalized using the cross sections calculated by the generator.  Modelling of the 
$s$-channel production of a single top quark, $t\bar{b} \rightarrow \ell\nu b\bar{b}$, and of the associated 
production of a top quark and a $W$ boson are performed using \acermc with \pythiasix in 7~\TeV~data 
and \mcatnlo with \herwig in 8~\TeV~data.  These samples are also normalized using the generator 
cross-section values.

Background from the \zjets ($Zjj$) process, which contributes when one of the leptons is not reconstructed 
and the \met is large, is modelled using \sherpa and the CT10 PDF set.  For the background with jets 
from QCD radiation, an inclusive Drell--Yan sample is produced at NLO\,\cite{sherpanlo} and merged 
with the leading-order (LO) production of additional partons (up to five).  The background with jets 
produced purely through the electroweak interaction is modelled at leading order.  This combination 
of samples is also used to model the $W(\rightarrow \tau\nu)$ + 2 jets background; the 7~\TeV~sample 
includes $WW$ and $WZ$ production.  The interference between the electroweak and QCD production of 
jets for these small backgrounds has a negligible impact on the measurements and is not modelled.

The diboson background processes $WW/WZ\rightarrow \ell\nu q\bar{q}^{(')}$ and 
$ZZ\rightarrow \ell\ell q\bar{q}$ provide only a small contribution at high dijet 
mass since the distribution peaks at the mass of the $W$ or $Z$ boson.  The 
interference between the single and pair production of electroweak bosons is 
negligible for the mass range selected by the analysis.  The diboson processes are 
modelled at leading order with \herwigpp and normalized to the NLO cross 
section\,\cite{dibosonxsec}.  The generation uses the CTEQ6L1 PDF set.  In 
7~\TeV~samples, $W\rightarrow \tau\nu$ decays are removed since they are included 
in the \wjets samples.

%% file: multijet.tex
Multijet production constitutes a background to the \wjets process when one of 
the jets is misidentified as a lepton and significant \met arises from either a 
momentum mismeasurement or the loss of particles outside the detector acceptance.  
Due to the very small fraction of multijet events with both of these properties, 
and their relatively poor modelling in simulation, a purely data-driven method is 
used to estimate this background.  The method inverts certain lepton identification 
criteria (described below) to obtain a multijet-dominated sample for modelling 
kinematic distributions.  The \met distribution is then fit to obtain a multijet 
normalization factor; this fit is performed separately in the signal, control, and 
validation regions.  Systematic uncertainties are estimated by modifying the fit 
distribution and the identification criteria, and by propagating detector and 
theoretical uncertainties.  

Modifications to the lepton identification criteria which enhance the multijet 
contribution are based on isolation and either the impact parameter with respect 
to the primary vertex (for muons) or the shower and track properties (for electrons).  
For the 7~TeV analysis, the impact parameter significance requirement is inverted 
in the muon channel ($|d_0|/\sigma_{d_0} > 3$).  This preferentially selects muons 
from heavy-flavour hadron decays, a dominant source of muons in multijet events.  
For the 8 TeV analysis, no requirement on impact parameter significance is made and 
instead a track isolation requirement is applied orthogonal to the requirement for 
selected muons ($0.15 < \pTisothree / \pT < 0.35$).

For the electron channel in $\sqrt{s}=7$~TeV data, triggers requiring a loose 
electron candidate are used to obtain a multijet modelling sample.  The electron 
candidate must satisfy medium criteria on track hit multiplicity and track--shower 
matching in $\eta$, but must fail to satisfy at least one of the tight shower-based 
criteria.  It also must not be isolated in the calorimeter: $\ETisothree / \ET > 0.2$.
In $\sqrt{s}=8$~TeV data, electron candidates must satisfy medium selection criteria 
consistent with the trigger used in the analysis.  As in the muon channel, a 
track isolation window is applied orthogonal to the requirement for selected 
electrons ($0.05 < \pTisotwo / \pT < 0.1$).

To normalize the multijet-dominated samples to the expected contribution with nominal 
lepton criteria, a fit to the \met distribution is performed.  The fit simultaneously 
determines the multijet and strong \wjets normalizations in the region where the nominal 
lepton criteria are applied, taking the multijet distribution from the sample with 
inverted lepton identification criteria.  Other contributions are fixed to their SM 
predictions, and the data are consistent with the post-fit distribution within 
uncertainties.  The strong \wjets normalization is consistent with that found in the 
fit to the dijet mass distribution described in Section~\ref{sec:xsec}.

Systematic uncertainties in the multijet normalization arise from uncertainties in 
the kinematic modelling and in jet, lepton, and \met reconstruction.  The modelling 
uncertainties dominate and are estimated using three methods: (1) modifying the 
lepton candidate selection for the kinematic distributions; (2) using \mT as an 
alternative fit distribution; and (3) varying the kinematic range of the fit.  For 
each method, the largest change in the normalization is taken as a systematic 
uncertainty and added in quadrature with reconstruction and modelling uncertainties 
for processes modelled with Monte Carlo simulation.  The leading uncertainty arises 
from the change in multijet 
normalization when fitting the \mT distribution instead of the \met distribution.  
The next largest uncertainty results from variations of the isolation and impact 
parameter requirements in the lepton selection used for the kinematic distributions.  
The total relative systematic uncertainty of the multijet normalization in the muon 
(electron) channel is 28\%~(67\%) for the $\sqrt{s}=7$~TeV analysis, and 36\%~(38\%) 
for the $\sqrt{s}=8$~TeV analysis.  The relatively large uncertainty in the 
$\sqrt{s}=7$~TeV electron channel results from a larger dependence on the fit 
distribution and range than in the other multijet fits.

%% file: plots.tex
\begin{figure}[htbp]
\centering
\includegraphics[width=0.48\textwidth]{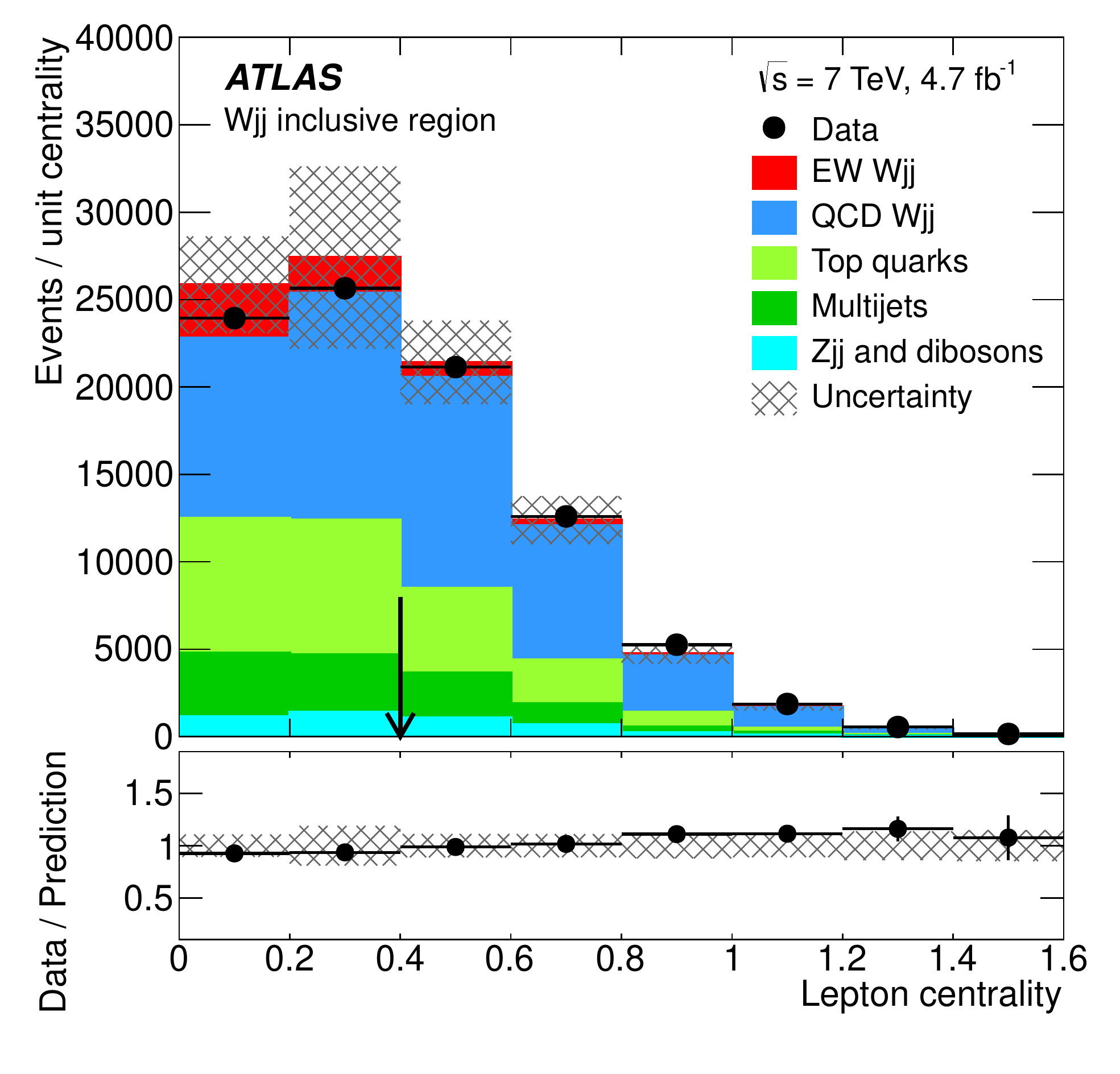}
\includegraphics[width=0.48\textwidth]{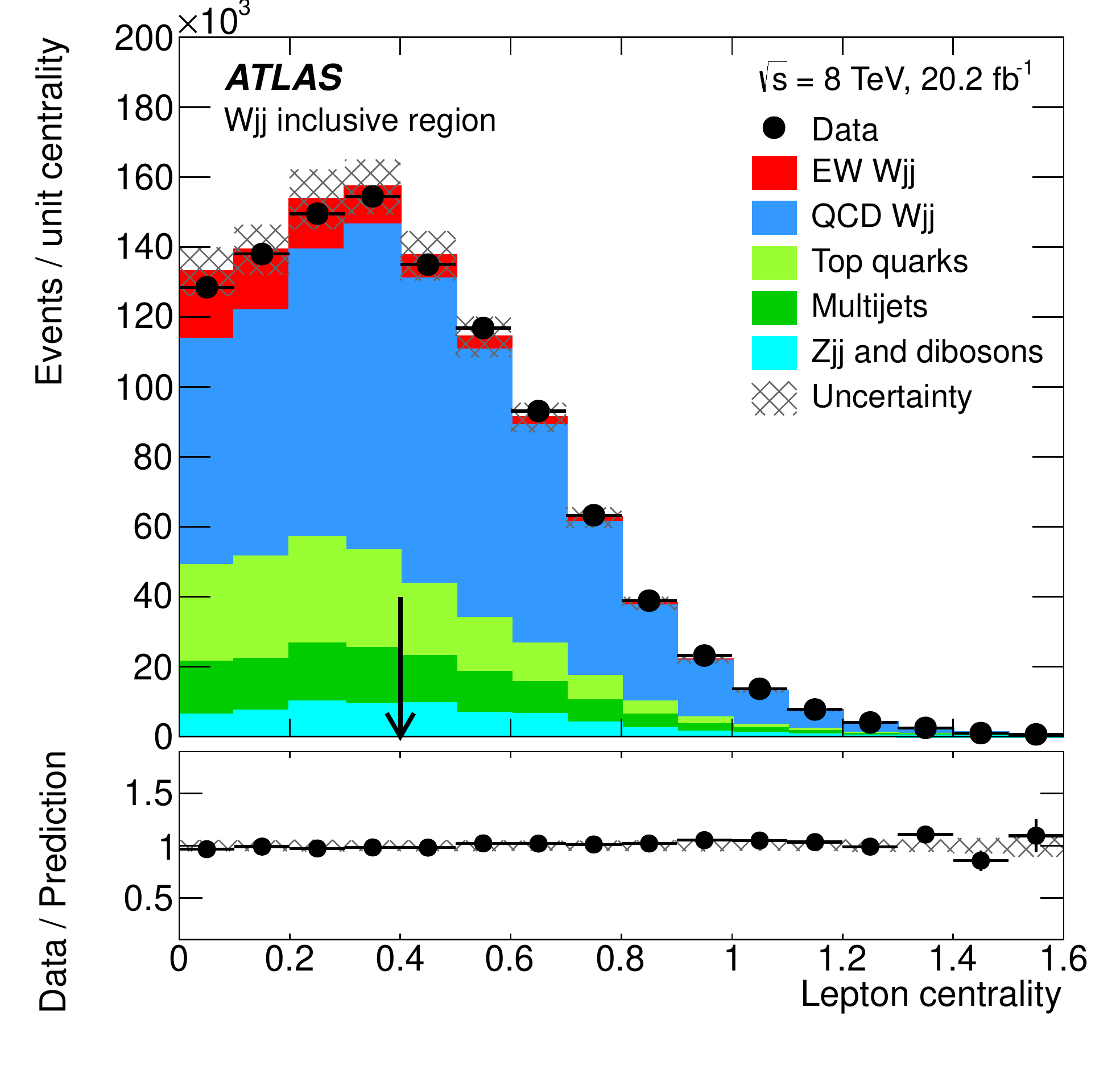}
\includegraphics[width=0.48\textwidth]{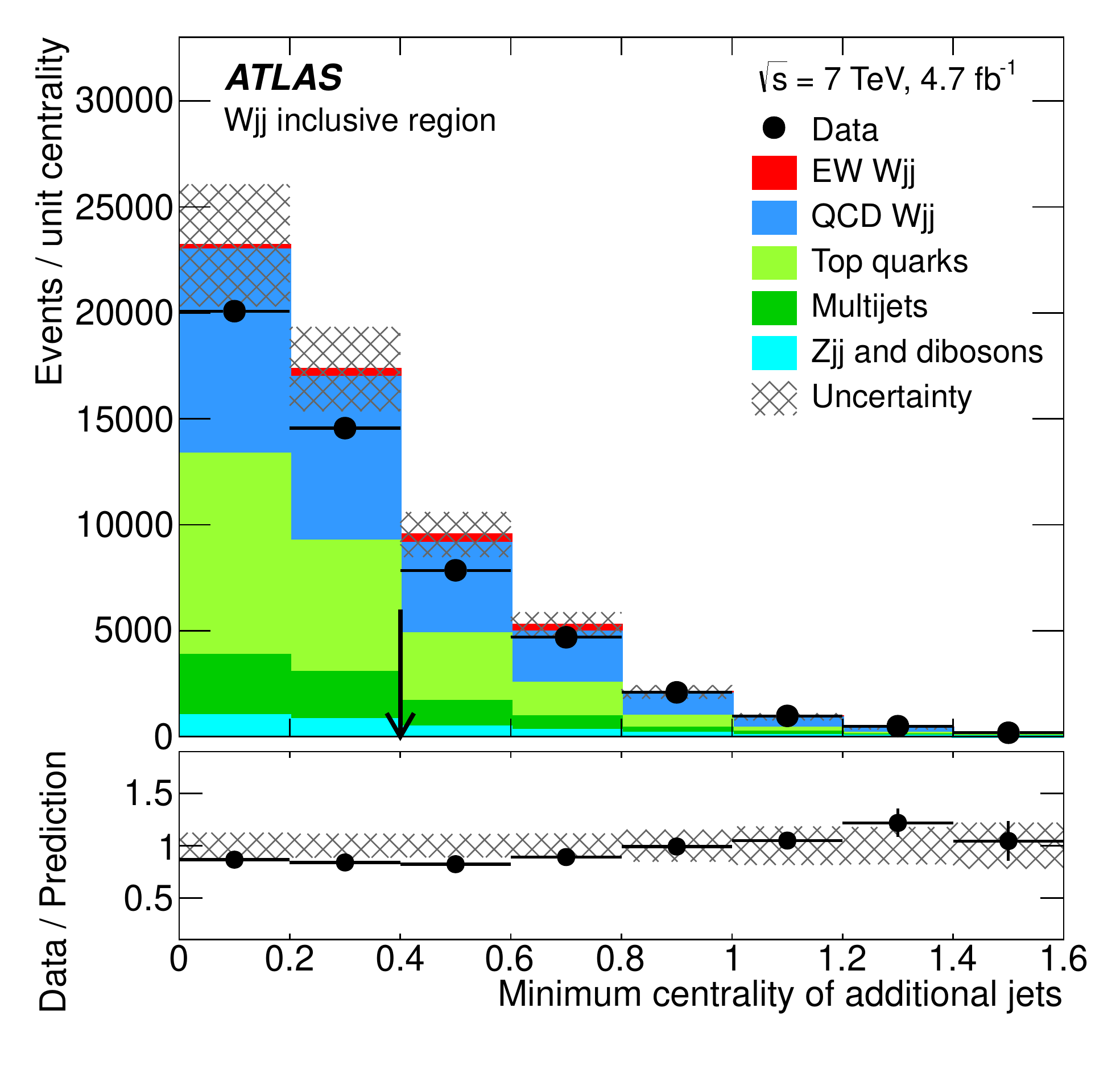}
\includegraphics[width=0.48\textwidth]{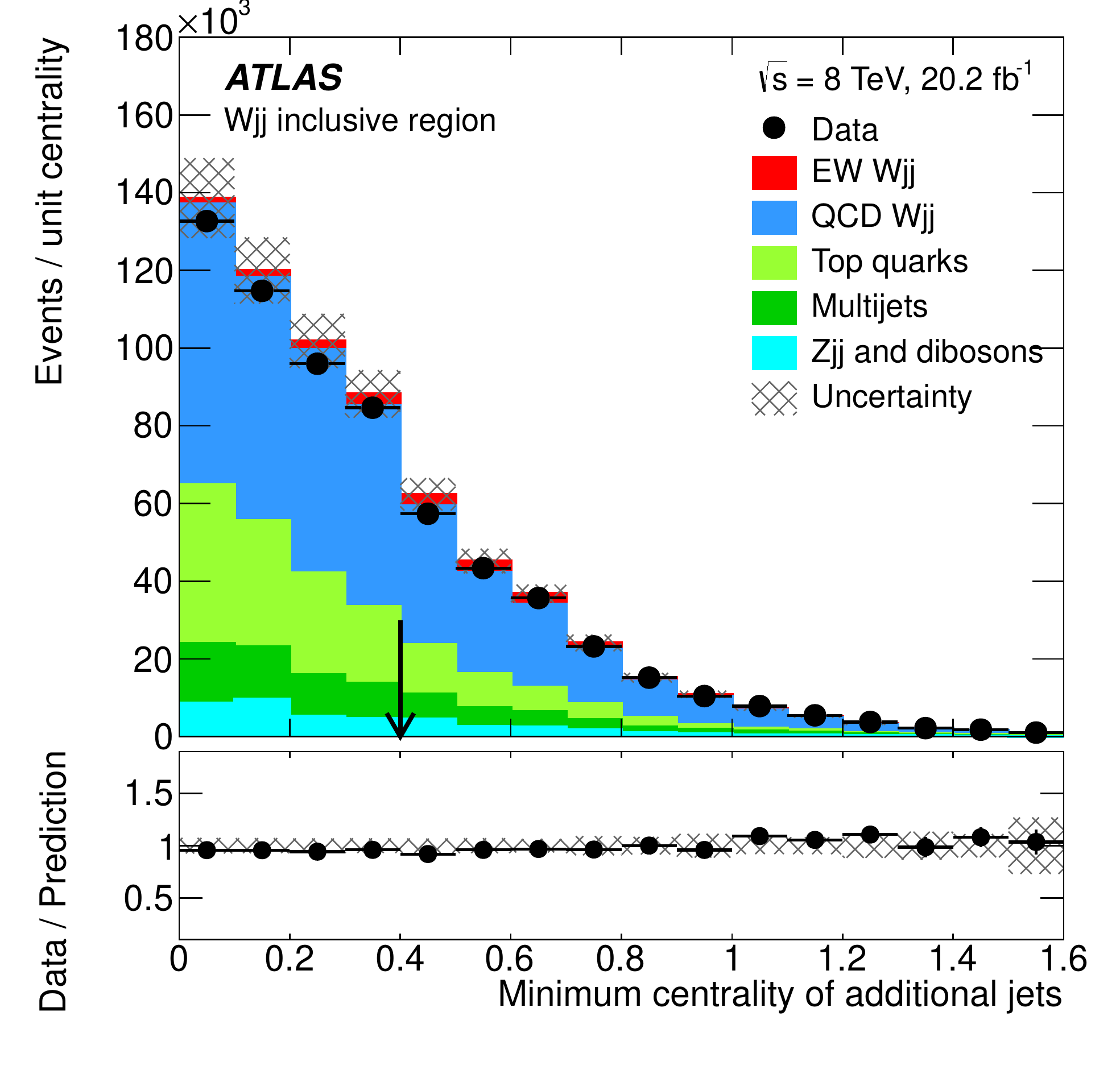}
\caption{Predicted and observed distributions of the lepton centrality (top) and the minimum centrality 
of additional jets (bottom) for events in the inclusive fiducial region (i.e. after preselection) in 
7~\TeV~(left) and 8~\TeV~(right) data.  The arrows in the lepton-centrality distributions separate the 
signal-region selection (to the left) from the control-region selection (to the right).  The 
arrows in the jet-centrality distributions separate the signal-region selection (to the right) 
from the validation-region selection (to the left).  The bottom panel in each distribution shows the 
ratio of data to the prediction.  The shaded band represents the statistical and experimental 
uncertainties summed in quadrature. }
\label{evtsel:controlInclusive}
\end{figure}
\begin{figure}[htbp]
\centering
\includegraphics[width=0.48\textwidth]{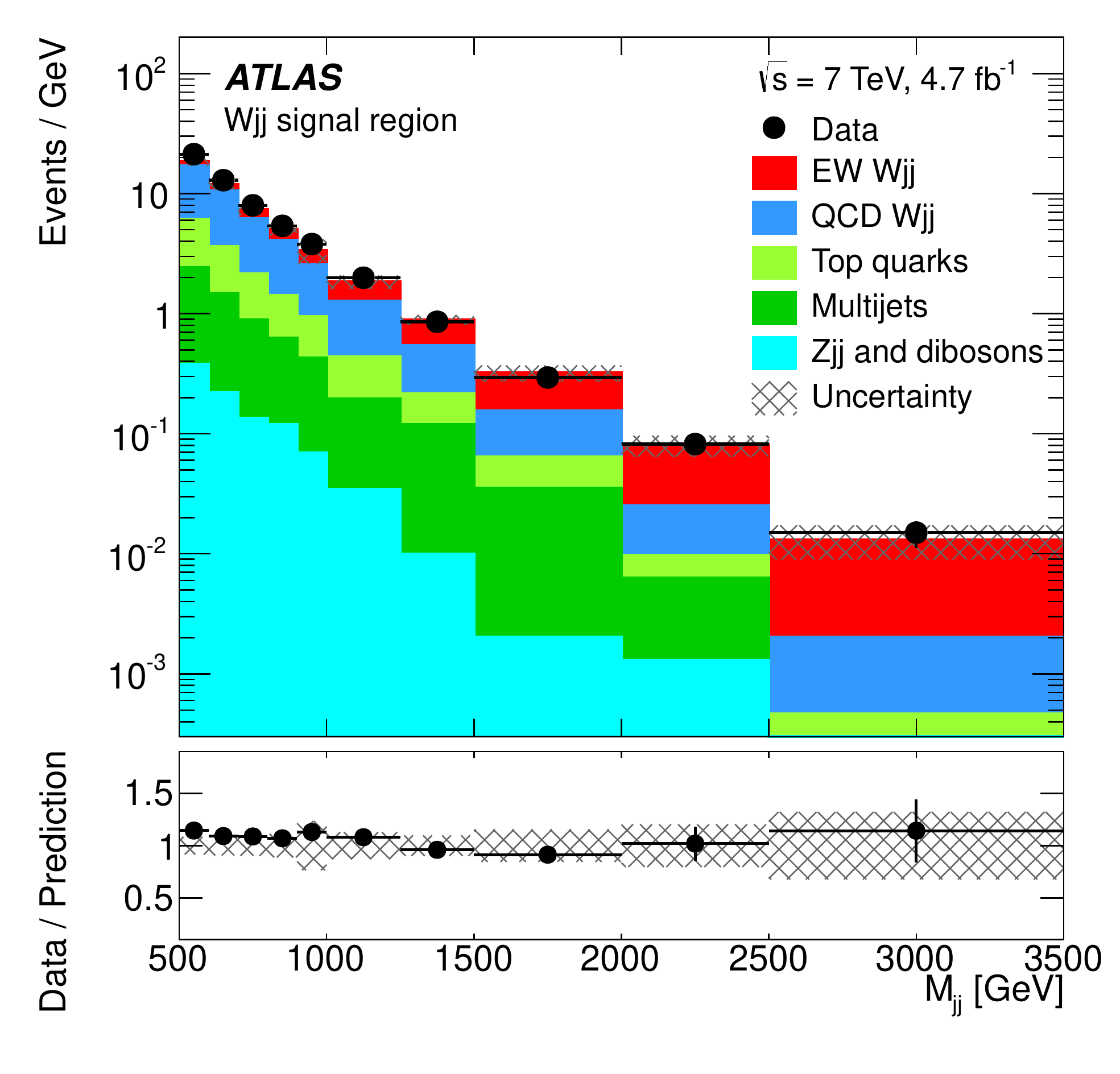}
\includegraphics[width=0.48\textwidth]{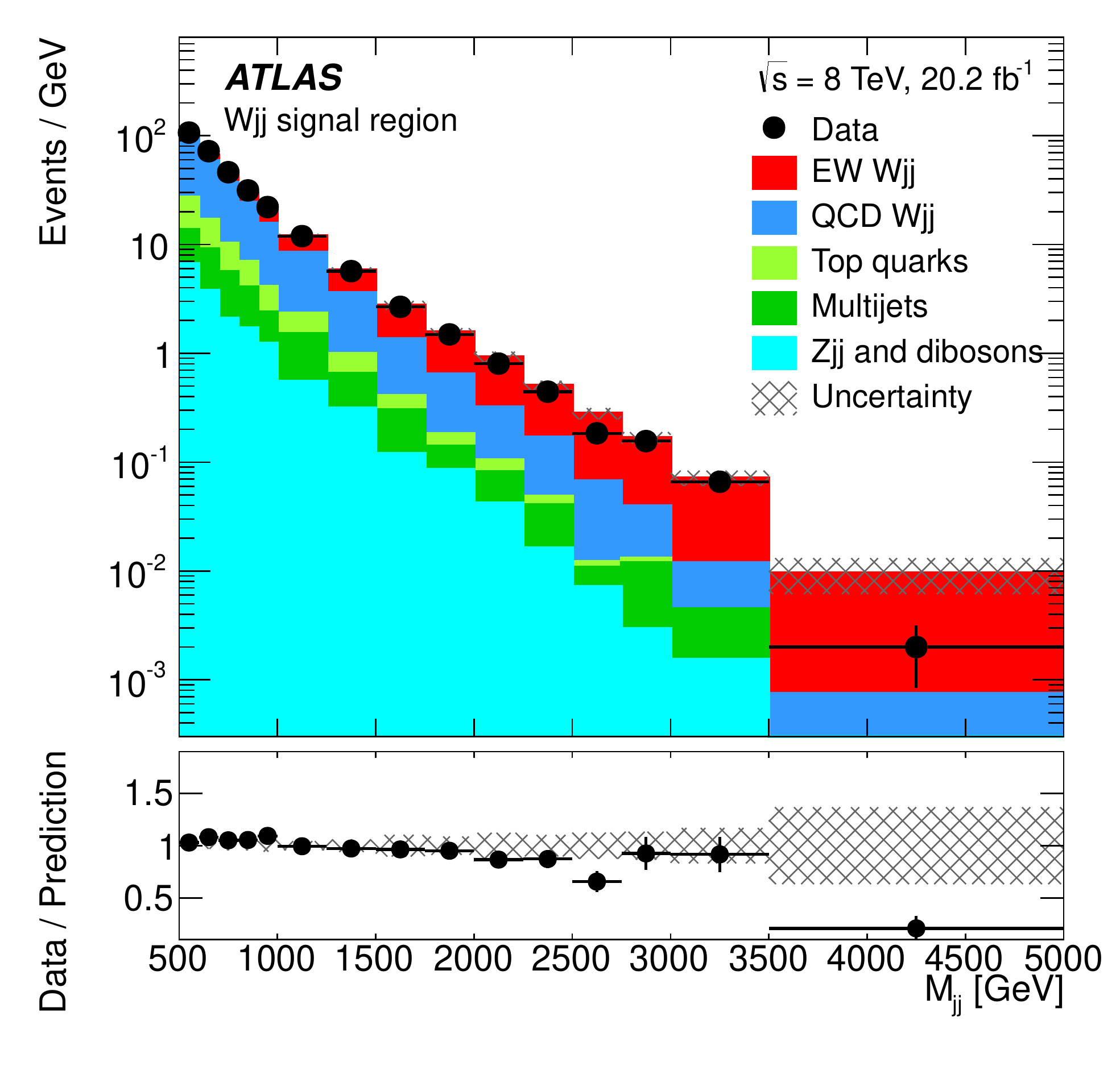}
\includegraphics[width=0.48\textwidth]{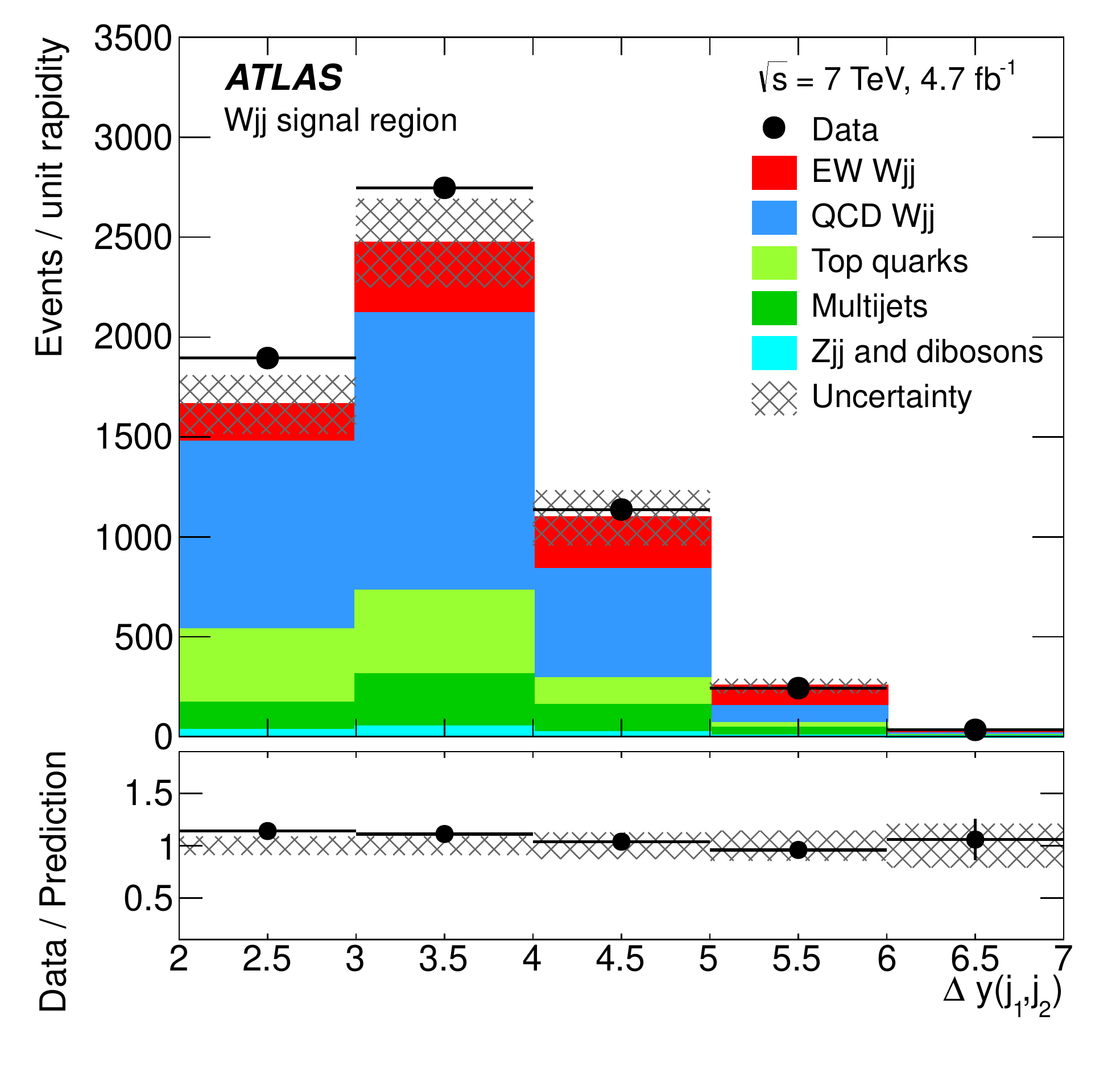}
\includegraphics[width=0.48\textwidth]{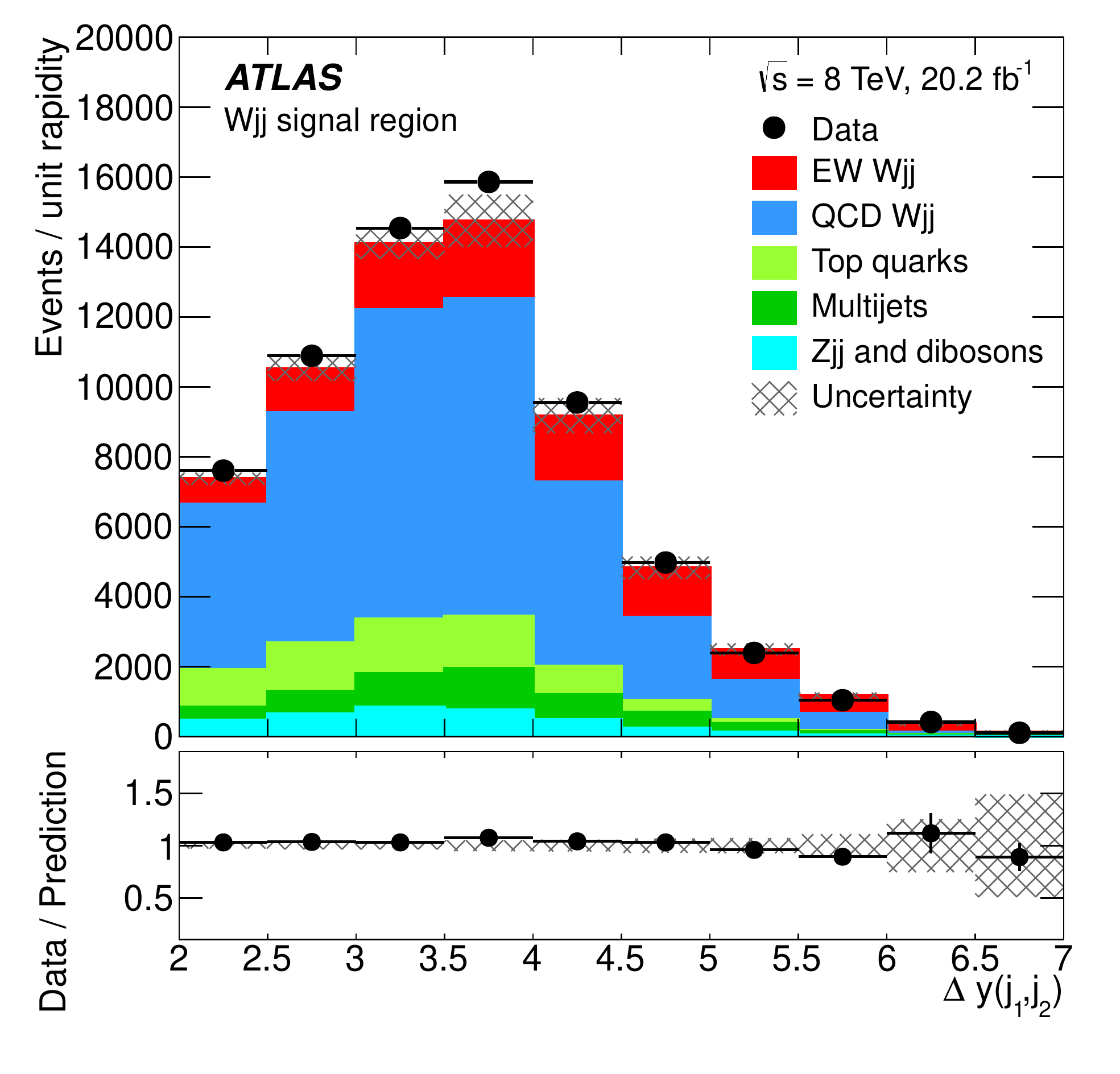}
\caption{Predicted and observed distributions of the dijet invariant mass (top) and $\dyjj$ (bottom) 
for events in the signal region in 7~\TeV~(left) and 8~\TeV~(right) data.  The bottom panel in each 
distribution shows the ratio of data to the prediction.  The shaded band represents the statistical 
and experimental uncertainties summed in quadrature. }
\label{evtsel:controlSignal}
\end{figure}

The distributions of lepton centrality and the minimum centrality of additional jets, 
which are used to separate signal, control, and validation regions, are shown in 
Figure~\ref{evtsel:controlInclusive} for the 7 and 8 TeV data and the corresponding SM 
predictions after the preselection.  The comparisons of the SM predictions to data show 
general agreement within the estimated uncertainties.  The predictions include correction 
factors for lepton identification and triggering, and the bands correspond to the 
combination of statistical and experimental uncertainties.  The signal-region dijet mass 
distributions, used to fit for the signal yield in the fiducial and total cross-section 
measurements, are shown in Figure~\ref{evtsel:controlSignal} for both data sets.  The 
figure also shows the dijet rapidity difference, which is correlated with dijet mass and 
demonstrates an enhancement in signal at high values.  Table~\ref{tab:signalyield} details 
the data and SM predictions for the individual processes in the signal region, and 
Table~\ref{tab:yields} shows the total predictions and the observed data in each of the 
fiducial regions defined in Section~\ref{sec:evtsel}.

\begin{table*}[tb!]
\caption{
Observed data and predicted SM event yields in the signal region.  The MC predictions are normalized to 
the theoretical cross sections in Table~\ref{tbl:mc}.  The relative uncertainty of the total SM prediction 
is ${\cal{O}}$(10\%).}
\label{tab:signalyield}
\begin{center}
\begin{tabular}{lrr}
\toprule
Process & 7 TeV & 8 TeV \\ 
\midrule
$\wjets$ (EW) & 920 & 5600 \\
$\wjets$ (QCD) & 3020 & 19600 \\
Multijets & 500 & 2350 \\
$t\bar{t}$ & 430 & 1960 \\
Single top & 244 & 1470 \\
$Zjj$ (QCD) & 470 & 1140 \\
Dibosons & 126 & 272 \\
$Zjj$ (EW) & 5 & 79 \\
\hline
Total SM & 5700 & 32500 \\
\hline
Data & 6063 & 33719 \\
\bottomrule
\end{tabular}
\end{center}
\end{table*}

\begin{table*}[htbp]
\caption{
Observed data and total predicted SM event yields in each measurement region.  The MC predictions are 
normalized to the theoretical cross sections times branching ratios in Table~\ref{tbl:mc}.  The relative 
uncertainty of the total SM prediction is ${\cal{O}}$(10\%). }
\label{tab:yields}
\begin{center}
\begin{tabular*}{0.925\textwidth}{lrrrr}
\toprule
Region name & \multicolumn{2}{c}{7 TeV} & \multicolumn{2}{c}{8 TeV} \\ 
            & SM prediction & ~~Data~~ & SM prediction & ~~Data~~ \\ 
\midrule
Fiducial and differential measurements & & & & \\
~~Signal region                        & 5700 & 6063 & 32500 & 33719 \\
~~Forward-lepton control region        & 5000 & 5273 & 29400 & 30986 \\
~~Central-jet validation region        & 2170 & 2187 & 12400 & 12677 \\
\hline
Differential measurement only & & & &  \\
~~Inclusive region, $\mjj>500$~GeV & - & - & 106000 & 107040 \\
~~Inclusive region, $\mjj>1$~TeV & - & - & 17400 & 16849 \\
~~Inclusive region, $\mjj>1.5$~TeV & - & - & 3900 & 3611 \\
~~Inclusive region, $\mjj>2$~TeV & - & - & 1040 & 890 \\
~~Forward-lepton/central-jet region & - & - & 12000 & 12267 \\
~~High-mass signal region & - & - & 6100 & 6052 \\
\hline
Anomalous coupling measurements only & & \\
~~High-$q^2$ region & - & - & 39 & 30 \\
\bottomrule
\end{tabular*}
\end{center}
\end{table*}

%% file: fidxsec.tex
The measurement of the fiducial EW \wjets cross section in the signal region uses 
a control-region constraint to provide a precise determination of the electroweak 
production cross section for $W$ bosons produced in association with dijets at 
high invariant mass.  The measurement is performed with an extended joint binned 
likelihood fit~\cite{roofit} of the \mjj distribution for the normalization factors 
of the QCD \wjets and EW \wjets \powheg+ \pythia predictions, \muqcd and \muew 
respectively, defined as follows:
\begin{eqnarray}
(\sigma_i^{\ell \nu jj} \times {\cal{A}}_i)^\mathrm{meas} & = & 
\mu_i \cdot (\sigma_i^{\ell \nu jj} \times {\cal{A}}_i)^\mathrm{theo} \nonumber \\
 & = & \frac{N_i}{{\cal{C}}_i {\cal{L}}}, 
\nonumber 
\end{eqnarray}

\noindent
where $\sigma^{\ell \nu jj}_i$ is the cross section of process $i$ (QCD \wjets or EW 
\wjets production in a single lepton channel), ${\cal{A}}_i$ is the acceptance for events 
to pass the signal selection at the particle level (see Table~\ref{tab:selection}), $N_i$ 
is the number of measured events, ${\cal{L}}$ is the integrated luminosity, and 
${\cal{C}}_i$ is the ratio of reconstructed to generated events passing the selection 
and accounts for experimental efficiencies and resolutions.  The fit includes a 
Gaussian constraint for all non-\wjets backgrounds, and accounts only for statistical 
uncertainties in the expected yield.  The fit result for \muew is translated into a 
fiducial cross section by multiplying \muew by the predicted fiducial cross section 
from \powheg + \pythia.  In addition, the total cross section for jets with 
$\pt > 20$~\GeV~is calculated by dividing the fiducial cross section by ${\cal{A}}$ for 
the EW \wjets process. 

The dijet mass provides the discriminating fit distribution.  The region at relatively low 
invariant mass ($\approx 500$--$1000$ \GeV) has low signal purity and primarily determines 
\muqcd, while events with higher invariant mass have higher signal purity and mainly 
determine \muew.  The interference between the processes is not included in the fit, and 
is instead taken as an uncertainty based on SM predictions.  

The uncertainty in the shape of the QCD \wjets distribution dominates the measurement, 
but is reduced by using the forward-lepton control region to correct the modelling of 
the \mjj shape.  This control region is defined in Table~\ref{tab:selection} and uses the 
same selection as the signal region, except for the inversion of the central-lepton 
requirement.  This section describes the application of the control-region constraint, 
the uncertainties in the measurement, and the results of the fit.

%% file: cr.tex
The SM prediction of the dijet mass distribution receives significant uncertainties from 
the experimental jet energy scale and resolution.  These uncertainties are constrained 
with a correction to the predicted distribution derived using data in a control region 
where the signal contribution is suppressed.  This forward-lepton control 
region is selected using the lepton centrality distribution.  Residual uncertainties 
arise primarily from differences in the dijet mass spectrum between the control region 
and the signal region.  

\begin{figure}[tbp!]
\centering
\includegraphics[width=0.48\textwidth]{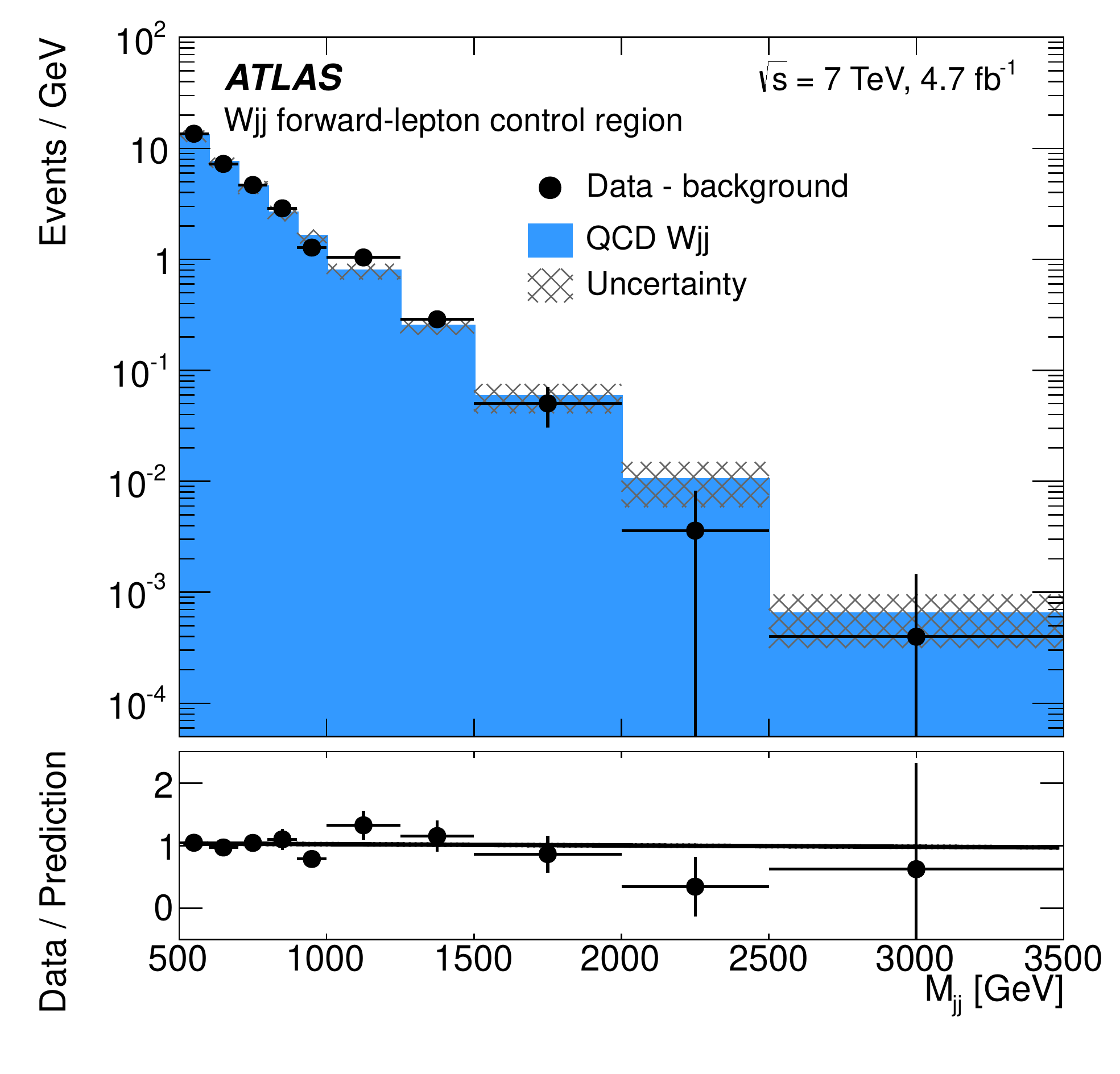}
\includegraphics[width=0.48\textwidth]{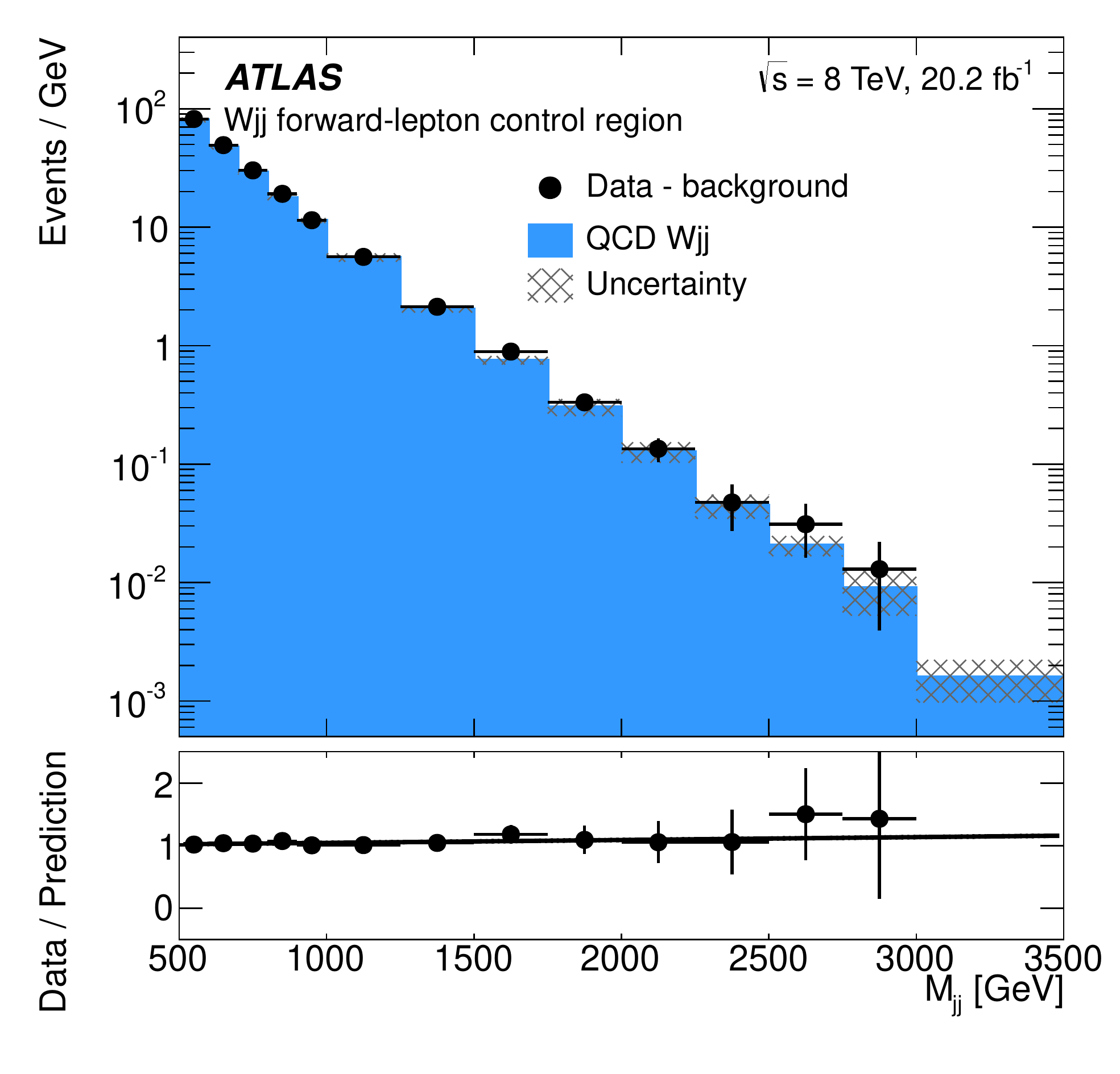}
\caption{Comparison of the predicted QCD \wjets dijet mass distribution to data with background 
processes subtracted, for events in the forward-lepton control region in 7~\TeV~(left) and 
8~\TeV~(right) data.  The bottom panel in each distribution shows the ratio of data to the QCD~\wjets 
prediction, and the result of a linear fit to the ratio.  The error bars represent statistical and 
experimental uncertainties summed in quadrature. }  
\label{fig:crmasssubtracted}
\end{figure}

To derive the \mjj correction, all processes other than strong~\wjets production are subtracted 
from the data and the result is compared to the prediction~(Figure~\ref{fig:crmasssubtracted}).  
The correction is then determined with a linear fit to the ratio of the subtracted data to the 
\wjets prediction.  The slopes of the fits in 7 and 8~\TeV~data are consistent with zero; they 
are $(0.2 \pm 1.1)$\%/ \TeV~and $(0.28 \pm 0.43)$\%/ \TeV, respectively, where the uncertainties 
are statistical only.  The effect of a slope correction of 1\%/ \TeV~is approximately 0.1 in the 
measured $\muew$.

\begin{figure}[tbp!]
\centering
\includegraphics[width=0.48\textwidth]{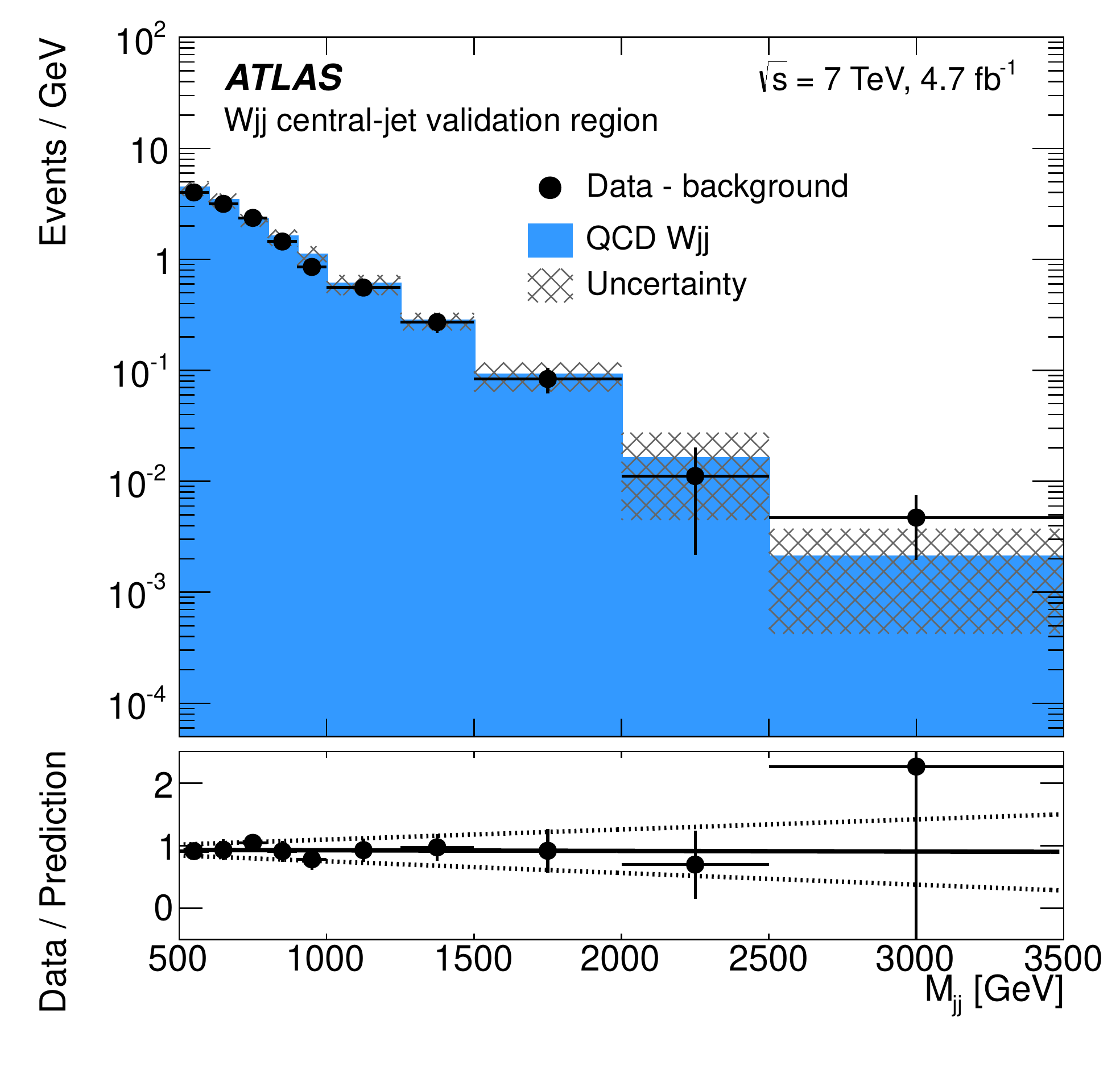}
\includegraphics[width=0.48\textwidth]{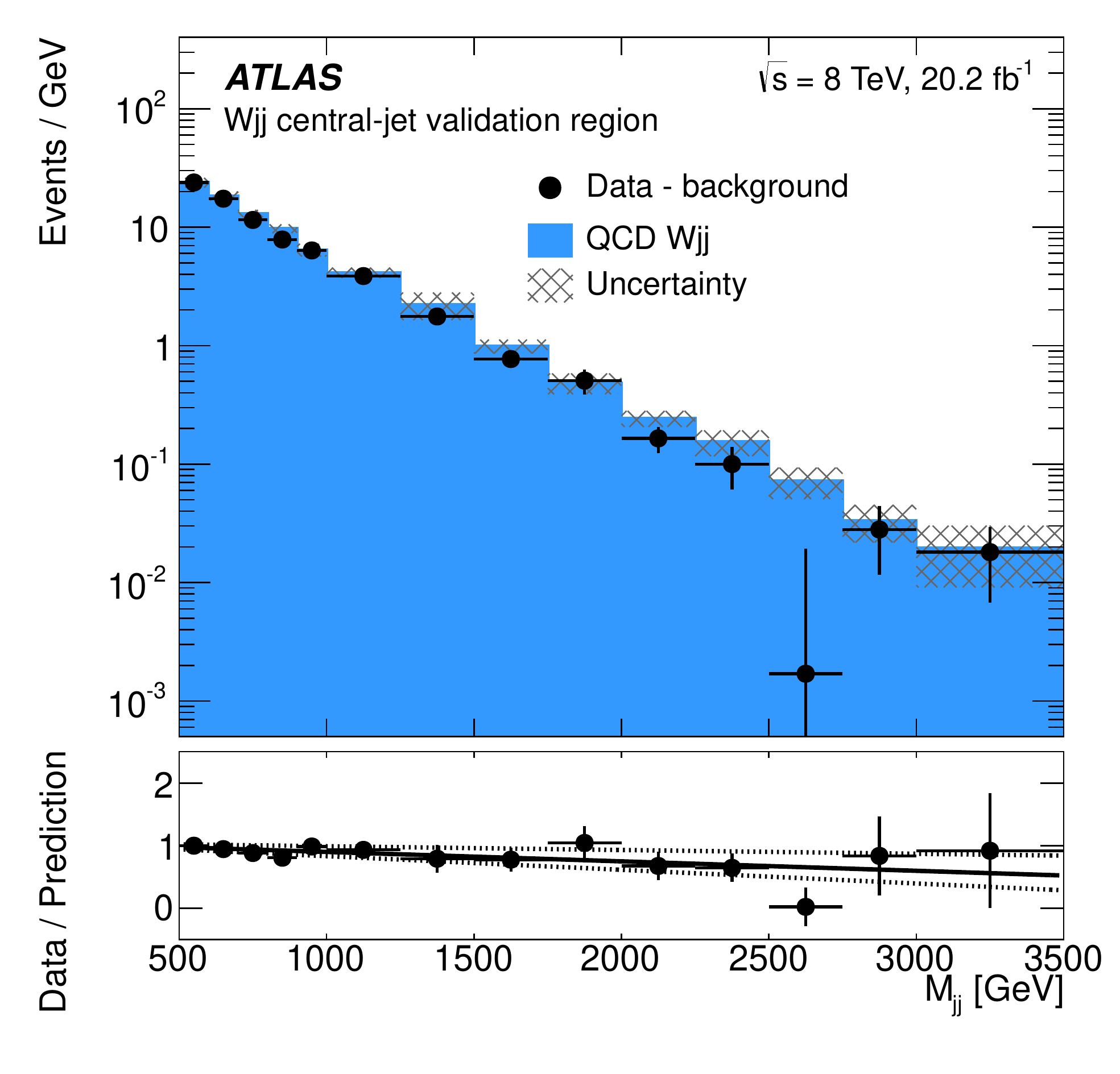}
\caption{Comparison of the corrected QCD \wjets background dijet mass distribution to data with 
background processes subtracted, for events in the central-jet validation region in 7~\TeV~(left) 
and 8~\TeV~(right) data.  The bottom panel in each subfigure shows the ratio of data to prediction, 
and the result of a linear fit to the ratio (solid line).  The error bars represent statistical and 
experimental uncertainties summed in quadrature.  The dotted lines show the fit with slope adjusted 
up and down by statistical and experimental uncertainties. }
\label{fig:vrmasssubtracted}
\end{figure}

Systematic uncertainties in the corrected dijet mass distribution in the signal and validation 
regions are estimated by varying each source of uncertainty up or down by $1\sigma$ and calculating 
the corresponding slope correction in the control region in the simulation.  This correction 
is applied to the prediction in the signal region and the fit performed on pseudodata 
derived from the nominal prediction.  The resulting change in $\muew$ is taken as the 
corresponding systematic uncertainty.  The method is illustrated in the central-jet validation 
region in Figure~\ref{fig:vrmasssubtracted}, where the background-subtracted and corrected \wjets 
dijet mass distribution is compared to data.  The ratio of subtracted data to the corrected \wjets 
prediction is consistent with a line of zero slope when considering statistical and experimental 
uncertainties (the dotted lines in the figure).  

%% file: fidxsuncertainties.tex
Uncertainties in $\muew$ consist of: statistical uncertainties in the fit to 
the normalizations of the signal and background~\wjets processes in the signal 
region; the statistical uncertainty of the correction from the control region;
 and experimental and theoretical uncertainties affecting the signal and 
background predictions.  Table~\ref{tbl:fidxsuncertainties} summarizes the 
uncertainties in the measurement of \muew. 

\begin{table}[htbp]
\caption{
The statistical and systematic uncertainty contributions to the measurements 
of $\muew$ in 7 and 8 TeV data.
}
\label{tbl:fidxsuncertainties}
\begin{center}
\begin{tabular}{lllrr}
\toprule
\multicolumn{3}{l}{Source}
& \multicolumn{2}{c}{Uncertainty in $\muew$}
\\
& & &
\multicolumn{1}{r}{7 TeV} & \multicolumn{1}{r}{8 TeV}
\\
\midrule
\multicolumn{3}{l}{Statistical                     } & & \\
\multicolumn{3}{l}{\quad Signal region             } & 0.094 & 0.028 \\
\multicolumn{3}{l}{\quad Control region            } & 0.127 & 0.044 \\
\\
\multicolumn{2}{l}{Experimental                    } & & \\
\multicolumn{3}{l}{\quad Jet energy scale ($\eta$ intercalibration) } & 0.124 & 0.053 \\
\multicolumn{3}{l}{\quad Jet energy scale and resolution (other)    } & 0.096 & 0.059 \\
\multicolumn{3}{l}{\quad Luminosity                                 } & 0.018 & 0.019 \\
\multicolumn{3}{l}{\quad Lepton and \met reconstruction             } & 0.021 & 0.012 \\
\multicolumn{3}{l}{\quad Multijet background                        } & 0.064 & 0.019 \\
\\
\multicolumn{2}{l}{Theoretical                                      } & & \\
\multicolumn{3}{l}{\quad MC statistics (signal region)              } & 0.027 & 0.026 \\
\multicolumn{3}{l}{\quad MC statistics (control region)             } & 0.029 & 0.019 \\
\multicolumn{3}{l}{\quad EW \wjets (scale and parton shower)        } & 0.012 & 0.031 \\
\multicolumn{3}{l}{\quad QCD \wjets (scale and parton shower)       } & 0.043 & 0.018 \\
\multicolumn{3}{l}{\quad Interference (EW and QCD \wjets)           } & 0.037 & 0.032 \\
\multicolumn{3}{l}{\quad Parton distribution functions              } & 0.053 & 0.052 \\
\multicolumn{3}{l}{\quad Other background cross sections            } & 0.002 & 0.002 \\
\multicolumn{3}{l}{\quad EW \wjets cross section                    } & 0.076 & 0.061 \\
\hline
\multicolumn{3}{l}{Total                                            } & 0.26~~ & 0.14~~ \\
\bottomrule
\end{tabular}
\end{center}
\end{table}

The total statistical uncertainty in $\muew$ of the joint likelihood fit is 0.16 (0.052) 
in 7~(8)~\TeV~data, where the leading uncertainty is the statistical uncertainty of the 
data in the control region rather than in the signal region.

Systematic uncertainties affecting the MC prediction are estimated by varying each 
uncertainty source up and down by $1\sigma$ in all MC processes, fitting the ratio 
of the varied QCD $\wjets$ prediction to the nominal prediction in the control region, 
and performing the signal region fit using the varied samples as pseudodata and 
the nominal samples as the templates.  The largest change in $\mu$ from the up 
and down variations is taken as a symmetric uncertainty.  The dominant experimental 
uncertainty in $\muew$ is due to the calibration of the $\eta$ dependence of the jet 
energy scale, and is $0.124$~(0.053) in 7~(8)~\TeV~data.  Other uncertainties in the 
jet energy scale (JES) and resolution (JER) are of similar size when combined, with 
the largest contribution coming from the uncertainty in modelling the ratio of responses
to quarks and gluons.  Uncertainties due to multijet modelling are estimated by separately 
varying the normalization and distribution of the multijet background in each phase-space 
region and combining the effects in quadrature.

Theoretical uncertainties arise from the statistical uncertainty on the MC predictions; 
the lack of interference between signal and background \wjets processes in the MC modelling; 
\wjets renormalization and factorization scale variations and parton-shower modelling, which 
affect the acceptance of the jet centrality requirement; parton distribution functions; and 
cross-section uncertainties.  The uncertainty due to MC statistics is 0.040~(0.032) in 
7~(8)~\TeV~data.  The interference uncertainty is estimated by including the \sherpa 
leading-order interference model as part of the background \wjets process and affects the 
measurement of $\muew$ by 0.037~(0.032) in 7~(8)~\TeV~data.  Uncertainties due to PDFs are 
0.053~(0.052) for 7~(8)~\TeV~data.  Scale and parton-shower uncertainties are $\approx 0.04$ 
in both the 7~and 8~\TeV~measurements.  The scale uncertainty in EW~\wjets production is 
larger at $\sqrt{s}=8$~TeV than at 7~\TeV~because of the increasing uncertainty with dijet 
mass and the higher mean dijet mass at 8~\TeV.  The scale uncertainty in QCD~\wjets 
production is larger at $\sqrt{s}=7$~\TeV~because the data constraint has less statistical 
power than at 8~\TeV.  

Finally, a 0.076 (0.061) uncertainty in the signal cross section at 7~(8)~\TeV~due to 
higher-order QCD corrections and non-perturbative modelling is estimated using scale 
and parton-shower variations, affecting the measurement of $\muew$ but not the 
extracted cross sections.

%% file: fidxsresults.tex
\begin{figure}[htbp]
\centering
\includegraphics[width=0.48\textwidth]{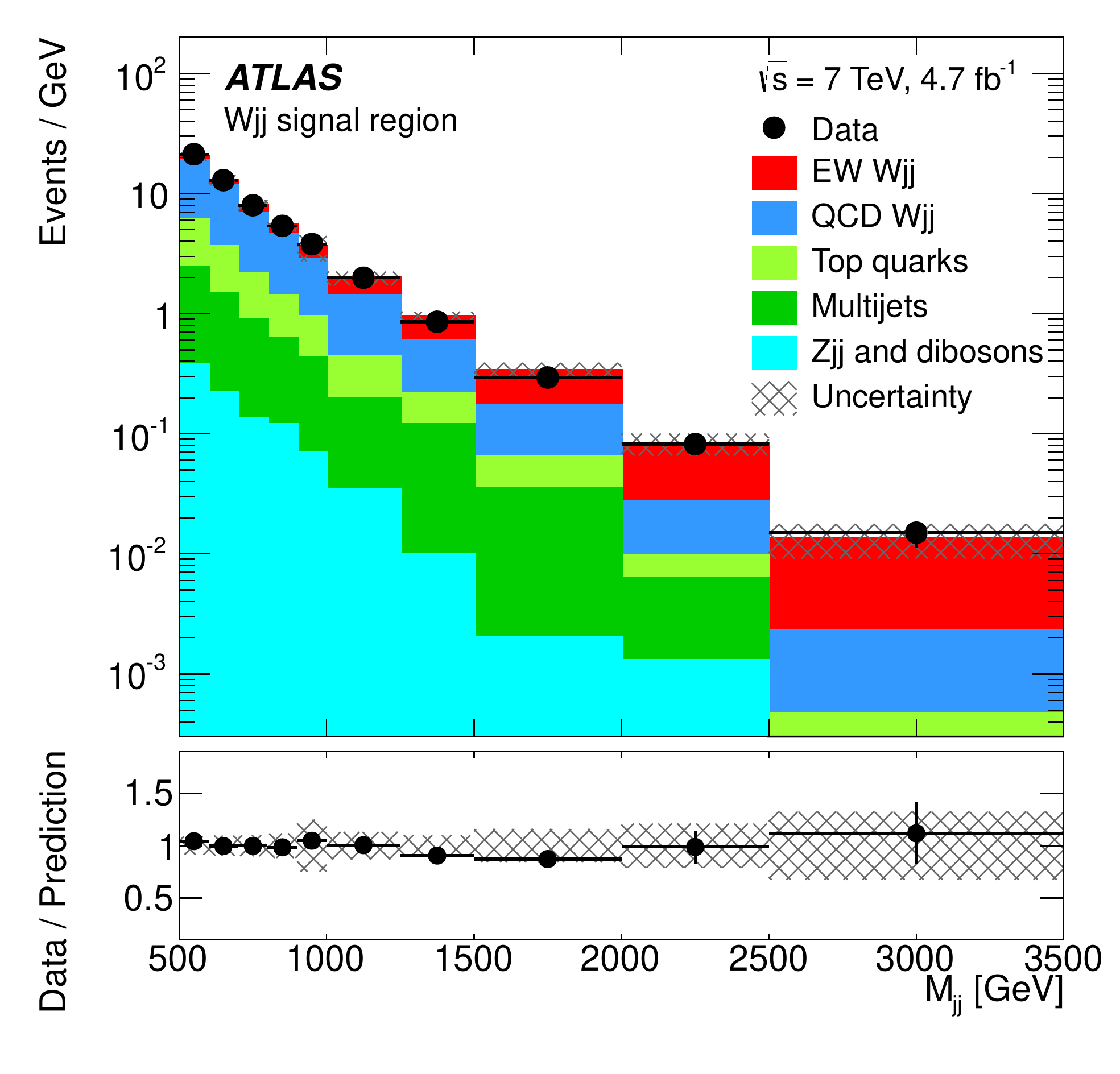}
\includegraphics[width=0.48\textwidth]{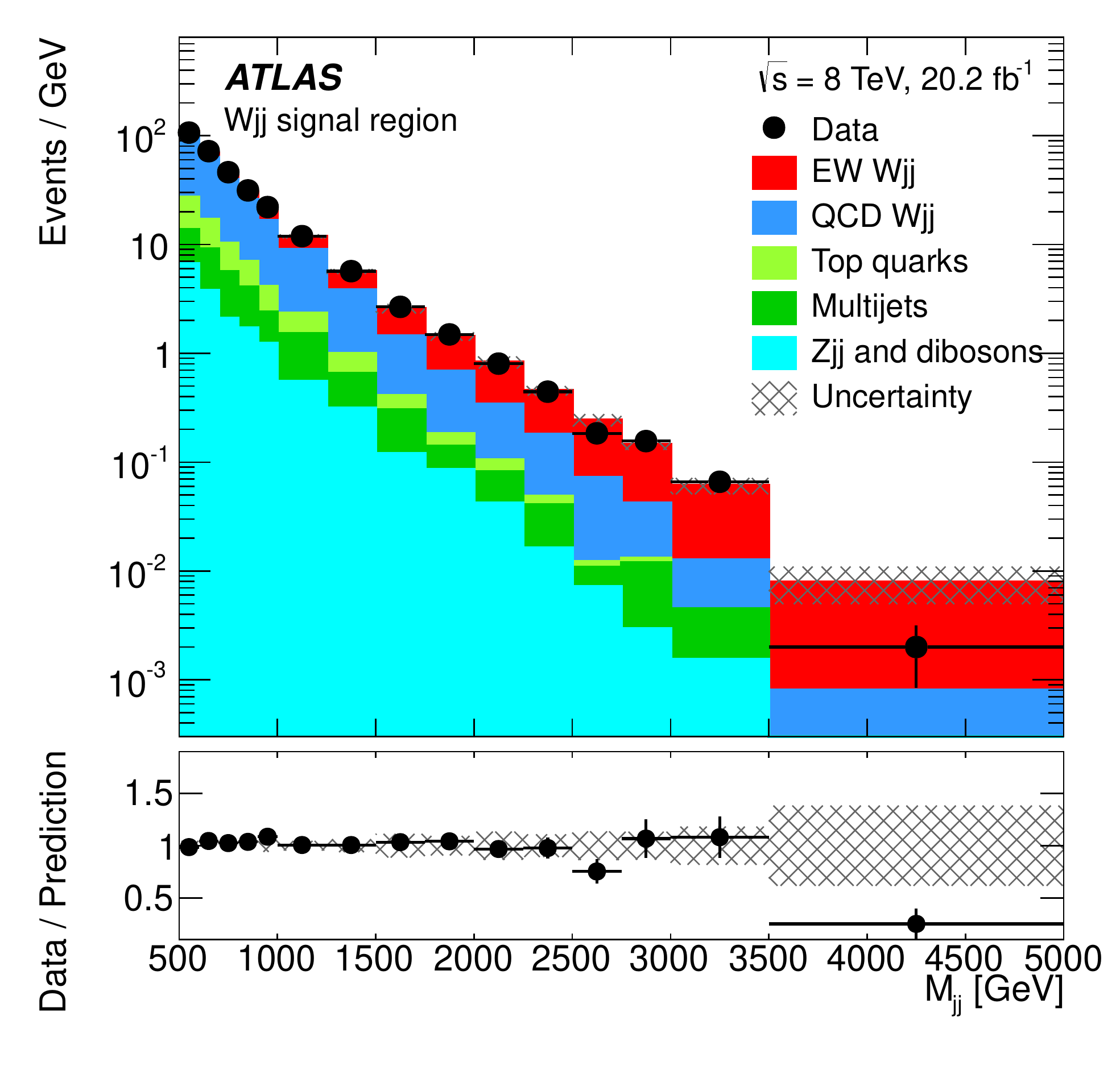}
\caption{Distributions of the dijet invariant mass for events in the signal region in 7~\TeV~(left) 
and 8~\TeV~(right) data, after fitting for the yields of the individual \wjets processes.  The 
bottom panel in each distribution shows the ratio of data to predicted signal-plus-background yields.  
The shaded band centred at unity represents the statistical and experimental uncertainties 
summed in quadrature. }
\label{fig:mjjfit}
\end{figure}

The dijet mass distributions in 7 and 8~\TeV~data after fitting for \muew and \muqcd are 
shown in Figure~\ref{fig:mjjfit}.  There is good overall agreement between the normalized 
distributions and the data.  The fit results for \muqcd are $1.16 \pm 0.07$ for 7~\TeV~data, 
and $1.09 \pm 0.05$ for 8~\TeV~data.  The measured values of \muew are consistent between 
electron and muon channels, with the following combined results:
\begin{eqnarray}
\muew~(7~\mathrm{{\TeV}}) & = & 1.00 \pm 0.16~\mathrm{(stat)}~\pm 0.17~\mathrm{(exp)}~\pm 0.12~\mathrm{(th)}, \nonumber \\ 
\muew~(8~\mathrm{{\TeV}}) & = & 0.81 \pm 0.05~\mathrm{(stat)}~\pm 0.09~\mathrm{(exp)}~\pm 0.10~\mathrm{(th)}. \nonumber 
\end{eqnarray}

\noindent
The measured value of $\muew$ has a total uncertainty of 0.26 (0.14) in 7~(8)~\TeV~data, 
and differs from the SM prediction of unity by $<0.1\sigma$~($1.4\sigma$).  In the absence 
of a control region, the uncertainty would increase to 0.37 (0.18) in 7~(8)~\TeV~data.

The fiducial signal region is defined by the selection in Table~\ref{tab:selection} using 
particle-level quantities after parton showering.  The measured and predicted cross sections 
times branching ratios in this region are shown in Table~\ref{tab:fidxs}.  The acceptance is 
calculated using \powheg + \pythia with a dominant uncertainty due to the parton-shower 
modelling which is estimated by taking the difference between \powheg + \pythia and \powheg + \herwigpp.  
The uncertainty in the predicted fiducial cross section at $\sqrt{s} = 8$~\TeV~includes a 4~fb 
contribution from scale variations and an 11~fb contribution from parton-shower modelling.

\begin{figure}[tbp!]
\centering
\includegraphics[width=0.69\textwidth]{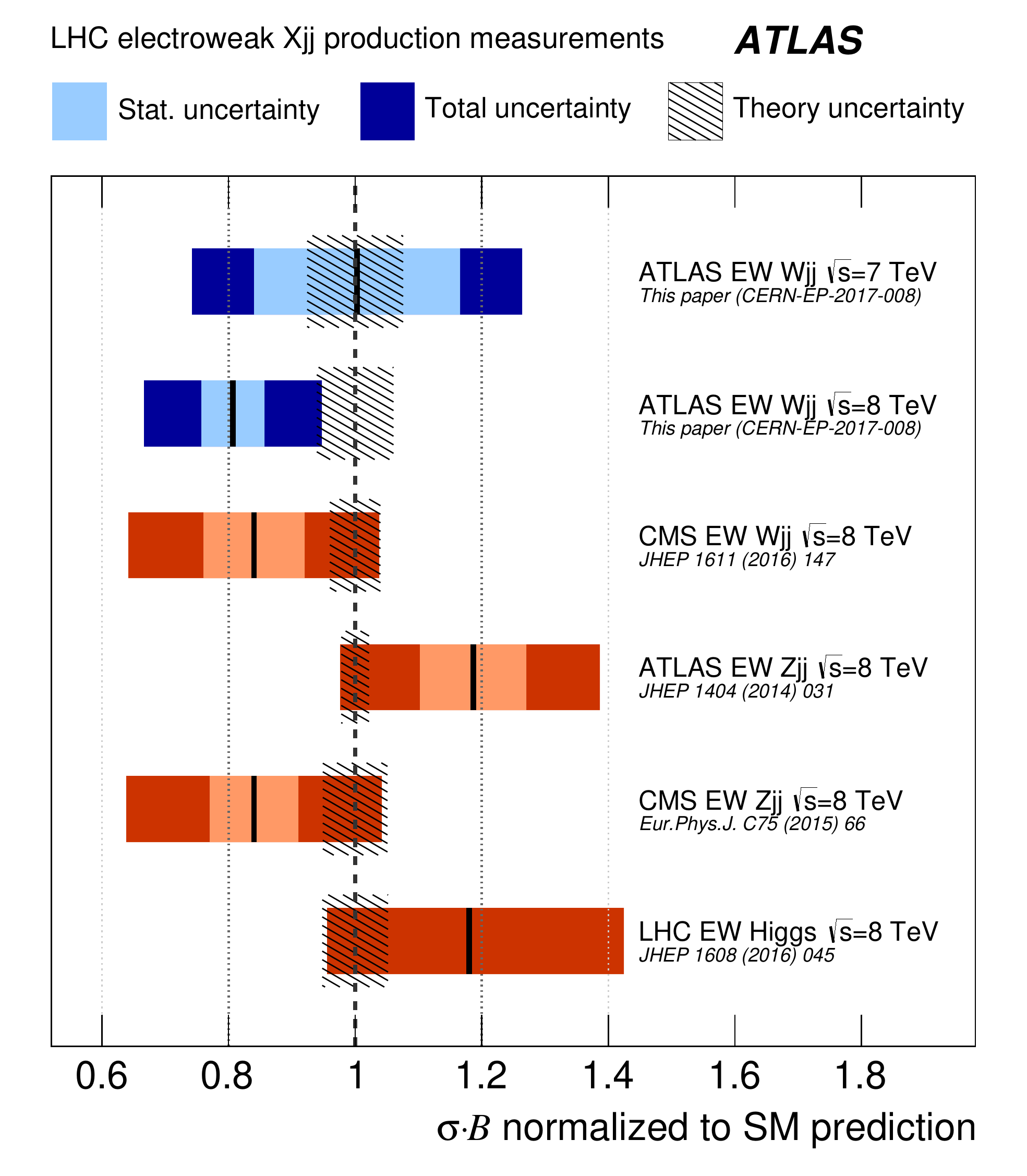}
\caption{Measurements of the cross section times branching fractions of electroweak production
of a single $W$, $Z$, or Higgs boson at high dijet invariant mass, divided by the SM predictions 
(\powheg+\pythia for ATLAS, \madgraph+\pythia for CMS, and \powheg+\pythia for the LHC combination). 
The lighter shaded band (where shown) represents the statistical uncertainty of the measurement, 
the outer darker band represents the total measurement uncertainty.  Theoretical uncertainties in 
the SM prediction are represented by the shaded region centred at unity.}
\label{fig:fidxsec:LHCEWKSF}
\end{figure}

A summary of this measurement and other measurements of boson 
production at high dijet invariant mass is shown in Figure~\ref{fig:fidxsec:LHCEWKSF}, normalized 
to SM predictions.  The measurement with the smallest relative uncertainty is the 8 TeV $\wjets$ 
measurement presented here.

\begin{table*}[tbp!]
\caption{
Measured fiducial cross sections of electroweak \wjets production in a single lepton channel, 
compared to predictions from \powheg + \pythia.  The acceptances and the inclusive measured 
production cross sections with $\pT > 20$~GeV jets are also shown.
}
\label{tab:fidxs}
\begin{center}
\begin{tabular}{ccccc}
\toprule
$\sqrt{s}$ & $\sigma^\mathrm{fid}_\mathrm{meas}$ [fb] & $\sigma^\mathrm{fid}_\mathrm{SM}$ [fb] & Acceptance ${\cal{A}}$ & 
$\sigma^\mathrm{inc}_\mathrm{meas}$ [fb] \\
\midrule
7 \TeV & $144 \pm 23~\mathrm{(stat)}~\pm 23~\mathrm{(exp)}~\pm 13~\mathrm{(th)}$ & $144 \pm 11$ & $0.053 \pm 0.004$ & 
$2760 \pm 670$ \\
8 \TeV & $159 \pm 10~\mathrm{(stat)}~\pm 17~\mathrm{(exp)}~\pm 15~\mathrm{(th)}$ & $198 \pm 12$ & $0.058 \pm 0.003$ & 
$2890 \pm 510$ \\
\bottomrule
\end{tabular}
\end{center}
\end{table*}

%% file: diffxsecintro.tex
Differential cross section measurements provide valuable information on the observed kinematic
properties of a process, testing the theoretical predictions and providing model-independent
results to probe for new physics.  This section presents differential measurements in the
$\sqrt{s}=8$~\TeV~data that discriminate EW \wjets from QCD \wjets production, after first introducing
the unfolding procedure, uncertainties, and the fiducial measurement regions.  The large event
yields allow more precise tests of these distributions than other VBF measurements and provide
the most comprehensive tests of predictions in VBF-fiducial regions.  Distributions sensitive
to anomalous triple gauge couplings are also presented and extend to values of momentum transfer
approaching 1~\TeV, directly probing these energies for the presence of new interactions.
Additional distributions are provided in Appendix~\ref{sec:appendix}, and the complete set of
measurements is available in \textsc{hepdata}\,\cite{HEPDATA}.

All differential production cross sections are measured both as absolute cross sections and as
distributions normalized by the cross section of the measured fiducial region ($\sigma^\mathrm{fid}_W$).
The normalizations are performed self-consistently, i.e. data measurements are normalized by the 
total fiducial data cross section and MC predictions are normalized by the corresponding MC cross 
section.  Many sources of uncertainty are reduced for normalized distributions, allowing 
higher-precision tests of the modelling of the shape of the measured observables.  

Unfolded differential cross-section measurements are performed for both QCD+EW~\wjets and 
EW~\wjets production and compared to theoretical predictions from the \powheg + \pythia, 
\sherpa, and \textsc{hej} event generators, which are described in
Section~\ref{sec:mc}.  The reported cross sections are for a single lepton flavour and are 
normalized by the width of the measured bin interval.

%% file: unfolding.tex
The MC simulations are used to correct the cross sections for detector and 
event selection inefficiencies, and for the effect of detector resolutions.  
An implementation\,\cite{Adye:2011gm} of a Bayesian iterative unfolding 
technique\,\cite{DAgostini2010} is used to perform these corrections.
The unfolding is based on a response matrix from the simulated events which 
encodes bin-to-bin migrations between a particle-level differential distribution 
and the equivalent reconstruction-level distribution.  The matrix gives transition 
probabilities from particle level to reconstruction level, and Bayes' theorem is 
employed to calculate the inverse probabilities.  These probabilities are used in 
conjunction with a prior particle-level signal distribution, which is taken from the 
\powheg + \pythia simulations, to unfold the background-subtracted reconstruction-level 
data distributions.  After this first unfolding iteration the unfolded data distribution 
is used as the new prior and the process repeated for another iteration.  The unfolding 
procedure is validated by unfolding the \sherpa simulation using the \powheg + \pythia 
response matrix.  For all distributions the unfolded and initial particle-level \sherpa 
predictions agree within the unfolding uncertainty assigned. Bin boundaries in unfolded 
distributions are chosen to ensure that $>66\%$ of particle-level events remain within 
the same interval at reconstruction level.

The sources of uncertainty discussed in Section~\ref{sec:xsec} are assessed for the 
unfolded differential production cross sections.  Figures are shown with statistical 
uncertainties as inner bars and total uncertainties as the outer bars.  Statistical 
uncertainties are estimated using pseudoexperiments, with correlations between bins 
determined using a bootstrap method\,\cite{Hayes:1988xc}.  The $W\to e\nu$ and 
$W\to\mu\nu$ channels are found to be statistically compatible, and are combined.  
Theoretical uncertainties include the effects of scale and PDF variations on the prior 
distribution and on the response matrix.  For unfolding EW \wjets production, additional 
theoretical uncertainties arise from modelling the QCD \wjets contribution subtracted 
from the data, and from the neglect of interference between the strong and electroweak 
\wjets processes.  The interference uncertainty is estimated using the same procedure as 
for the fiducial measurement (Section~\ref{sec:xsec}), i.e. by adding the \sherpa 
interference model to the background prediction.  The interference uncertainty is 
shown explicitly as a shaded area in each bin of the measured distributions.  
An uncertainty in the unfolding procedure is estimated by reweighting the simulation 
such that the distributions match the unfolded data, and then unfolding the data with 
the reweighted simulation; the change in the unfolded measurement is symmetrized and 
taken as an uncertainty.  Experimental uncertainties are assessed by unfolding the data 
distributions using a modified response matrix and prior incorporating the change in 
detector response.

\begin{figure}[!tbp]
\centering
\includegraphics[width=0.49\textwidth]{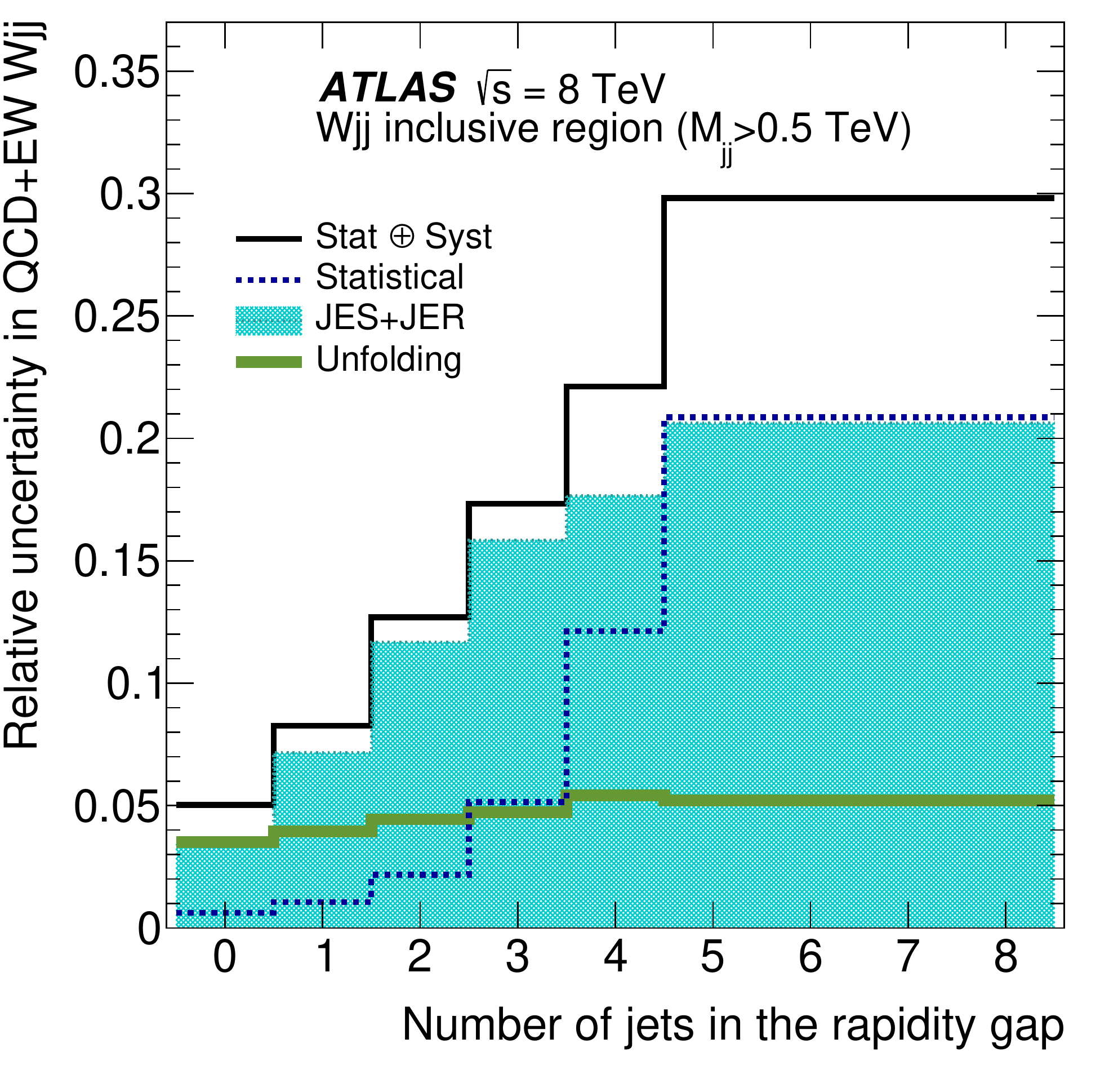}
\includegraphics[width=0.49\textwidth]{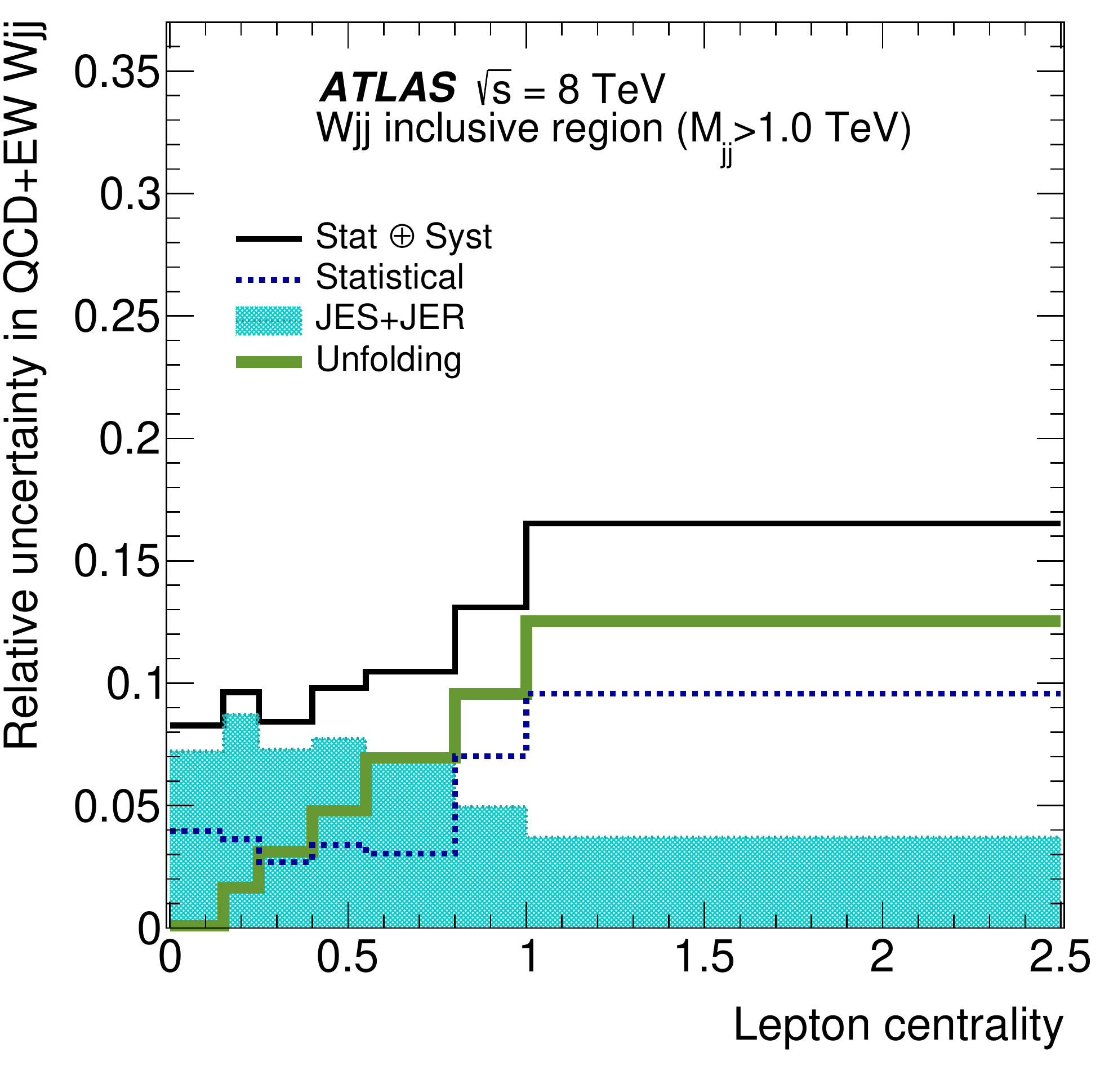}
\includegraphics[width=0.49\textwidth]{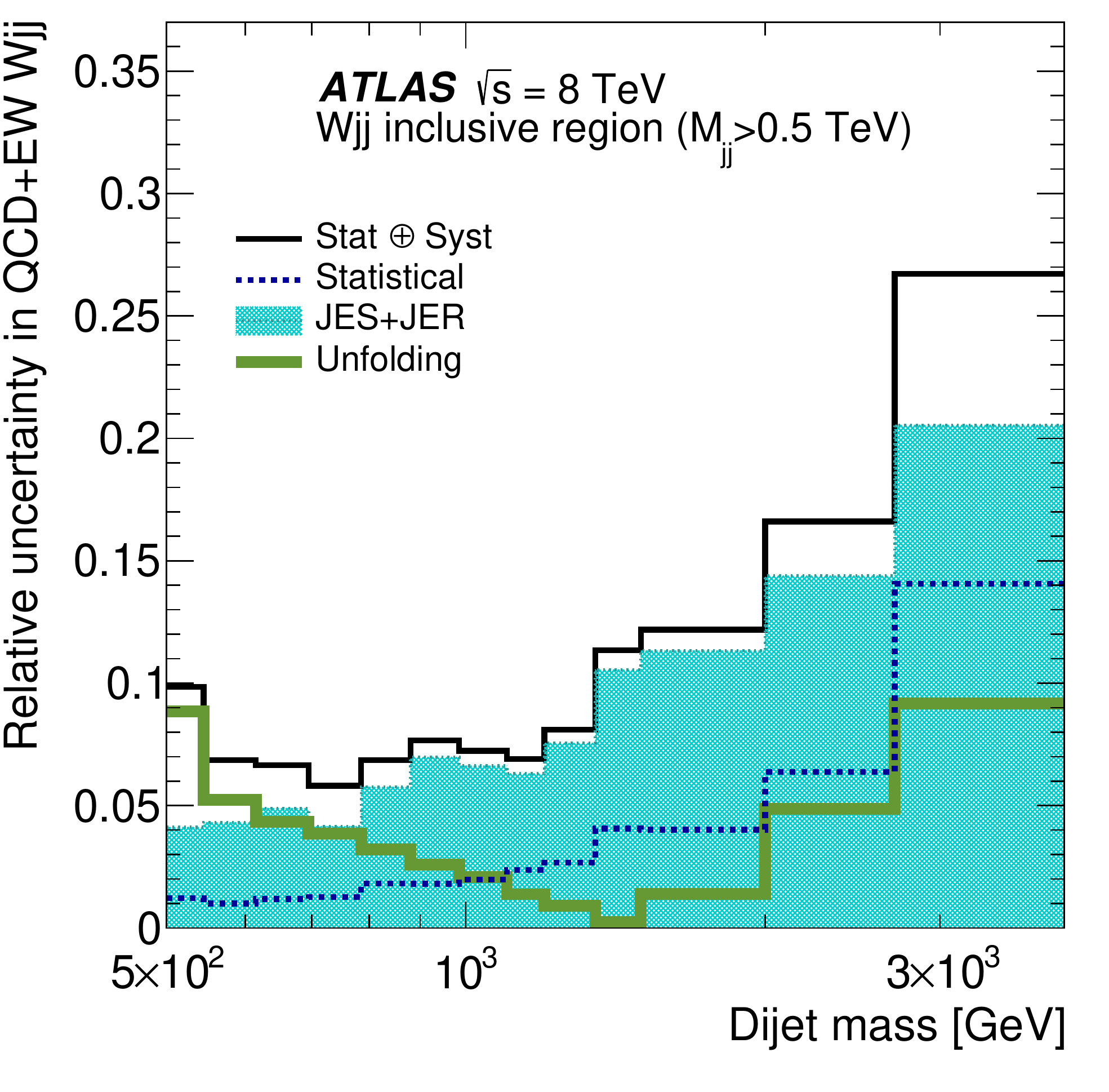}
\includegraphics[width=0.49\textwidth]{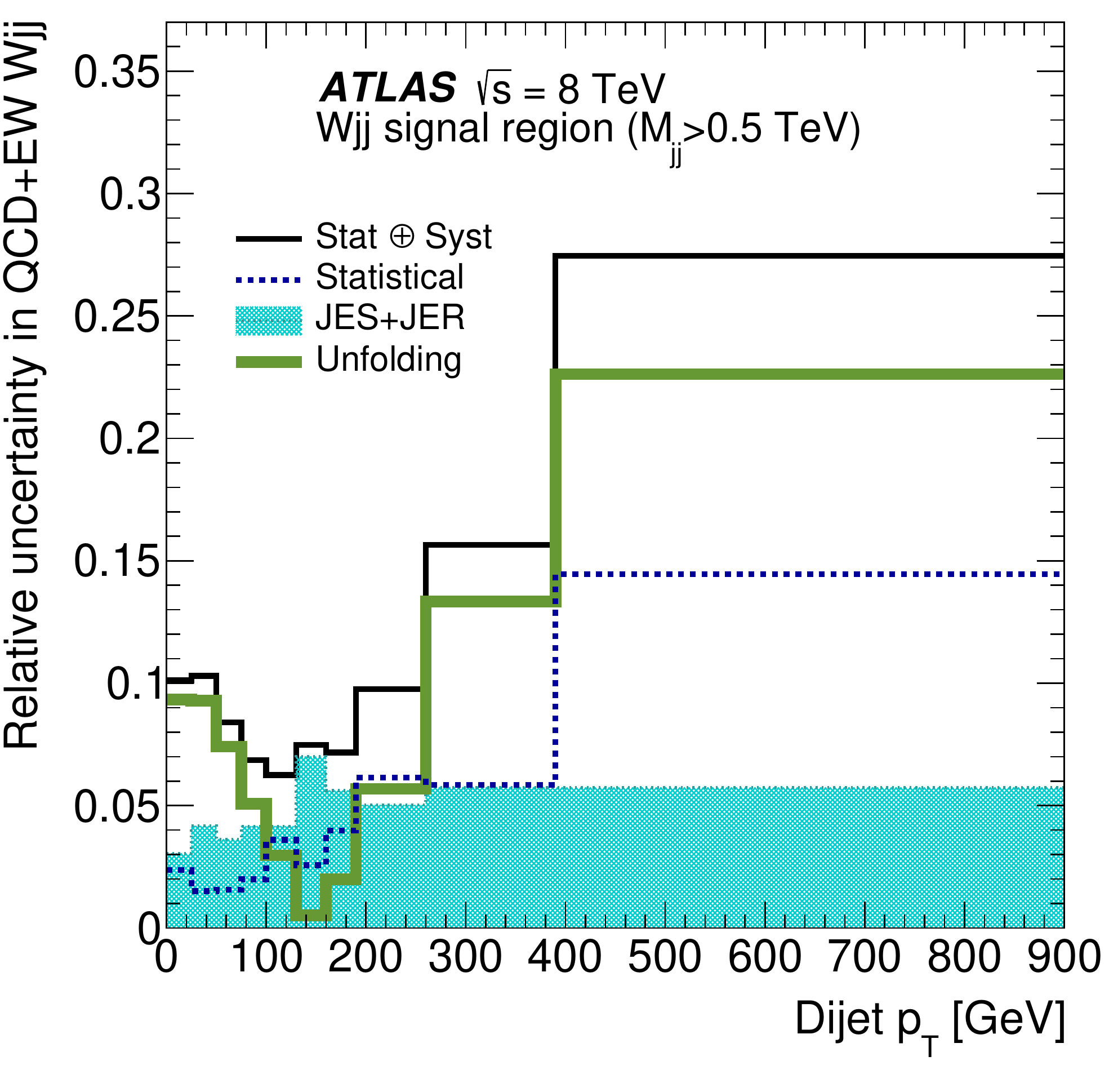}
\caption{Relative uncertainties in example unfolded differential cross sections for the combined 
QCD+EW~\wjets processes.  The examples are: the number of jets in the rapidity gap between the 
two highest-$\pt$ jets in the inclusive region (top left); the lepton centrality distribution in 
the inclusive $\mjj>1$~TeV region (top right); $\mjj$ in the inclusive region (bottom left); and 
the dijet $\pt$ in the signal region (bottom right). Dominant contributions to the total systematic 
uncertainty are highlighted separately.}
\label{unfolding:systSummary_examples}
\end{figure}

\begin{figure}[htbp]
\centering
\includegraphics[width=0.49\textwidth]{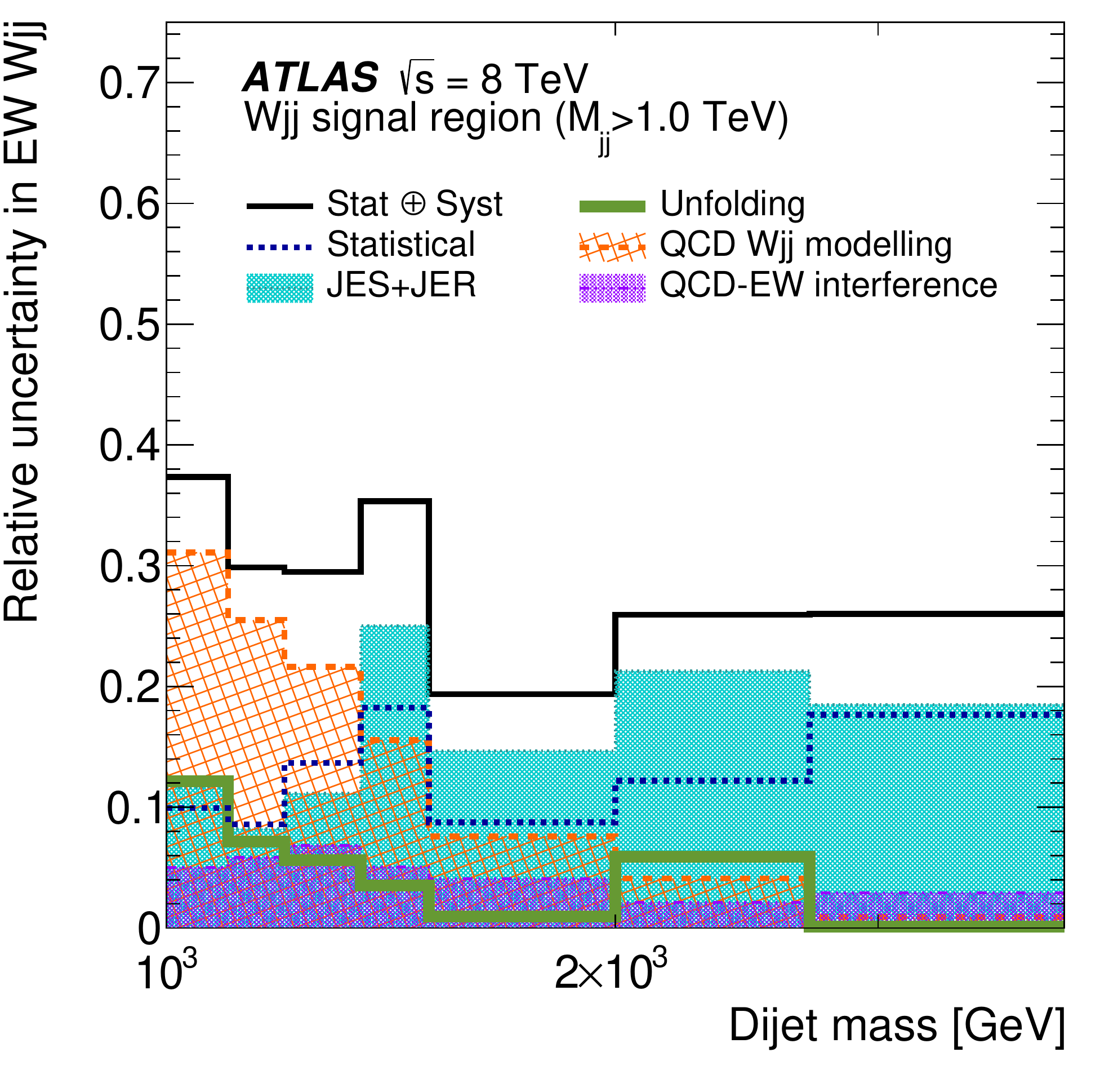}
\includegraphics[width=0.49\textwidth]{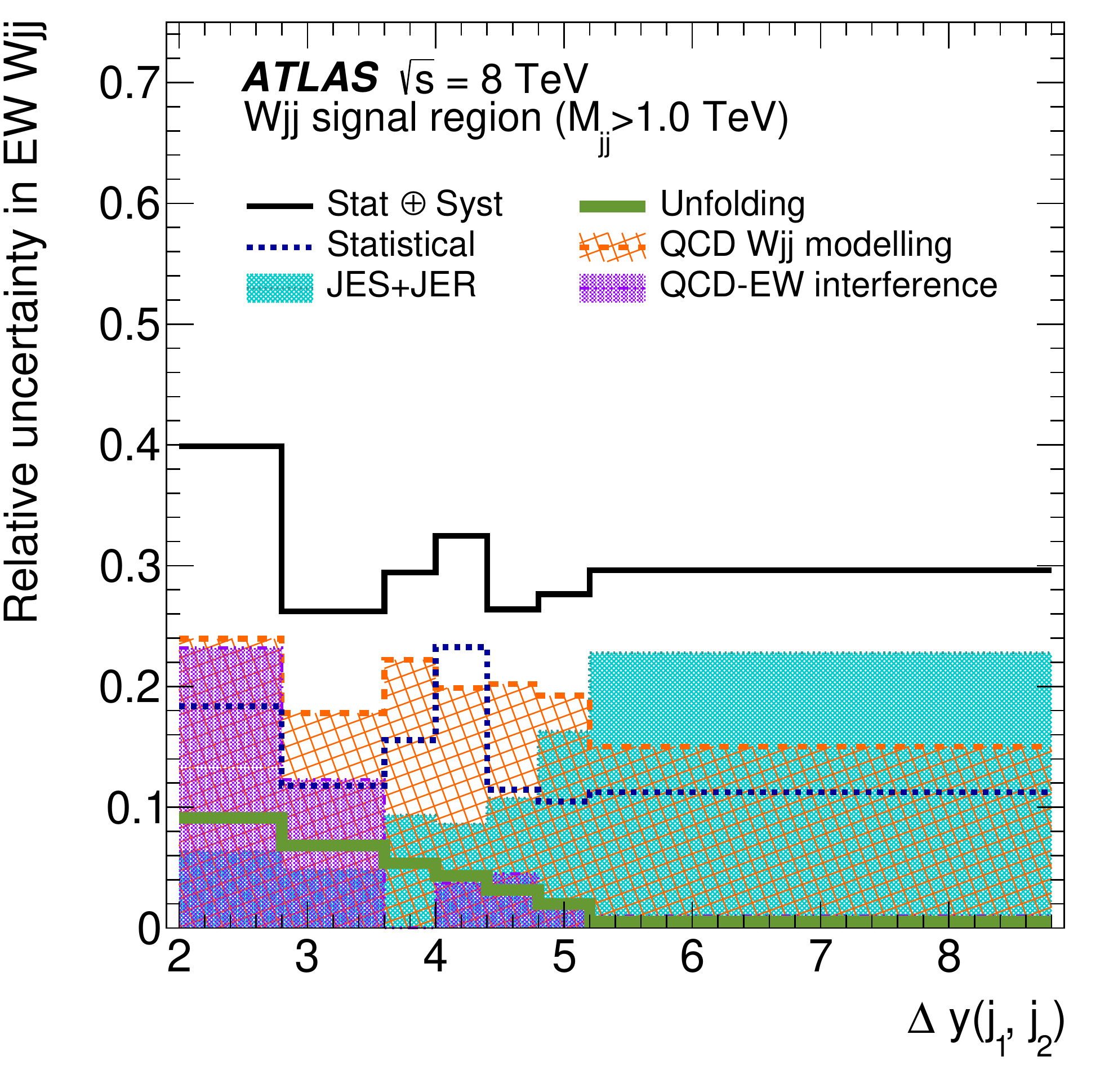}
\includegraphics[width=0.49\textwidth]{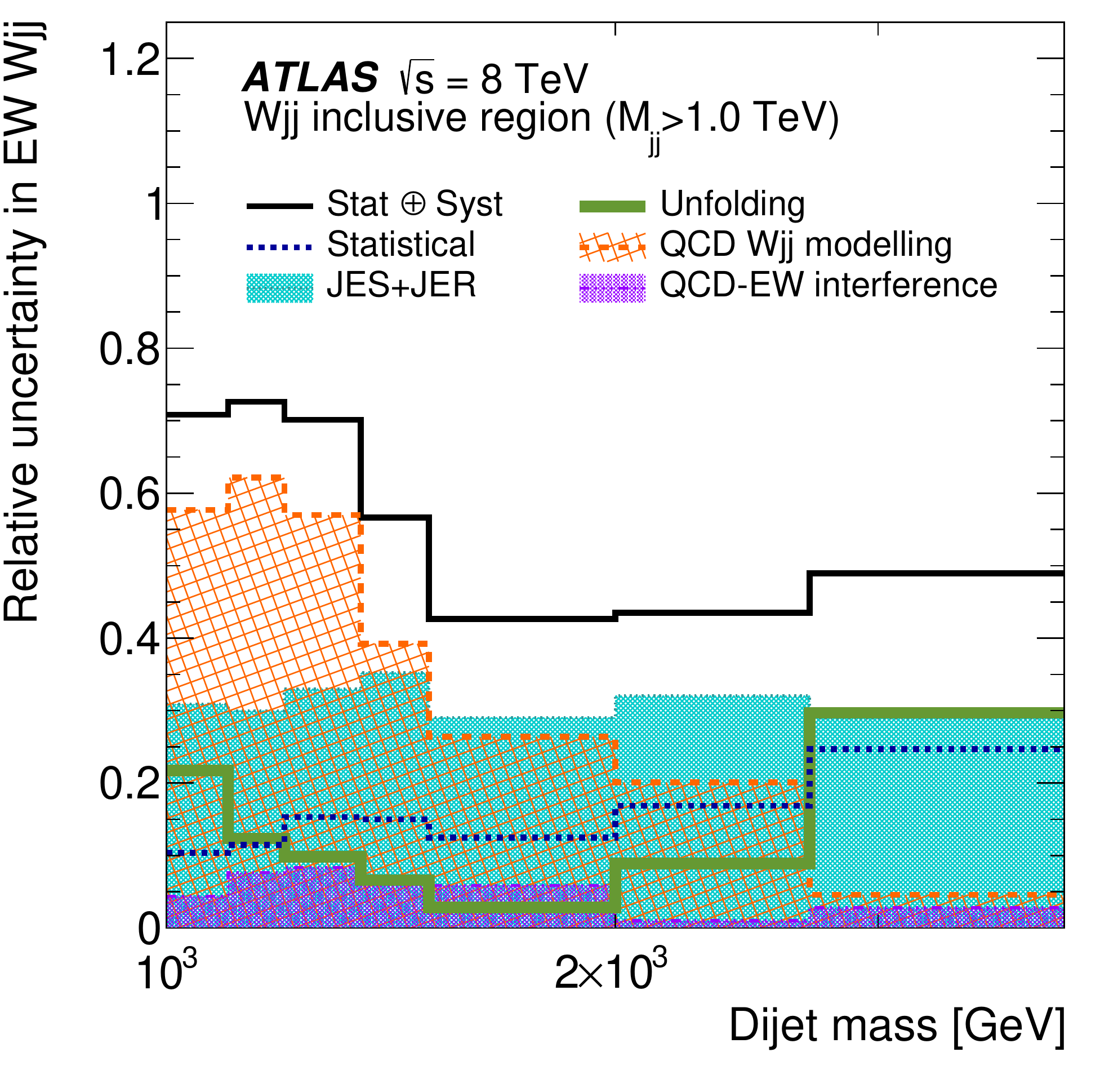}
\includegraphics[width=0.49\textwidth]{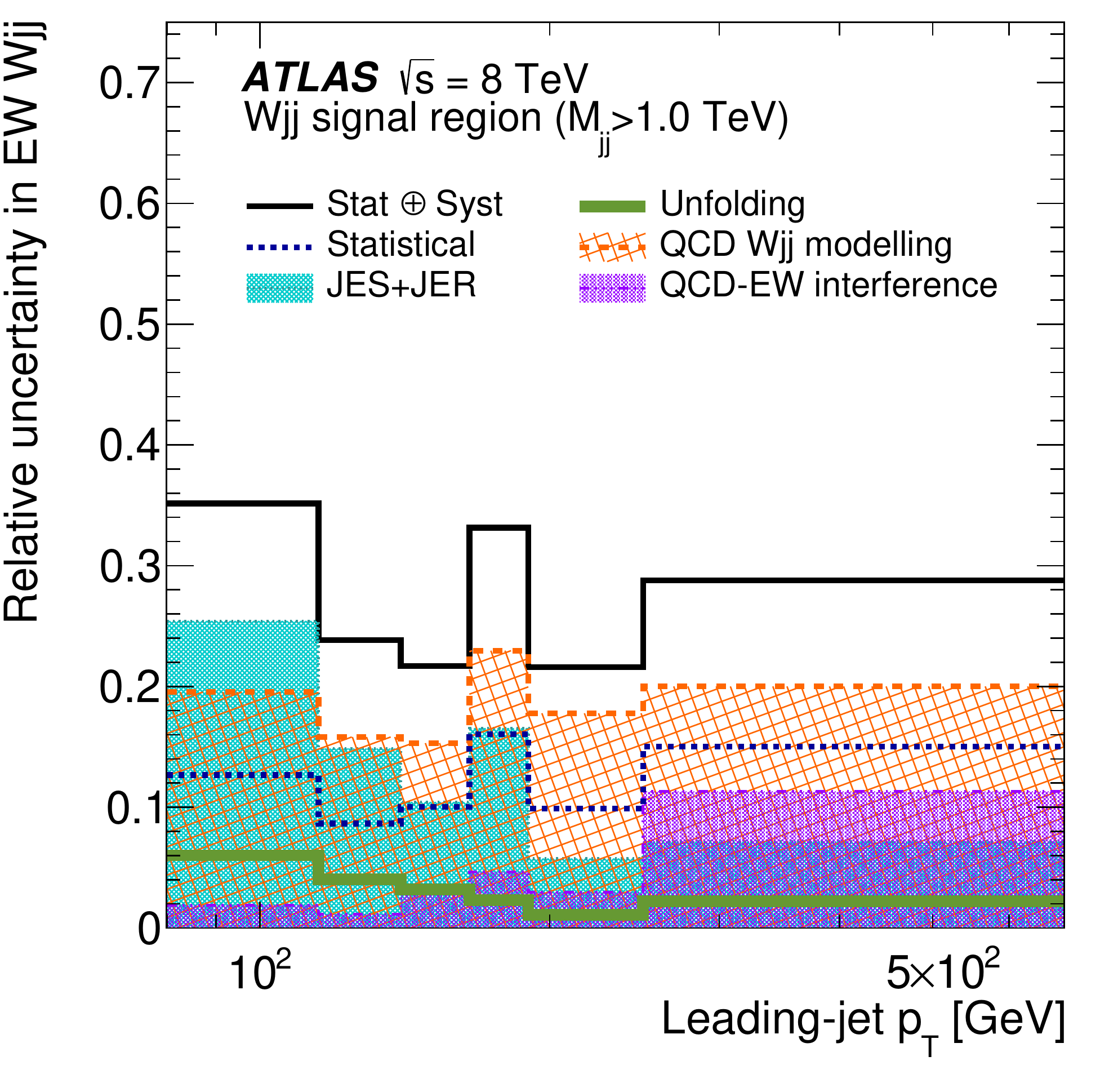}
\caption{Relative uncertainties in example unfolded differential cross sections for the EW \wjets 
processes. The examples are $\mjj$ (top left) and $\dyjj$ (top right) in the high-mass signal 
region; $\mjj$ in the $\mjj>1$~TeV inclusive region (bottom left); and leading-jet $\pt$ in the 
high-mass signal region (bottom right).  Dominant contributions to the total systematic uncertainty 
are highlighted separately. }
\label{unfolding:systSummaryEWK_examples}
\end{figure}

Figures~\ref{unfolding:systSummary_examples} and~\ref{unfolding:systSummaryEWK_examples} 
summarize the uncertainty contributions to example unfolded data distributions for 
QCD+EW~\wjets and EW~\wjets production, respectively.  For measurements of combined 
QCD+EW~\wjets production, the jet energy scale and resolution uncertainties dominate 
the total uncertainty except in regions where statistical uncertainties are significant.  
The unfolding uncertainty is typically relevant in these regions and in regions dominated 
by QCD \wjets production where the statistical uncertainties are small.
In measurements of EW~\wjets production, uncertainties in the modelling of strong \wjets 
production are particularly important at low dijet invariant mass, where the EW~\wjets 
signal purity is lowest.  Interference uncertainties become dominant at low dijet rapidity 
separation but are otherwise not the leading contribution to the total uncertainty.  A 
recent study\,\cite{Denner:2014ina} of interference in $Z$+jets VBF topologies, 
incorporating NLO electroweak corrections, predicted similar behaviour.  For the bulk of the 
EW \wjets distributions, the leading sources of uncertainty are statistical, QCD \wjets 
modelling, and jet energy scale and resolution, and contribute roughly equally.

%% file: diffregions.tex
The differential cross sections of the combined \wjets processes are measured in the following nine 
fiducial regions:
\begin{itemize}
\item the four mutually orthogonal fiducial regions defined in Figure~\ref{evtsel:phasespaceIllustration}, 
three of which are electroweak-suppressed ($<5\%$ contribution) and one electroweak-enhanced ($15$--$20\%$ 
contribution);
\item an additional electroweak-enhanced signal region with $\mjj>1.0$~\TeV~($35$--$40\%$ electroweak 
\wjets contribution); and 
\item four inclusive fiducial regions defined by the preselection requirements in Table~\ref{tab:selection} 
with $\mjj > 0.5,~1.0,~1.5$~and~2.0~\TeV.
\end{itemize}

The inclusive fiducial regions probe the observables used to distinguish EW and QCD~\wjets production, 
namely lepton and jet centrality, and the number of jets radiated in the rapidity gap between 
the two leading jets.  The four successively higher invariant mass thresholds increasingly enhance the 
EW~\wjets purity of the differential distributions, without lepton and jet topology requirements. 

The combined QCD+EW \wjets production is measured in all regions to test the modelling of QCD \wjets 
production in a VBF topology.  In regions sensitive to EW \wjets contributions, the prediction for 
QCD \wjets only is shown along with the combined QCD+EW \wjets prediction in order to indicate the 
effect of the EW \wjets process.  Differential measurements of EW \wjets production are performed in 
regions with $\mjj>1.0$~\TeV, where the expected EW~\wjets fraction is $>20\%$.  The QCD \wjets 
background is subtracted using the multiplicative normalization factor of $\muqcd = 1.09 \pm 0.02$~(stat) 
determined from the fits in Section~\ref{sec:xsec}.  This substantially reduces the normalization 
uncertainty, confining theoretical uncertainties to the shapes of the background distributions.

Performing a complete unfolding of the EW~\wjets signal process leads to better precision on the unfolded data, 
particularly in the case of normalized distributions, than could be achieved by subtracting the particle-level 
QCD~\wjets production background from unfolded QCD+EW~\wjets production data.  All EW~\wjets differential 
measurements are nonetheless also performed as combined QCD+EW \wjets production measurements so that such a 
subtraction could be performed with other QCD \wjets predictions.

\begin{figure}[!ptb]
\centering
\includegraphics[width=0.875\textwidth]{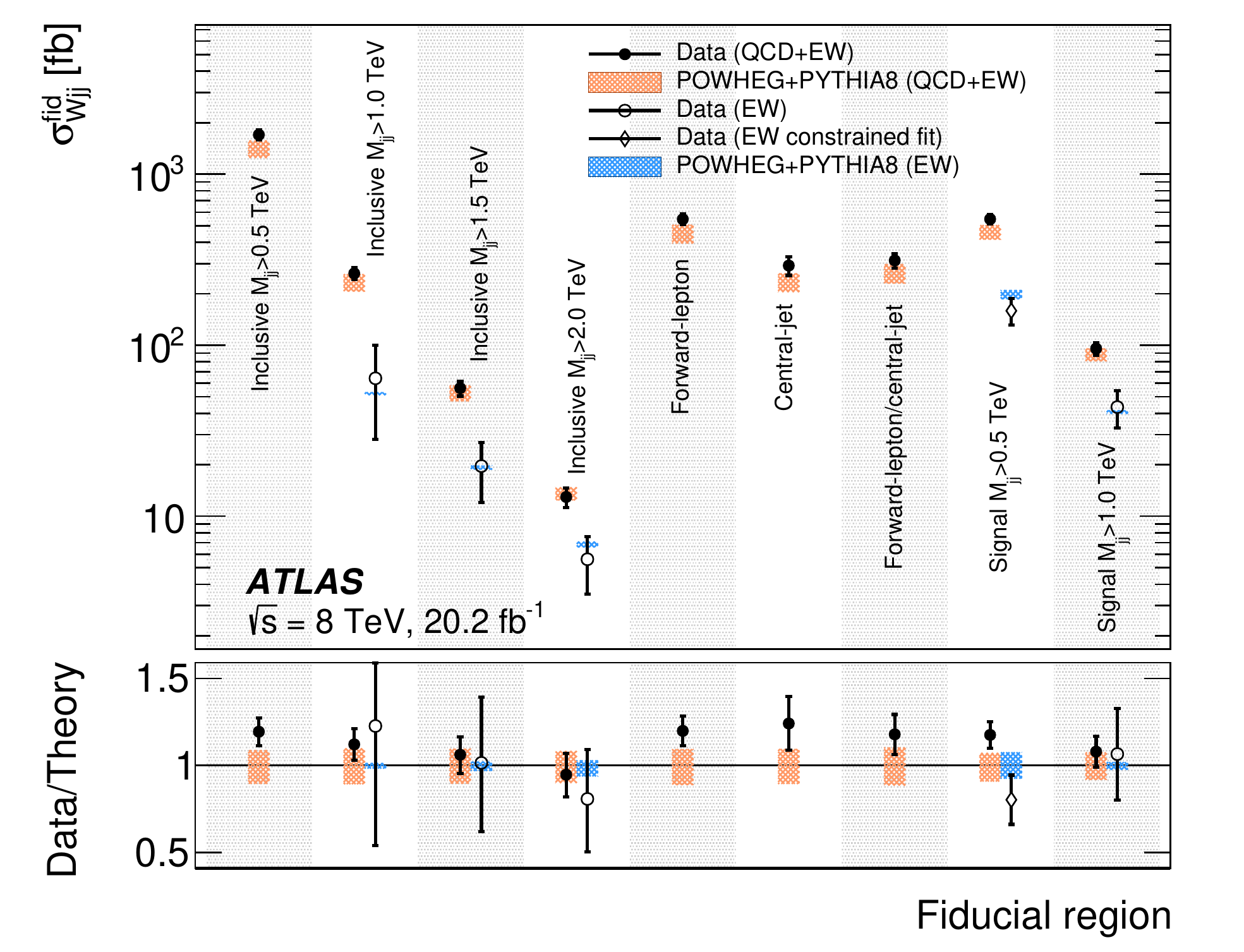}
\caption{Integrated production cross sections QCD+EW \wjets (solid data points) and EW \wjets (open data points) 
production in each measured particle-level fiducial region in a single lepton channel; EW \wjets production is 
only measured in fiducial regions where there is sufficient purity.  For each measurement the error bar 
represents the statistical and systematic uncertainties summed in quadrature.  Comparisons are made to predictions 
from \powheg + \pythia and the bottom pane shows the ratio of data to these predictions.}
\label{unfolding:intxsec_public}
\end{figure}

Integrated cross sections for \wjets production are determined in each fiducial region.  
Table~\ref{unfolding:intxsec_public_table} and Figure~\ref{unfolding:intxsec_public} show the measured integrated 
production cross sections for a single lepton flavour ($\sigma^\mathrm{fid}_{W}$) for QCD+EW \wjets production 
and, in high dijet invariant-mass regions, for EW \wjets production.  Also shown is the value of the EW \wjets 
cross section extracted from the constrained fit described in Section~\ref{sec:fidxsresults}.  All measurements 
are broadly compatible with predictions from \powheg + \pythia.  In fiducial regions dominated by QCD \wjets 
production the measured cross sections are approximately 15--20\% higher than predictions.  The integrated 
EW \wjets production cross sections have larger relative uncertainties than the precisely constrained fiducial 
EW \wjets cross-section measurement.

\begin{table*}[!ptb]
\caption{Integrated fiducial cross sections for QCD+EW and EW \wjets production and the equivalent predictions 
from \powheg + \pythia.  The uncertainties displayed are the values of the statistical and systematic uncertainties 
added in quadrature.}
\label{unfolding:intxsec_public_table}
\begin{center}
\begin{tabular}{lcccc}
\toprule
Fiducial region & \multicolumn{4}{c}{$\sigma^\mathrm{fid}_{W}$ [fb]} \\
                & \multicolumn{2}{c}{QCD+EW} & \multicolumn{2}{c}{EW} \\
                & Data & \powheg + \pythia & Data & \powheg + \pythia \\
\midrule
Inclusive $\mjj>0.5$~\TeV & $1700\pm 110$ & $1420\pm 150$ & --- & --- \\[0.1ex]
Inclusive $\mjj>1.0$~\TeV & $263\pm 21$ & $234\pm 26$ & $64\pm 36$ & $52\pm 1$\\[0.1ex]
Inclusive $\mjj>1.5$~\TeV & $56\pm 5$ & $53\pm 5$ & $20\pm 8$ & $19\pm 0.5$ \\[0.1ex]
Inclusive $\mjj>2.0$~\TeV & $13\pm 2$ & $14\pm 1$ & $5.6\pm 2.1$ & $6.9\pm 0.2$\\[0.1ex]
Forward-lepton & $545\pm 39$ & $455\pm 51$ & --- & ---\\[0.1ex]
Central-jet & $292\pm 36$ & $235\pm 28$ & --- & ---\\[0.1ex]
Forward-lepton/central-jet & $313\pm 30$ & $265\pm 32$ & --- & ---\\[0.1ex]
Signal $\mjj>0.5$~\TeV & $546\pm 35$ & $465\pm 39$ & $159\pm 27$ & $198\pm 12$\\[0.1ex]
Signal $\mjj>1.0$~\TeV & $96\pm 8$ & $89\pm 7$ & $43\pm 11$ & $41\pm 1$\\
\bottomrule
\end{tabular}
\end{center}
\end{table*}

\begin{figure}[!hptb]
\centering
\includegraphics[width=0.97\textwidth]{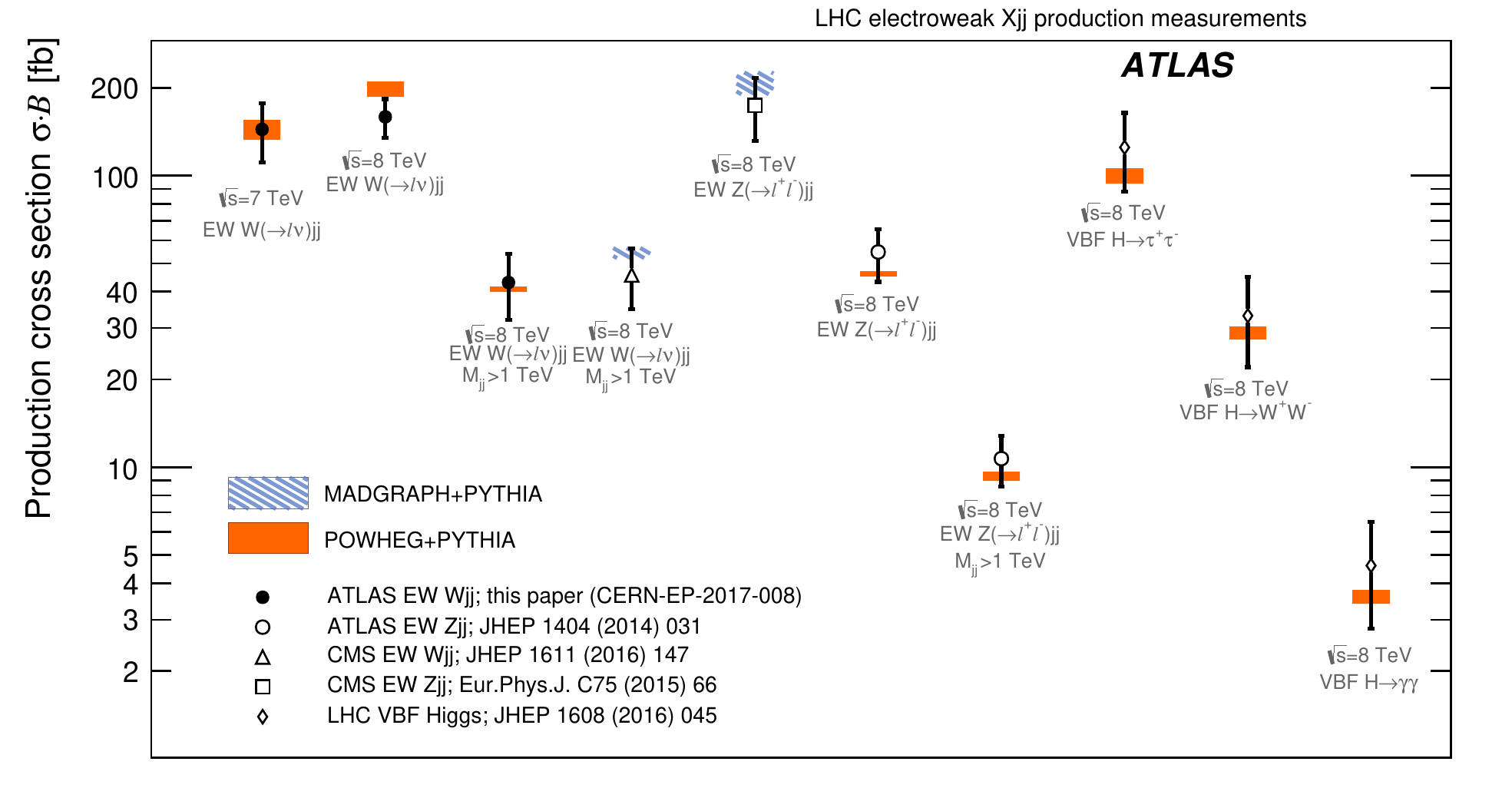}
\caption{Measurements of the cross sections times branching fractions of electroweak production of a single $W$, $Z$, 
or Higgs boson with two jets at high dijet invariant mass and in fiducial measurement regions.  For each measurement 
the error bar represents the statistical and systematic uncertainties summed in quadrature.  Shaded bands represent 
the theory predictions.  The \mjj threshold defining the fiducial $Zjj$ region differs between ATLAS and CMS, leading 
to different inclusive cross sections. }
\label{fig:LHCEWKHiggsxsec}
\end{figure}

The measurements of electroweak $\wjets$ fiducial cross sections are compared to measurements of electroweak $Zjj$ 
production and VBF Higgs boson production in Figure~\ref{fig:LHCEWKHiggsxsec}.  These other measurements are 
extrapolated to lower dijet mass (for $Zjj$ production) or to inclusive production (for Higgs boson
production) so their apparent cross sections are generally increased relative to the \wjets fiducial cross 
sections.

%% file: diffqcdvsew.tex
Differential measurements are performed in the following distributions that provide 
discrimination between strong and electroweak \wjets production:
\begin{itemize}
\item $\mjj$, the invariant mass of the two highest-$\pt$ jets;
\item $\dyjj$, the absolute rapidity separation between the two highest-$\pt$ jets;
\item $C_\ell$, lepton centrality, the location in rapidity of the lepton relative to 
the average rapidity of the two highest-\pt jets, defined in Eq.~(\ref{eqn:lepcentrality});
\item $C_j$, jet centrality, the location in rapidity of any additional jet relative to 
the average rapidity of the two highest-\pt jets, defined in Eq.~(\ref{eqn:lepcentrality}); and 
\item $\ngapjet$, the number of additional jets in the rapidity gap bounded by the two 
highest-$\pt$ jets (i.e., jets with $C_j < 0.5$).
\end{itemize}

\begin{figure}[htbp]
\centering
\includegraphics[width=0.45\textwidth]{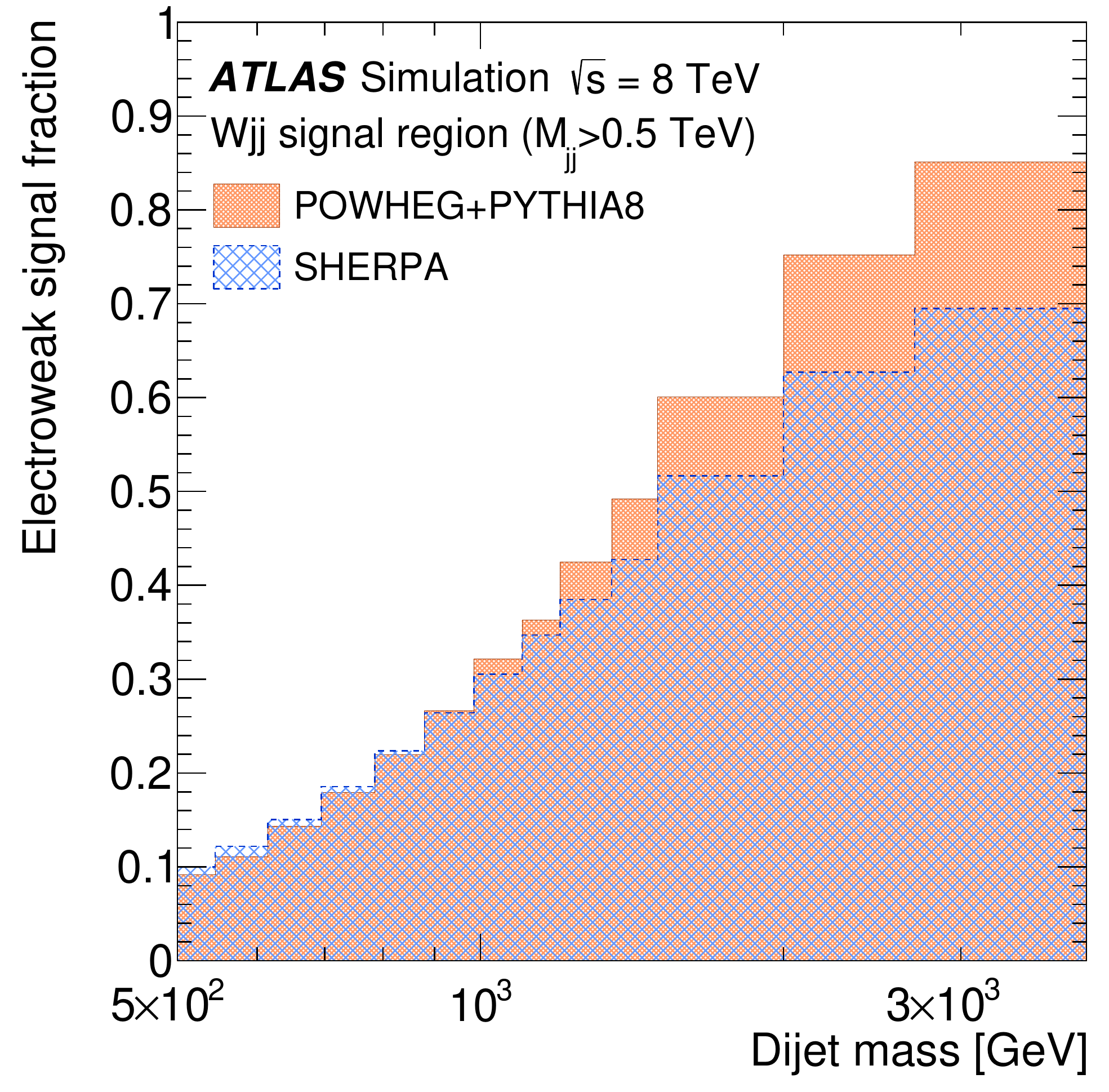}
\includegraphics[width=0.45\textwidth]{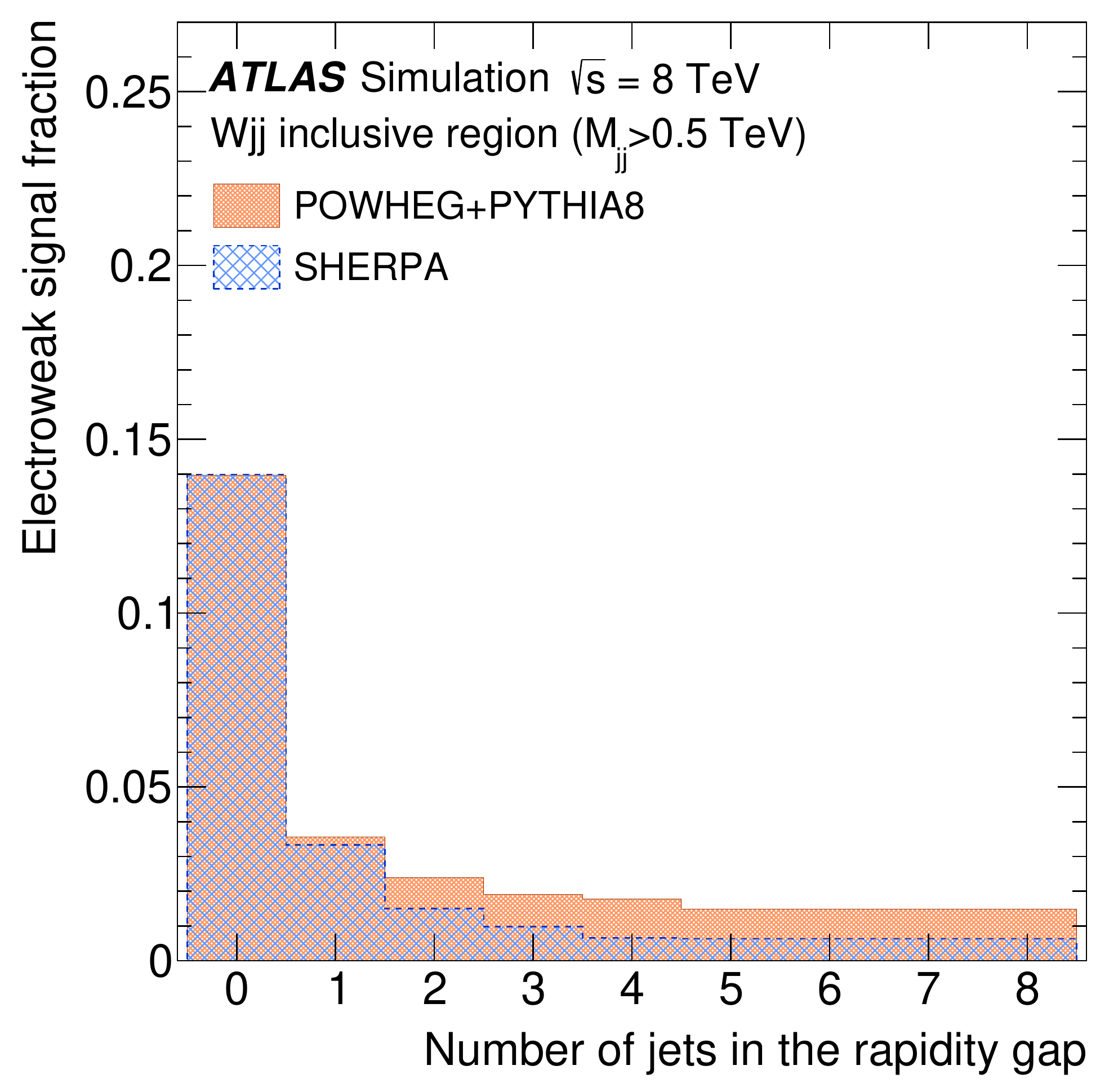}
\caption{Fraction of EW~\wjets signal relative to the combined QCD+EW~\wjets production,
predicted by \powheg + \pythia and \sherpa simulations for observables in the signal (left) 
and inclusive (right) fiducial regions.}
\label{unfolding:EWKfraction}
\end{figure}

The first two observables use the dijet system to distinguish the $t$-channel VBF topology 
from the background. The remaining observables use the rapidity of other objects relative 
to the dijet rapidity gap, exploiting the colourless gauge boson exchange to distinguish the 
EW \wjets signal from the QCD \wjets background.  Figure~\ref{unfolding:EWKfraction} shows 
the \powheg + \pythia and \sherpa predictions of the fraction of \wjets events produced via 
electroweak processes, as a function of the dijet invariant mass in the signal fiducial region 
and the number of jets emitted in the dijet rapidity gap for the inclusive fiducial region with 
$\mjj>0.5$~\TeV.  

%% file: diffdijet.tex
\begin{figure}[!tbp]
\centering
\includegraphics[width=0.48\textwidth]{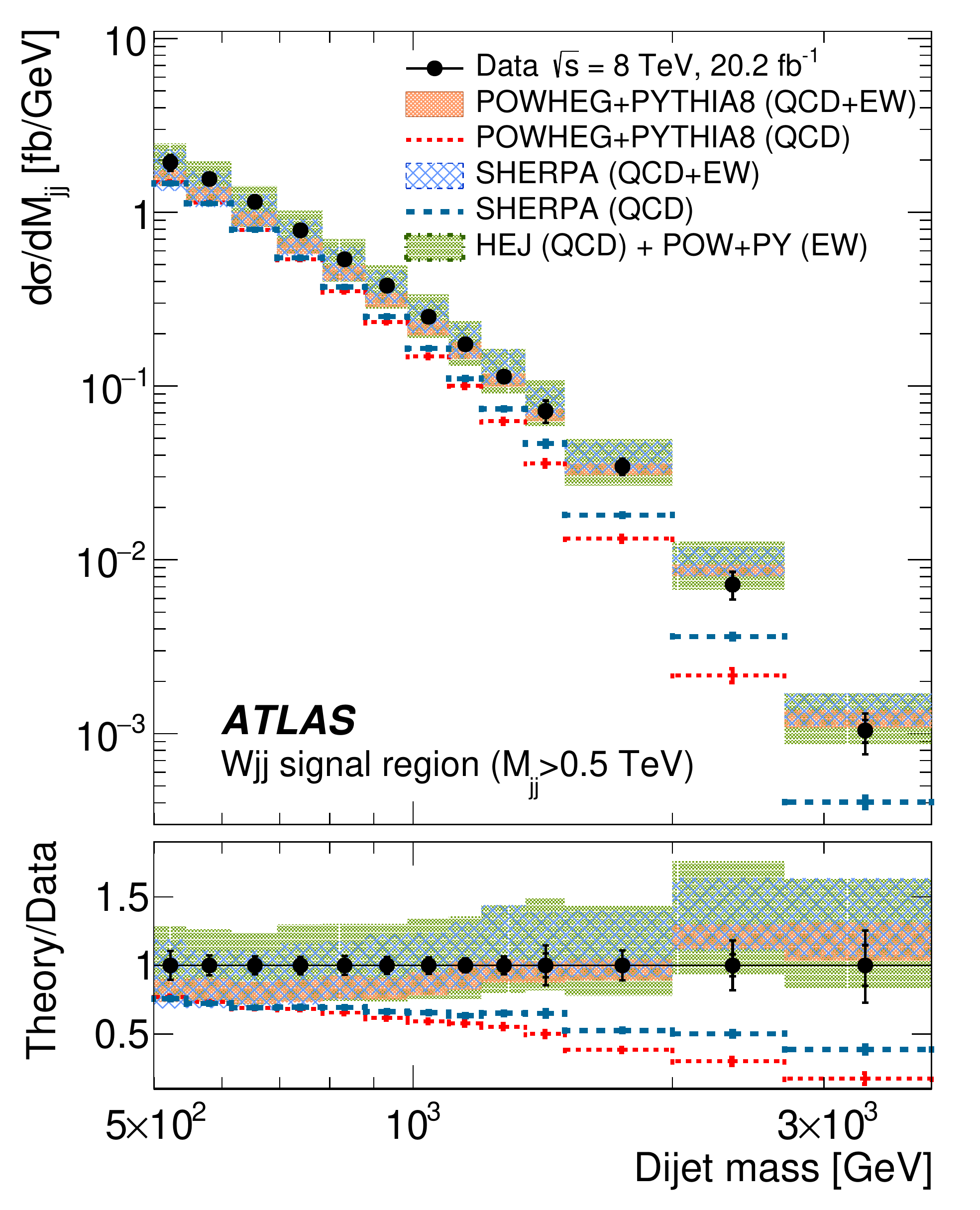}
\includegraphics[width=0.48\textwidth]{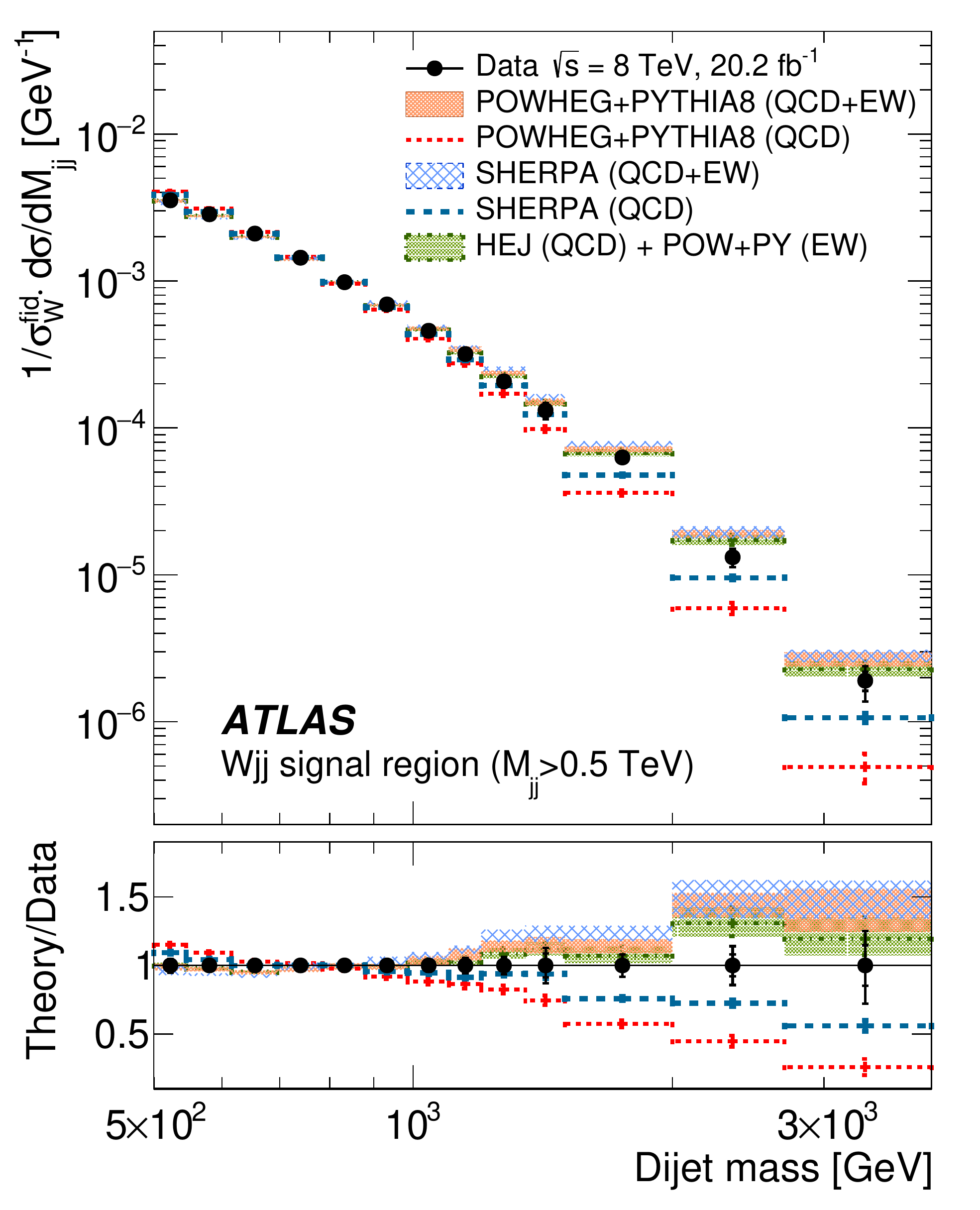}
\includegraphics[width=0.48\textwidth]{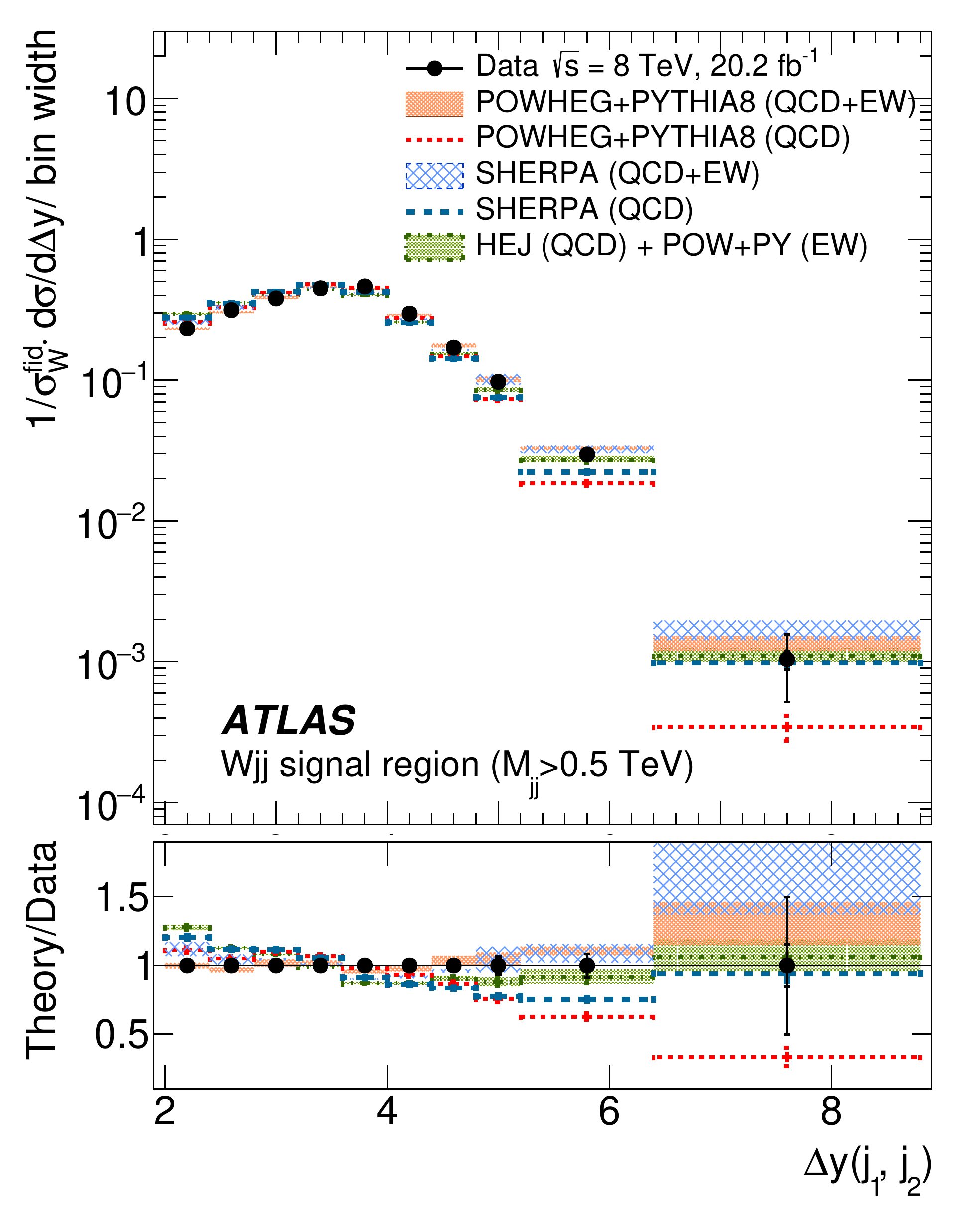}
\includegraphics[width=0.48\textwidth]{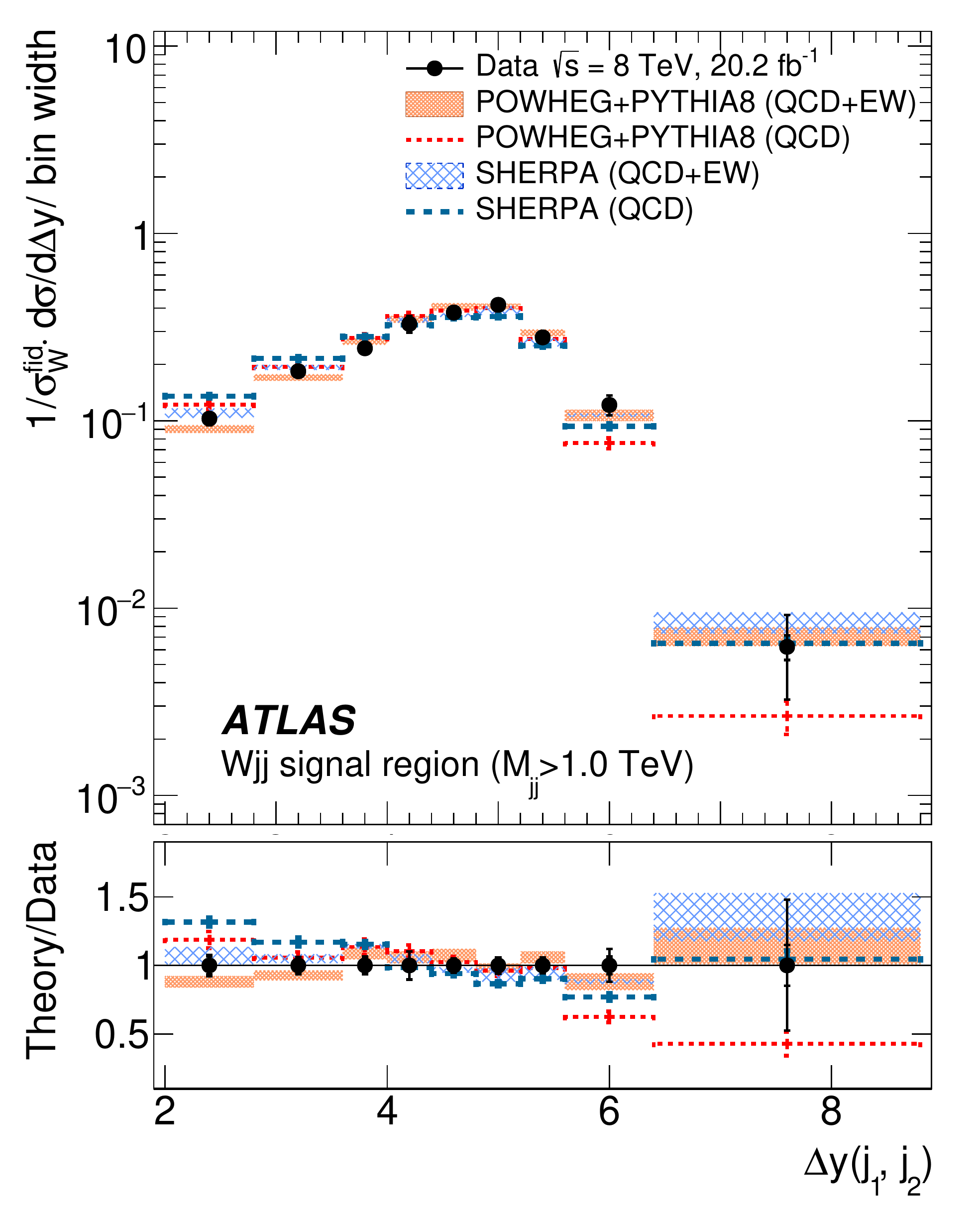}
\caption{Top: Unfolded absolute (left) and normalized (right) differential \wjets production cross 
sections as a function of dijet mass for the signal fiducial region.  Bottom: Unfolded normalized production 
cross sections as a function of $\dyjj$ for the signal regions with $\mjj > 0.5$~\TeV~(left) and 
$\mjj > 1.0$~\TeV~(right).  Both statistical (inner bar) and total (outer bar) measurement uncertainties are 
shown, as well as ratios of the theoretical predictions to the data (the bottom panel in each distribution).}
\label{unfolding:combined_measurementdijetmass1Dsignal}
\end{figure}

The best discrimination between QCD and EW \wjets production is provided by the dijet mass distribution, 
as demonstrated in the top plots of Figure~\ref{unfolding:combined_measurementdijetmass1Dsignal}.  
The distribution of dijet rapidity separation is correlated with this distribution but is purely 
topological.  The discrimination provided by $\dyjj$ is shown in the bottom plots of the figure for 
$\mjj >0.5$~and~1~\TeV.  

\begin{figure}[!tbp]
\centering
\includegraphics[width=0.49\textwidth]{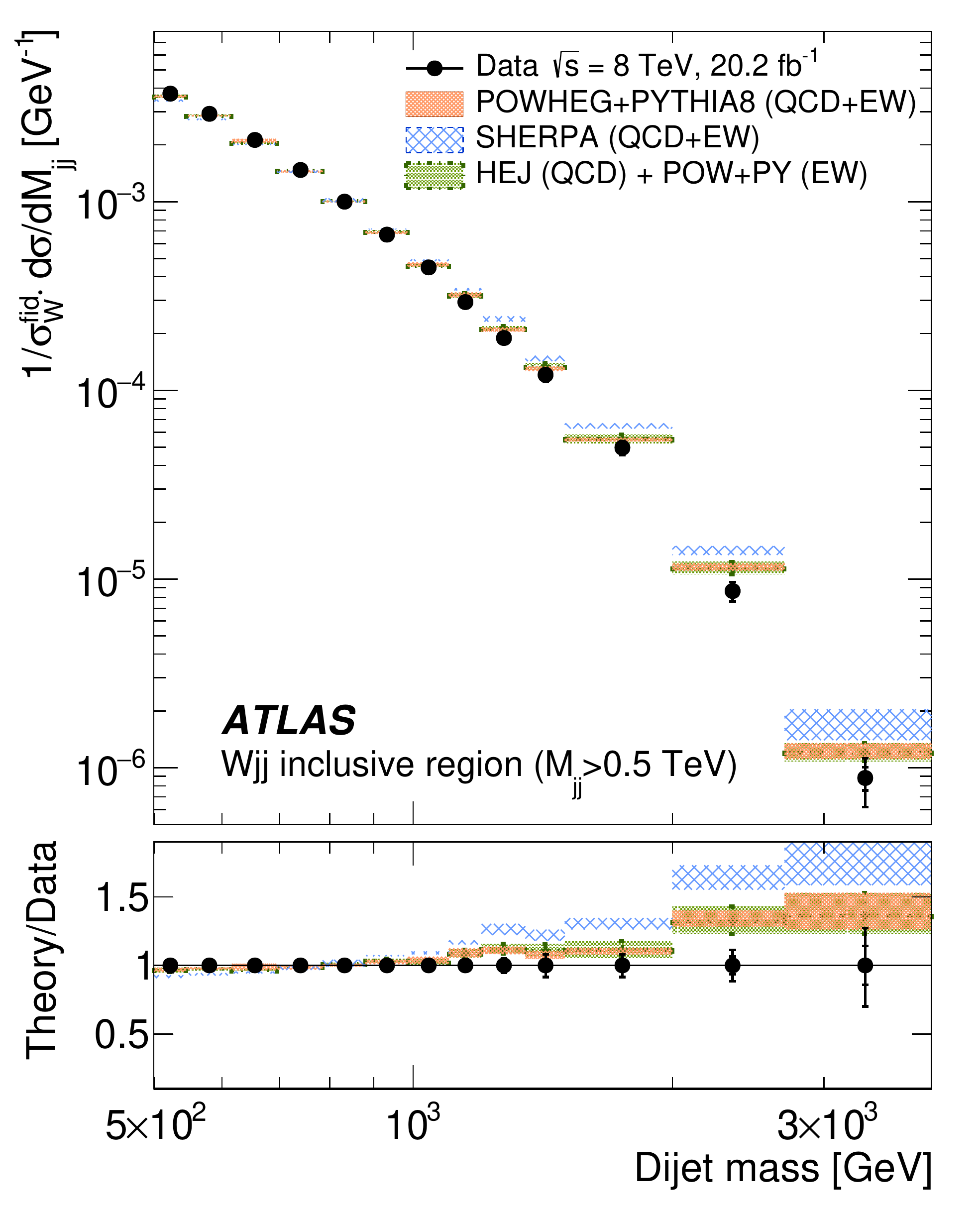}
\includegraphics[width=0.49\textwidth]{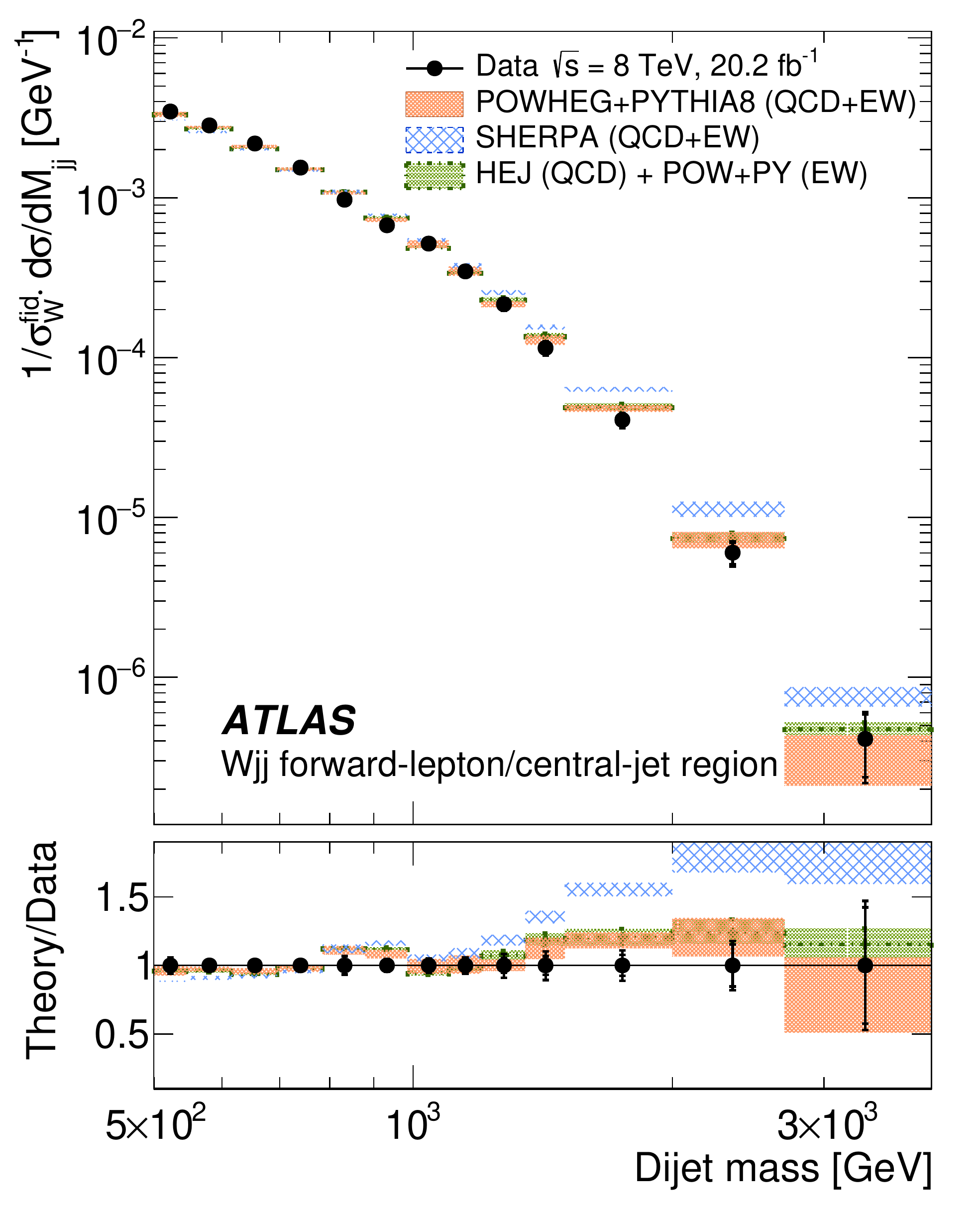}
\includegraphics[width=0.49\textwidth]{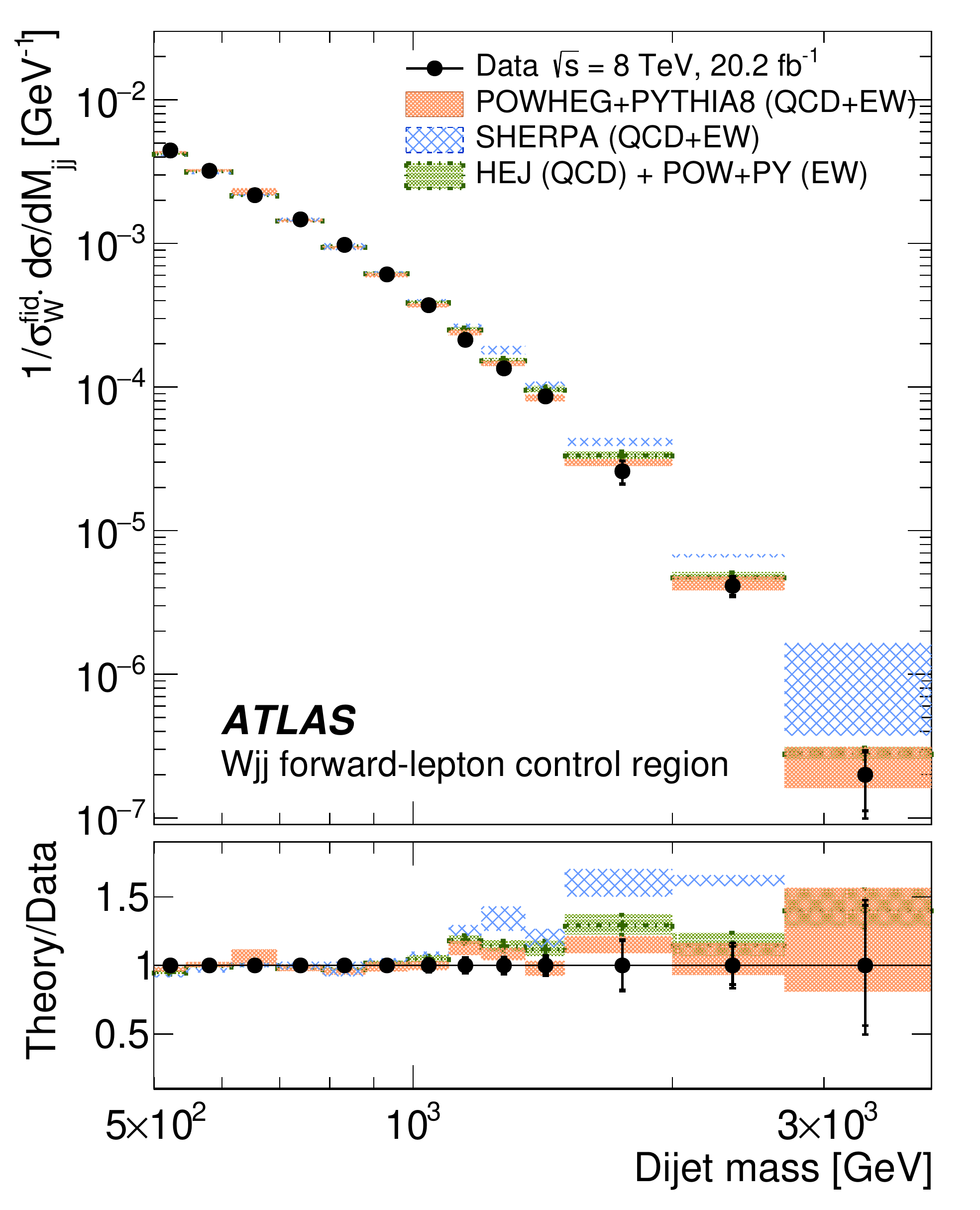}
\includegraphics[width=0.49\textwidth]{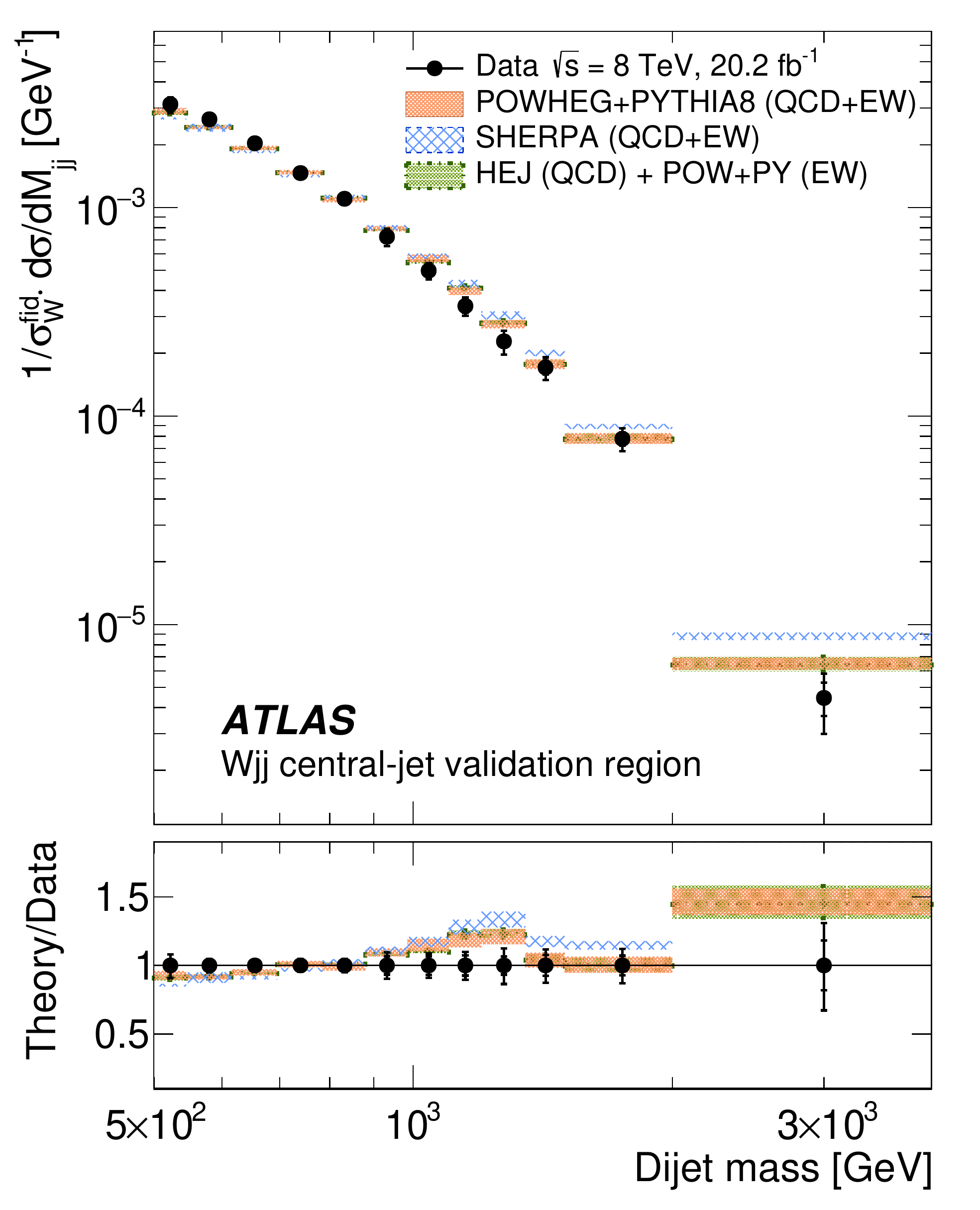}
\caption{Unfolded normalized differential \wjets production cross sections as a function of dijet invariant mass in the
inclusive, forward-lepton/central-jet, forward-lepton, and central-jet fiducial regions.  Both statistical
(inner bar) and total (outer bar) measurement uncertainties are shown, as well as ratios of the theoretical
predictions to the data (the bottom panel in each distribution). }
\label{unfolding:dijetmass1Dinclusive}
\end{figure}

\begin{figure}[htbp]
\centering
\includegraphics[width=0.49\textwidth]{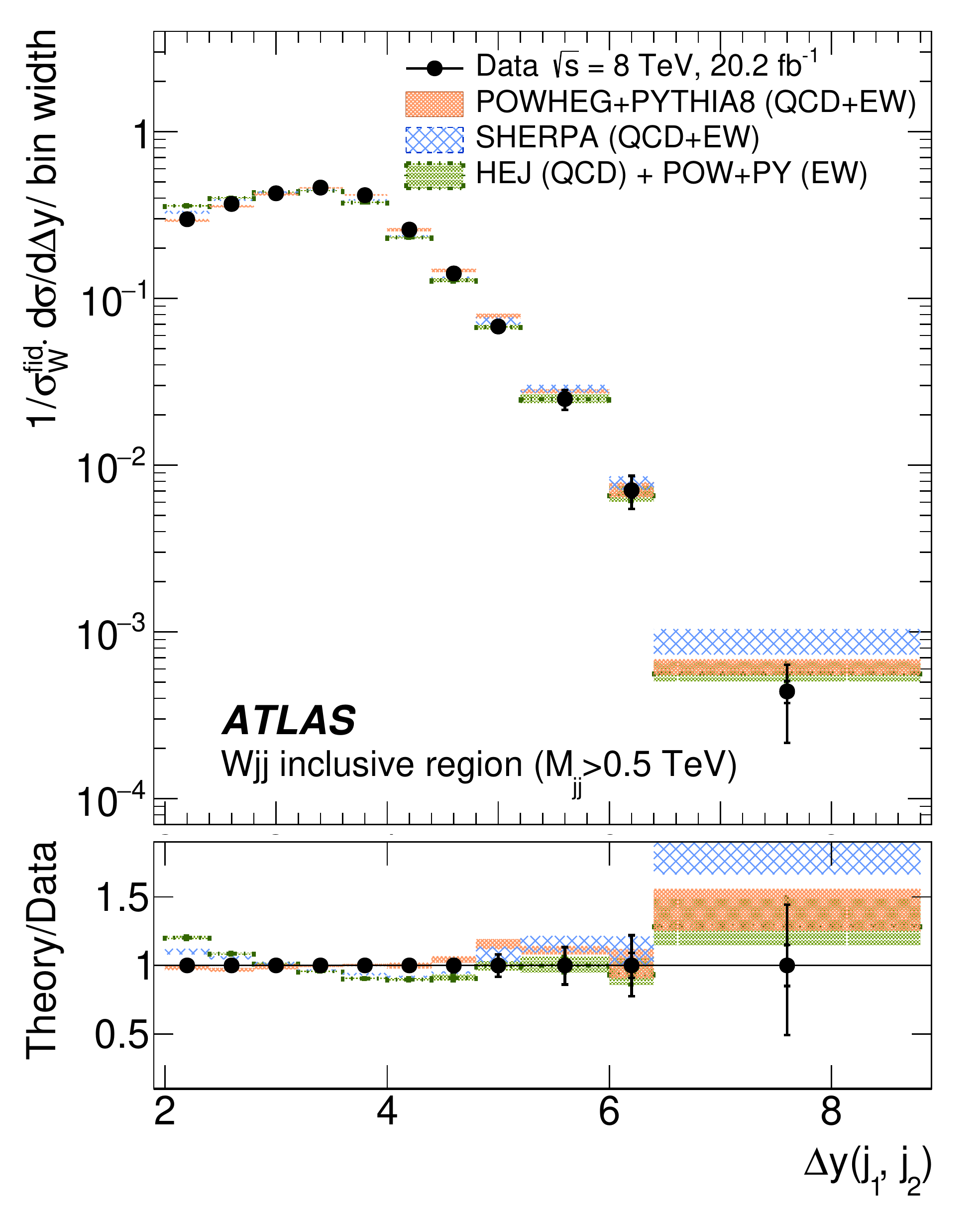}
\includegraphics[width=0.49\textwidth]{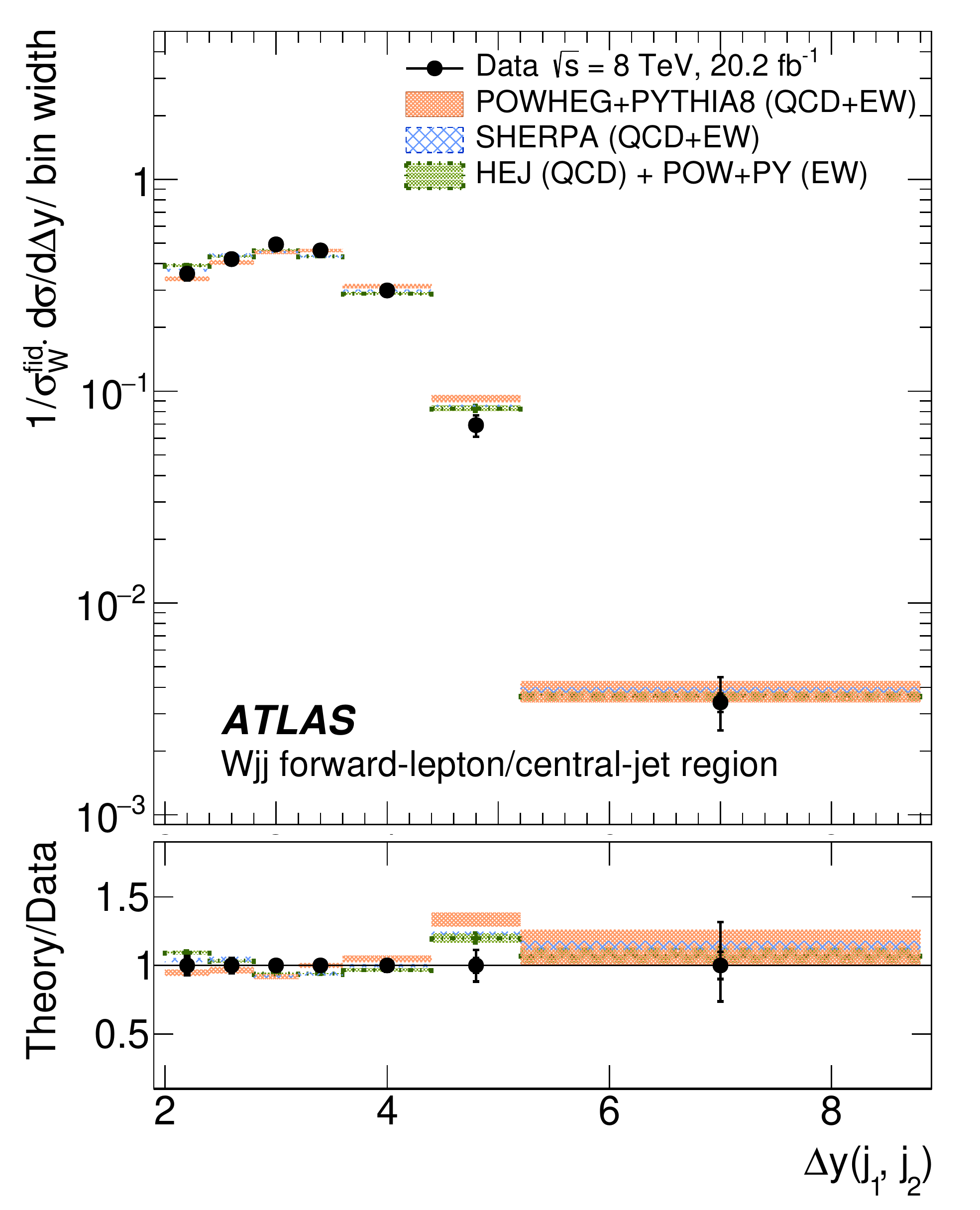}
\includegraphics[width=0.49\textwidth]{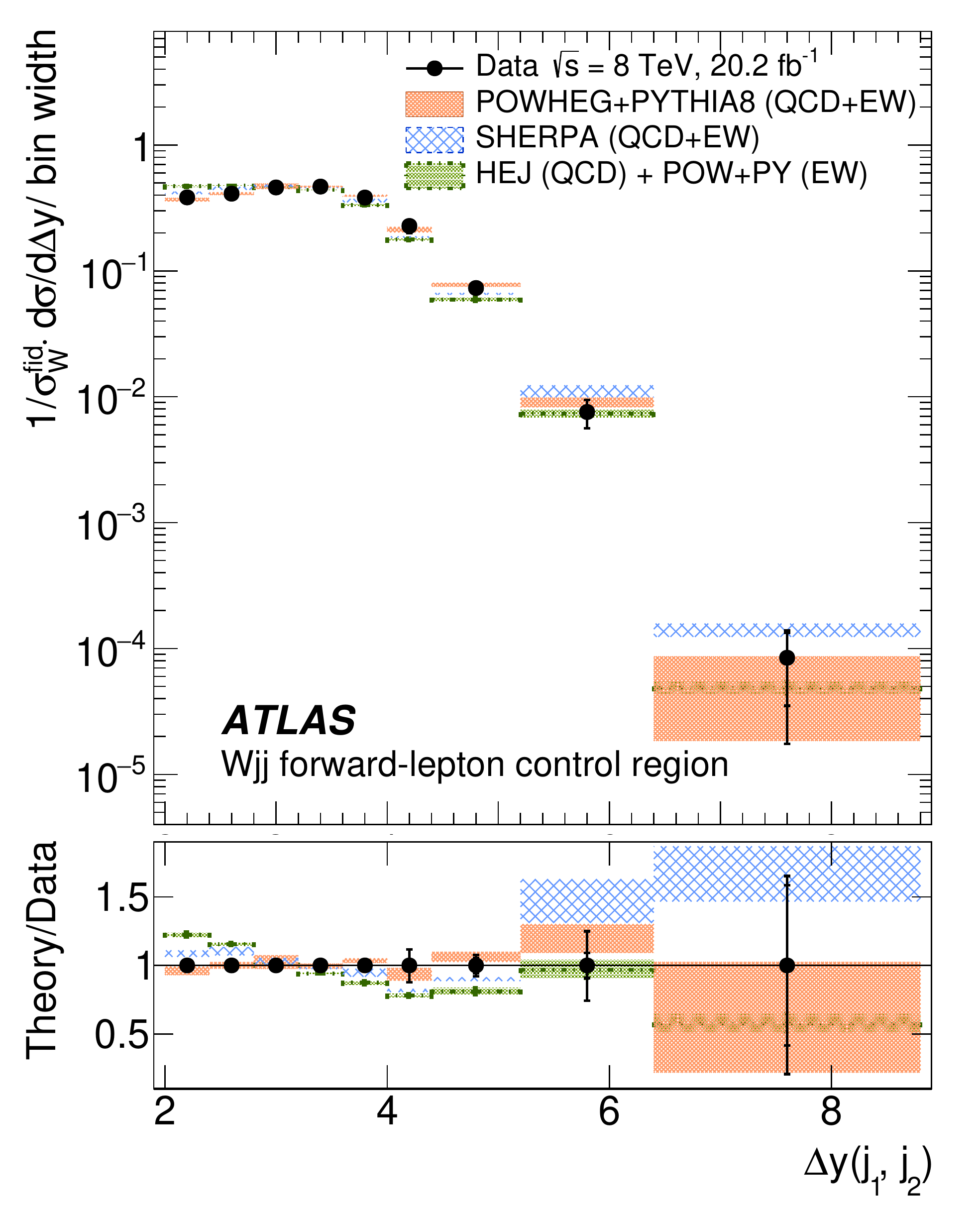}
\includegraphics[width=0.49\textwidth]{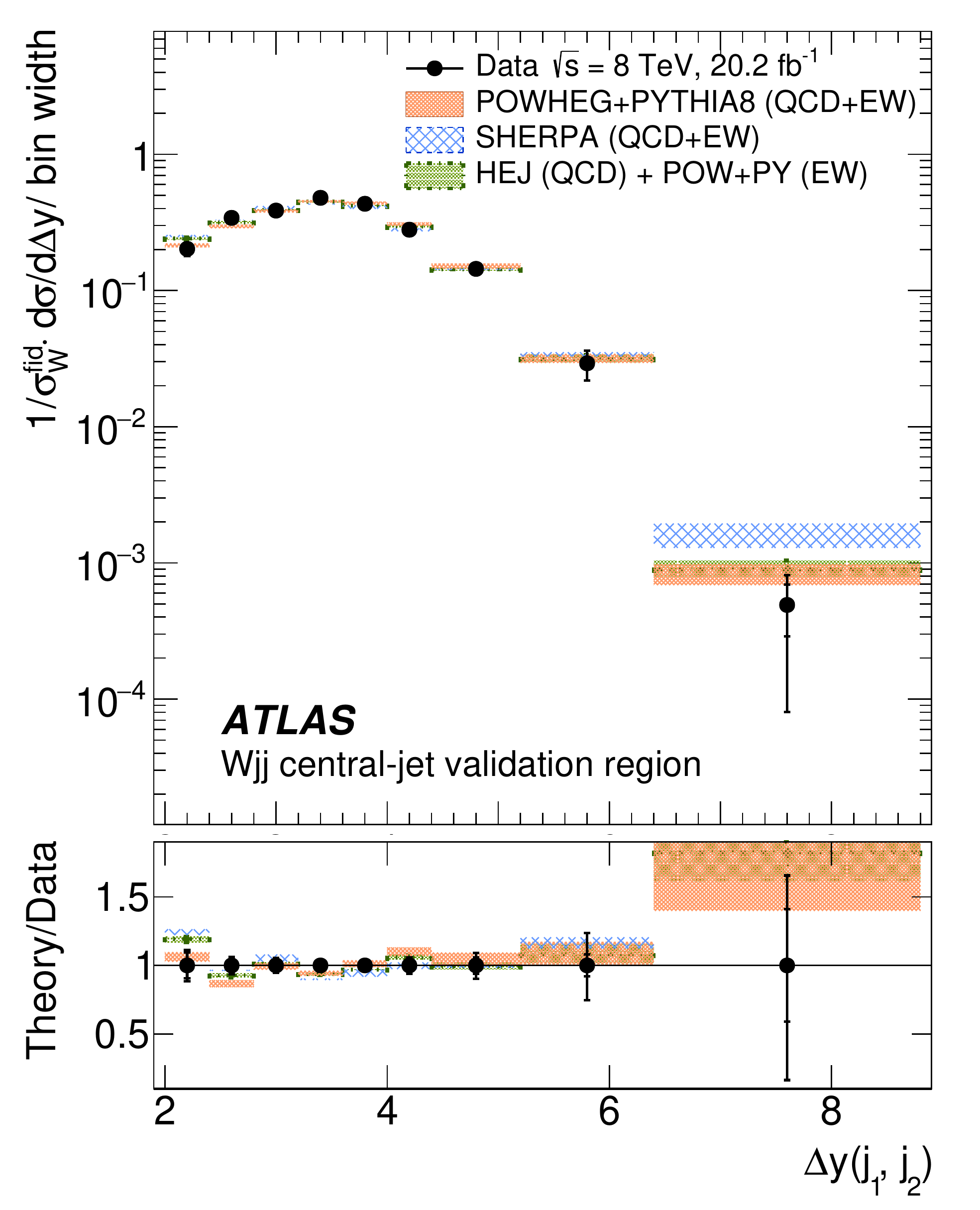}
\caption{Unfolded normalized differential \wjets production cross sections as a function of $\dyjj$ in the
inclusive, forward-lepton/central-jet, forward-lepton, and central-jet fiducial regions.  Both statistical
(inner bar) and total (outer bar) measurement uncertainties are shown, as well as ratios of the theoretical 
predictions to the data (the bottom panel in each distribution). }
\label{unfolding:dijetdyjj1Dinclusive}
\end{figure}

The QCD~\wjets modelling of the dijet distributions is important for extracting the cross section 
for EW~\wjets production.  The modelling of the \mjj distribution in regions dominated by 
QCD \wjets production is shown in Figure~\ref{unfolding:dijetmass1Dinclusive}.  Predictions 
from \textsc{hej}, which are expected to provide a good description at high dijet invariant mass 
where large logarithms contribute, are similar to the NLO predictions from \powheg + \pythia.  
\sherpa predicts more events at high dijet invariant mass than observed in data in these fiducial 
regions, whereas \powheg + \pythia and \textsc{hej} are in better agreement with data.  The dijet 
rapidity separation (Figure~\ref{unfolding:dijetdyjj1Dinclusive}) shows similar behavior, with 
\sherpa overestimating the rate at large separation.  The \textsc{hej} distributions have larger 
deviations from the data due to the reduced accuracy of resummation at small $\dyjj$. 

\begin{figure}[tbp]
\centering
\includegraphics[width=0.48\textwidth]{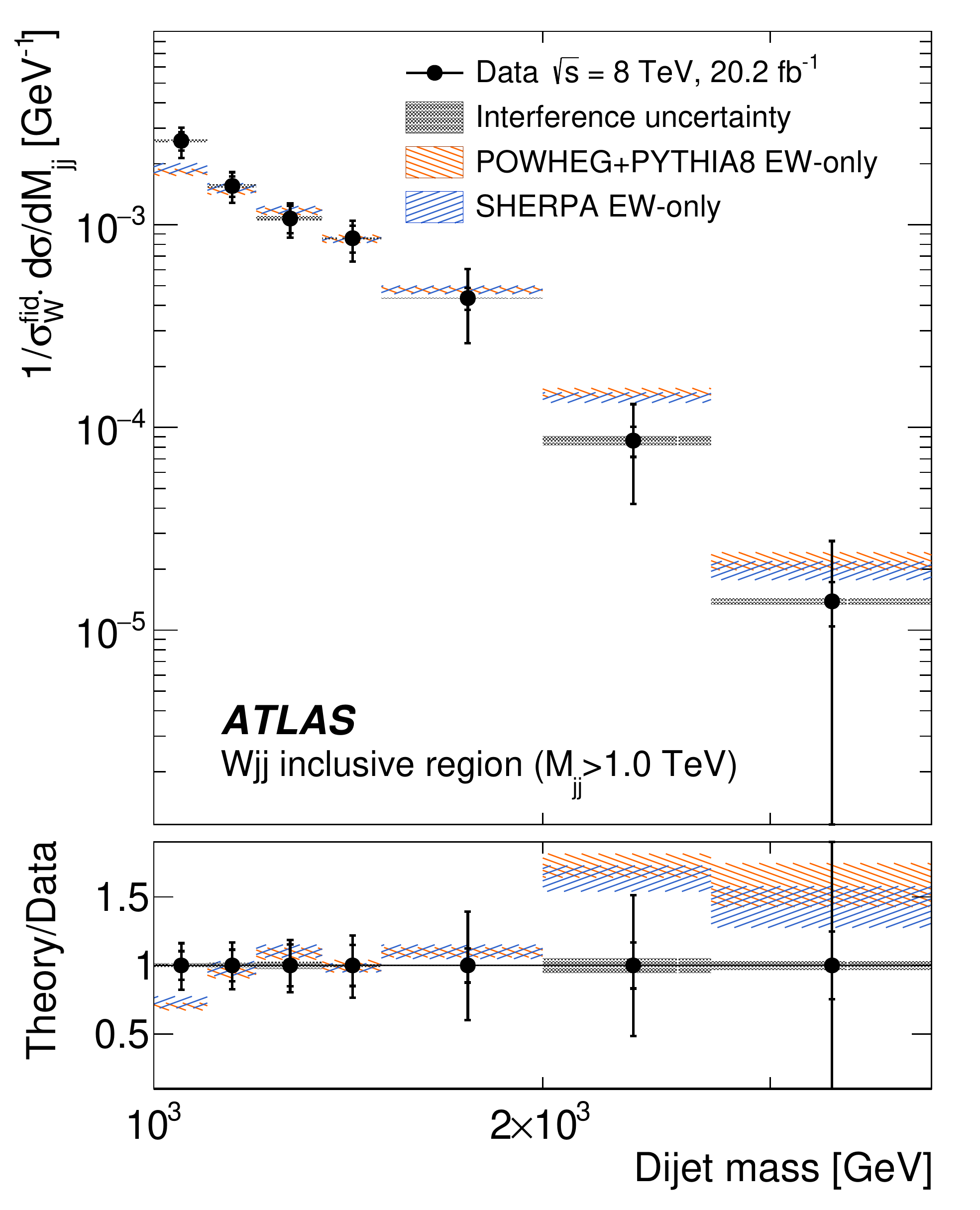}
\includegraphics[width=0.48\textwidth]{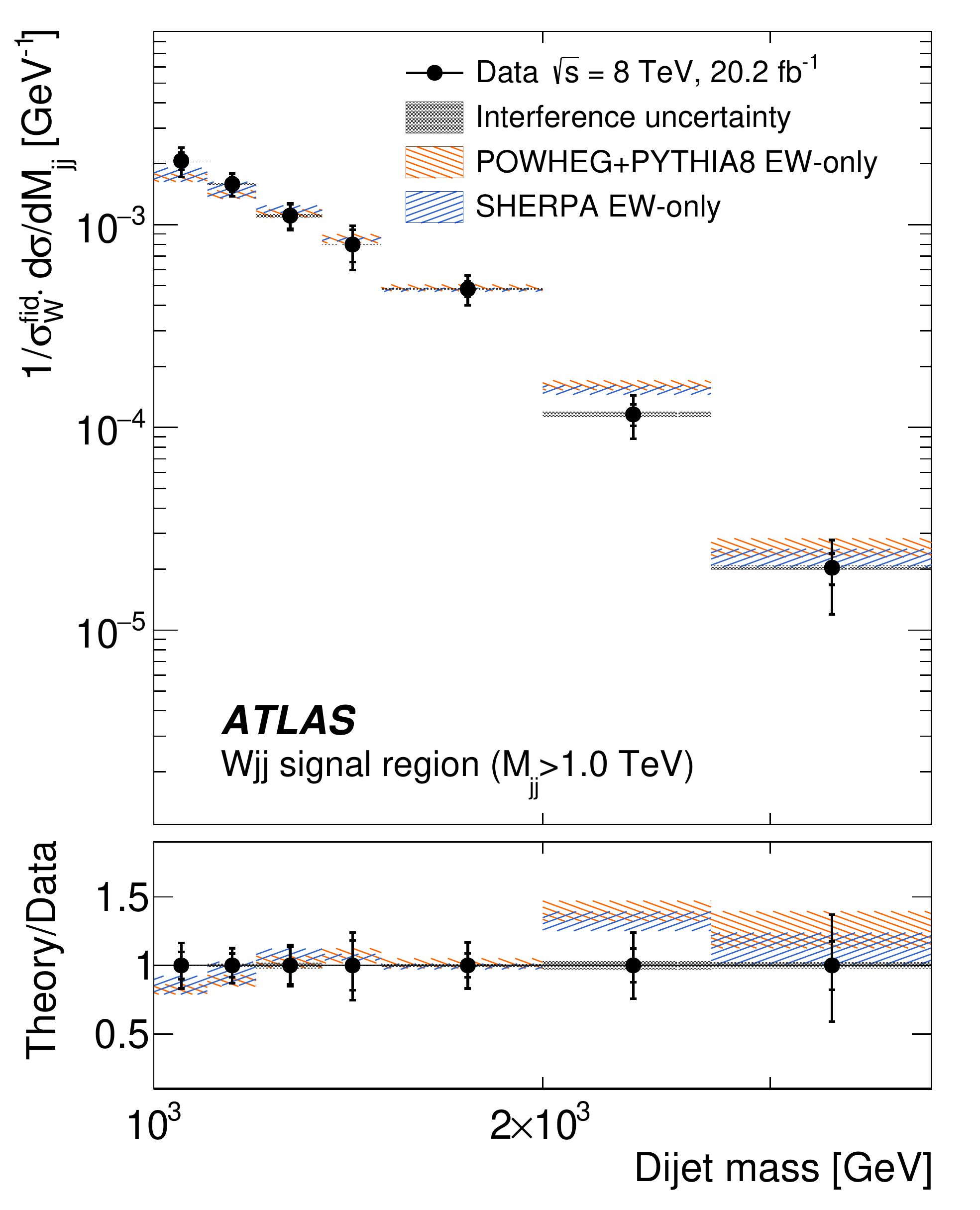}
\includegraphics[width=0.48\textwidth]{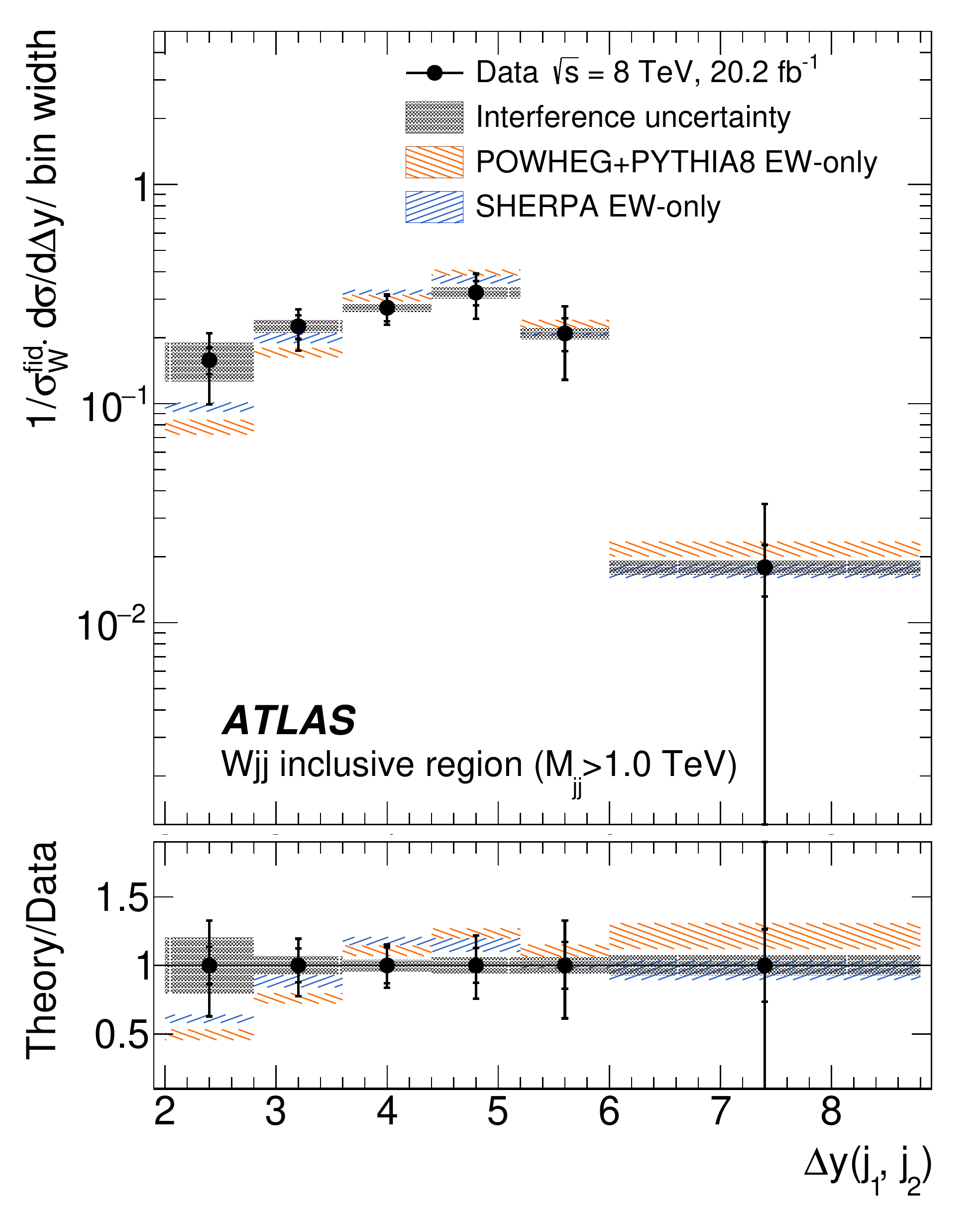}
\includegraphics[width=0.48\textwidth]{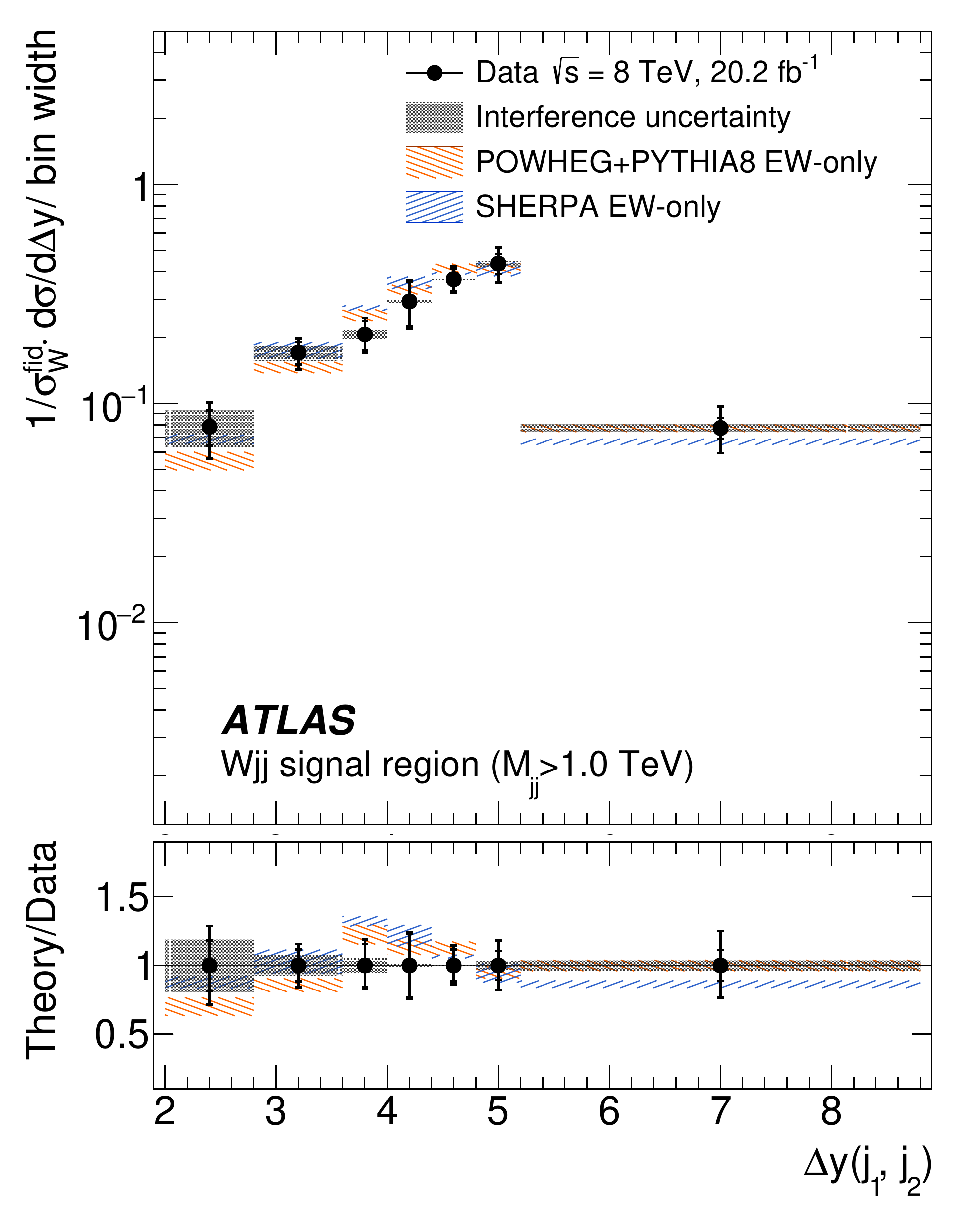}
\caption{Unfolded normalized differential EW~\wjets production cross sections as a function of the dijet
invariant mass (top) and $\dyjj$ (bottom) for the inclusive (left) and signal (right) fiducial regions with 
$\mjj > 1.0$~\TeV.  Both statistical (inner bar) and total (outer bar) measurement uncertainties are shown, 
as well as ratios of the theoretical predictions to the data (the bottom panel in each distribution). }
\label{unfolding:EWKcombined_measurementdijetmass1Dhighmass10}
\end{figure}

The dijet distributions are generally well modelled for the EW \wjets process, as shown in  
Figure~\ref{unfolding:EWKcombined_measurementdijetmass1Dhighmass10} for the inclusive and 
signal regions with $\mjj>1.0$~\TeV.  The reduced purity in the inclusive region causes larger 
measurement uncertainties, and the measurements have larger absolute discrepancies with 
respect to predictions.  The interference uncertainty is largest at low $\dyjj$, where the 
topology is less VBF-like.

%% file: difftopology.tex
The event topology distinguishes electroweak VBF production from other processes, in 
particular the lack of hadronic activity in the rapidity gap between the leading two 
jets and the tendency for the boson to be emitted within this gap.  These topological 
features are studied using the distributions of the jet multiplicity in the gap, the 
fraction of events with no jets with the gap, and the rapidity of the lepton and 
jets relative to the gap.  

\begin{figure}[!hptb]
\centering
\includegraphics[width=0.48\textwidth]{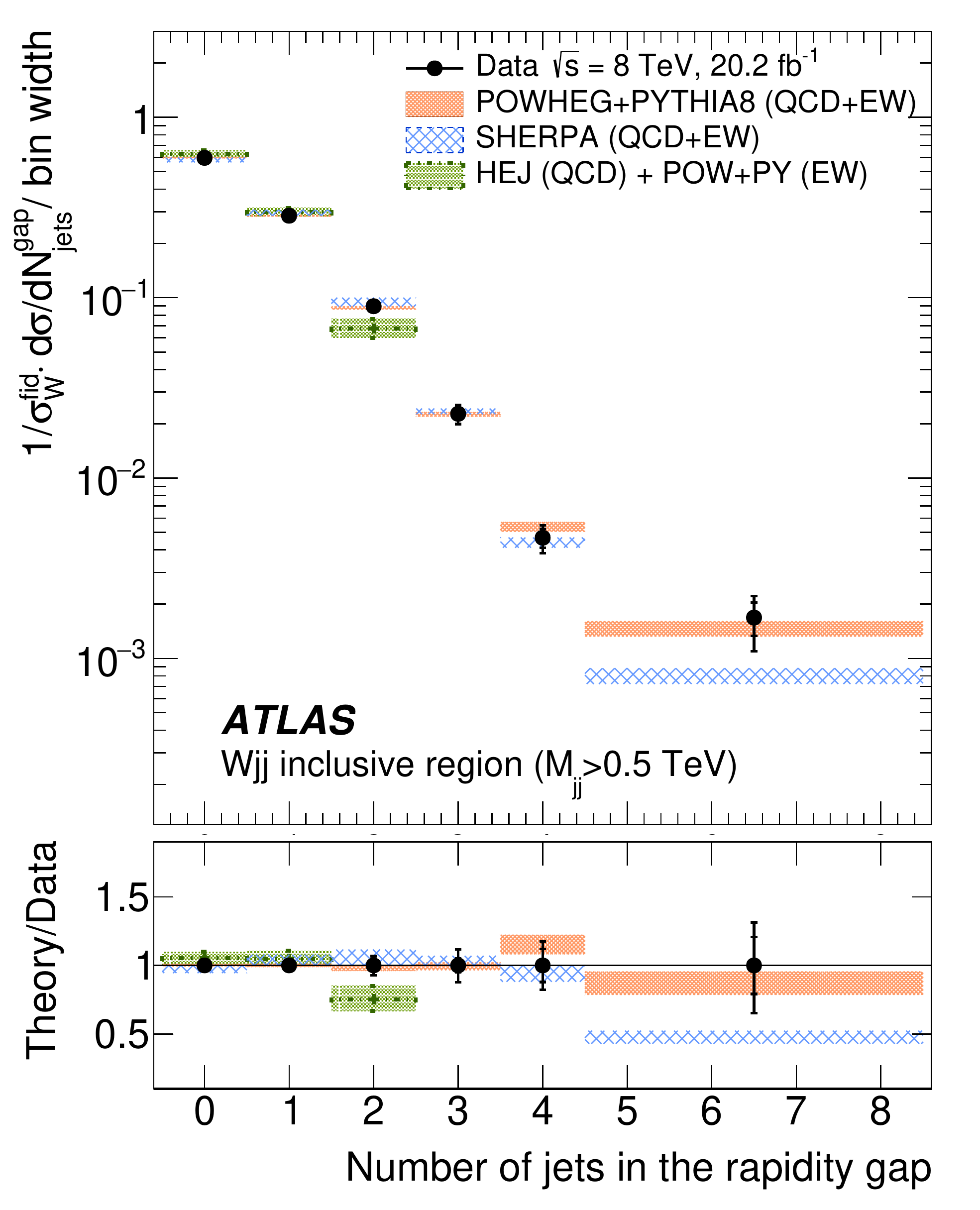}
\includegraphics[width=0.48\textwidth]{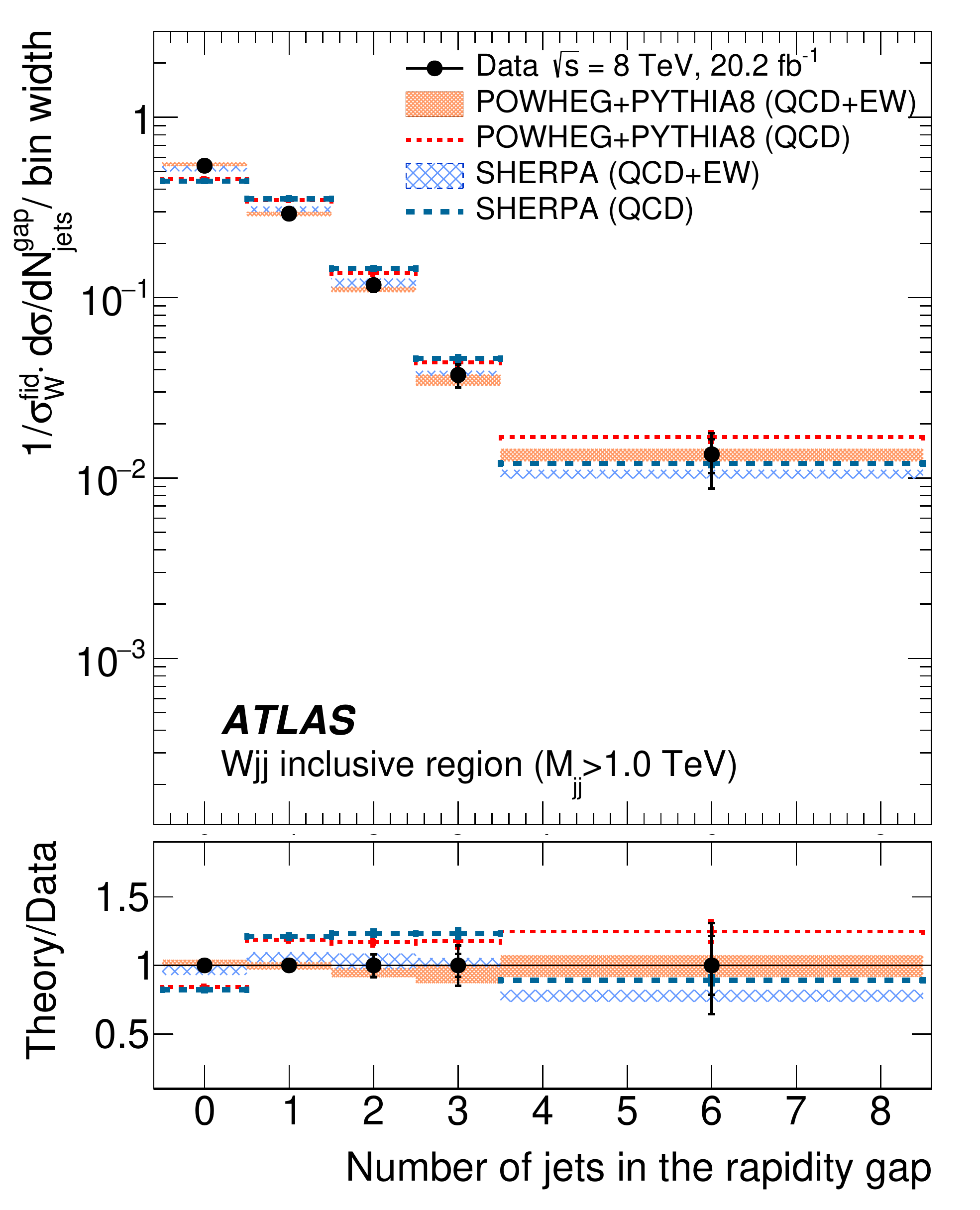}
\includegraphics[width=0.48\textwidth]{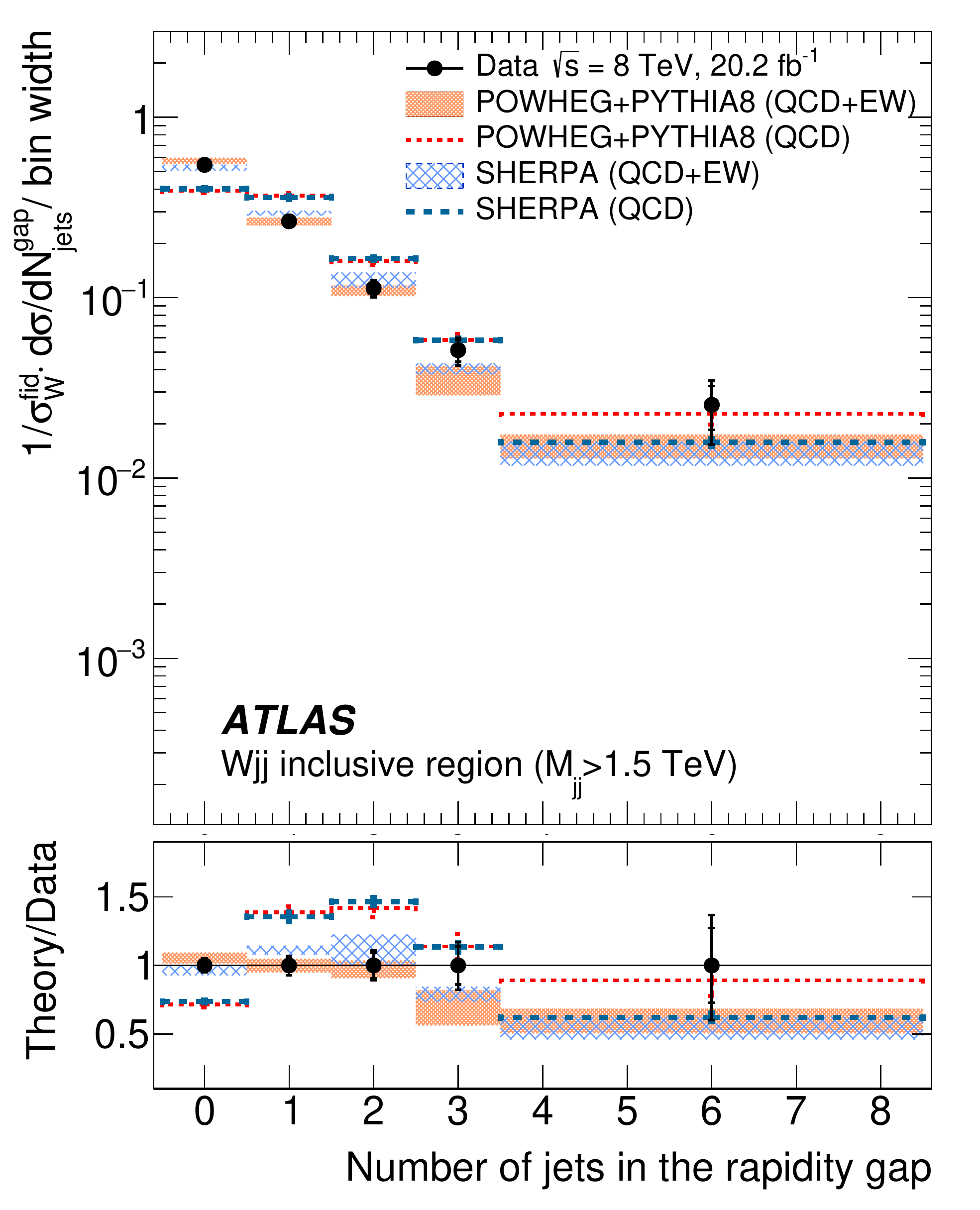}
\includegraphics[width=0.48\textwidth]{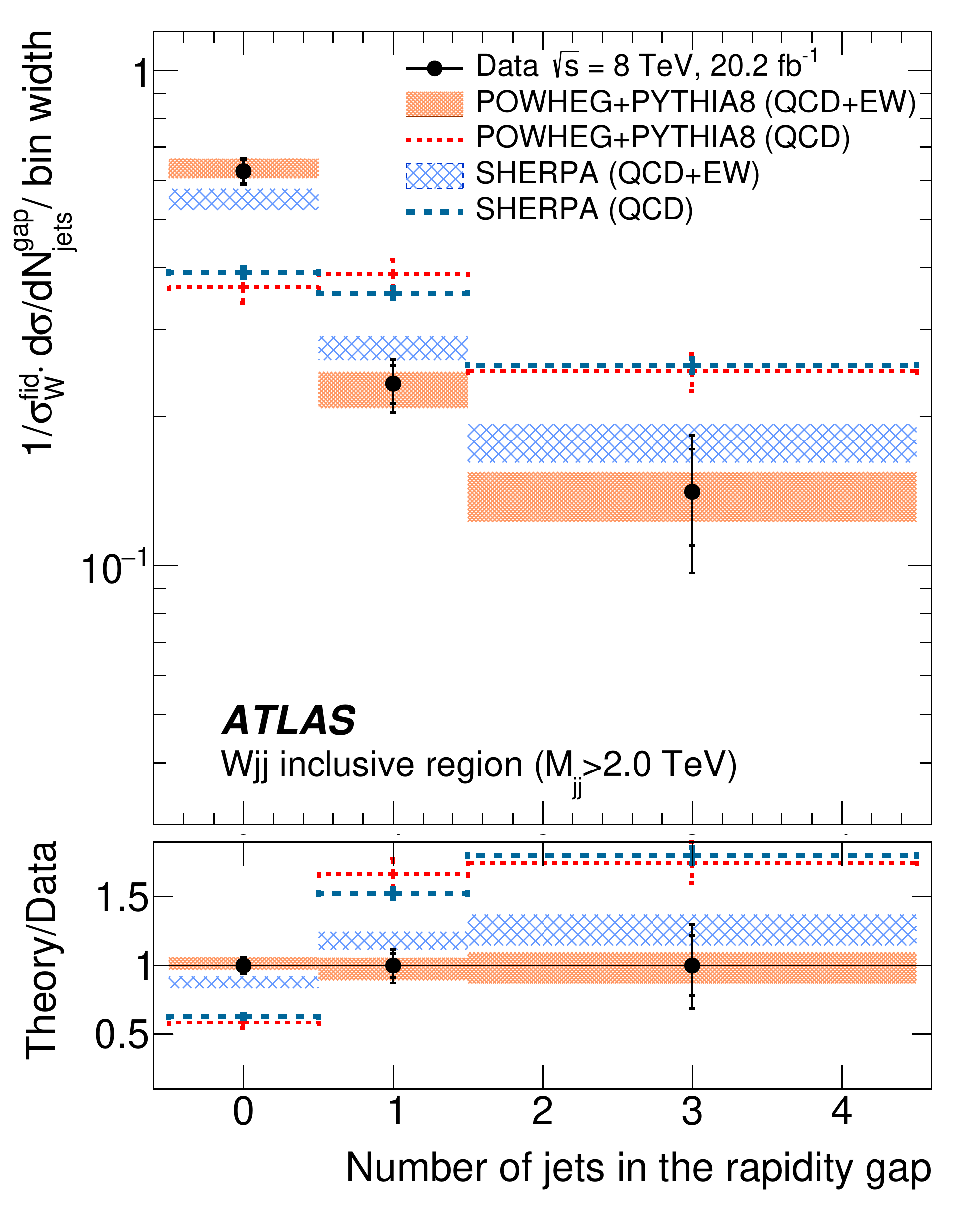}
\caption{Unfolded normalized distribution of the number of jets with $\pt>30$~\GeV~in the rapidity 
interval bounded by the two highest-$\pt$ jets in the inclusive fiducial region with $\mjj$ 
thresholds of 0.5~\TeV~(top left),~1.0~\TeV~(top right),~1.5~\TeV~(bottom left), and 2.0~\TeV~(bottom 
right).  Both statistical (inner bar) and total (outer bar) measurement uncertainties are shown, as 
well as ratios of the theoretical predictions to the data (the bottom panel in each distribution).  }
\label{unfolding:norm_measurementngapjets}
\end{figure}

\begin{figure}[!hptb]
\centering
\includegraphics[width=0.47\textwidth]{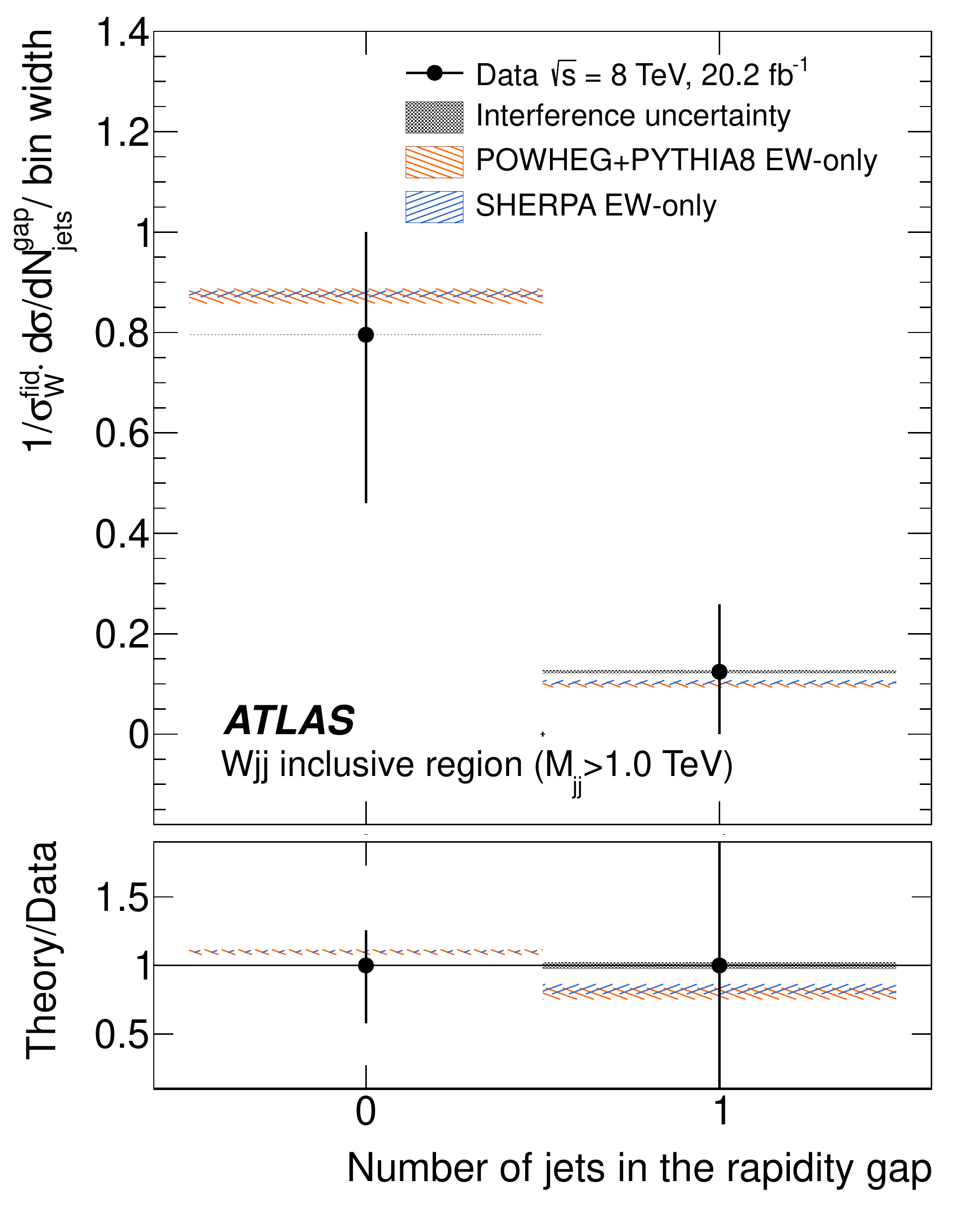}
\includegraphics[width=0.47\textwidth]{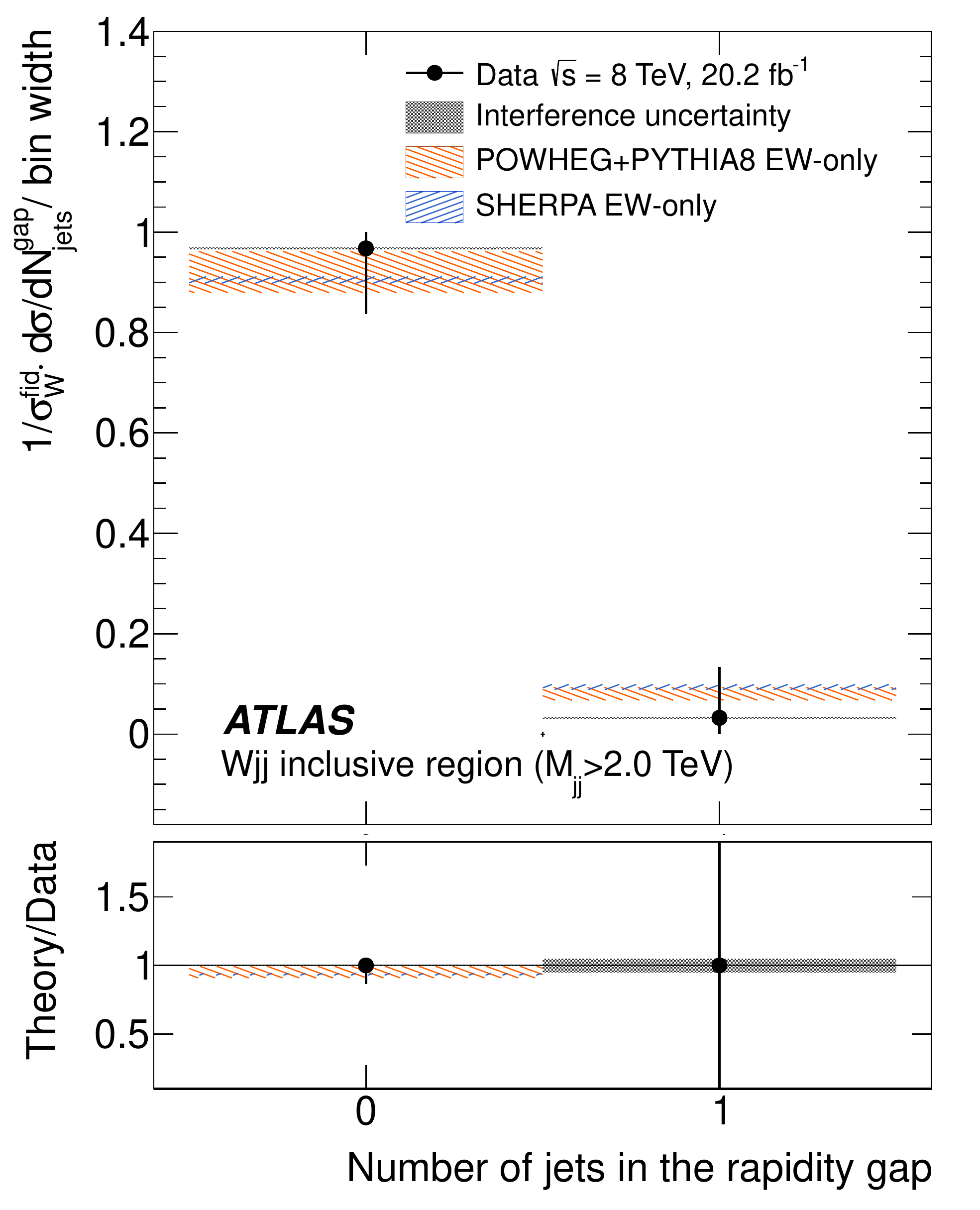}
\caption{Unfolded normalized differential EW~\wjets production cross sections as a function of the number of
jets with $\pt>30$~\GeV~in the rapidity interval bounded by the two highest-$\pt$ jets in the inclusive fiducial
region, with $\mjj> 1.0$~\TeV~(left) and $\mjj > 2.0$~\TeV~(right).  Both statistical (inner bar) and total (outer bar)
measurement uncertainties are shown, as well as ratios of the theoretical predictions to the data (the bottom panel in
each distribution). }
\label{unfolding:EWKcombined_measurementngapjets1Dhighmass10_15}
\end{figure}

Figure~\ref{unfolding:norm_measurementngapjets}~shows the normalized differential cross 
section as a function of the number of $\pt>30$~\GeV~jets emitted into the rapidity gap 
for progressively increasing \mjj thresholds.  In the lowest invariant-mass fiducial 
region, strong \wjets production dominates and predictions from \powheg + \pythia, \sherpa, 
and \textsc{hej} all describe the data well.  As the dijet invariant mass threshold is 
increased, the differences in shape between predictions with and without the EW~\wjets 
contribution become apparent.  The corresponding differential measurements for EW~\wjets 
production are shown in Figure~\ref{unfolding:EWKcombined_measurementngapjets1Dhighmass10_15} 
for the inclusive regions with $\mjj > 1.0$~\TeV~and $2.0$~\TeV.  The measured fraction of 
EW \wjets events with no additional central jets is higher than that of QCD+EW \wjets events, 
as also demonstrated in Table~\ref{tab:jetvetoeff}.  The table shows that the measured 
zero-jet fraction, frequently referred to as the jet-veto efficiency, is consistent with 
the \powheg + \pythia QCD+EW \wjets prediction for progressively increasing $\mjj$.  As 
$\mjj$ increases the relative contribution of the EW \wjets process increases substantially.

\begin{table*}[!hptb]
\caption{
Jet-veto efficiency for each $\mjj$ threshold compared to \powheg + \pythia QCD+EW and QCD \wjets 
simulations. The uncertainties comprise statistical and systematic components added in quadrature.}
\label{tab:jetvetoeff}
\begin{center}
\begin{tabular}{l|cccc}
\toprule
 & \multicolumn{4}{c}{Jet-veto efficiency} \\
& $\mjj>0.5$~TeV & $\mjj>1.0$~TeV &$\mjj>1.5$~TeV &$\mjj>2.0$~TeV \\
\midrule
Data                       & $0.596\pm 0.014$ & $0.54\pm 0.02$ & $0.55\pm 0.03$ & $0.63\pm 0.04$ \\
\powheg+\pythia (QCD+EW) & $0.597\pm 0.005$ & $0.55\pm 0.01$ & $0.57\pm 0.02$ & $0.63\pm 0.03$ \\
\powheg+\pythia (QCD) & $0.569\pm 0.002$ & $0.45\pm 0.01$ & $0.39\pm 0.01$ & $0.36\pm 0.03$ \\
\bottomrule
\end{tabular}
\end{center}
\end{table*}

\begin{figure}[!htpb]
\centering
\includegraphics[width=0.48\textwidth]{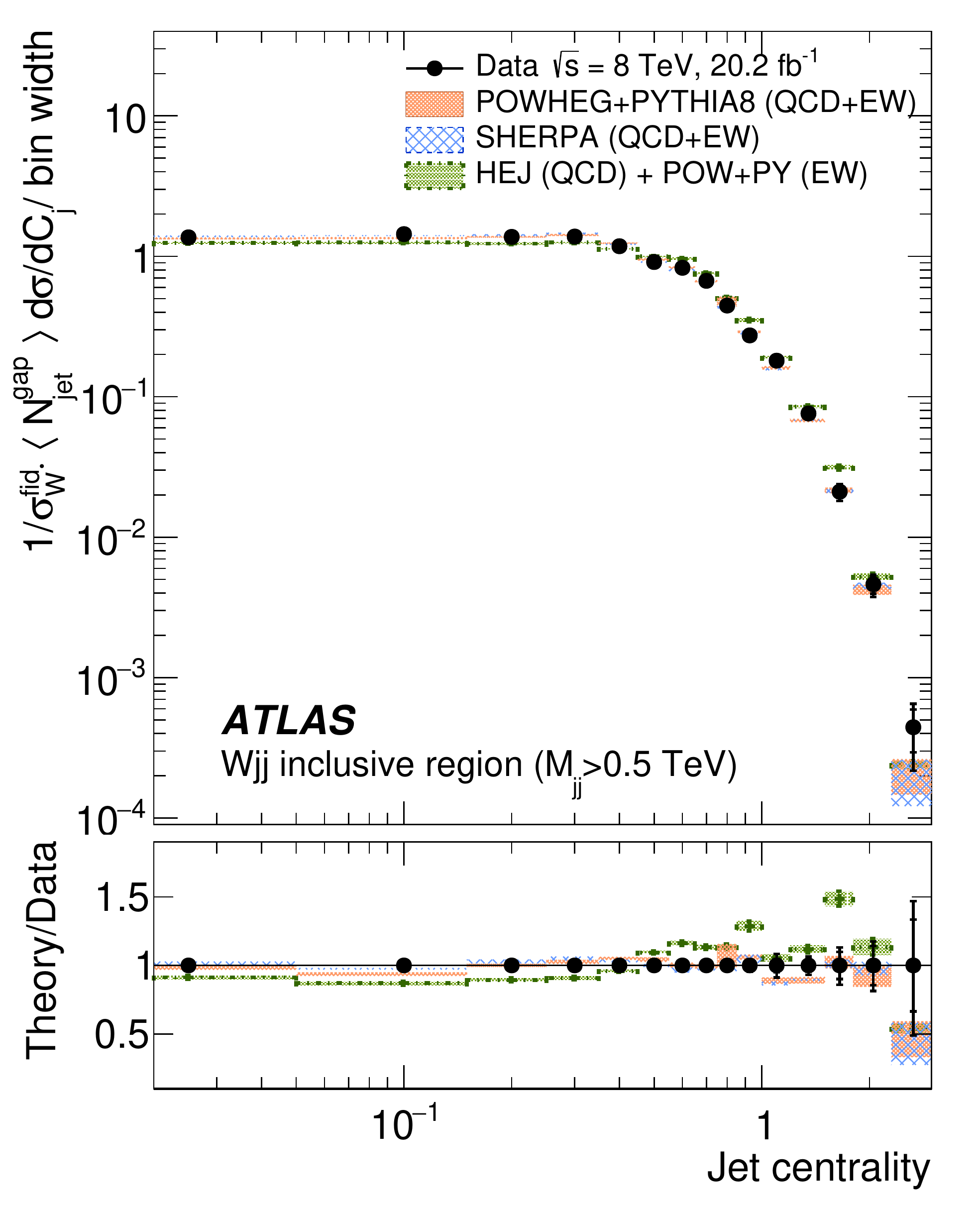}
\includegraphics[width=0.48\textwidth]{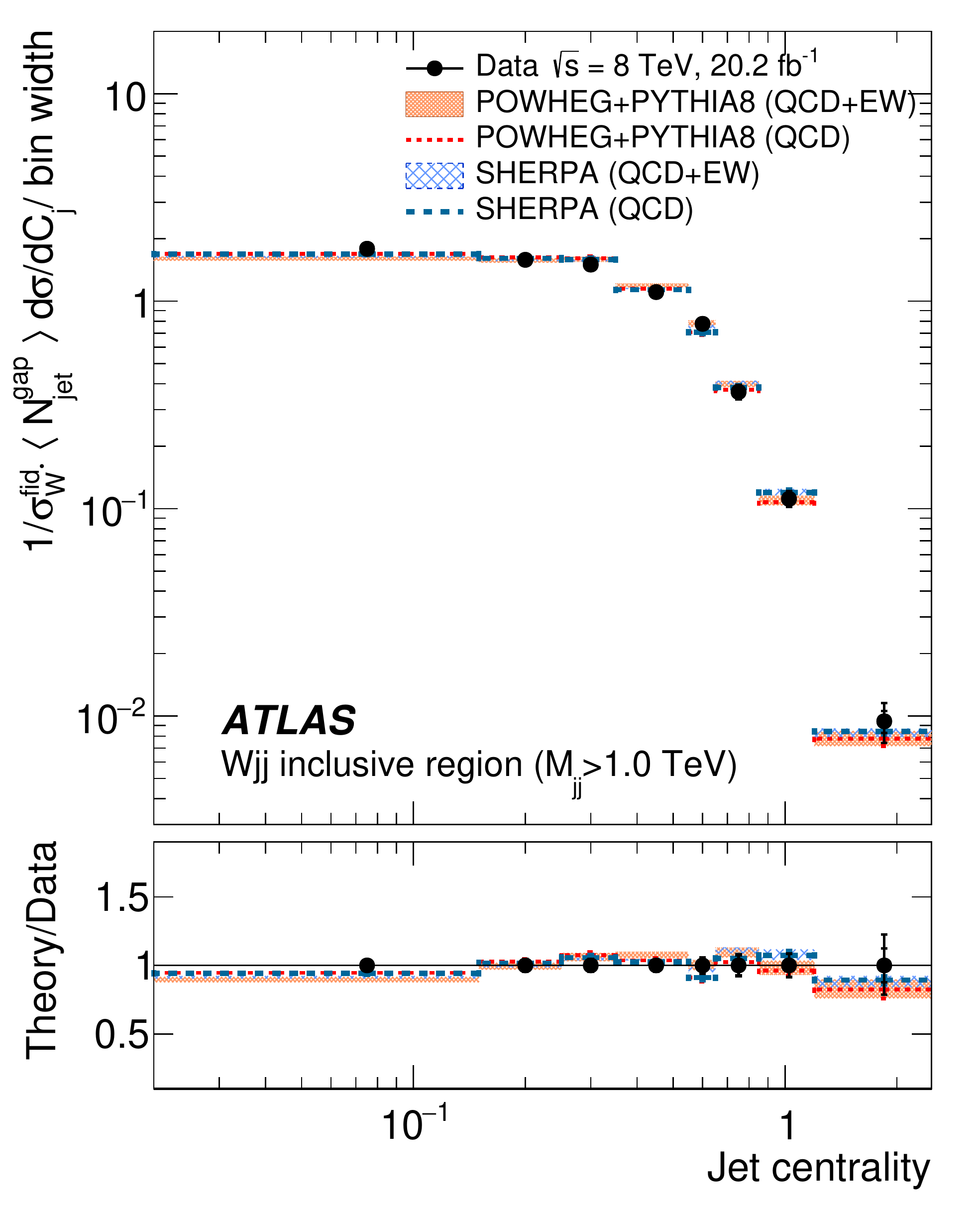}
\caption{Unfolded normalized differential QCD+EW \wjets production cross sections as a function of jet 
centrality for the inclusive fiducial region with $\mjj > 0.5$~\TeV~(left) and 1.0~\TeV~(right). 
Both statistical (inner bar) and total (outer bar) measurement uncertainties are shown, as well 
as ratios of the theoretical predictions to the data (the bottom panel in each distribution).}
\label{unfolding:combined_measurementJC1Dinclusive}
\end{figure}

Jet centrality is related to the number of jets in the rapidity gap, as events with $C_j < 0.5$ 
have a jet within the gap.  Figure~\ref{unfolding:combined_measurementJC1Dinclusive} shows good 
agreement between the predictions and data in the QCD+EW \wjets differential cross section weighted 
by the mean number of gap jets.  Since the rate for additional jet production is low in EW~\wjets 
production, there are too few events to perform a measurement of the jet centrality distribution 
for this process.

\begin{figure}[!hptb]
\centering
\includegraphics[width=0.48\textwidth]{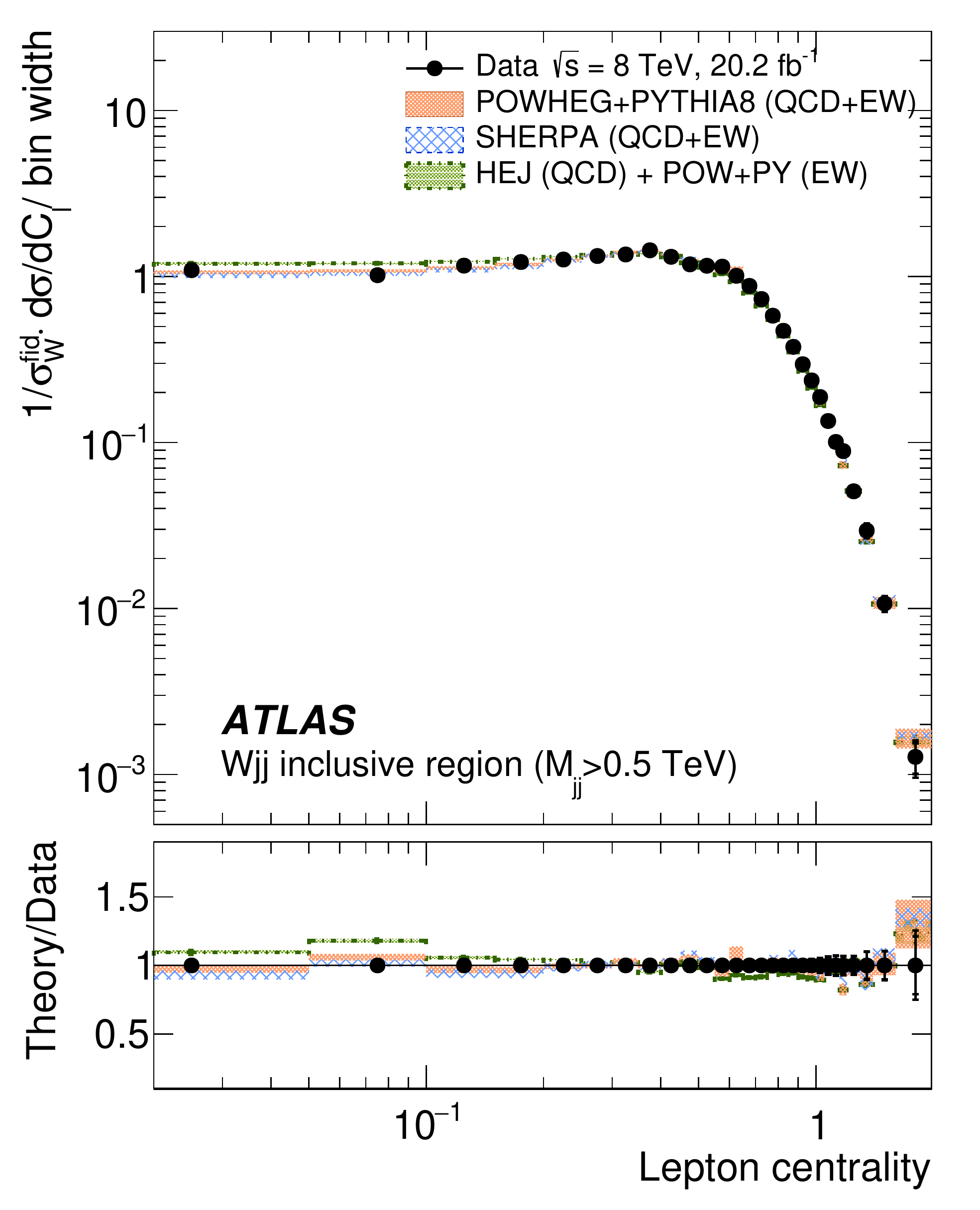}
\includegraphics[width=0.48\textwidth]{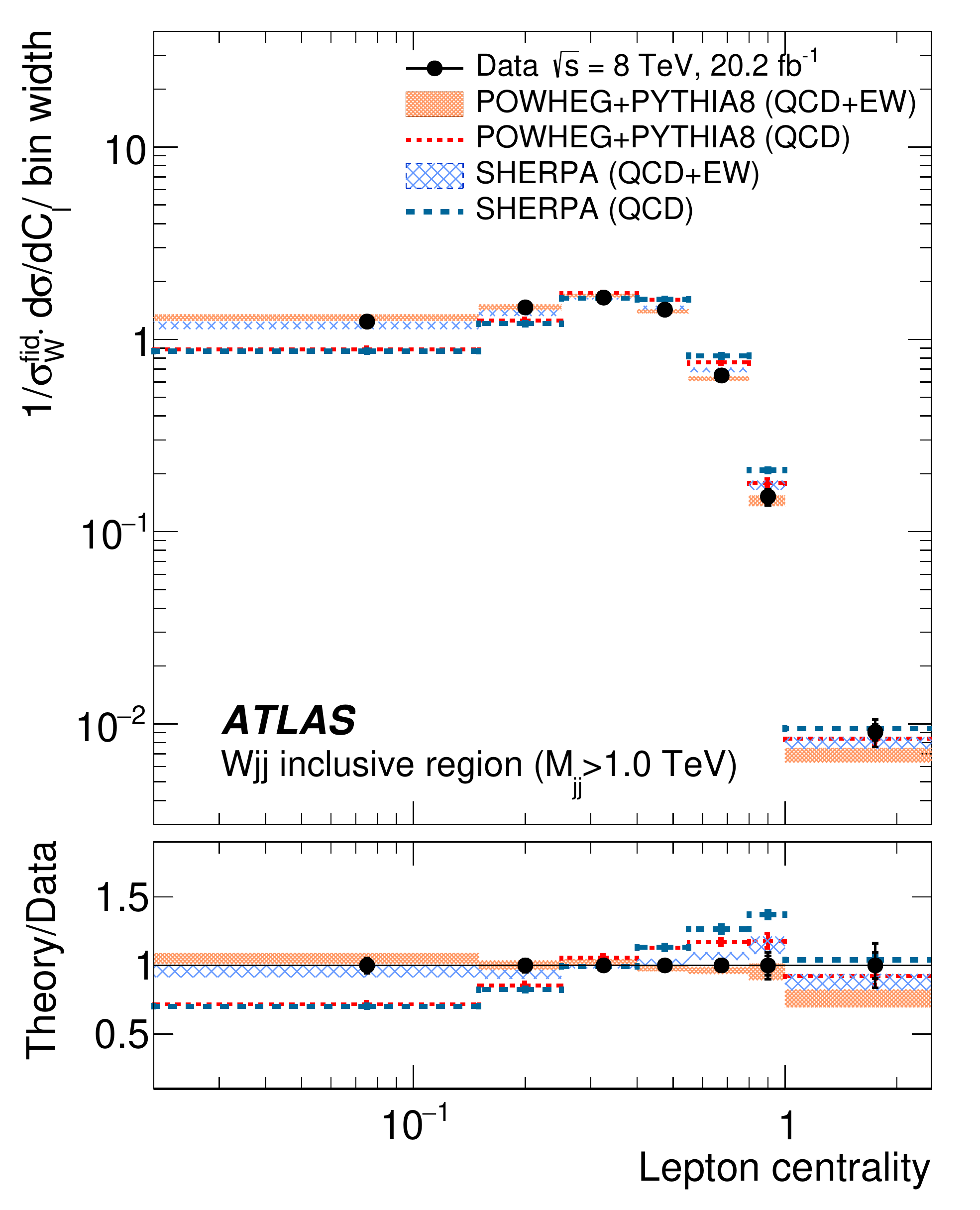}
\includegraphics[width=0.48\textwidth]{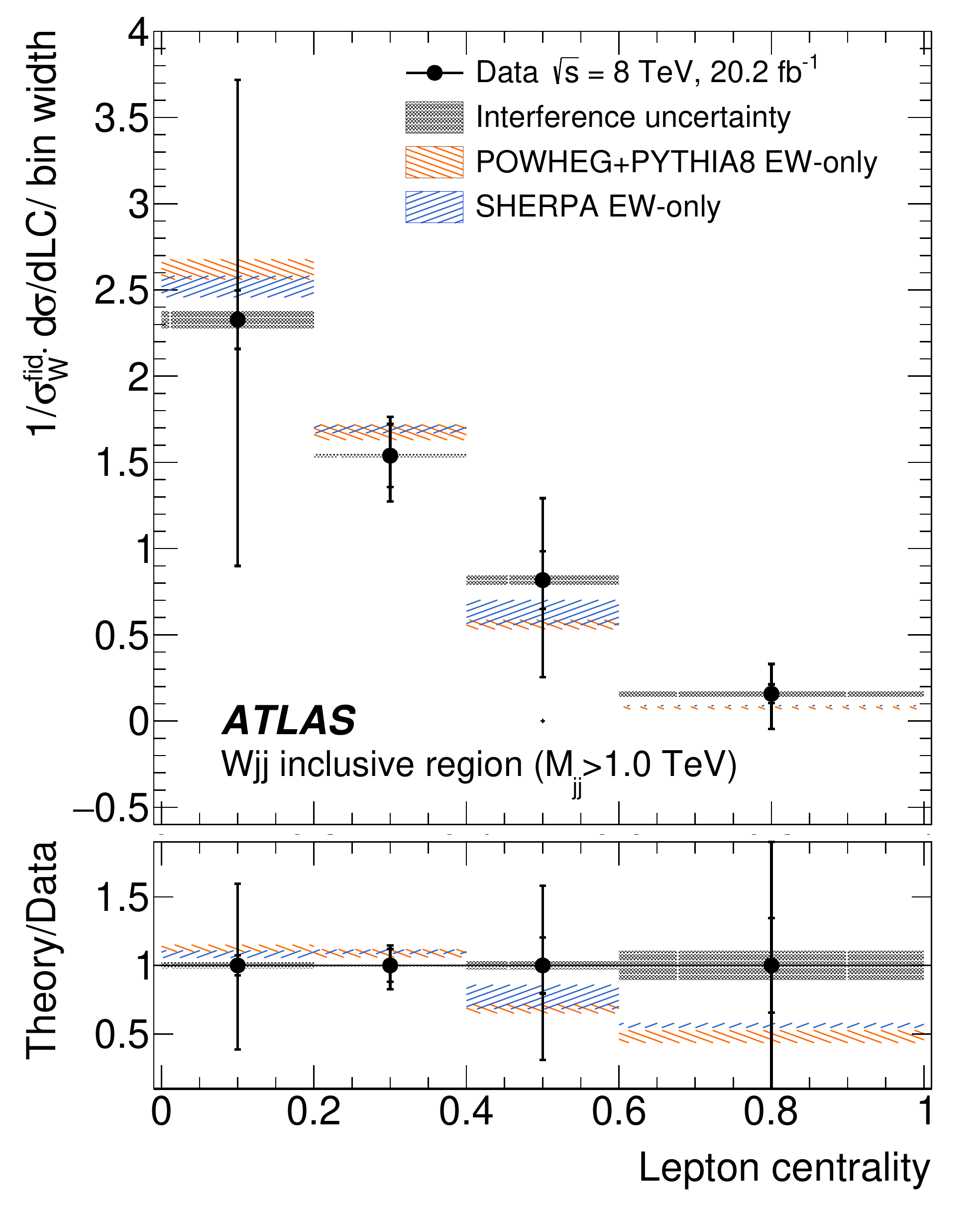}
\includegraphics[width=0.48\textwidth]{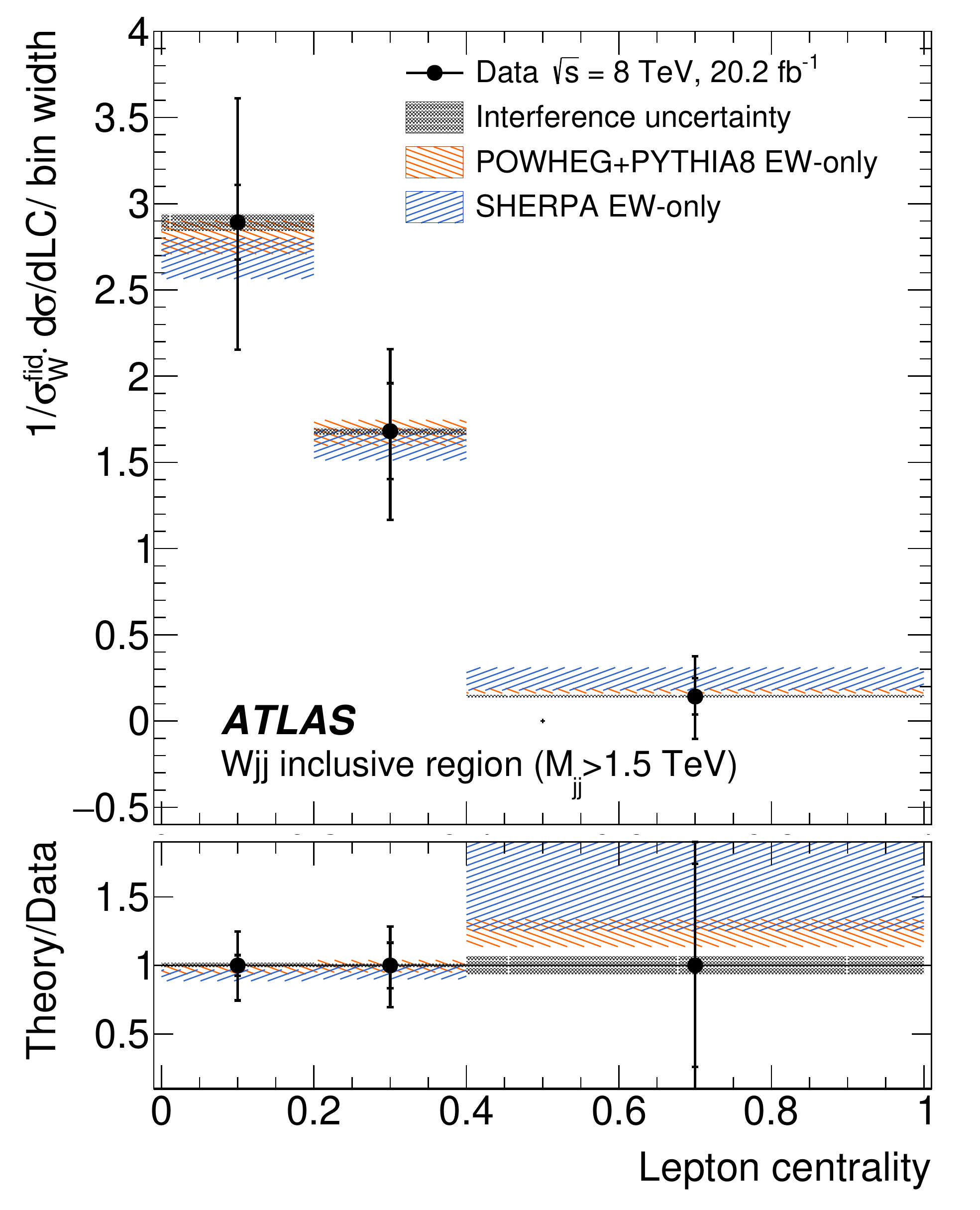}
\caption{Unfolded normalized differential QCD+EW \wjets (top) and EW (bottom) production cross sections 
as a function of lepton centrality for the inclusive fiducial region with 
$\mjj > 0.5$~\TeV~(top left), 1.0~\TeV~(top right and bottom left), and 1.5~\TeV~(bottom right). 
Both statistical (inner bar) and total (outer bar) measurement uncertainties are shown, as well as 
ratios of the theoretical predictions to the data (the bottom panel in each distribution).}
\label{unfolding:combined_measurementLC1Dinclusive}
\end{figure}

The lepton centrality distribution indirectly probes the rapidity of the $W$~boson relative to the 
dijet rapidity interval.  The differential cross section in the inclusive region as a function of 
lepton centrality is shown in Figure~\ref{unfolding:combined_measurementLC1Dinclusive} for three 
$\mjj$ thresholds.  All QCD+EW~\wjets predictions adequately describe the lepton centrality in the 
region with the lowest dijet mass threshold, which is dominated by QCD \wjets production.  As the 
$\mjj$ threshold is increased the differences between QCD and QCD+EW \wjets production become more 
apparent, particularly at low lepton centrality where EW~\wjets production is enhanced.  The 
measurement of this distribution for EW~\wjets production shows good agreement with the predictions.

%% file: diffatgc.tex
Differential measurements are performed in distributions that provide enhanced sensitivity to 
anomalous gauge couplings:
\begin{itemize}
\item $\ptjlead$, the $\pt$ of the highest-$\pt$ jet;
\item $\pt^{jj}$, the $\pt$ of the dijet system (vector sum of the $\pt$ of the two highest-\pt jets); and 
\item $\Delta \phi(j_1,j_2)$, the magnitude of the azimuthal angle between the two highest-$\pt$ jets,
\end{itemize}
where the last observable is sensitive to anomalous CP-violating couplings~\cite{vbfcp}.

\begin{figure}[htbp]
\centering
\includegraphics[width=0.48\textwidth]{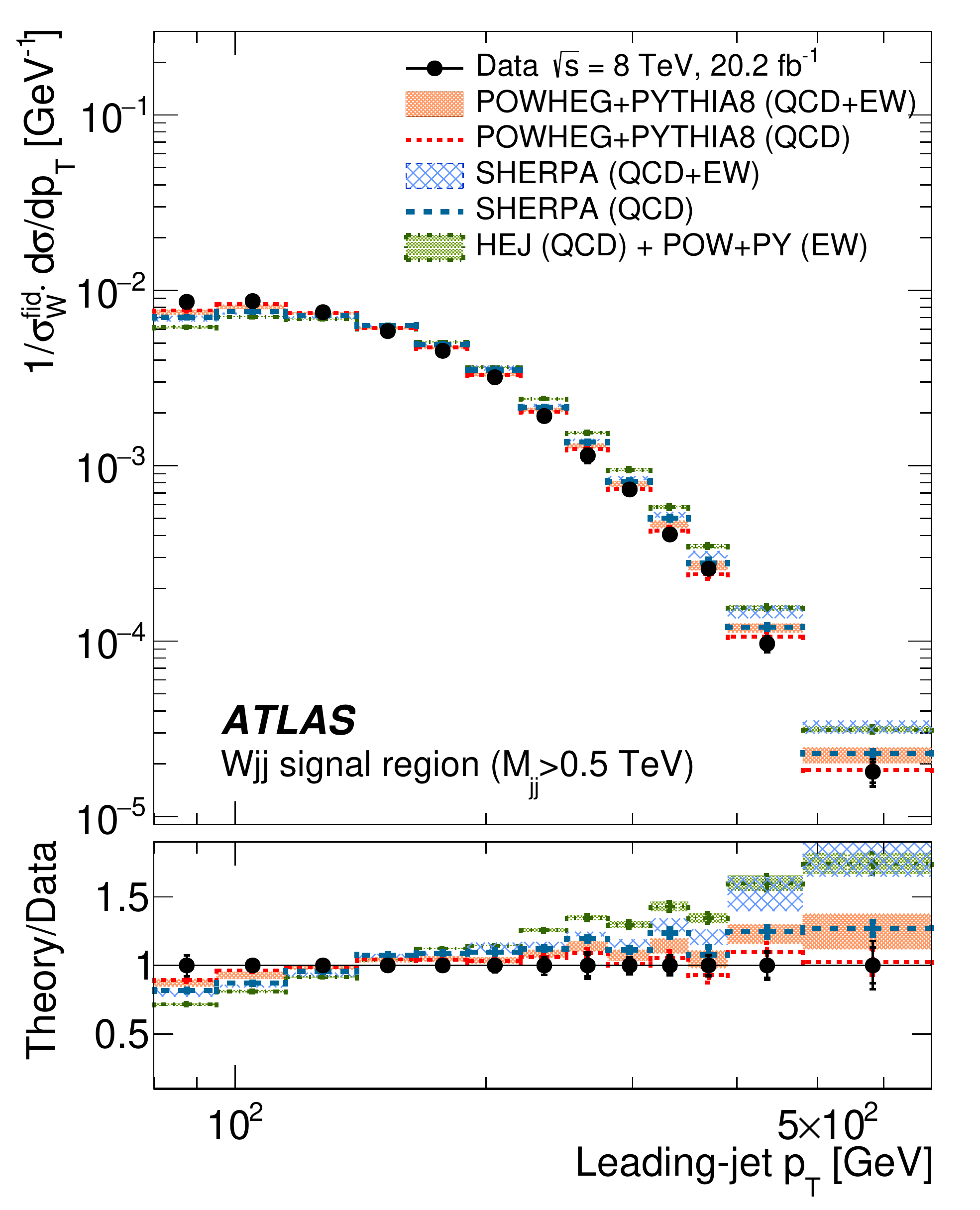}
\includegraphics[width=0.48\textwidth]{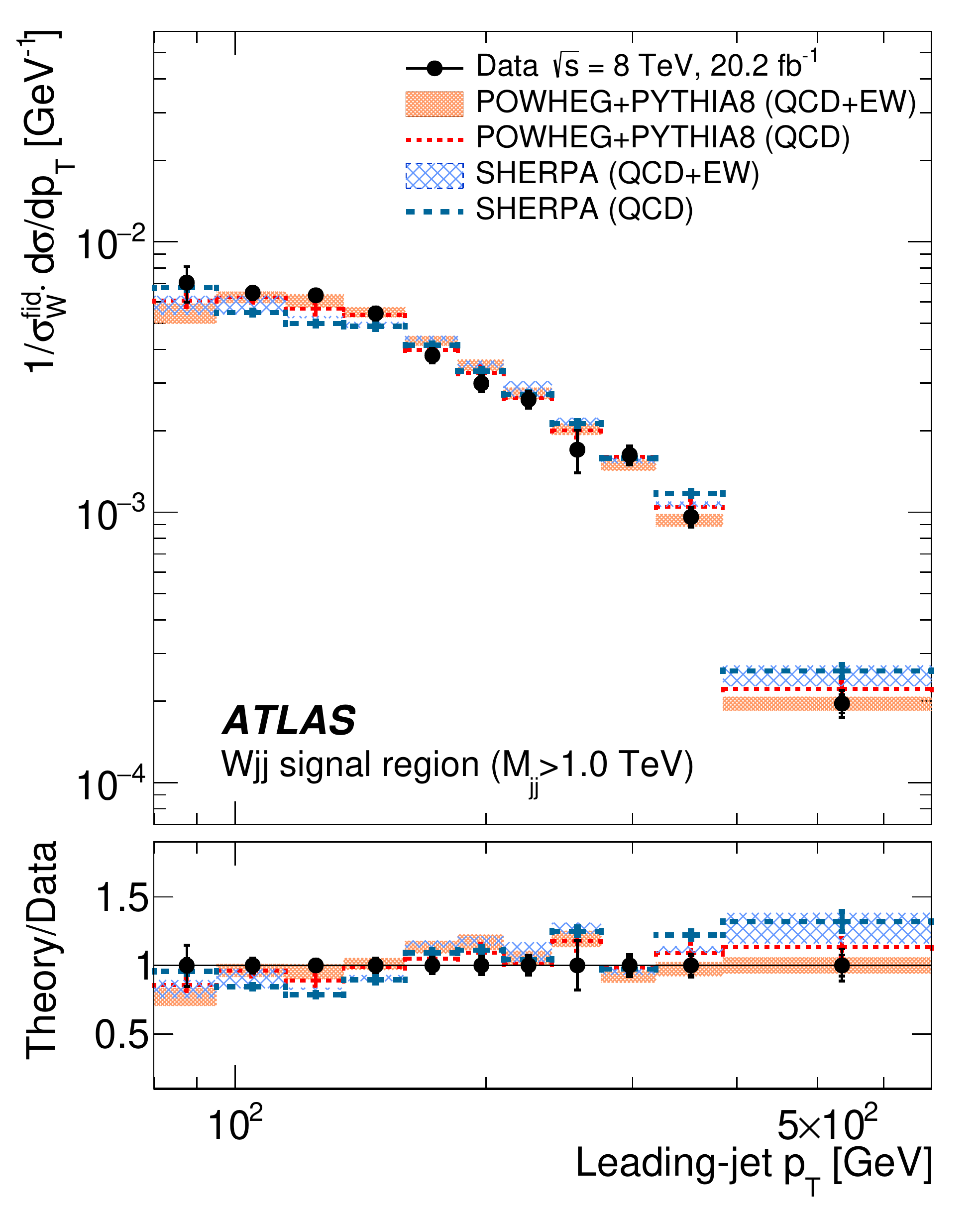}
\includegraphics[width=0.48\textwidth]{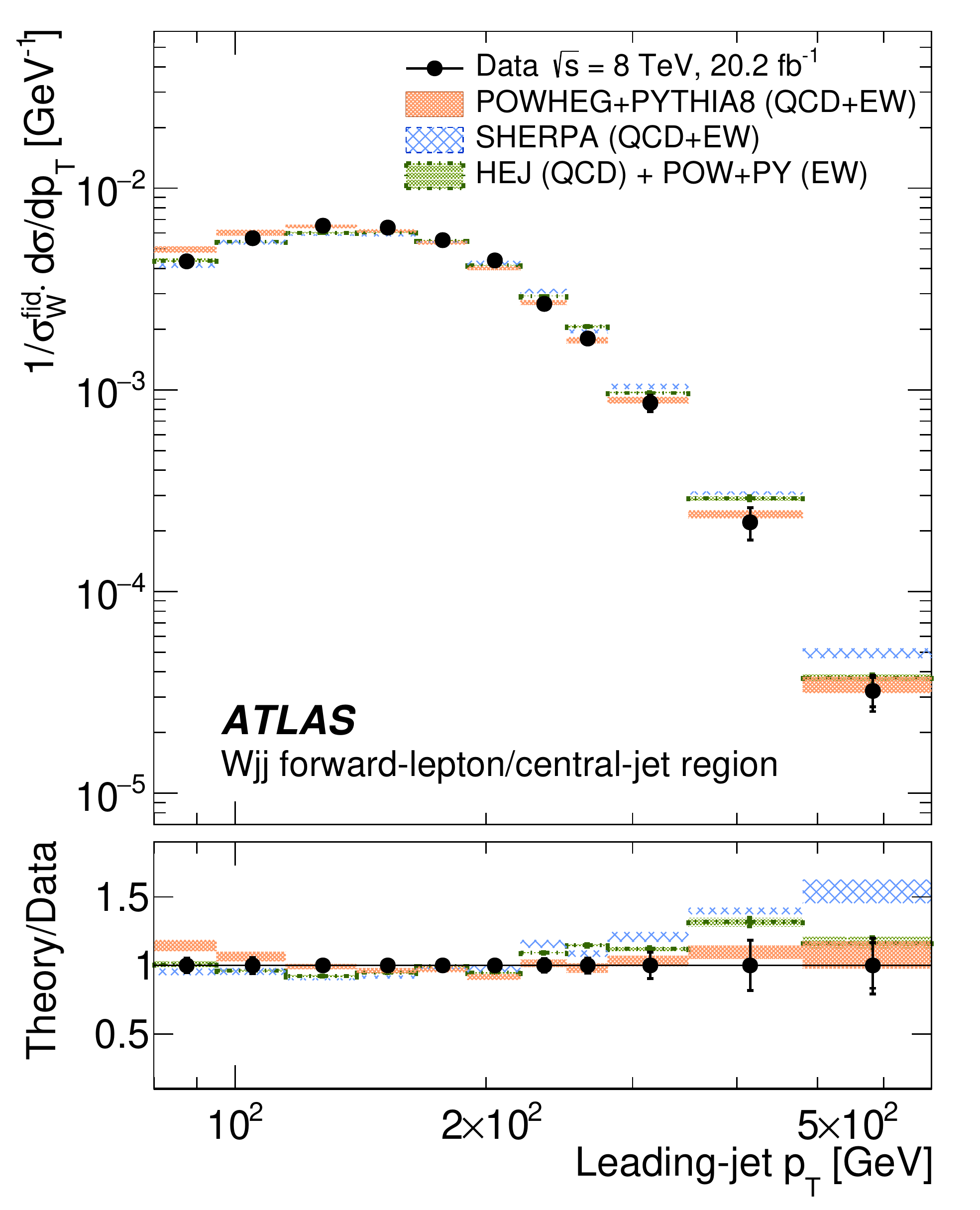}
\includegraphics[width=0.48\textwidth]{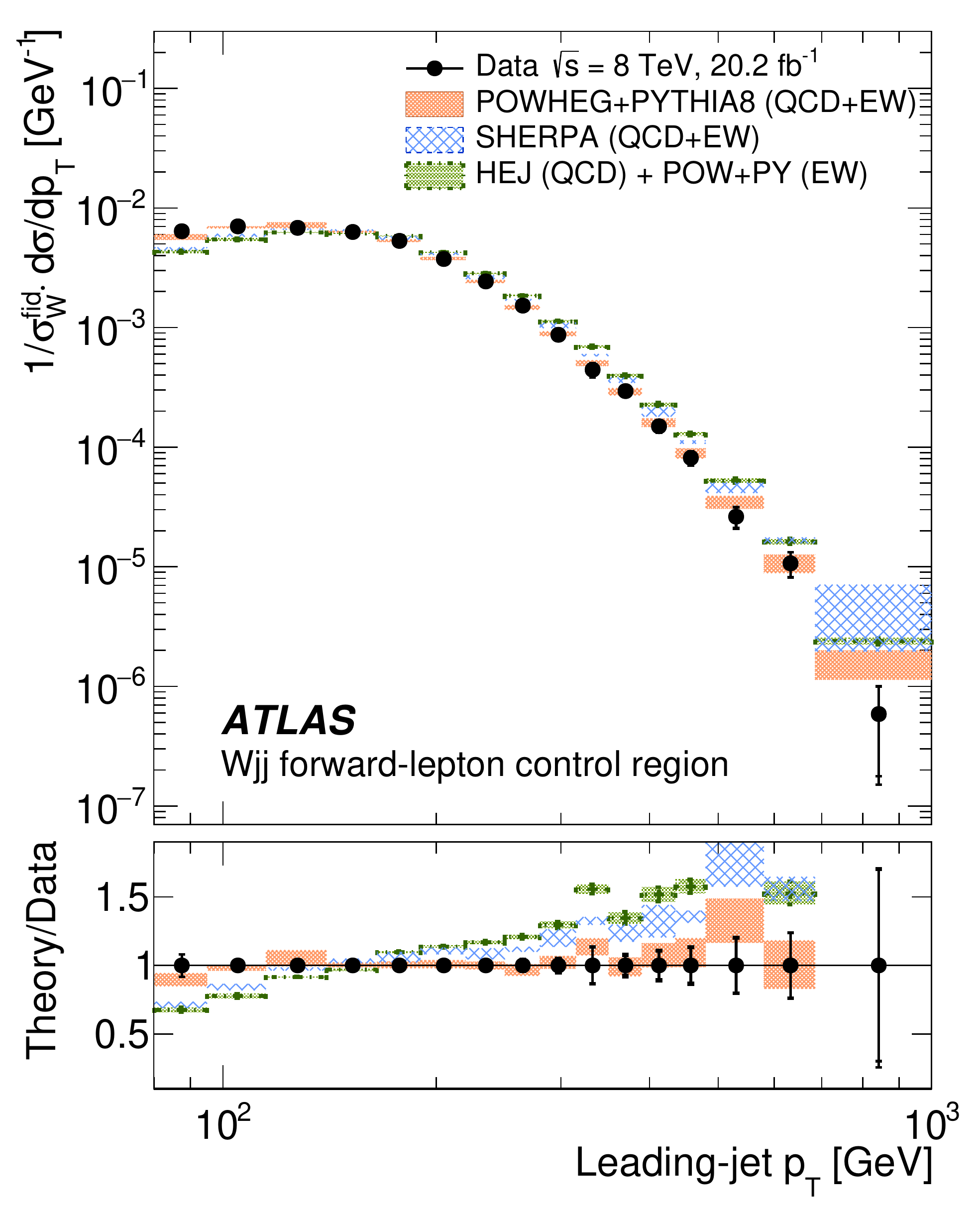}
\caption{Unfolded normalized differential \wjets production cross sections as a function of the leading-jet \pt
in the signal, high-mass signal, forward-lepton/central-jet, and forward-lepton fiducial regions.
Both statistical (inner bar) and total (outer bar) measurement uncertainties are shown, as well as ratios
of the theoretical predictions to the data (the bottom panel in each distribution).}
\label{unfolding:combinedj1pt}
\end{figure}

\begin{figure}[tbp]
\centering
\includegraphics[width=0.48\textwidth]{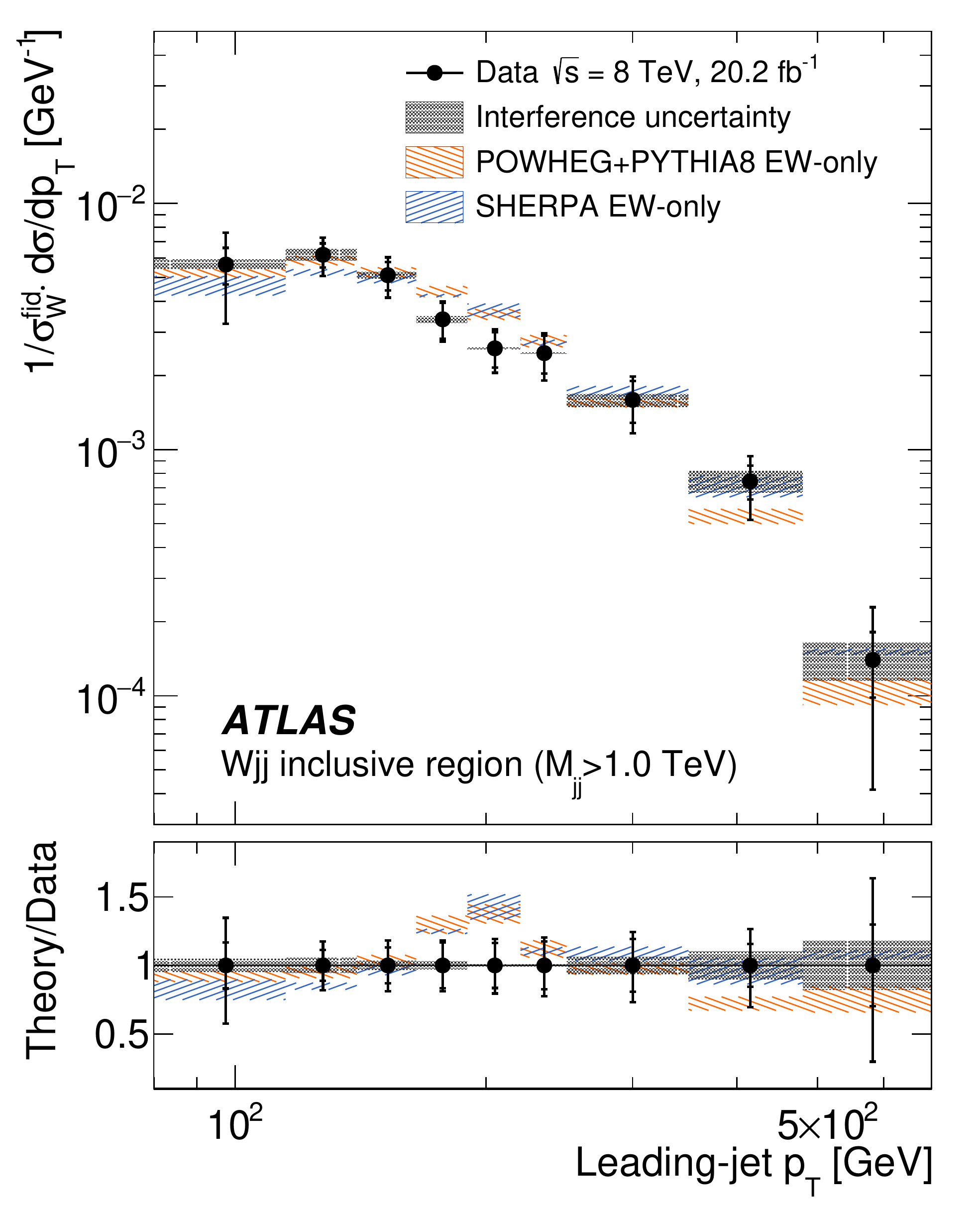}
\includegraphics[width=0.48\textwidth]{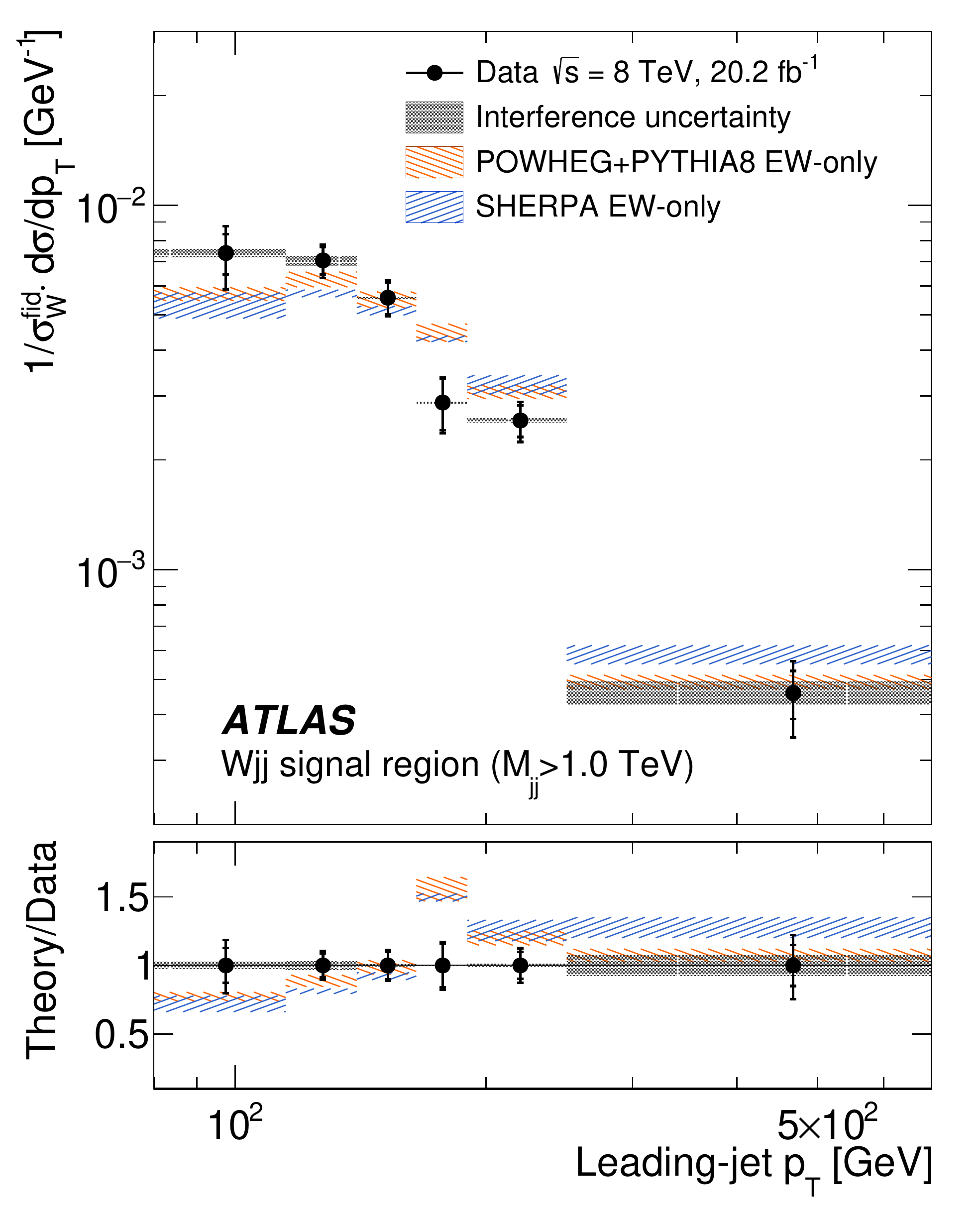}
\includegraphics[width=0.48\textwidth]{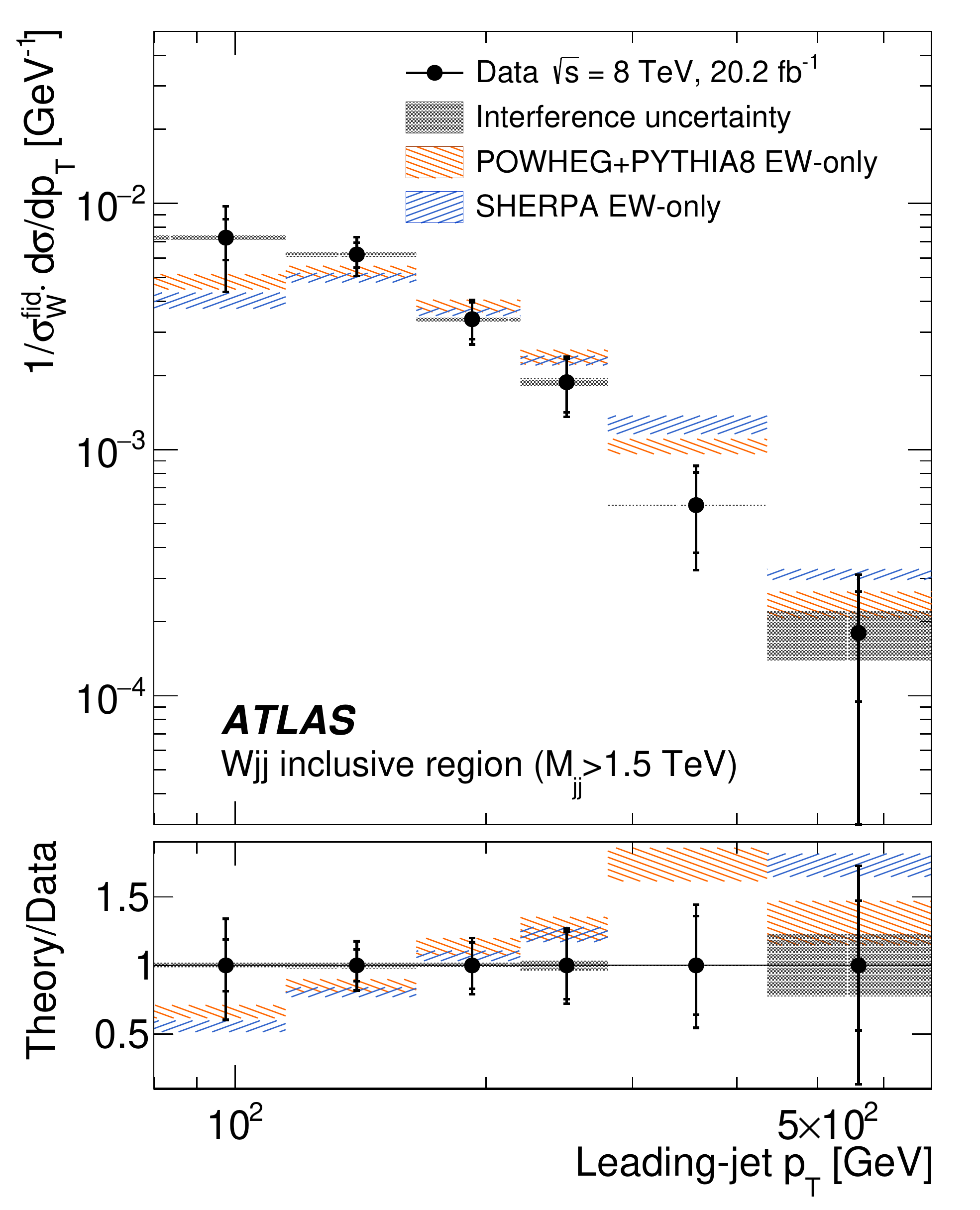}
\includegraphics[width=0.48\textwidth]{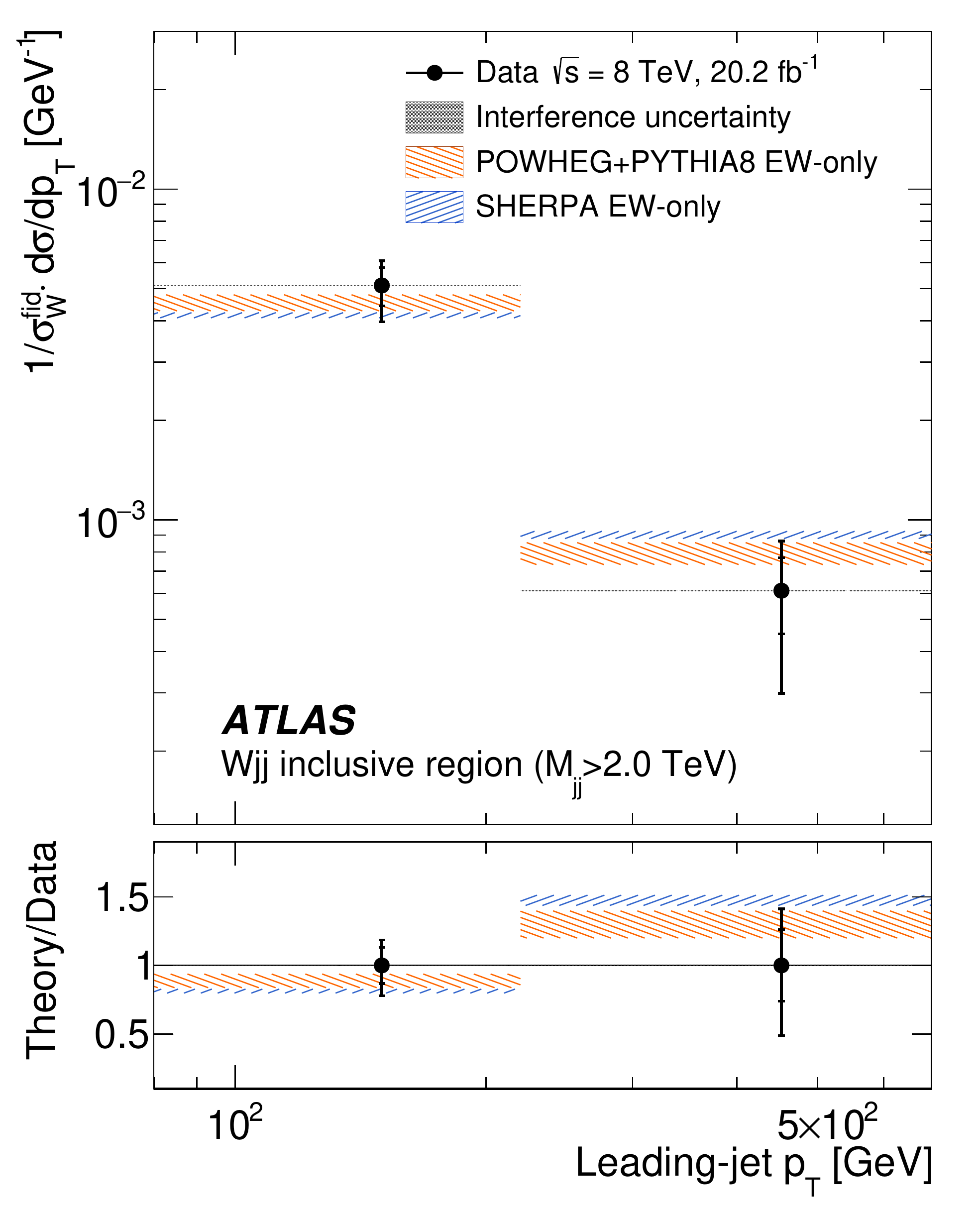}
\caption{Unfolded normalized differential EW~\wjets production cross sections as a function of the
leading-jet $\pt$ for the inclusive fiducial region with three thresholds on the dijet invariant mass (1.0~\TeV,
1.5~\TeV, and 2.0~\TeV), and for the signal-enriched fiducial region with a minimum dijet invariant mass of
1.0~\TeV.  Both statistical (inner bar) and total (outer bar) measurement uncertainties are shown, as well as ratios
of the theoretical predictions to the data (the bottom panel in each distribution). }
\label{unfolding:EWKcombined_measurementj1pt1Dhighmass10}
\end{figure}

The transverse momentum distribution of the leading jet, shown in Figure~\ref{unfolding:combinedj1pt}, 
has a substantial correlation with the momentum transfer in $t$-channel events.  The QCD+EW~\wjets 
measurements are globally well described by \powheg + \pythia, while predictions from \sherpa and 
\textsc{hej} both show a harder spectrum than observed in data.  For EW~\wjets production the 
\powheg + \pythia and \sherpa predictions give a harder spectrum than observed in the data, particularly 
in the higher purity regions (Figure~\ref{unfolding:EWKcombined_measurementj1pt1Dhighmass10}).  The 
overestimation of rates at high jet \pt may be reduced by the inclusion of NLO electroweak 
corrections\,\cite{openloops1}.

\begin{figure}[htbp]
\centering
\includegraphics[width=0.48\textwidth]{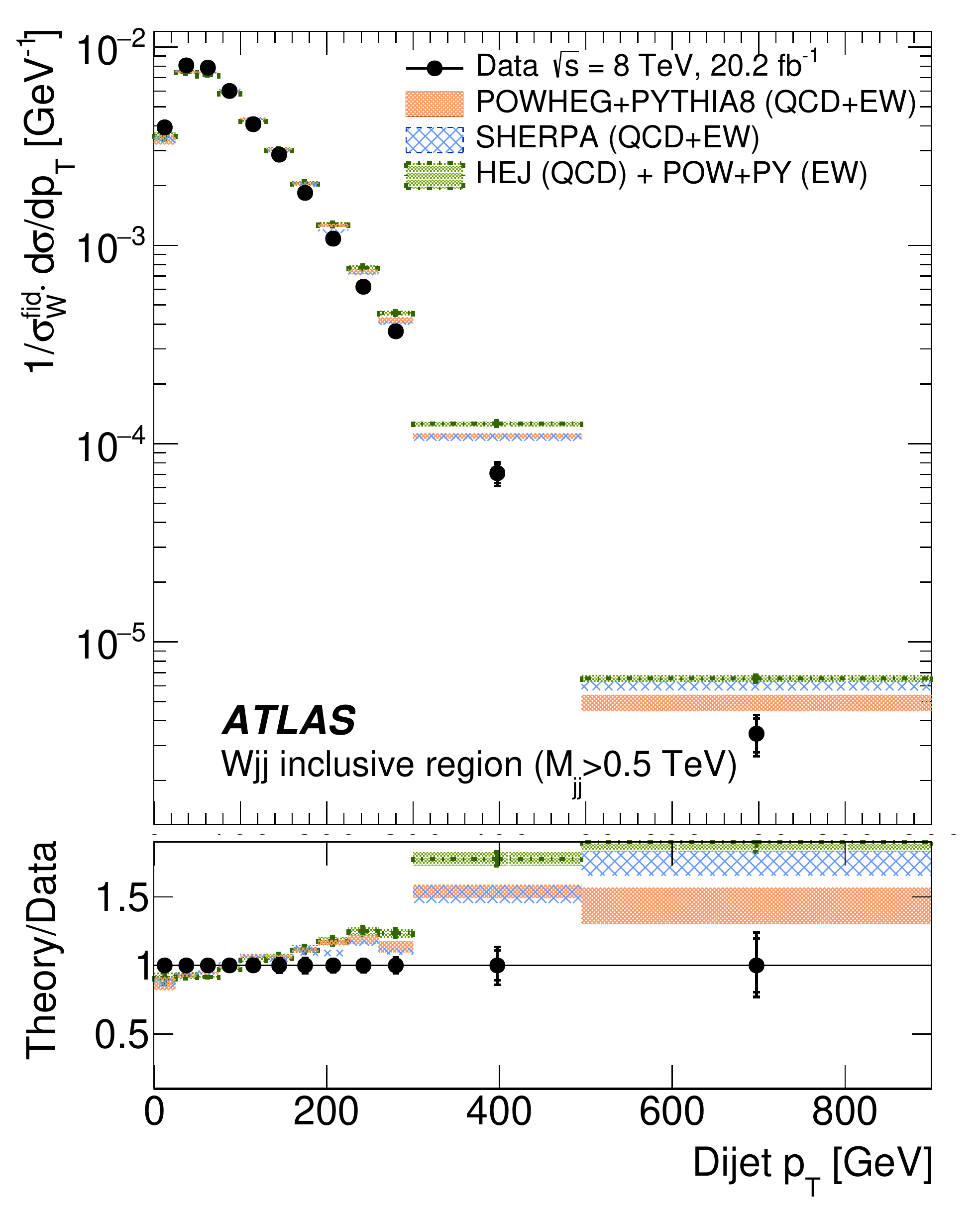}
\includegraphics[width=0.48\textwidth]{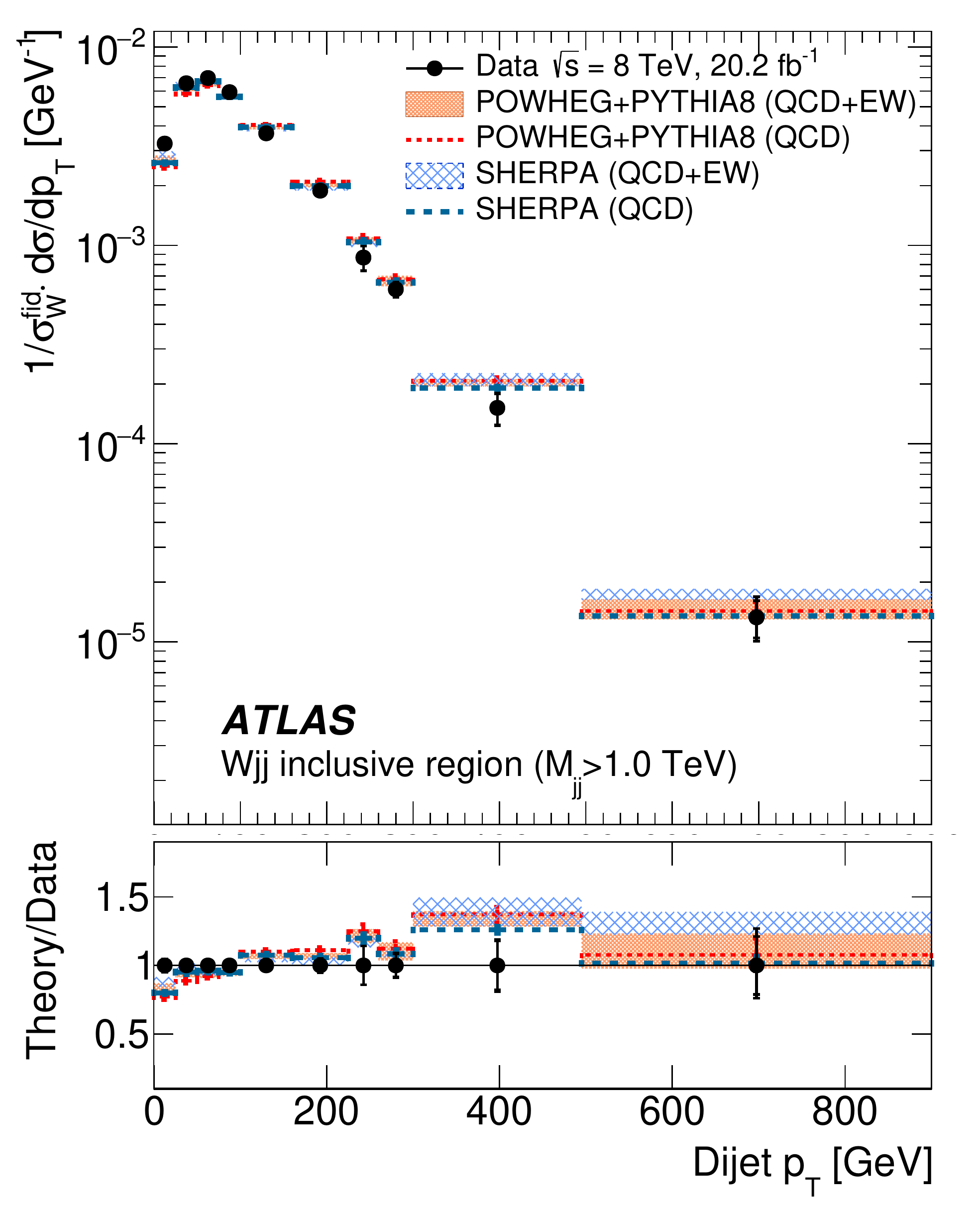}
\includegraphics[width=0.48\textwidth]{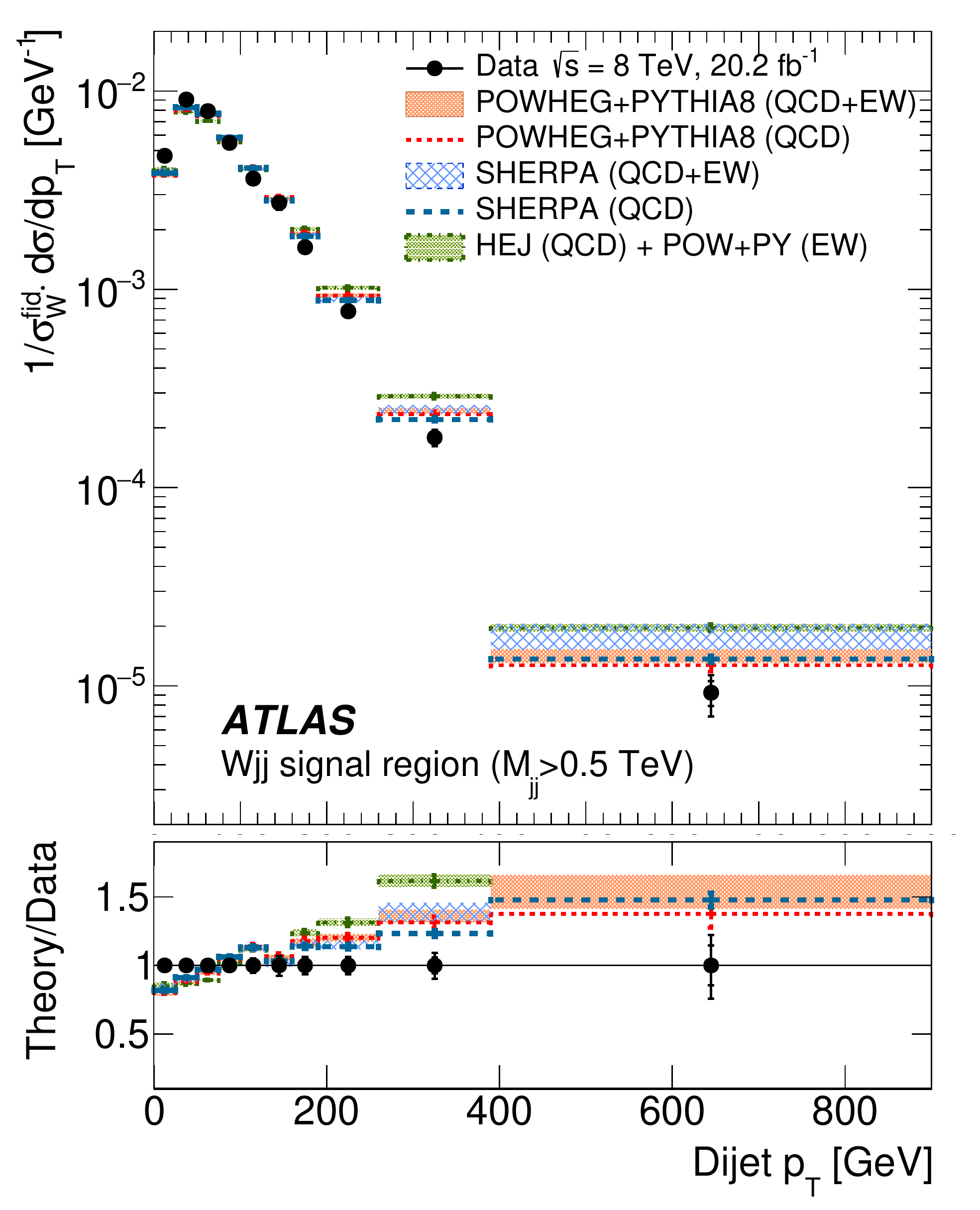}
\includegraphics[width=0.49\textwidth]{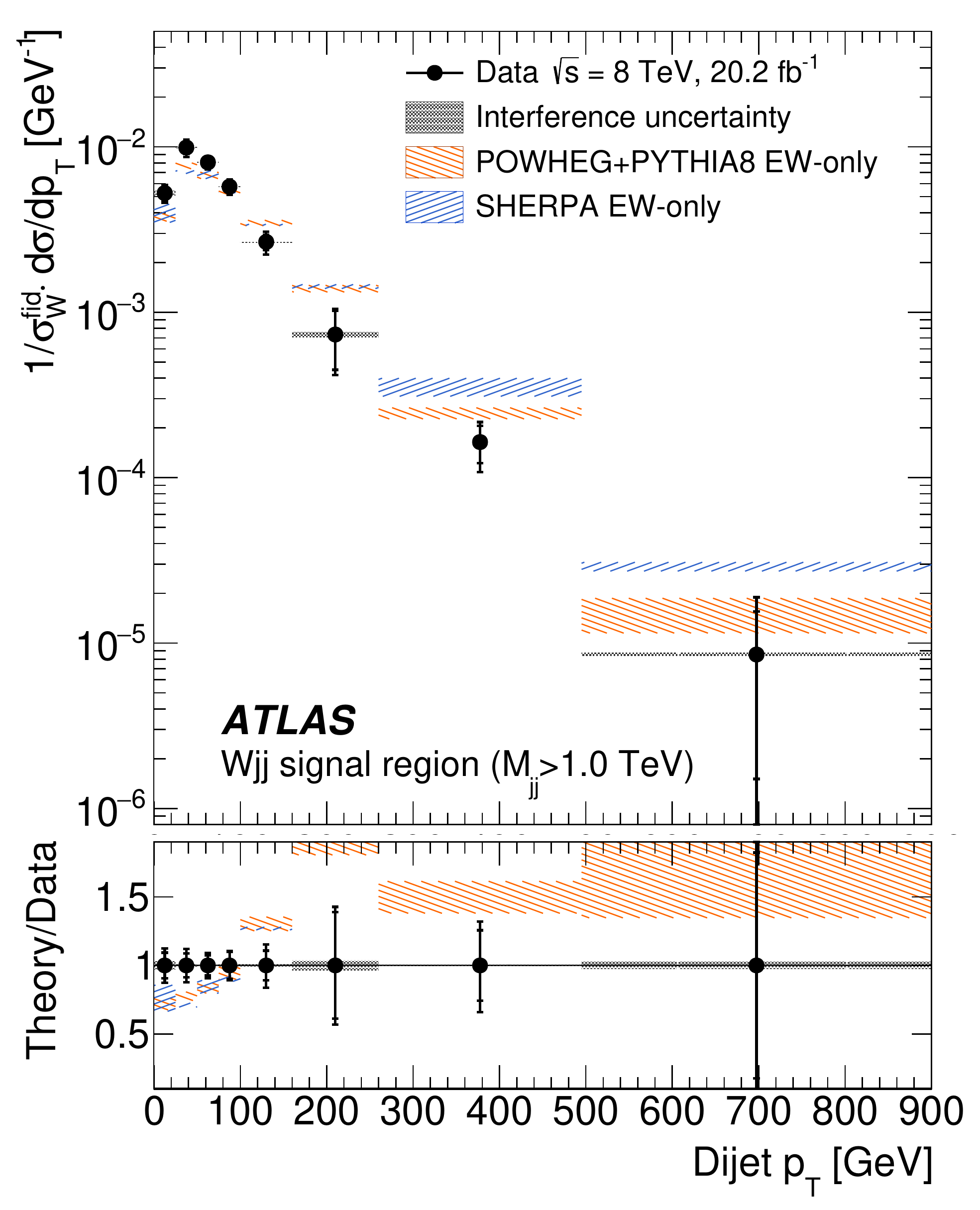}
\caption{Unfolded normalized differential \wjets production cross sections as a function of dijet $\pt$ for the
inclusive (top) and signal (bottom) regions with $\mjj>0.5$~\TeV~(left) and $\mjj > 1.0$~\TeV~(right).  The bottom 
right distribution shows EW~\wjets production and the other distributions show QCD+EW~\wjets production.  Both 
statistical (inner bar) and total (outer bar) measurement uncertainties are shown, as well as ratios of the
theoretical predictions to the data (the bottom panel in each distribution).}
\label{unfolding:combined_measurementdijetpt1Dinclusive}
\end{figure}

The transverse momentum of the dijet system is also correlated with the momentum transfer in $t$-channel 
events.  Figure~\ref{unfolding:combined_measurementdijetpt1Dinclusive} shows the measured normalized \pT 
distribution of the dijet system compared to the various predictions.  There is a trend for all 
predictions to overestimate the relative rate at high dijet $\pt$ in the inclusive and signal-enhanced 
regions, both for QCD+EW~\wjets and EW~\wjets production.  As in the case of the jet \pt distribution, 
the discrepancy could be due to missing NLO electroweak corrections, which reduce the predictions at high 
$W$-boson \pt\,\cite{openloops1}.

\begin{figure}[htbp]
\centering
\includegraphics[width=0.48\textwidth]{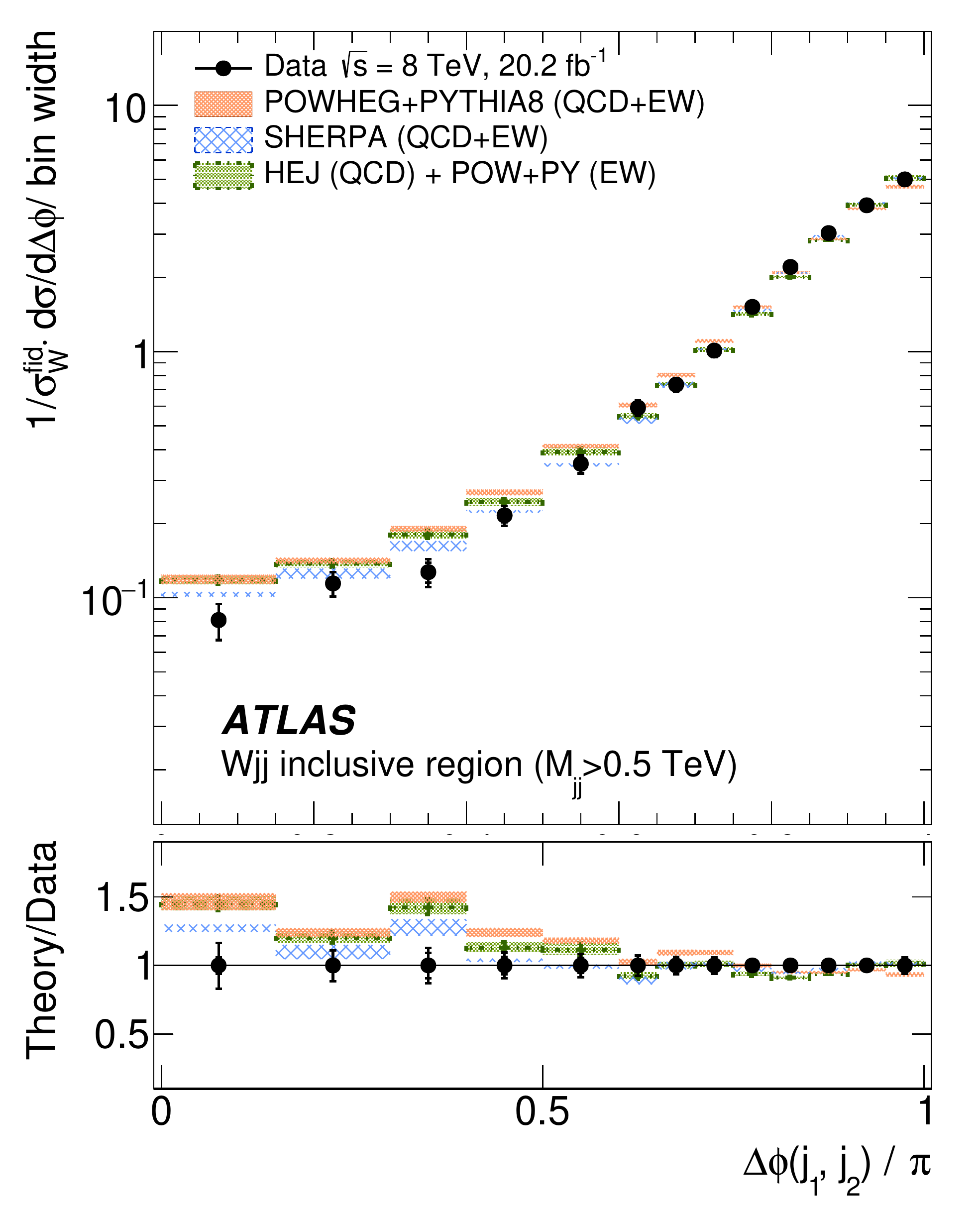}
\includegraphics[width=0.48\textwidth]{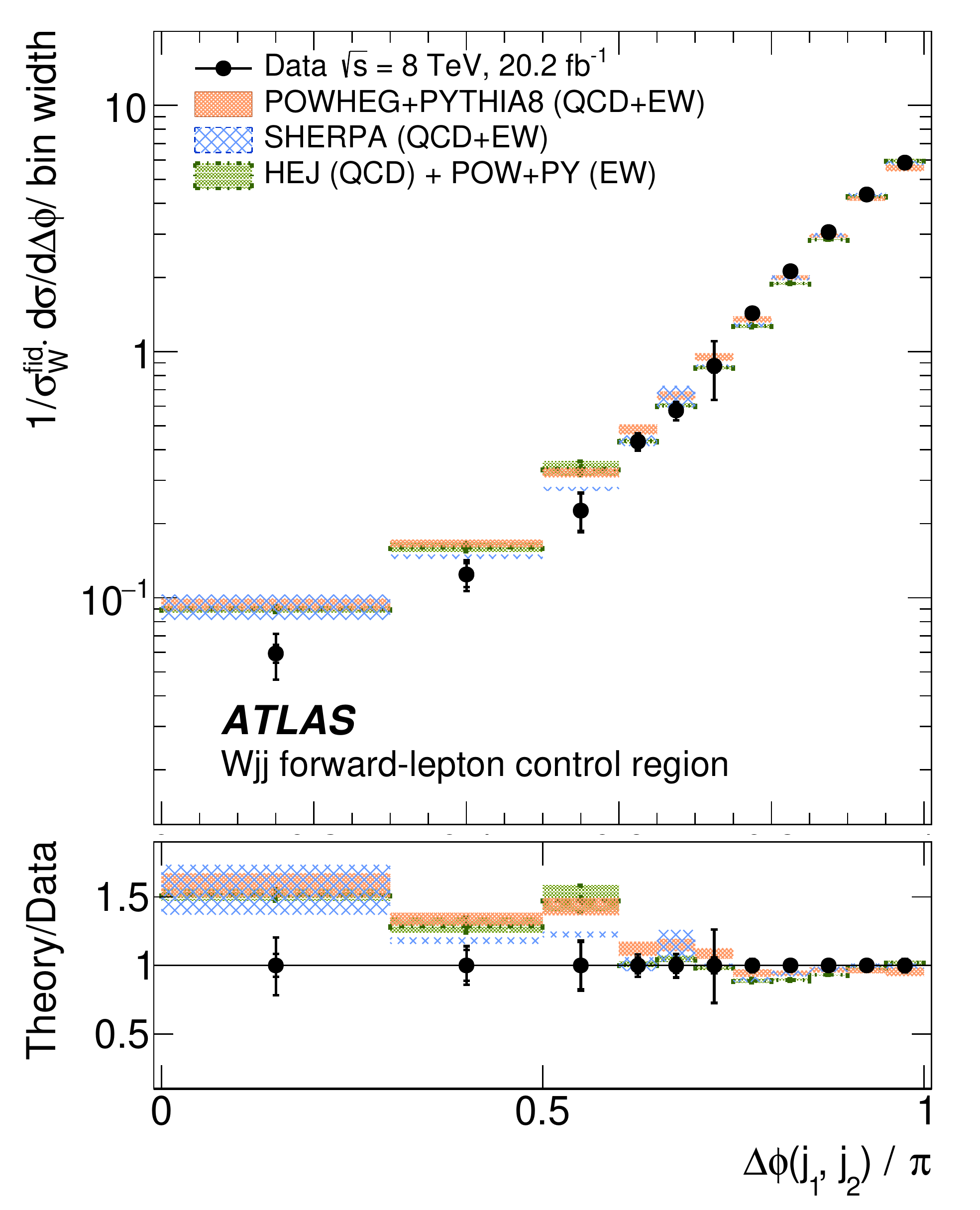}
\includegraphics[width=0.48\textwidth]{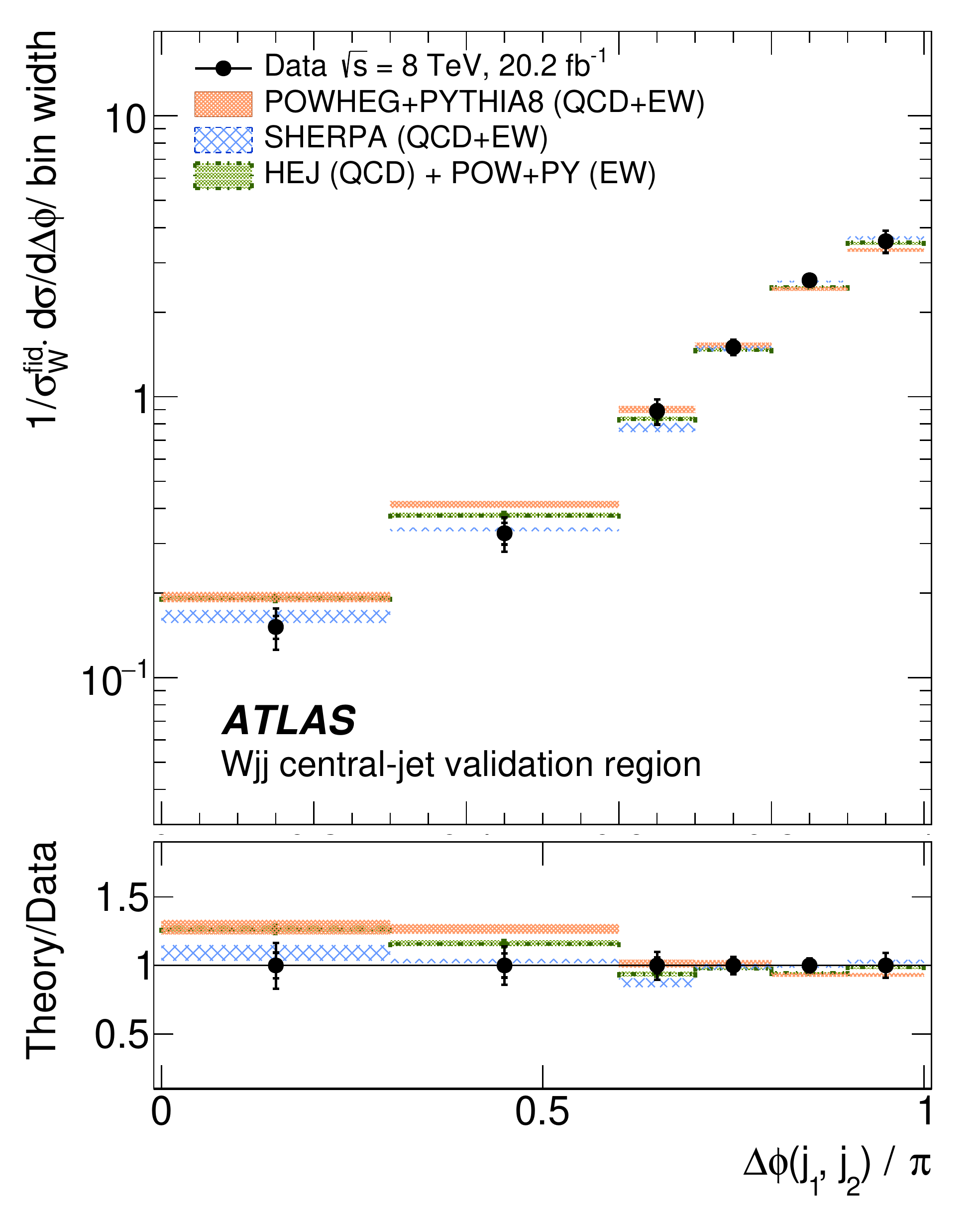}
\includegraphics[width=0.48\textwidth]{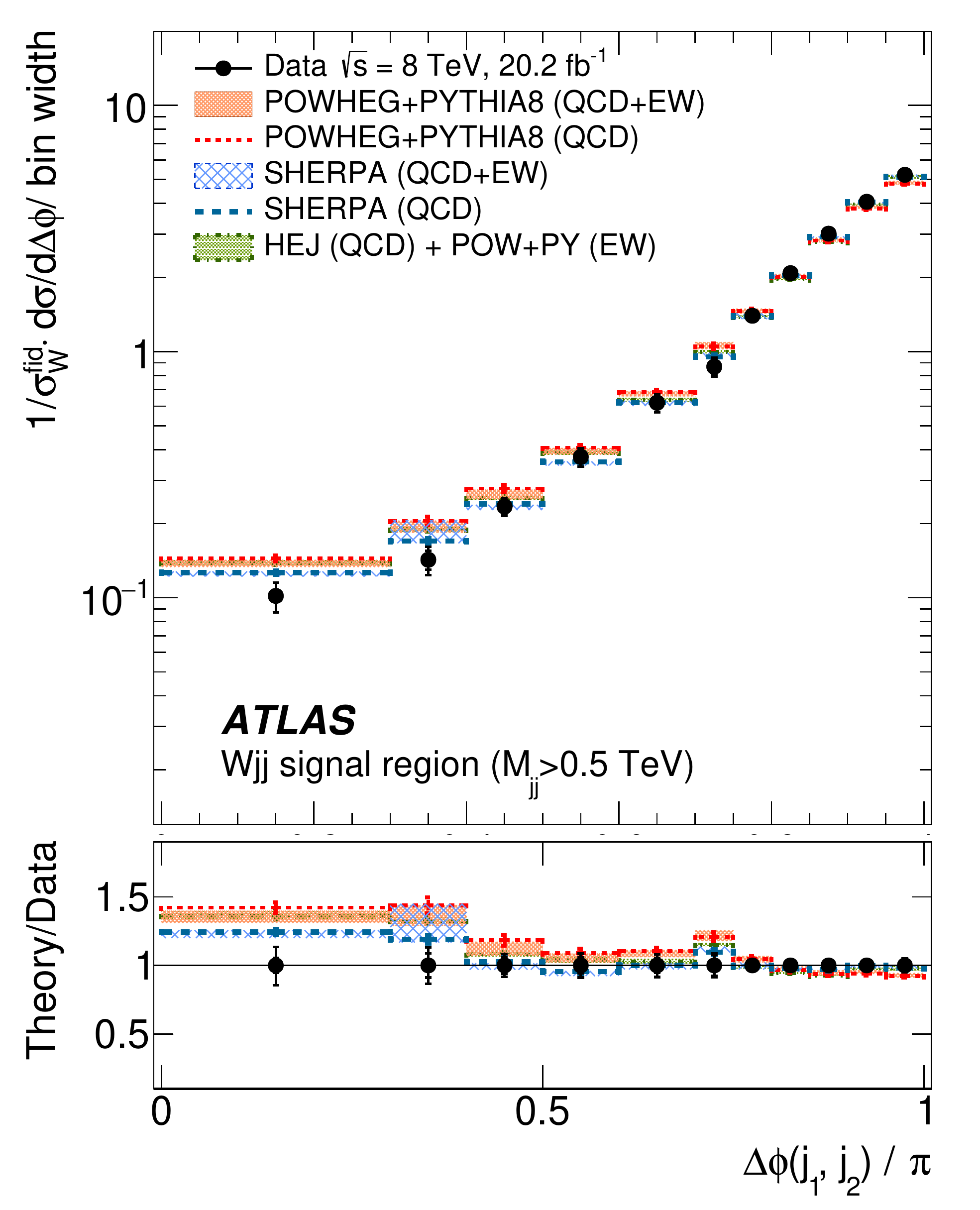}
\caption{Unfolded normalized differential \wjets production cross sections as a function of $\Delta\phi(j_1,j_2)$
for the inclusive, forward-lepton control, central-jet validation, and signal fiducial regions.  Both statistical
(inner bar) and total (outer bar) measurement uncertainties are shown, as well as ratios of the theoretical
predictions to the data (the bottom panel in each distribution). }
\label{unfolding:combined_measurementdphi121Dinclusive}
\end{figure}

\begin{figure}[tbp]
\centering
\includegraphics[width=0.48\textwidth]{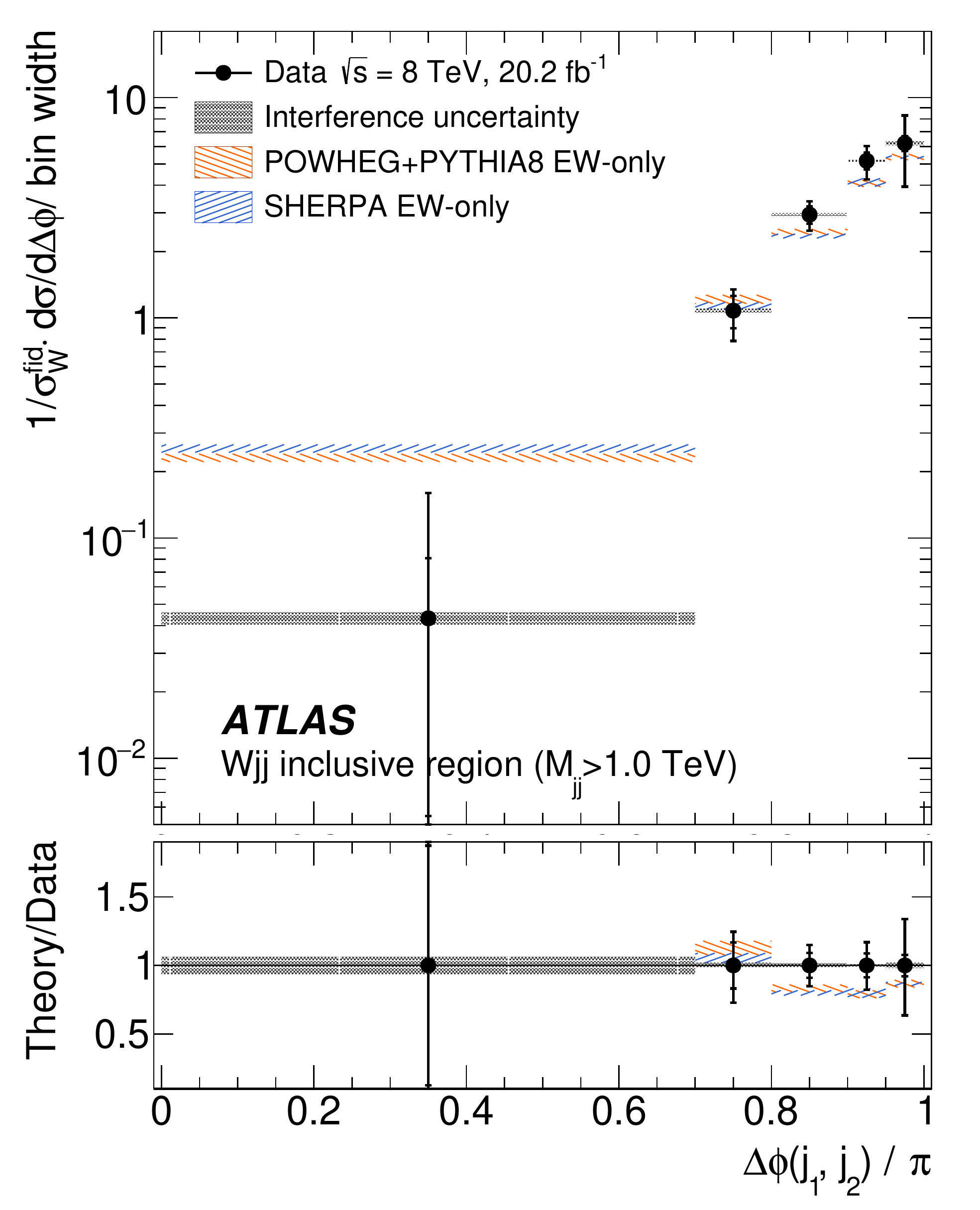}
\includegraphics[width=0.48\textwidth]{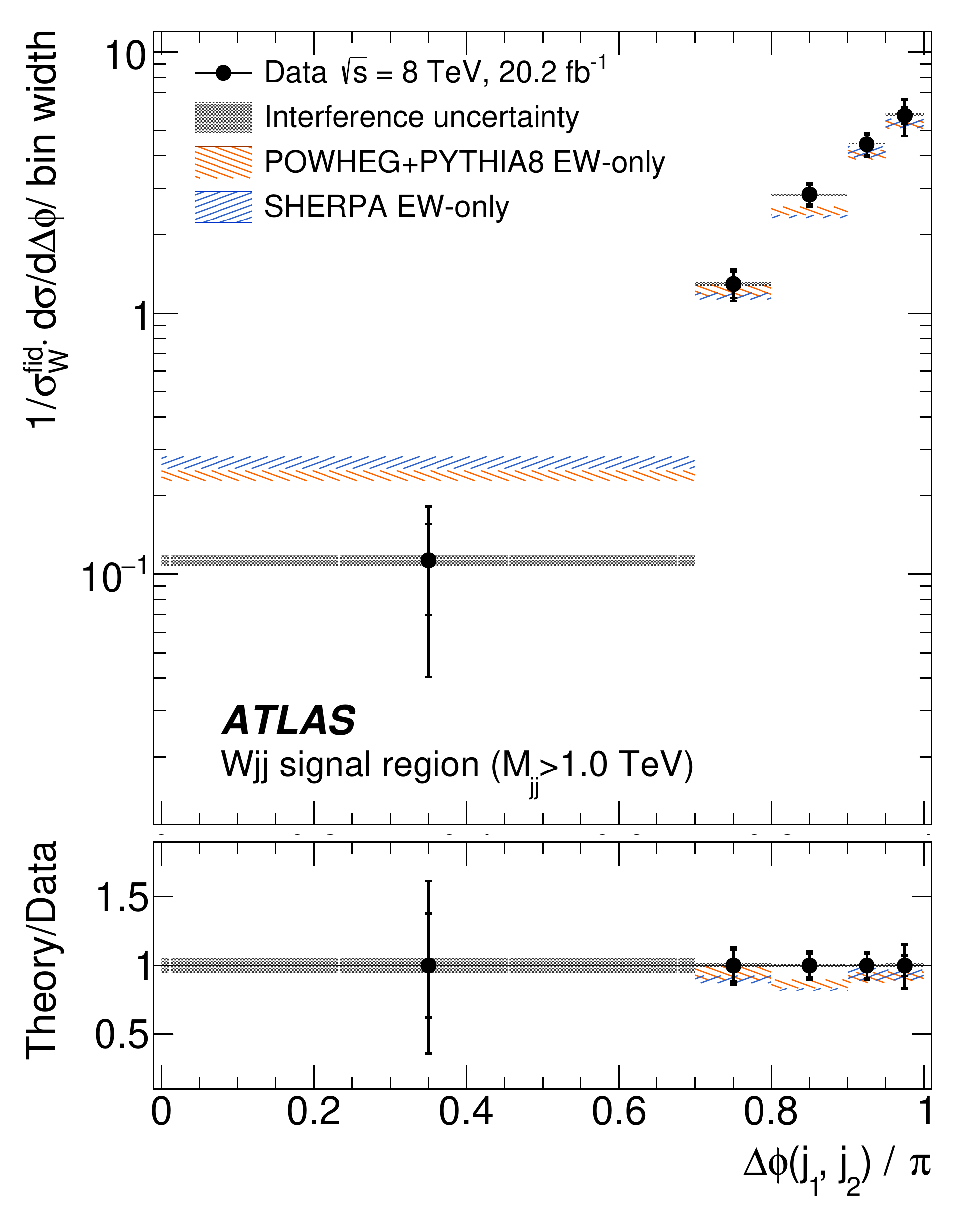}
\caption{Unfolded normalized differential EW~\wjets production cross sections as a function
of the azimuthal angle between the two leading jets, for the inclusive and signal fiducial regions with
$\mjj > 1.0$~\TeV.  Both statistical (inner bar) and total (outer bar) measurement uncertainties are shown,
as well as ratios of the theoretical predictions to the data (the bottom panel in each distribution).}
\label{unfolding:EWKcombined_measurementdphi1Dhighmass10}
\end{figure}

The azimuthal angle between the two leading jets can be used to probe for new CP-odd operators in 
VBF production.  The normalized differential cross sections for QCD+EW \wjets production as a function 
of this angle are shown in the inclusive, forward-lepton control, central-jet validation, and signal 
fiducial regions in Figure~\ref{unfolding:combined_measurementdphi121Dinclusive}.  Good agreement 
between the data and all predictions is seen, with a slight tendency for predictions to overestimate 
the relative rate at small angles in all fiducial regions.  
Figure~\ref{unfolding:EWKcombined_measurementdphi1Dhighmass10} shows the normalized EW~\wjets cross 
section as a function of the azimuthal angle between the two leading jets for the inclusive and signal 
fiducial regions with $\mjj>1.0$~\TeV.

%% file: atlas-aTGC.tex
\label{sec:aTGCs}
The triple-gauge-boson vertex is directly probed by the vector-boson-fusion process.
Non-SM couplings at this vertex would affect the production rates and distributions.  The 
couplings are constrained in the context of an aTGC or EFT framework, using the yield in 
the anomalous coupling signal region (Table~\ref{tab:selection}) to constrain the parameters.  
The results are complementary~\cite{Baur:1993fv} to those obtained in diboson production~\cite{WW_ATLAS}, which 
corresponds to the exchange of one off-shell boson in the $s$-channel rather than two in the 
$t$-channel.

\subsection{Theoretical overview}
The signal-region measurements are sensitive to the $WWV$ ($V$ = $Z$ or $\gamma$) couplings 
present in the $t$-channel production mode shown in Figure~\ref{intro:Fig:EWK2Jets}(a).   
These couplings can be characterized by an effective Lagrangian $\mathcal{L}^{WWV}_\mathrm{eff}$
including operators up to mass-dimension six~\cite{atgc}:
\begin{align}\label{aTGC:effLagr}
i\mathcal{L}^{WWV}_\mathrm{eff}&=g_{WWV}\, \biggl\{ \Bigl\lbrack g_1^V V^{\mu}(W^{-}_{\mu \nu}W^{+\nu} -W^{+}_{\mu \nu}W^{-\nu}) 
+ \kappa_{V} W^{+}_{\mu }W^{-}_{\nu}V^{\mu \nu} + \frac{\lambda_{V}}{m_{W}^{2}}V^{\mu \nu}W^{+ \rho}_{\nu}W^{-}_{\rho \mu} 
\Bigr\rbrack \nonumber \\  \notag
&\qquad - \Bigl\lbrack \frac{\tilde{\kappa}_{V}}{2} W^{-}_{\mu }W^{+}_{\nu} \epsilon^{\mu \nu \rho \sigma} V_{\rho \sigma} + 
\frac{\tilde{\lambda}_{V}}{2 m_{W}^{2}}  W^{-}_{\rho \mu }W^{+ \mu}_{\nu} \epsilon^{ \nu \rho \alpha \beta} 
V_{\alpha \beta} \Bigr\rbrack \biggr\}, 
\end{align}

where $W^{\pm}_{\mu \nu} = \partial_\mu W^{\pm}_\nu - \partial_\nu W^{\pm}_{\mu}$, with $W^{\pm}_\mu$ the 
$W^{\pm}$ field; $V_{\mu \nu} = \partial_\mu V_\nu - \partial_\nu V_\mu$, with $V_\mu$ the $Z$ or $\gamma$ 
field; $m_W$ is the $W$-boson mass; and the individual couplings have SM values $g_1^V = 1$, $\kappa_{V} = 1$, 
$\lambda_{V} = 0$, $\tilde{\kappa}_{V} = 0$, and $\tilde{\lambda}_{V} = 0$.   The overall coupling constants 
$g_{WWV}$ are given by $g_{WW\gamma} = -e$ and $g_{WWZ} = -e\cdot\cot(\theta_W)$, where $e$ is the 
electromagnetic coupling and $\theta_W$ is the weak mixing angle.  The terms in the first row of the Lagrangian 
conserve $C$, $P$, and $CP$, while those in the second violate $CP$.  Deviations of the $g_1^V$ and $\kappa_{V}$ 
parameters from the SM are denoted by $\Delta g_1^Z= g_1^Z-1$ and $\Delta\kappa_{V}=\kappa_{V}-1$, respectively.  
The requirement of gauge invariance at the level of dimension-six operators leads to the following 
relations~\cite{hisz}:
\begin{equation}
\Delta g_1^Z = \Delta \kappa_Z + \Delta \kappa_{\gamma} \tan^2\theta_W,  \quad
\lambda_{\gamma} = \lambda_Z \equiv \lambda_V, \quad g_1^{\gamma}=1, \quad
\tilde{\kappa}_{\gamma} = - \tilde{\kappa}_Z \cot^2\theta_W, \quad \mathrm{ and } \quad 
\tilde{\lambda}_{\gamma} = \tilde{\lambda}_Z \equiv \tilde{\lambda}_V.
\nonumber
\end{equation}

The presence of anomalous couplings leads to unphysically large cross sections when the square of the momentum 
transfer $(q^2)$ between the incoming partons is large.  To preserve unitarity, a form factor is introduced with 
a new-physics scale $\Lambda$ that suppresses the anomalous coupling at high energies:
\begin{equation}
\alpha(q^2) = \frac{\alpha}{(1+ q^2 / \Lambda^2)^2},
\label{aTGC:couplings}
\nonumber
\end{equation}
where $\alpha$ is the anomalous coupling of interest.  In the following, 95\% confidence-level intervals are set for 
a unitarization scale of $\Lambda = 4 \, \text{TeV}$ and for a scale that effectively removes the form factor (shown 
as $\Lambda = \infty$).  The scale $\Lambda = 4 \, \text{TeV}$ is chosen because it does not violate unitarity 
for any parameter in the expected range of sensitivity.  

An alternative to the use of a form factor is to employ an effective field theory, which is an 
expansion in inverse powers of the energy scale of new interactions assuming perturbative coupling 
coefficients.  An EFT allows the comprehensive investigation of a complete set of dimension-six 
operators in a Lagrangian with SM fields.  The dimension-six terms introduced in the EFT can be 
expressed as
\begin{equation}
\mathcal{L}_{\textrm{EFT}} = \sum_i \frac{c_i}{\Lambda^2}O_i, 
\nonumber
\end{equation}
where $O_i$ are field operators with dimension $6$, the scale of new physics is $\Lambda$, and $c_i$ are 
dimensionless coefficients.  The operators relevant to triple-gauge-boson couplings in the HISZ basis~\cite{hisz} are 
\begin{eqnarray}
O_B & = & (D_\mu H)^{\dagger} B^{\mu\nu}D_\nu H, \nonumber \\
O_W & = & (D_\mu H)^{\dagger} W^{\mu\nu}D_\nu H, \nonumber \\
O_{WWW} & = & \mathrm{Tr}[W_{\mu\nu} W^{\nu}_{\rho} W^{\rho\mu}], \nonumber \\ 
O_{\tilde{W}} & = & (D_\mu H)^{\dagger} \tilde{W}^{\mu\nu} D_\nu H, \nonumber \\ 
O_{\tilde{W}WW} & = & \mathrm{Tr}[W_{\mu\nu}W^{\nu}_{\rho} \tilde{W}^{\rho\mu}], 
\nonumber 
\end{eqnarray}

where $H$ is the Higgs-boson field, $B_{\mu \nu} = \partial_\mu B_\nu - \partial_\nu B_{\mu}$, $B^{\mu}$ is the 
U(1)$_\mathrm{Y}$ gauge field, and $\tilde{W}^{\mu\nu} = \frac{1}{2}\epsilon_{\mu\nu\rho\sigma}W^{\rho\sigma}$.  
The coefficients of these operators are related to the aTGC parameters via the following equations:
\begin{align}
\nonumber \frac{c_W}{\Lambda^2} &= \frac{2}{m^2_Z} (g_1^Z -1),\\
\nonumber \frac{c_B}{\Lambda^2} &= \frac{2}{\tan^2\theta_W m^2_Z}(g_1^Z -1)-\frac{2}{\sin^2\theta_W m^2_Z} (\kappa_{Z} -1),\\
\nonumber \frac{c_{WWW}}{\Lambda^2} &= \frac{2}{3 g^2 m^2_W} \lambda_V, \\
\nonumber \frac{c_{\tilde{W}}}{\Lambda^2} &= -\frac{2}{\tan^2\theta_W m^2_W} \tilde{\kappa}_Z, \\
\frac{c_{\tilde{W}WW}}{\Lambda^2} &= \frac{2}{3 g^2 m^2_W} \tilde{\lambda}_V, 
\nonumber 
\end{align}

where $g$ is the weak coupling, $m_Z$ is the $Z$-boson mass, and the aTGC parameters do not 
have any form-factor suppression.

\subsection{Experimental method}

The signal region defined to increase the sensitivity to anomalous triple-gauge-boson couplings requires 
$\mjj > 1$~\TeV~and leading-jet $\pt > 600$~\GeV~(Table~\ref{tab:selection}).  The leading-jet $\pt$ is 
chosen because it is highly correlated with the $q^2$ of the signal $t$-channel process.  The $\pt$ 
threshold is optimized to maximize sensitivity to anomalous couplings, considering both the statistical and 
systematic uncertainties.  The event yields in the reconstructed signal region used for setting the constraints 
are given in Table~\ref{tab:yields}.  The SM prediction is negligible for $\pT > 1$~\TeV, yielding an 
approximate lower bound for the validity of the EFT constraints.

The effects of anomalous couplings are modelled with \sherpa.  Each sample is normalized by a factor 
$k = \text{NLO}/\text{LO}$ given by the ratio of \powheg + \pythia to \sherpa SM predictions of 
electroweak \wjets production.  The number of events expected for a given parameter value is calculated as:
 \begin{equation}
N_{\text{reco}}=\mathcal{L}\times\sigma^{\ell \nu jj} \times {\cal{A}} \times {\cal{C}} 
\times k,
\label{tgcreco}
\nonumber 
 \end{equation}
where $\mathcal{L}$ is the integrated luminosity of the 8~\TeV~data, $\sigma^{\ell \nu jj}$ is the 
cross section for the corresponding anomalous-coupling variation, ${\cal{A}}$ is the selection acceptance 
at particle level, and ${\cal C}$ is the ratio of selected reconstruction-level events to the particle-level 
events in the fiducial phase-space region.  The factor containing the cross section and acceptance 
($\sigma^{\ell \nu jj} \times {\cal{A}}$) is parameterized as a quadratic function of each aTGC parameter, 
with a 10\% statistical uncertainty in the parameterization.   

Theoretical uncertainties due to missing higher orders, estimated with factors of 2 and 1/2 variations of the 
renormalization and factorization scales, are estimated to be 8\% of the strong \wjets yield and 14\% of the 
electroweak \wjets yield in the region with leading-jet $\pt>600$~\GeV.  Detector uncertainties are correlated 
between strong and electroweak production and are estimated to be 11\% of the combined yield.

\subsection{Confidence-level intervals for aTGC parameters}

Confidence-level (C.L.) intervals are calculated using a frequentist approach\,\cite{fc}.  A negative log-likelihood 
function is constructed based on the expected numbers of background and signal events, and the number of observed data 
events.  The likelihood is calculated as a function of individual aTGC parameter variations, with the other parameters 
set to their SM values.  To obtain 95\% confidence-level intervals, pseudoexperiments are produced with the number of 
pseudodata events drawn from a Poisson distribution, where the mean is given by the total SM prediction 
Gaussian-fluctuated according to theoretical and experimental uncertainties.  

\begin{table}[tb!]
 \caption{Expected and observed 95\% C.L. allowed ranges for all aTGC parameters considered with the other parameters set 
to their SM values.  A form factor with unitarization scale equal to 4 TeV enforces unitarity for all aTGC parameters.  
The results are derived from the high-$q^2$ region yields given in Table~\ref{tab:yields}.
}
\label{aTGC:LEPlimits}
\begin{center}
\begin{tabular} {ccccc} 
\toprule
& \multicolumn{2}{c}{$\Lambda$ = 4 \TeV} & \multicolumn{2}{c}{$\Lambda$ = $\infty$}  \\ 
& Expected & Observed & Expected & Observed  \\ 
\midrule
$\Delta g_1^{Z}$ & $[-0.39, 0.35]$ & $[-0.32, 0.28]$ & $[-0.16, 0.15]$ & $[-0.13, 0.12]$ \\ [1ex]
$\Delta \kappa_{Z}$ & $[-0.38, 0.51]$ & $[-0.29, 0.42]$ & $[-0.19, 0.19]$ & $[-0.15, 0.16]$ \\ [1ex]  
$\lambda_{V}$ &  $[-0.16, 0.12]$ & $[-0.13, 0.090]$ & $[-0.064, 0.054]$  & $[-0.053, 0.042]$ \\ [1ex] 
$\tilde \kappa_{Z}$ &  $[-1.7, 1.8]$ & $[-1.4, 1.4]$ & $[-0.70, 0.70]$ & $[-0.56, 0.56]$ \\ [1ex] 
$\tilde \lambda_{V}$ &  $[-0.13, 0.15]$ & $[-0.10, 0.12]$ & $[-0.058, 0.057]$ & $[-0.047, 0.046]$ \\ [1ex] 
\bottomrule
\end{tabular}
\end{center}
 \end{table}

\begin{table}[tb!]
 \caption{Expected and observed 95\% C.L. intervals for individual EFT coefficients divided by the square 
of the new physics scale $\Lambda$, with other coefficients set to zero.  Intervals are calculated using 
the high-$q^2$ region yields (Table~\ref{tab:yields}). }
\label{aTGC:EFTlimits}
\begin{center}
\begin{tabular} {ccc}
\toprule
Parameter & Expected [\TeV $^{-2}$] & Observed [\TeV $^{-2}$] \\
\midrule
$\frac{c_W}{\Lambda^2}$  &  $[-39, 37]$ & $[-33, 30]$ \\ [1ex] 
$\frac{c_B}{\Lambda^2}$  &  $[-200, 190]$ & $[-170, 160]$ \\ [1ex] 
$\frac{c_{WWW}}{\Lambda^2}$ & $[-16, 13]$ & $[-13, 9]$ \\ [1ex] 
$\frac{c_{\tilde W}}{\Lambda^2}$ & $[-720, 720]$ & $[-580, 580]$ \\ [1ex] 
$\frac{c_{\tilde WWW}}{\Lambda^2}$ & $[-14, 14]$ & $[-11, 11]$ \\ [1ex] 
\bottomrule
\end{tabular}
\end{center}
 \end{table}

\begin{figure}[htbp]
\centering

    \includegraphics[width=0.45\textwidth]{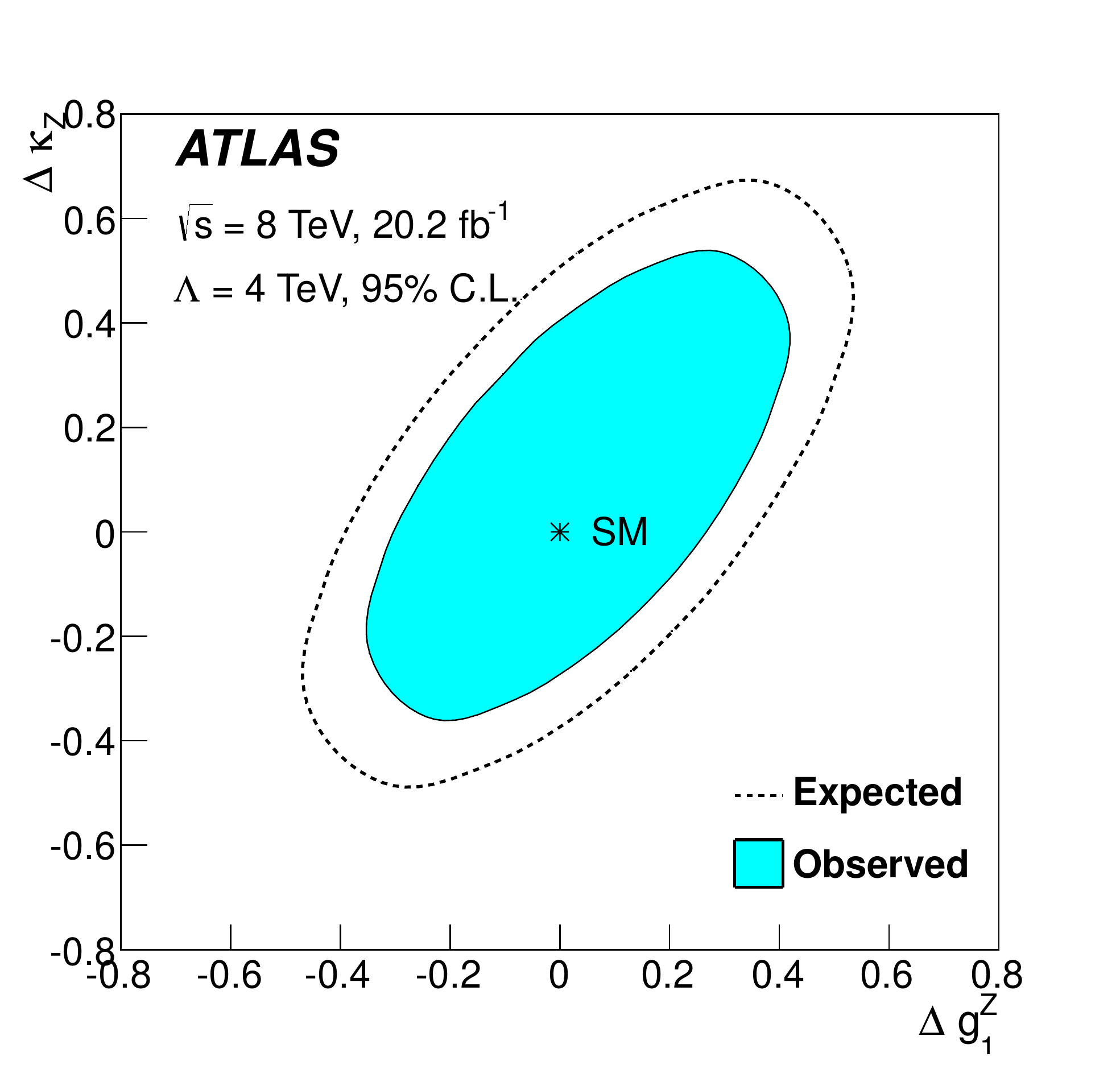}
    \includegraphics[width=0.45\textwidth]{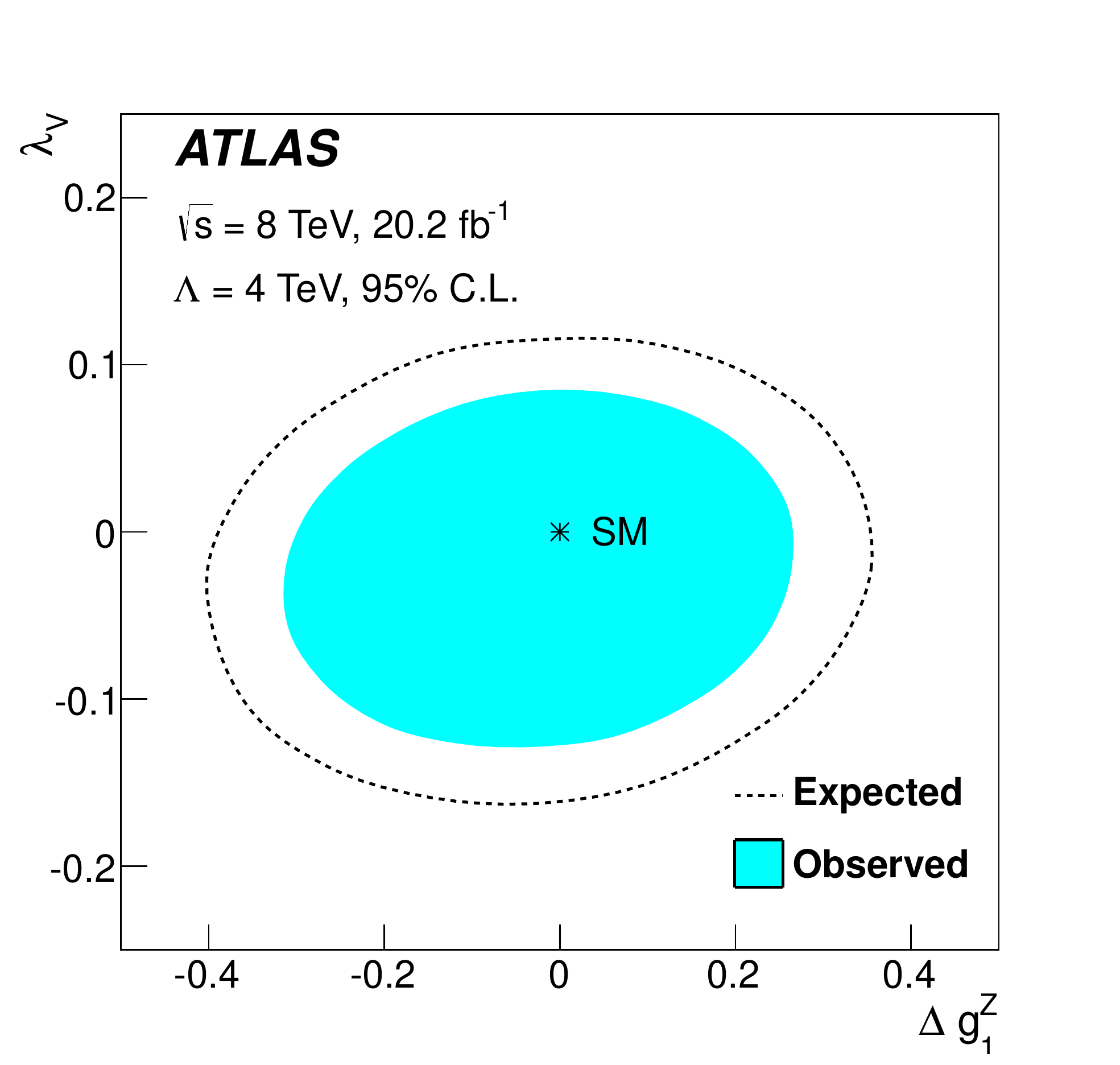}
    \includegraphics[width=0.45\textwidth]{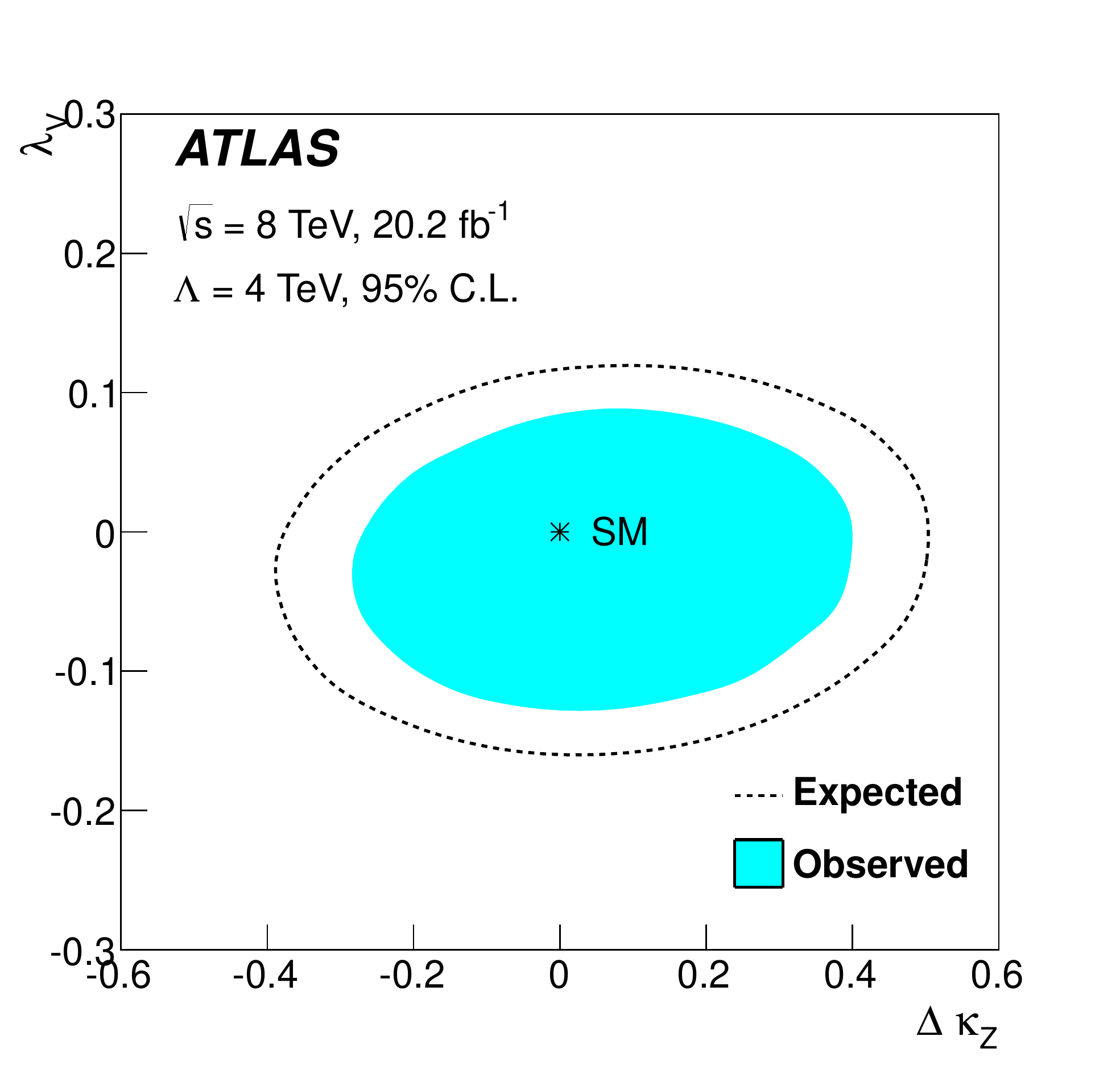}
  \caption{The observed (solid blue) and expected (open dashed) 95\% C.L. allowed regions in two-parameter planes for 
$\Lambda = 4$~\TeV.  The regions are derived using a single measured yield and therefore reduce to the corresponding 
one-parameter interval when the other parameter is set to zero.  Constraints on $\tilde{\lambda}_V$ are similar to those 
on $\lambda_V$. }
  \label{fig:aTGC:2dim_limits}
\end{figure}

Tables~\ref{aTGC:LEPlimits}~and~\ref{aTGC:EFTlimits} give the expected and observed 95\% C.L. interval for each 
parameter probed, with the other parameters set to their SM values.  All observed intervals are narrower than 
the expected intervals due to a slight deficit of data events compared with the SM prediction 
(Table~\ref{tab:yields}).  The $\lambda_V$ intervals are competitive with those derived from $WW$ 
production~\cite{WW_ATLAS}.  The 95\% C.L. regions in planes with two parameters deviating from their SM values 
are shown in Figure~\ref{fig:aTGC:2dim_limits}.  Since the regions are determined using a single measured yield, 
only the size of the region is constrained and not its shape.  Thus, along an axis where one parameter is equal 
to zero, the corresponding one-parameter C.L. interval is recovered.  The constraints on $\tilde{\lambda}_V$ are 
similar to $\lambda_V$ since the sensitivity is dominated by the square of the anomalous-coupling amplitude rather 
than its interference with the SM amplitude.

%% file: summary.tex
Measurements of the fiducial and differential cross sections of electroweak production 
of $W$ bosons in association with two jets have been performed using the lepton decay 
channel and events with high dijet invariant mass.  The measurements use data collected 
by the ATLAS detector from proton--proton collisions at the LHC at centre-of-mass 
energies of $\sqrt{s}=7$~and~8~\TeV, corresponding to 4.7 and 20.2~fb$^{-1}$ of 
integrated luminosity, respectively.  The cross sections in a fiducial region with a 
signal purity of ${\cal{O}}$(15\%) are 
\begin{eqnarray}
\sigma^\mathrm{fid}_{\mathrm{EW}~\ell\nu jj}~(7~\mathrm{TeV}) & = & 
144 \pm 23~\mathrm{(stat)}~\pm 23~\mathrm{(exp)}~\pm 13~\mathrm{(th)~fb}, \nonumber \\
\sigma^\mathrm{fid}_{\mathrm{EW}~\ell\nu jj}~(8~\mathrm{TeV}) & = & 
159 \pm 10~\mathrm{(stat)}~\pm 17~\mathrm{(exp)}~\pm 15~\mathrm{(th)~fb}, \nonumber
\end{eqnarray}

\noindent
corresponding to a deviation of $<0.1\sigma~(1.4\sigma)$ from the SM prediction of 
$144 \pm 11$ ($198 \pm 12$)~fb at $\sqrt{s}=7~(8)$~\TeV.  The large sample size of 
the 8~\TeV~measurement yields the smallest relative uncertainty of existing fiducial 
cross-section measurements of electroweak boson production in a VBF topology.

Differential cross sections of the $\sqrt{s}=8$~\TeV~electroweak \wjets production 
process are measured in a high-purity region with $\mjj>1$~\TeV.  The cross 
sections are measured as a function of dijet mass, dijet rapidity separation, 
dijet azimuthal angular separation, dijet $\pt$, leading-jet~$\pt$, the number 
of jets within the dijet rapidity gap, and lepton and jet centralities.  
Additionally, differential cross sections are measured in various fiducial 
regions for the combined electroweak and strong \wjets production with high 
dijet invariant mass.  The differential measurements are integrated in each 
fiducial region to obtain additional fiducial cross-section measurements.  The most 
inclusive region, where $\mjj>0.5$~\TeV, $\dyjj > 2$, $\ptjlead > 80$~\GeV, and 
$\ptjsub > 60$~\GeV, has a measured QCD+EW fiducial cross section at $\sqrt{s}=8$~\TeV~of 
$\sigma^\mathrm{fid}_{\mathrm{QCD+EW}~\ell\nu jj} =1700 \pm 110$~fb.

The region of increased purity for electroweak production of~\wjets ($\mjj>1$~\TeV) 
is used to constrain dimension-six triple-gauge-boson operators motivated by an 
effective field theory.  To improve the sensitivity to high-scale physics affecting the 
triple-gauge-boson vertex, events with leading-jet~$\pt >600$~\GeV~are also used to 
constrain CP-conserving and CP-violating operators in the HISZ scenario, both with 
and without a form-factor suppression.  A 95\% C.L. range of $[-0.13,0.09]$ is 
determined for $\lambda_V$ with a suppression scale of 4~\TeV~and the other parameters 
set to their SM values.  Limits are also set on the parameters of an effective field theory. 
The operator coefficient $c_{WWW}/\Lambda^2$ is proportional to $\lambda_V$ and is 
constrained to $[-13, 9]/{\mathrm{\TeV}}^2$ at 95\% C.L.  Constraints on CP-violating 
operators are similar to those on the CP-conserving operators.

%% file: Acknowledgements.tex

We thank CERN for the very successful operation of the LHC, as well as the
support staff from our institutions without whom ATLAS could not be
operated efficiently.

We acknowledge the support of ANPCyT, Argentina; YerPhI, Armenia; ARC, Australia; BMWFW and FWF, Austria; ANAS, Azerbaijan; SSTC, Belarus; CNPq and FAPESP, Brazil; NSERC, NRC and CFI, Canada; CERN; CONICYT, Chile; CAS, MOST and NSFC, China; COLCIENCIAS, Colombia; MSMT CR, MPO CR and VSC CR, Czech Republic; DNRF and DNSRC, Denmark; IN2P3-CNRS, CEA-DSM/IRFU, France; SRNSF, Georgia; BMBF, HGF, and MPG, Germany; GSRT, Greece; RGC, Hong Kong SAR, China; ISF, I-CORE and Benoziyo Center, Israel; INFN, Italy; MEXT and JSPS, Japan; CNRST, Morocco; NWO, Netherlands; RCN, Norway; MNiSW and NCN, Poland; FCT, Portugal; MNE/IFA, Romania; MES of Russia and NRC KI, Russian Federation; JINR; MESTD, Serbia; MSSR, Slovakia; ARRS and MIZ\v{S}, Slovenia; DST/NRF, South Africa; MINECO, Spain; SRC and Wallenberg Foundation, Sweden; SERI, SNSF and Cantons of Bern and Geneva, Switzerland; MOST, Taiwan; TAEK, Turkey; STFC, United Kingdom; DOE and NSF, United States of America. In addition, individual groups and members have received support from BCKDF, the Canada Council, CANARIE, CRC, Compute Canada, FQRNT, and the Ontario Innovation Trust, Canada; EPLANET, ERC, ERDF, FP7, Horizon 2020 and Marie Sk{\l}odowska-Curie Actions, European Union; Investissements d'Avenir Labex and Idex, ANR, R{\'e}gion Auvergne and Fondation Partager le Savoir, France; DFG and AvH Foundation, Germany; Herakleitos, Thales and Aristeia programmes co-financed by EU-ESF and the Greek NSRF; BSF, GIF and Minerva, Israel; BRF, Norway; CERCA Programme Generalitat de Catalunya, Generalitat Valenciana, Spain; the Royal Society and Leverhulme Trust, United Kingdom.

The crucial computing support from all WLCG partners is acknowledged gratefully, in particular from CERN, the ATLAS Tier-1 facilities at TRIUMF (Canada), NDGF (Denmark, Norway, Sweden), CC-IN2P3 (France), KIT/GridKA (Germany), INFN-CNAF (Italy), NL-T1 (Netherlands), PIC (Spain), ASGC (Taiwan), RAL (UK) and BNL (USA), the Tier-2 facilities worldwide and large non-WLCG resource providers. Major contributors of computing resources are listed in Ref.~\cite{ATL-GEN-PUB-2016-002}.

%% file: appendix.tex
\begin{figure}[htbp]\centering
\includegraphics[width=0.4\textwidth]{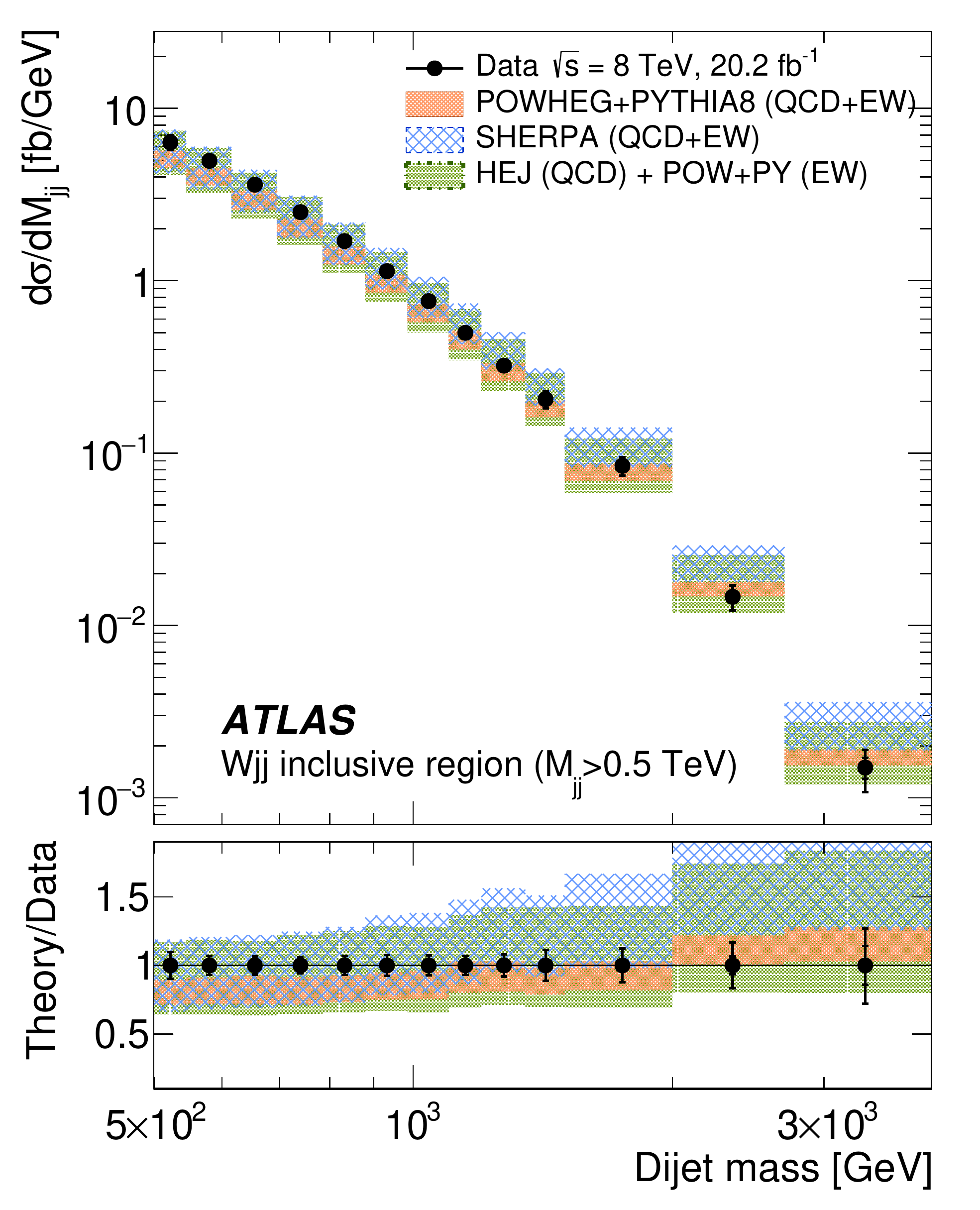}
\includegraphics[width=0.4\textwidth]{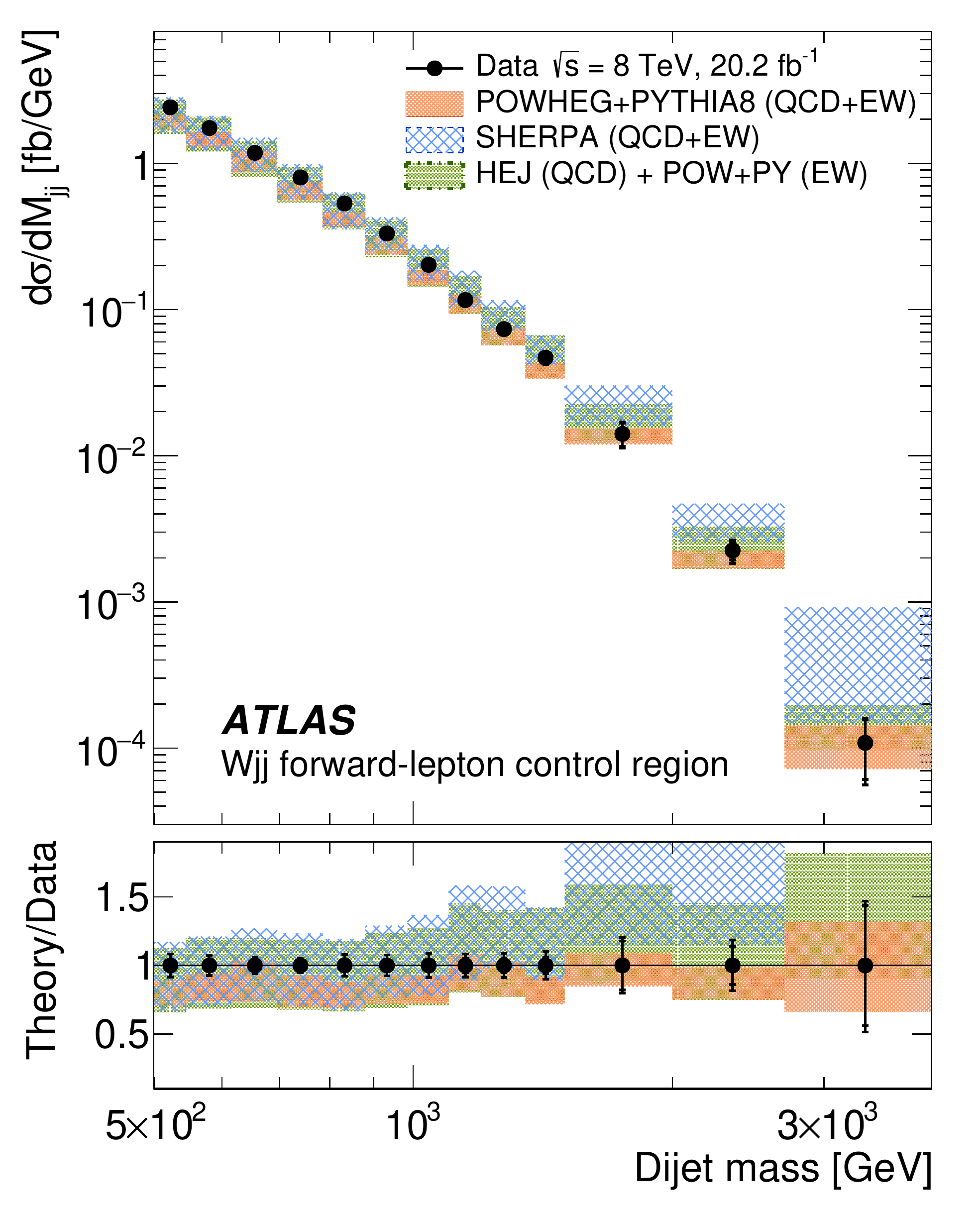}
\includegraphics[width=0.4\textwidth]{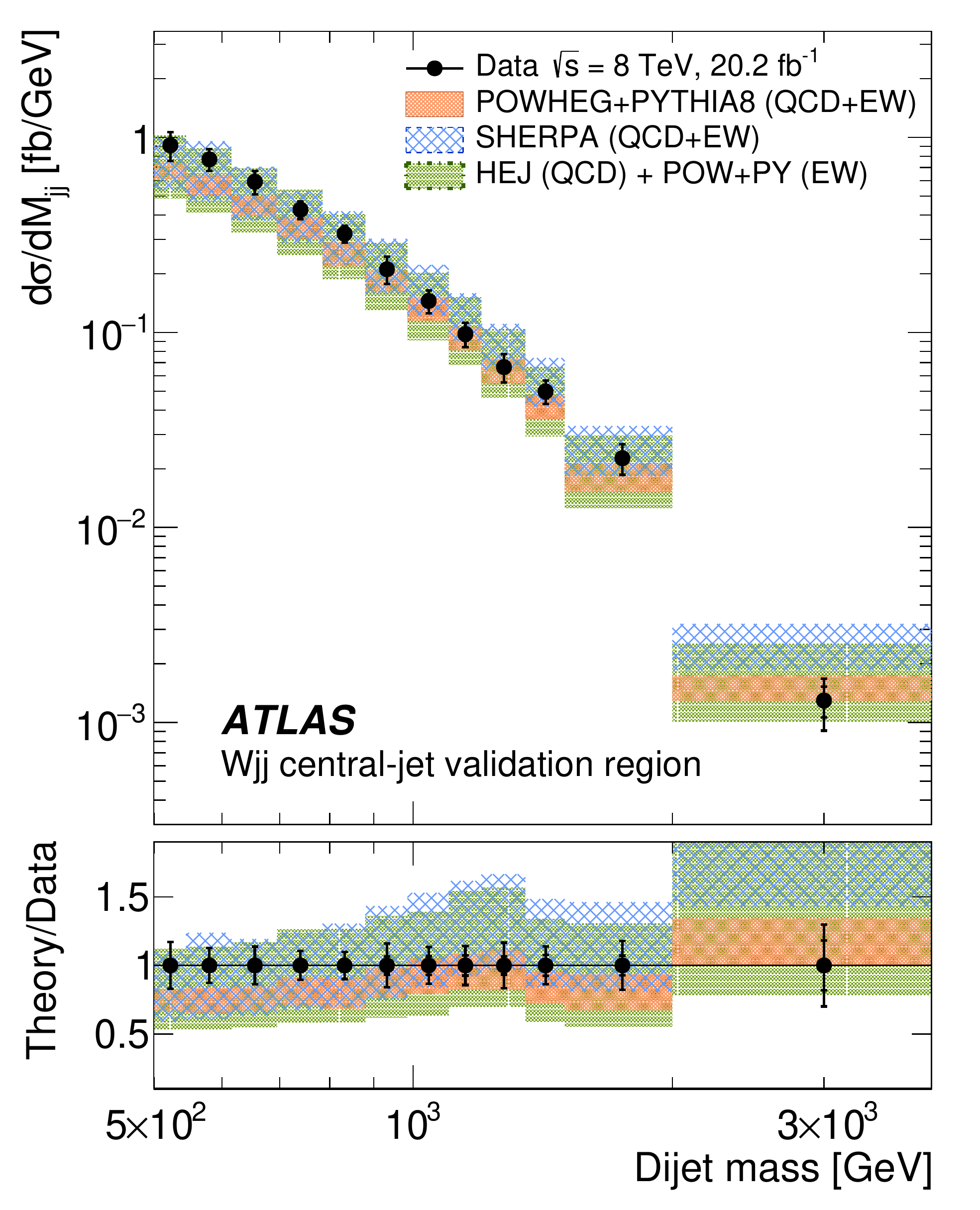}
\includegraphics[width=0.4\textwidth]{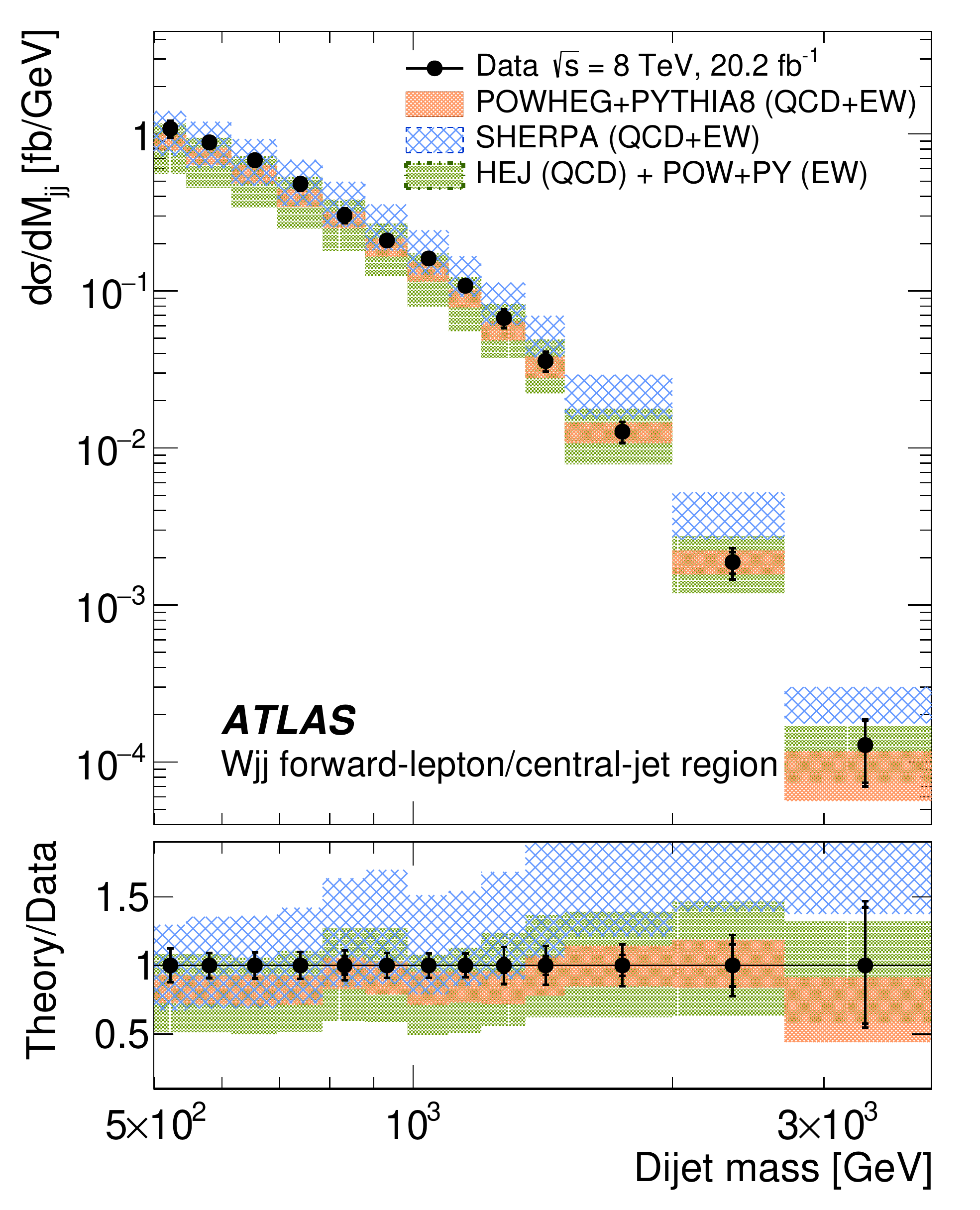}
\caption{Unfolded differential \wjets production cross sections as a function of dijet mass
for the inclusive (top left), forward-lepton (top right), central-jet (bottom left), and
forward-lepton/central-jet (bottom right) fiducial regions, which are enriched in strong
\wjets production.  Both statistical (inner bar) and total (outer bar) measurement uncertainties
are shown, as well as ratios of the theoretical predictions to the data (the bottom panel in each
distribution).}
\label{unfolding:combined_measurementdijetmass1Dinclusive}
\end{figure}
\begin{figure}[htbp]
\centering
\includegraphics[width=0.49\textwidth]{figures/unfolding/normalisedXsec/measurement_combined_dy12_1D_inclusive.pdf}
\includegraphics[width=0.49\textwidth]{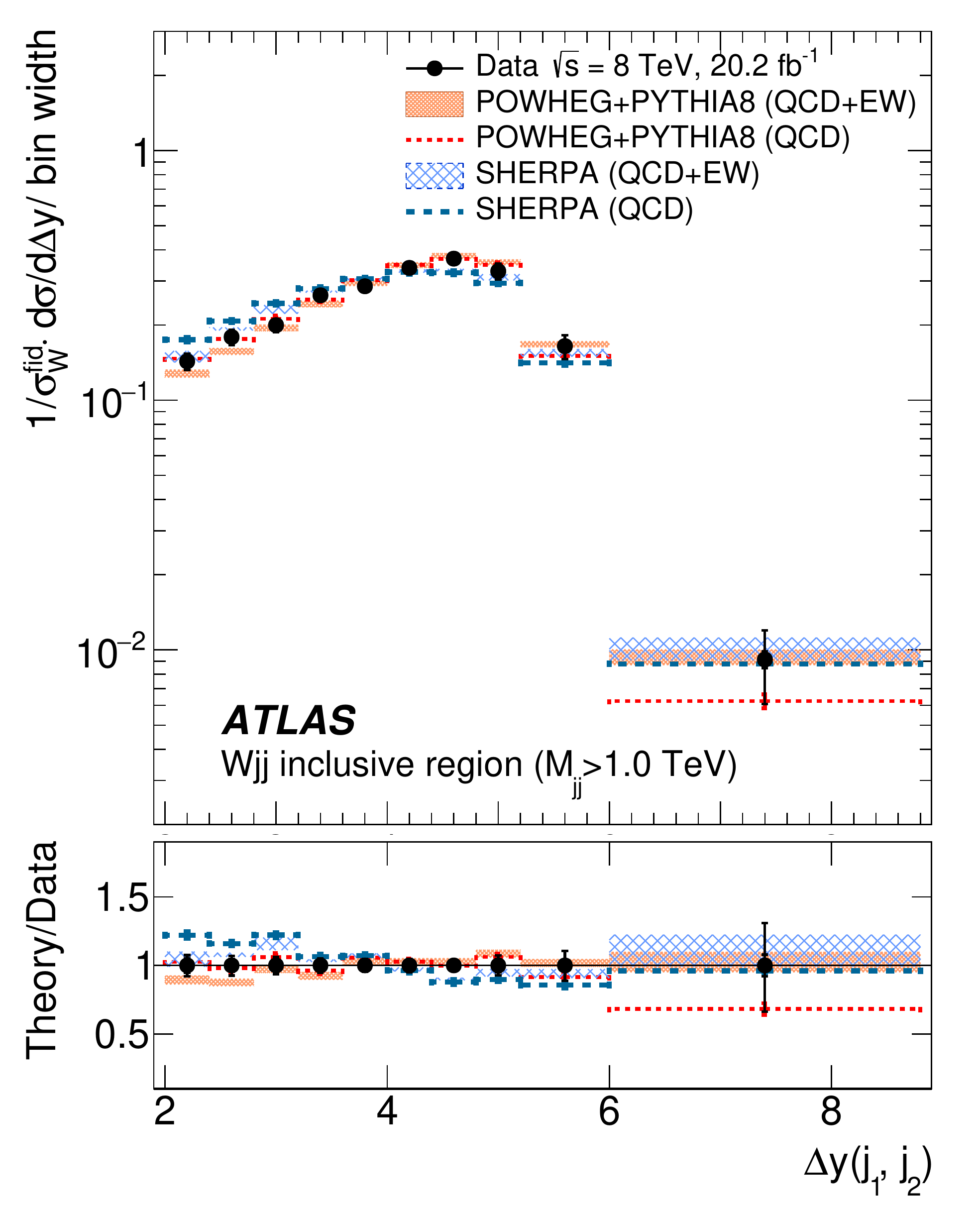}
\includegraphics[width=0.49\textwidth]{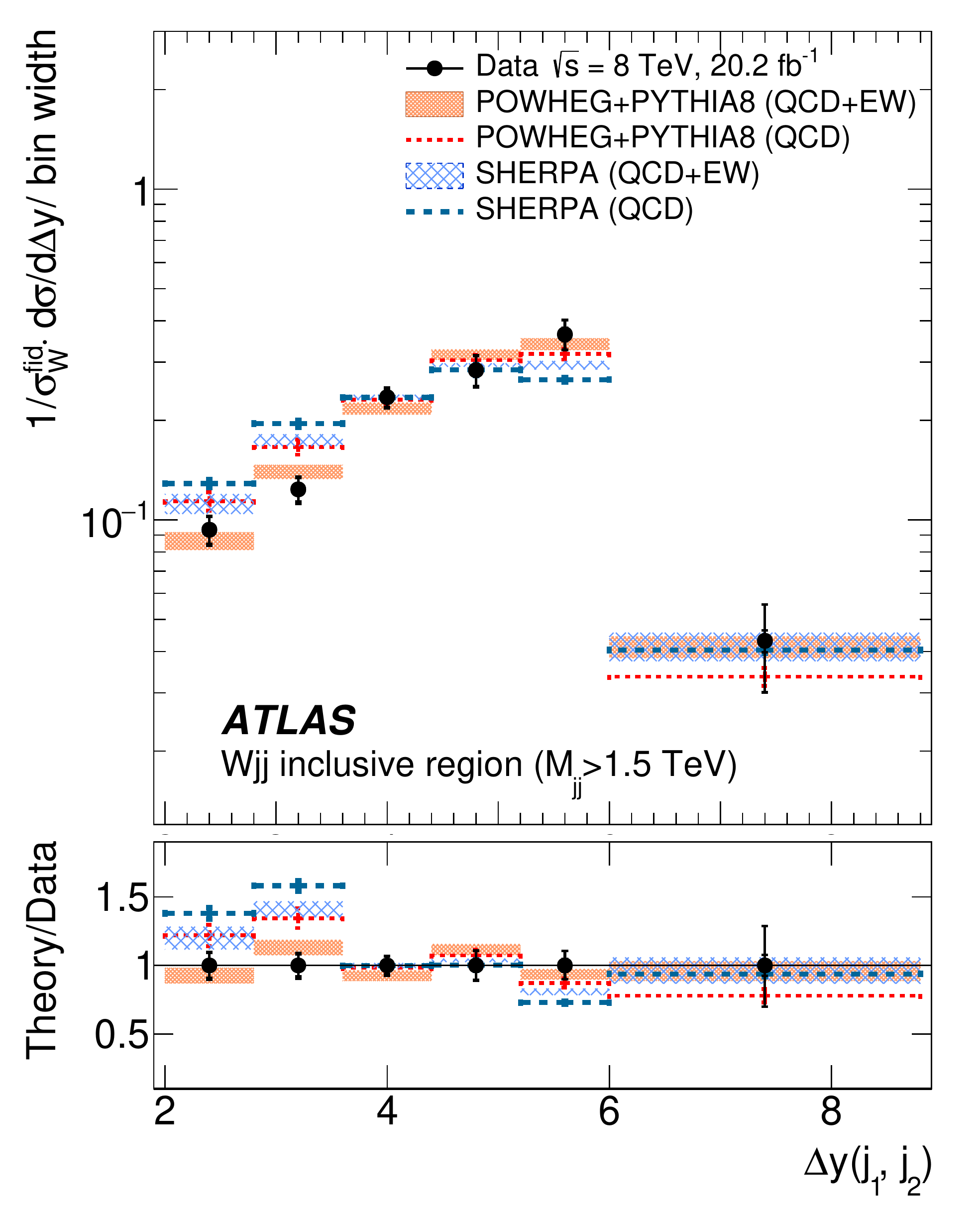}
\includegraphics[width=0.49\textwidth]{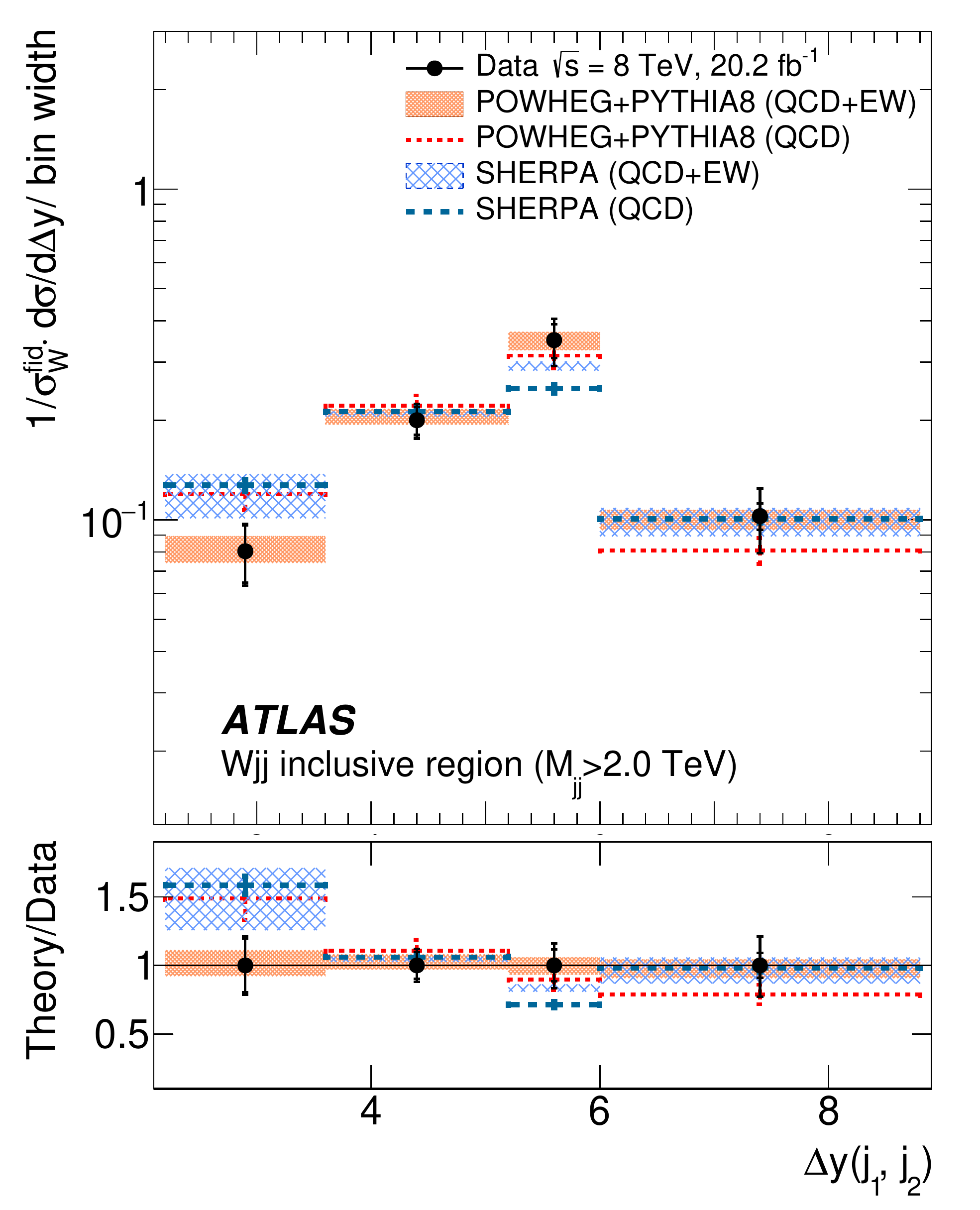}
\caption{Unfolded normalized differential \wjets production cross sections as a function of $\dyjj$ in the inclusive 
fiducial region with four thresholds on the dijet invariant mass (0.5~\TeV, 1.0~\TeV, 1.5~\TeV, and 2.0~\TeV). 
Both statistical (inner bar) and total (outer bar) measurement uncertainties are shown, as well as ratios
of the theoretical predictions to the data (the bottom panel in each distribution).}
\label{unfolding:aux:AUX9}
\end{figure}
\begin{figure}[htbp]
\centering
\includegraphics[width=0.48\textwidth]{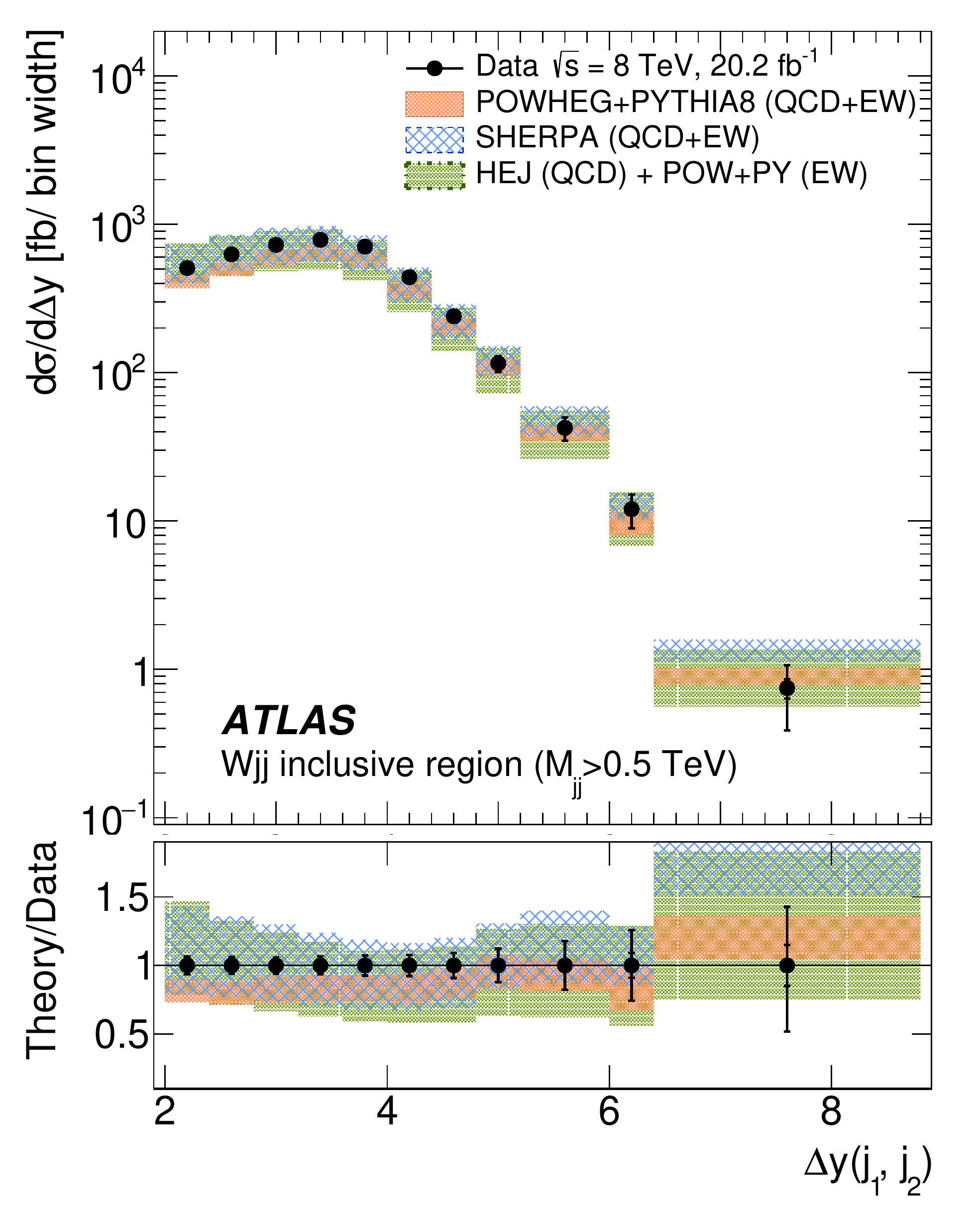}
\includegraphics[width=0.48\textwidth]{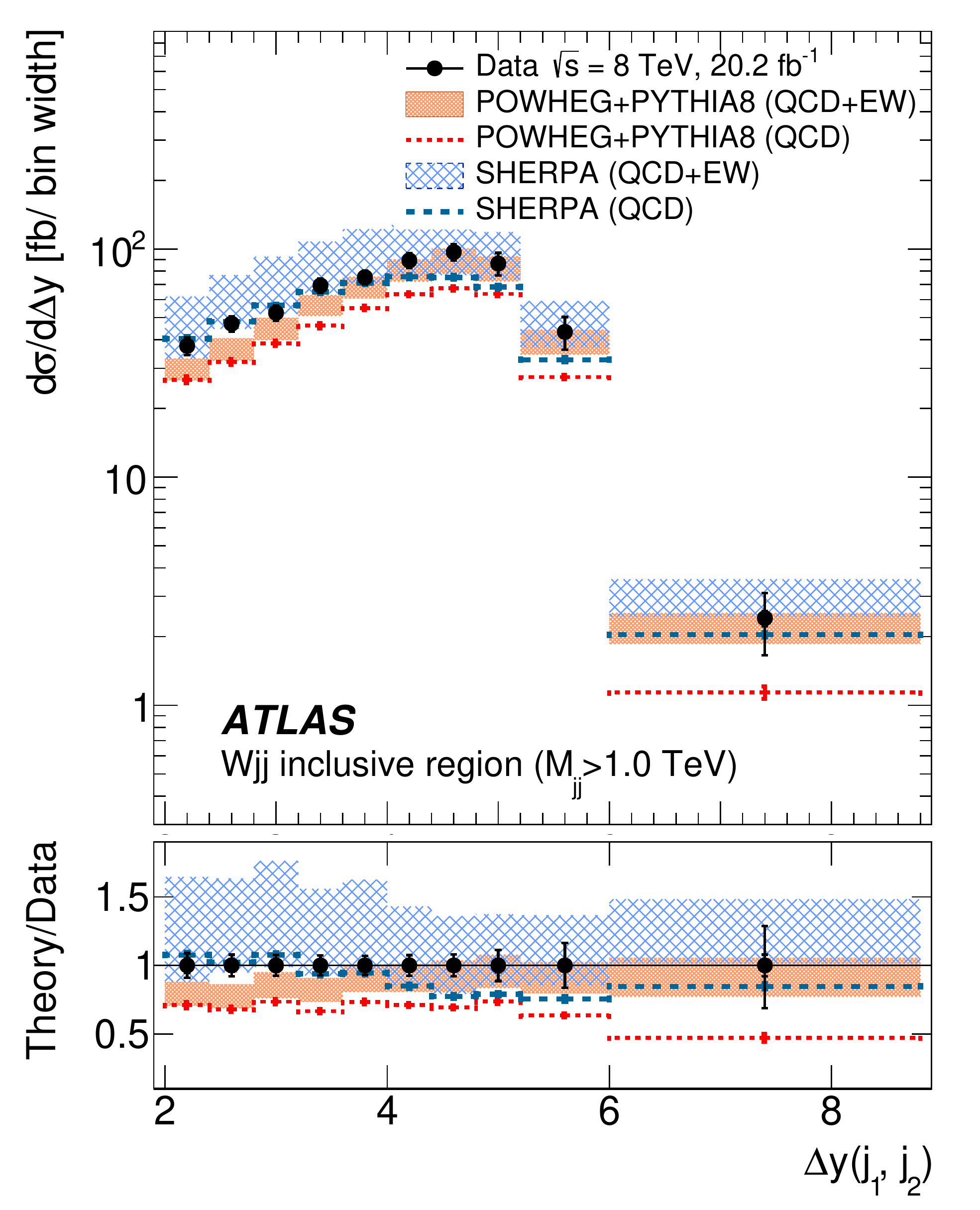}
\includegraphics[width=0.48\textwidth]{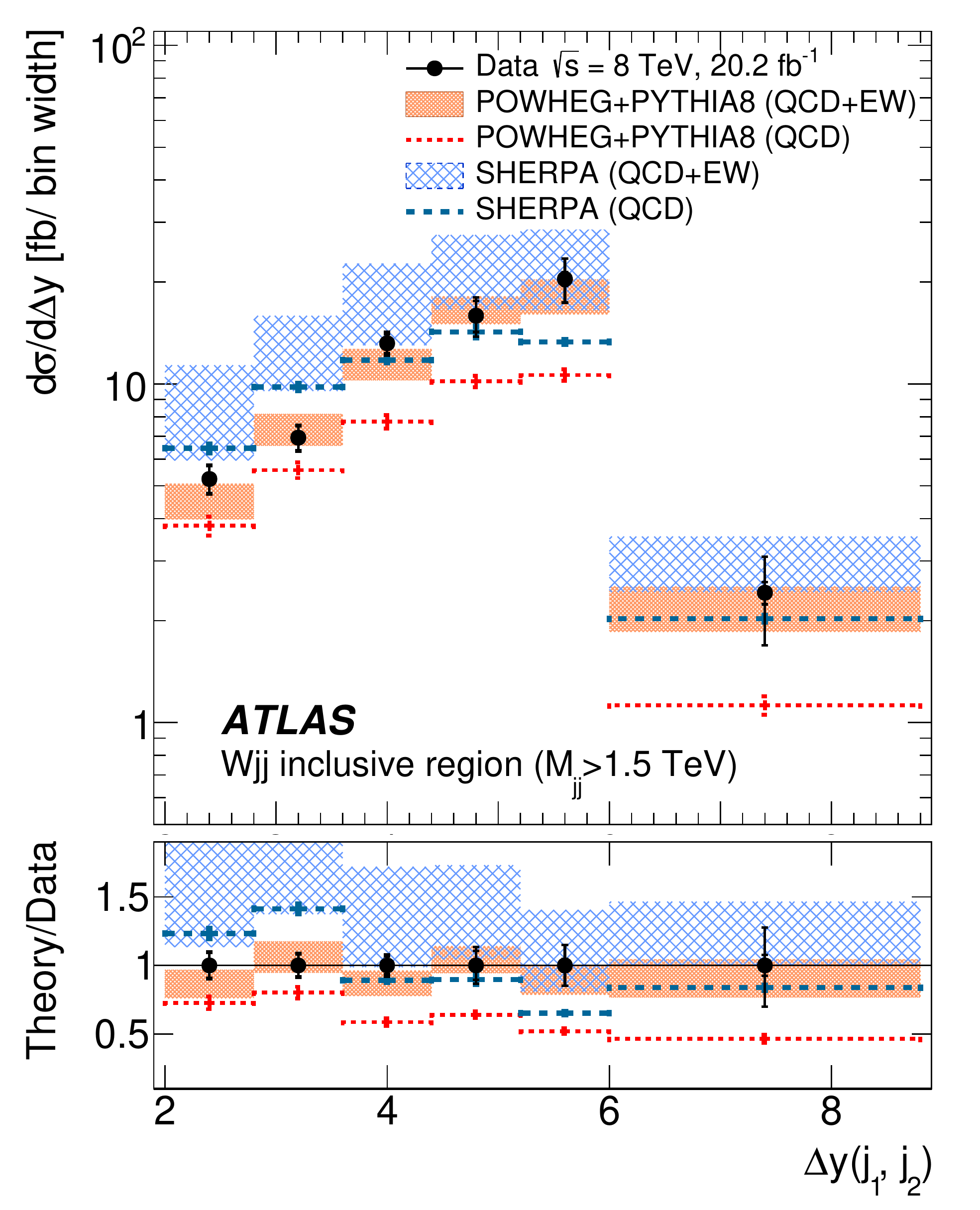}
\includegraphics[width=0.48\textwidth]{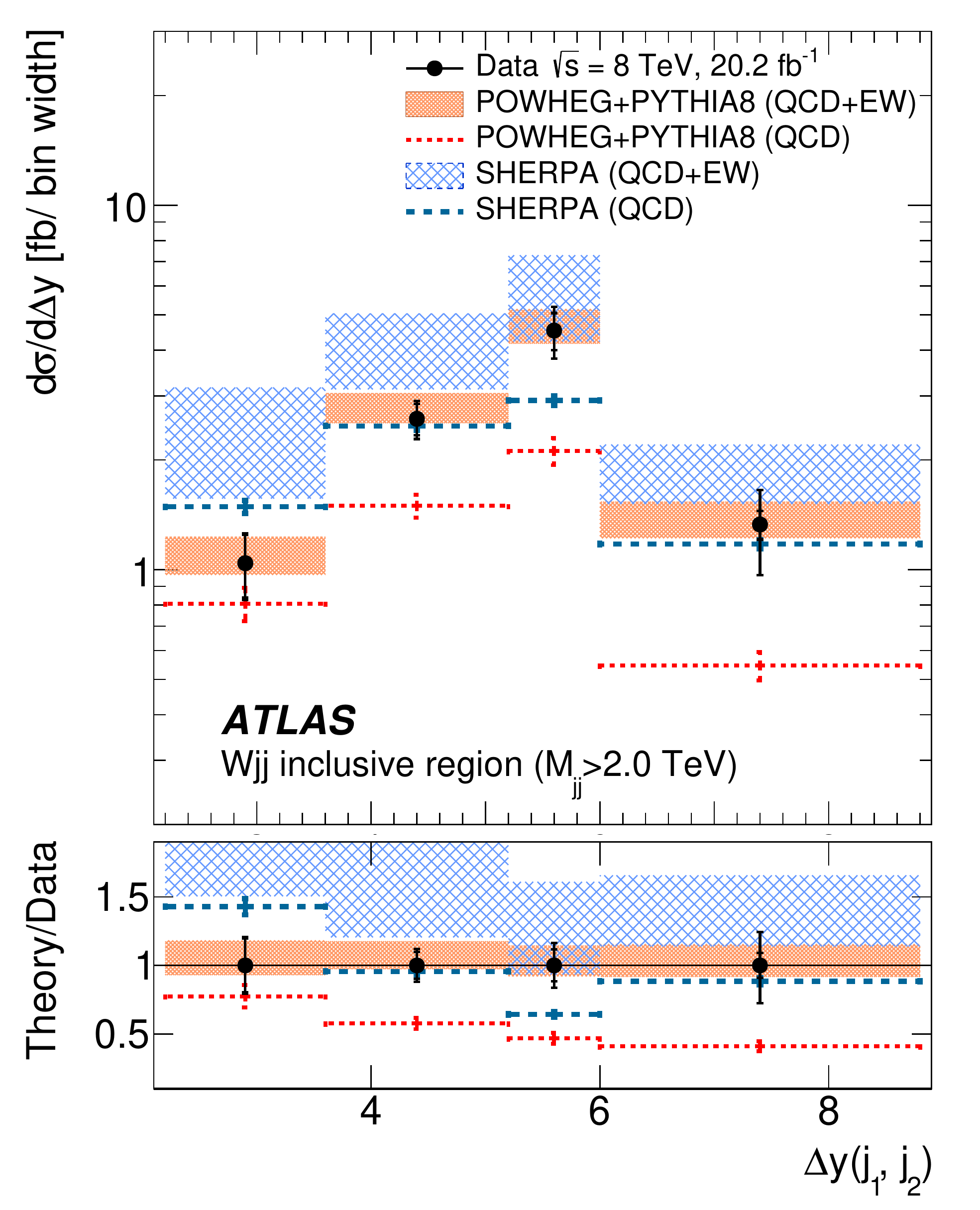}
\caption{Unfolded absolute differential \wjets production cross sections as a function of
$\dyjj$ for the inclusive fiducial region with progressively increasing dijet mass thresholds.
Both statistical (inner bar) and total (outer bar) measurement uncertainties are shown, as well
as ratios of the theoretical predictions to the data (the bottom panel in each distribution).}
\label{unfolding:combined_measurementdy121Dinclusive}
\end{figure}
\begin{figure}[htbp]
\centering
\includegraphics[width=0.35\textwidth]{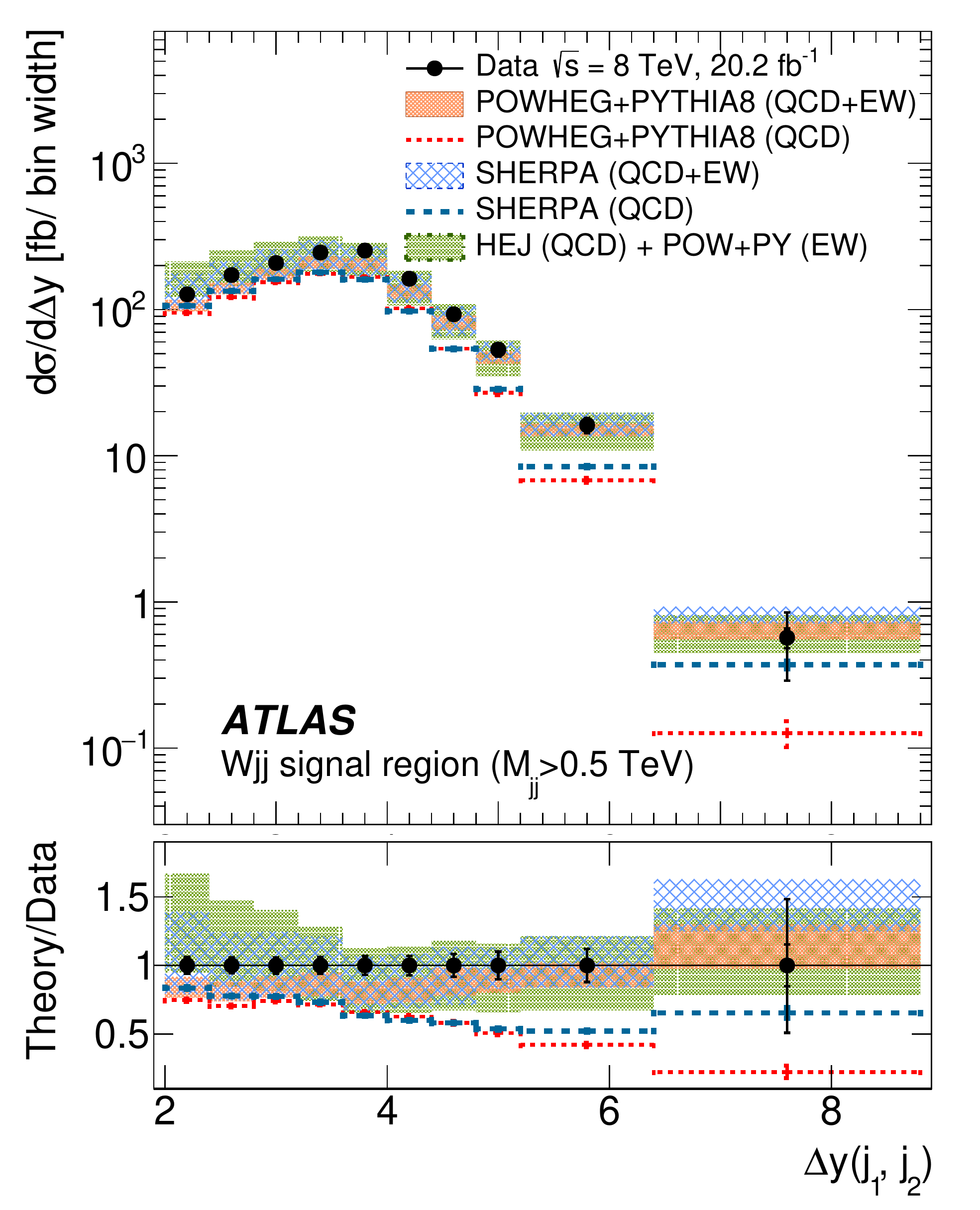}
\includegraphics[width=0.35\textwidth]{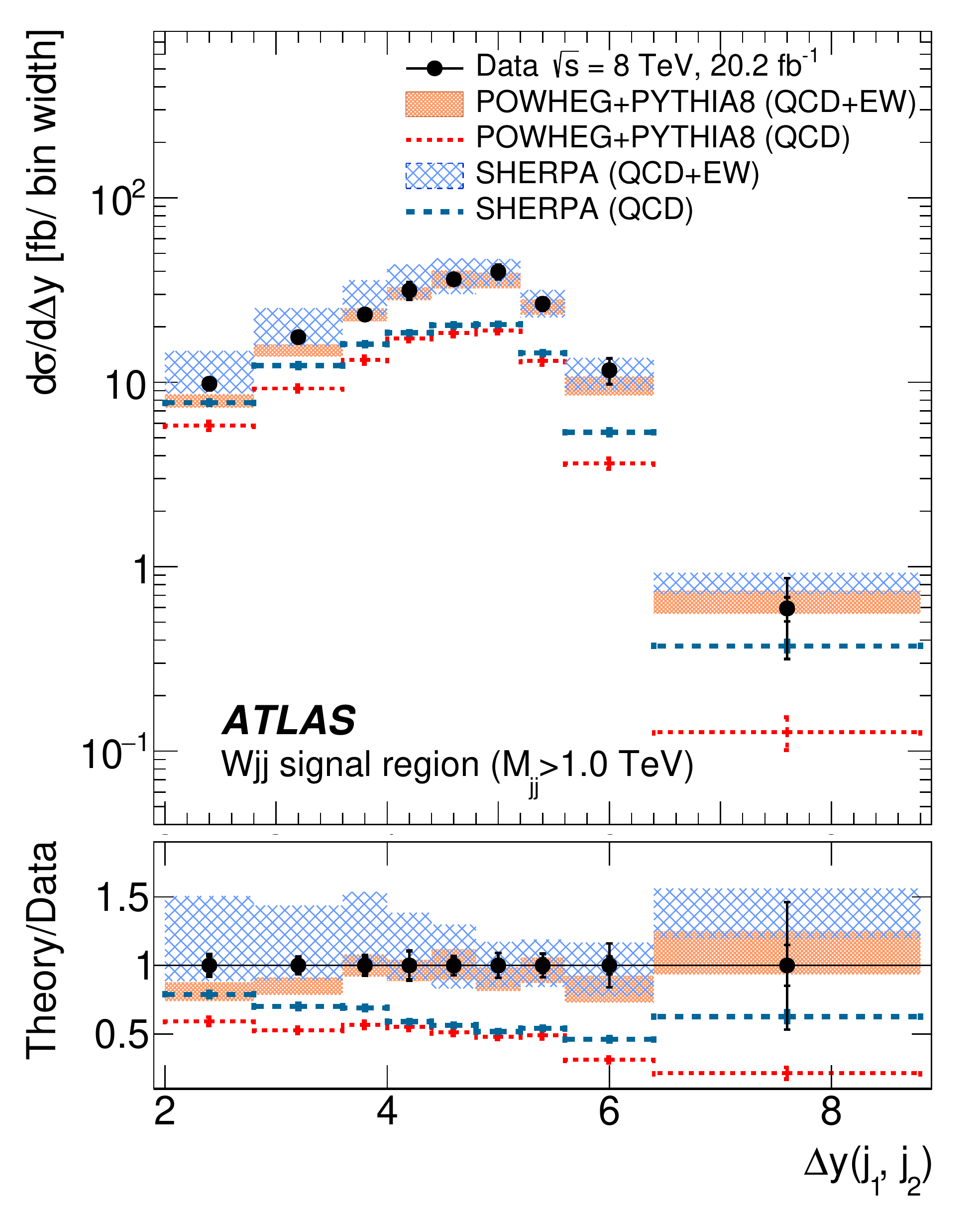}
\includegraphics[width=0.35\textwidth]{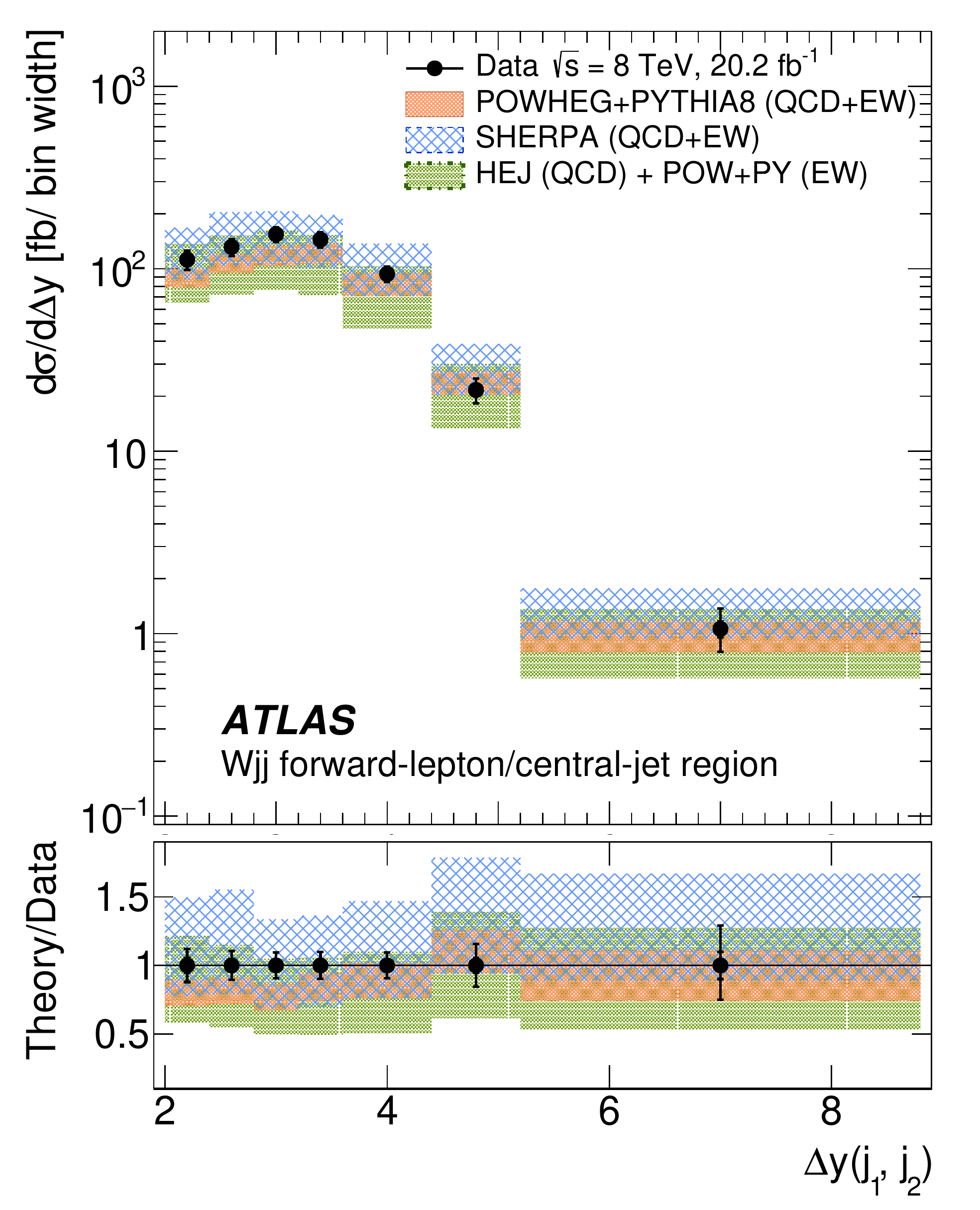}
\includegraphics[width=0.35\textwidth]{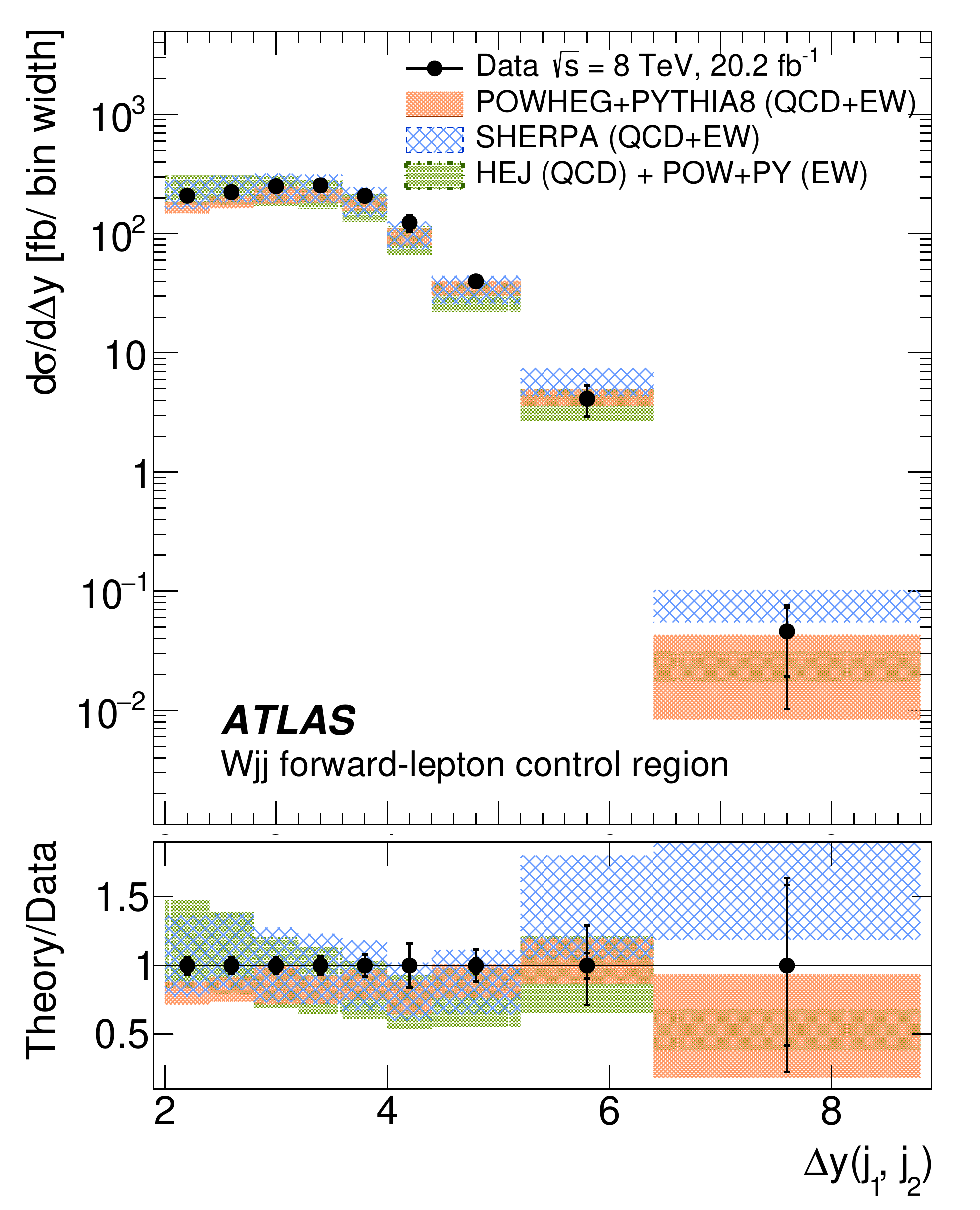}
\includegraphics[width=0.35\textwidth]{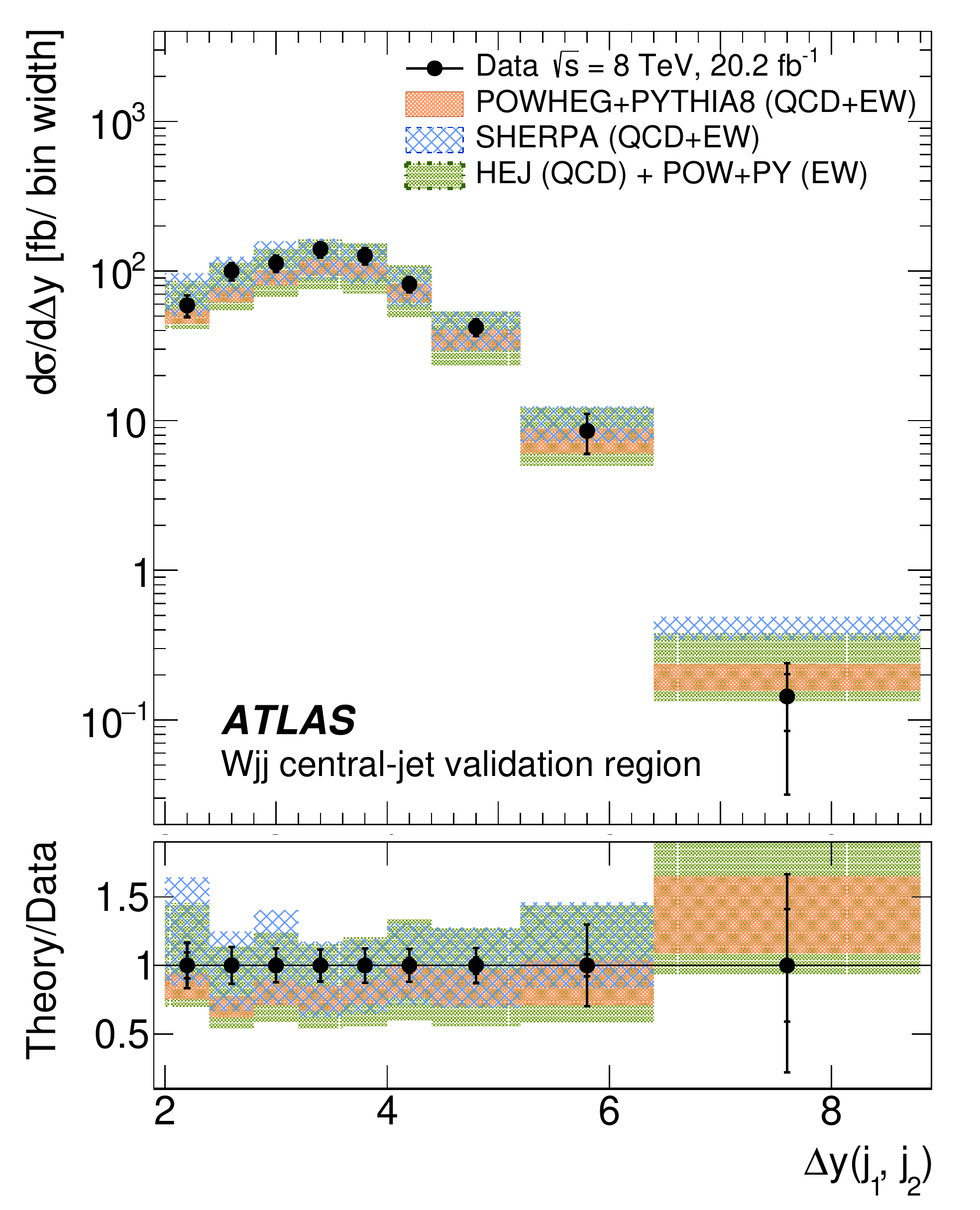}
\caption{Differential \wjets production cross sections as a function of $\dyjj$ in the signal and high-mass signal 
fiducial regions, and in the forward-lepton, central-jet validation, and forward-lepton/central-jet fiducial regions.  
Both statistical (inner bar) and total (outer bar) measurement uncertainties are shown, as well as ratios
of the theoretical predictions to the data (the bottom panel in each distribution). }
\label{unfolding:aux:AUX18}
\end{figure}
\begin{figure}[htbp]
\centering
\includegraphics[width=0.49\textwidth]{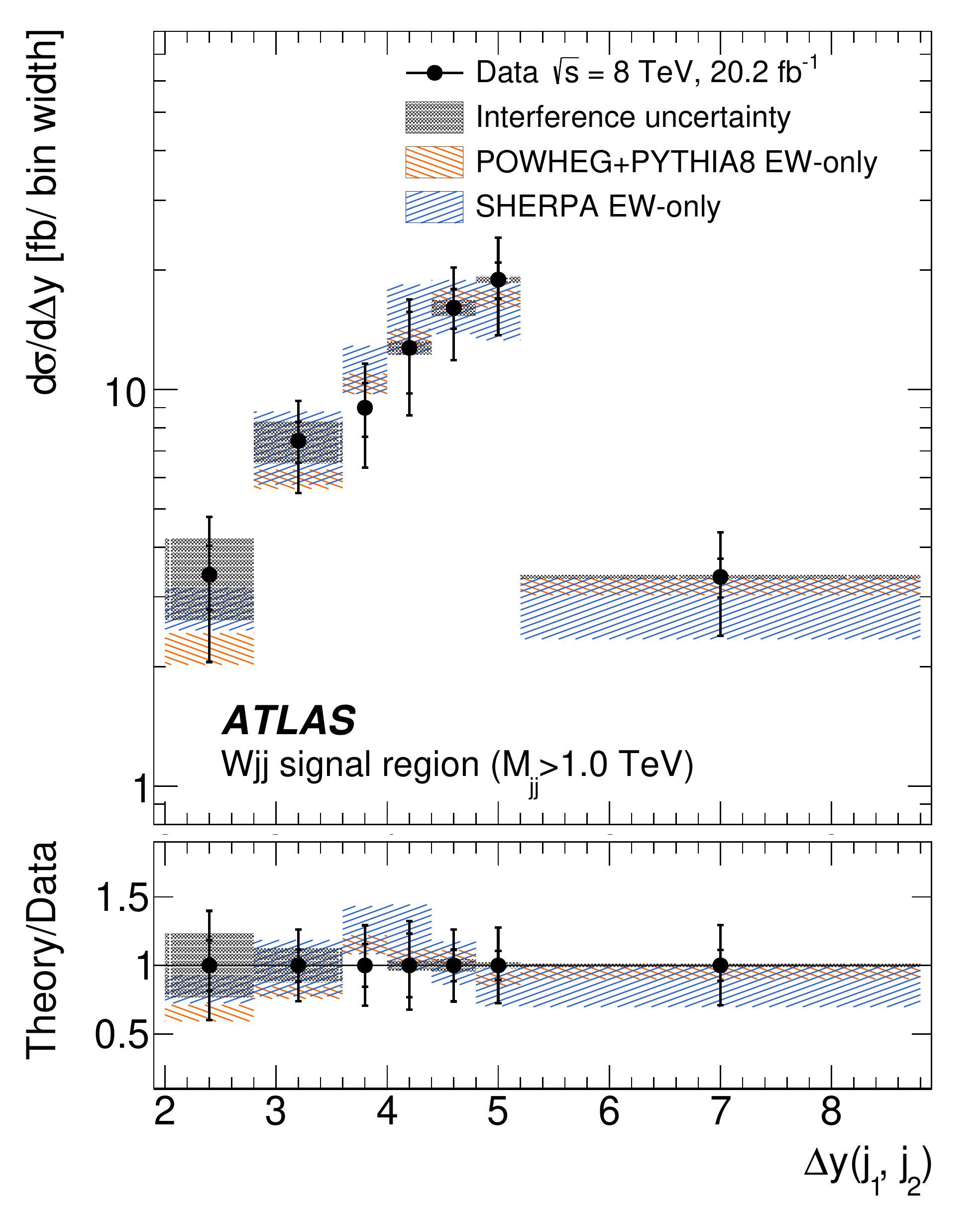}
\includegraphics[width=0.49\textwidth]{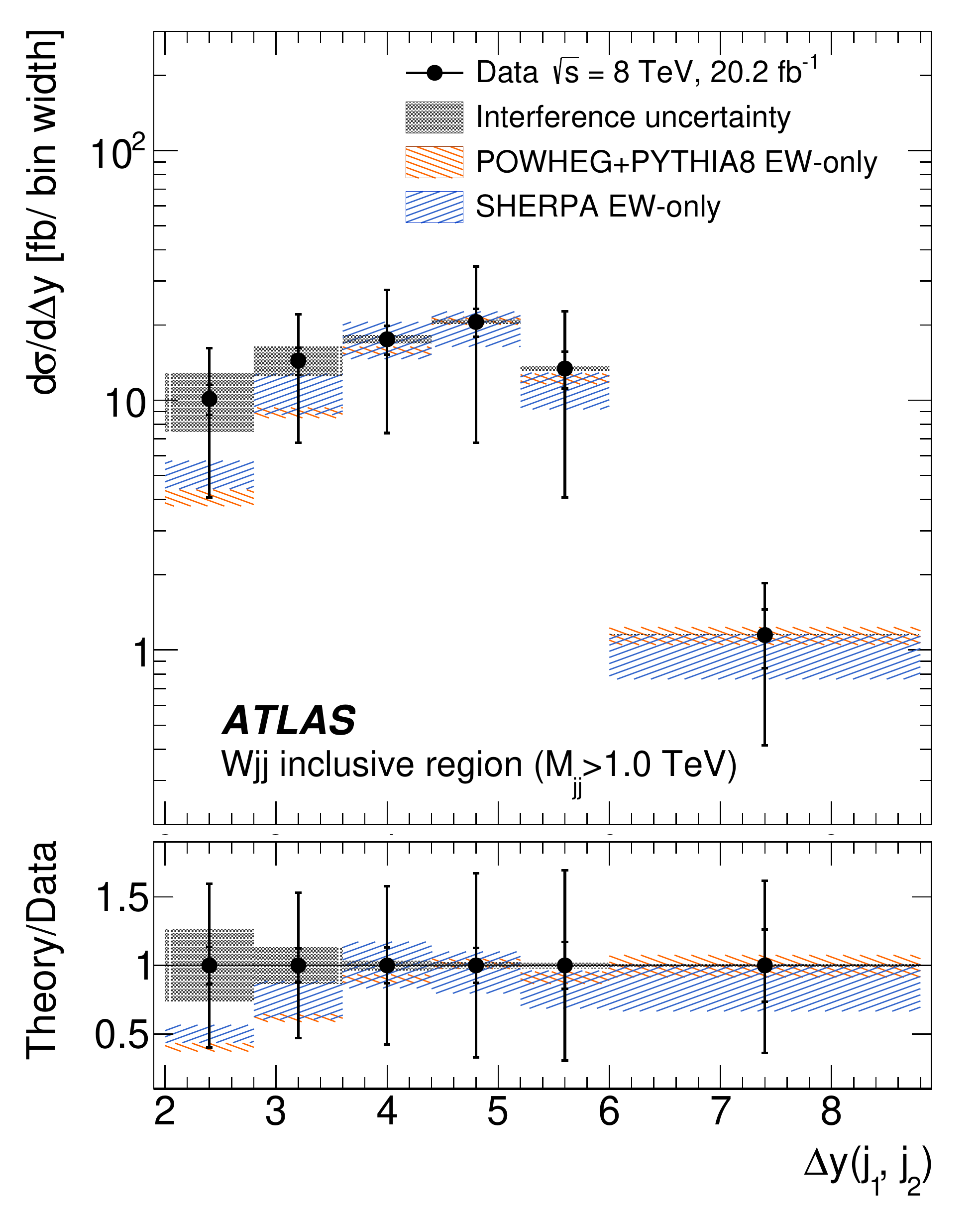}
\includegraphics[width=0.49\textwidth]{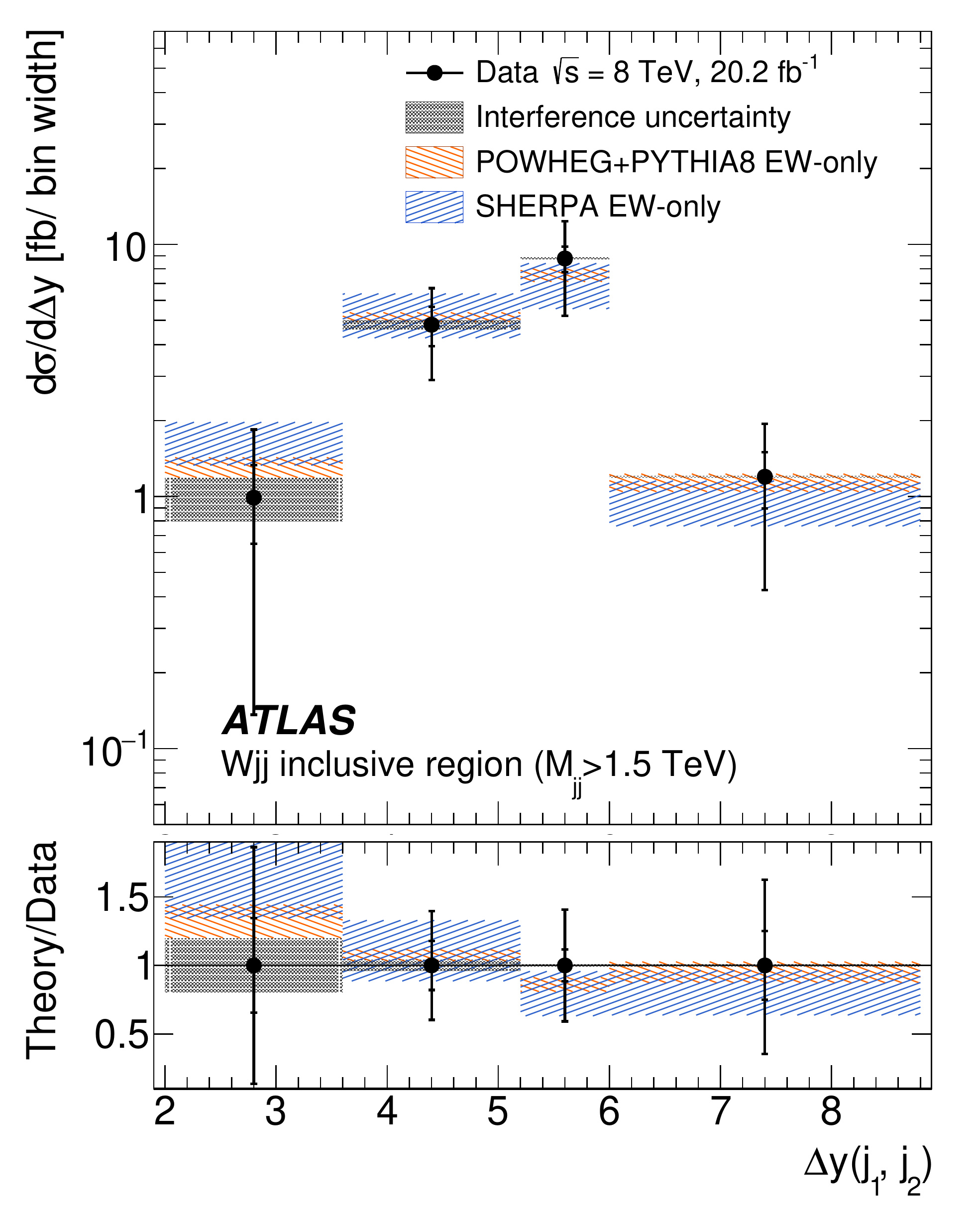}
\includegraphics[width=0.49\textwidth]{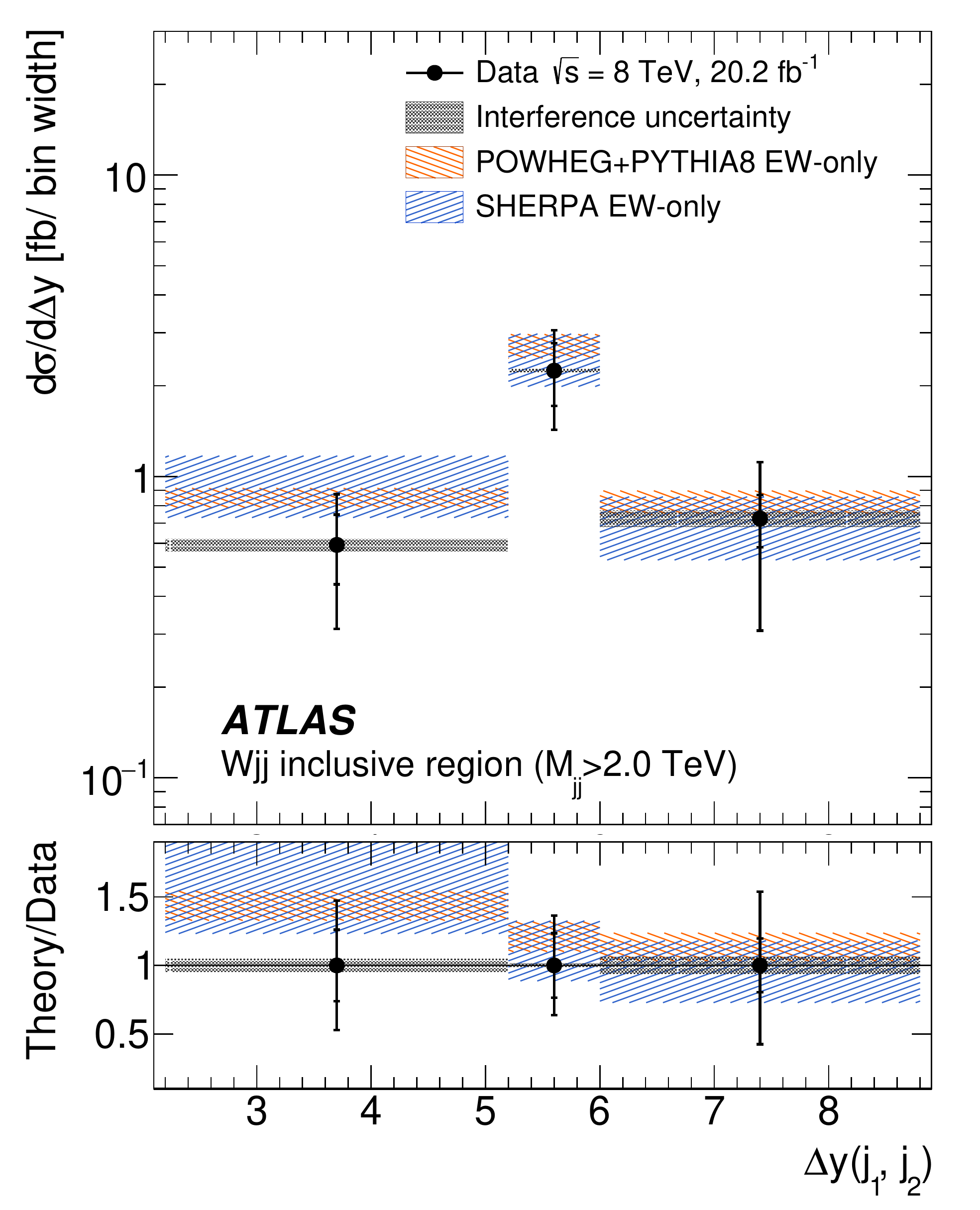}
\caption{Differential electroweak \wjets production cross sections as a function of $\dyjj$ in the high-mass signal 
region and the inclusive fiducial region with three thresholds on the dijet invariant mass (1.0~\TeV, 1.5~\TeV, and 
2.0~\TeV).  Both statistical (inner bar) and total (outer bar) measurement uncertainties are shown, as well as ratios
of the theoretical predictions to the data (the bottom panel in each distribution). }
\label{unfolding:aux:AUX5}
\end{figure}
\begin{figure}[htbp]
\centering
\includegraphics[width=0.49\textwidth]{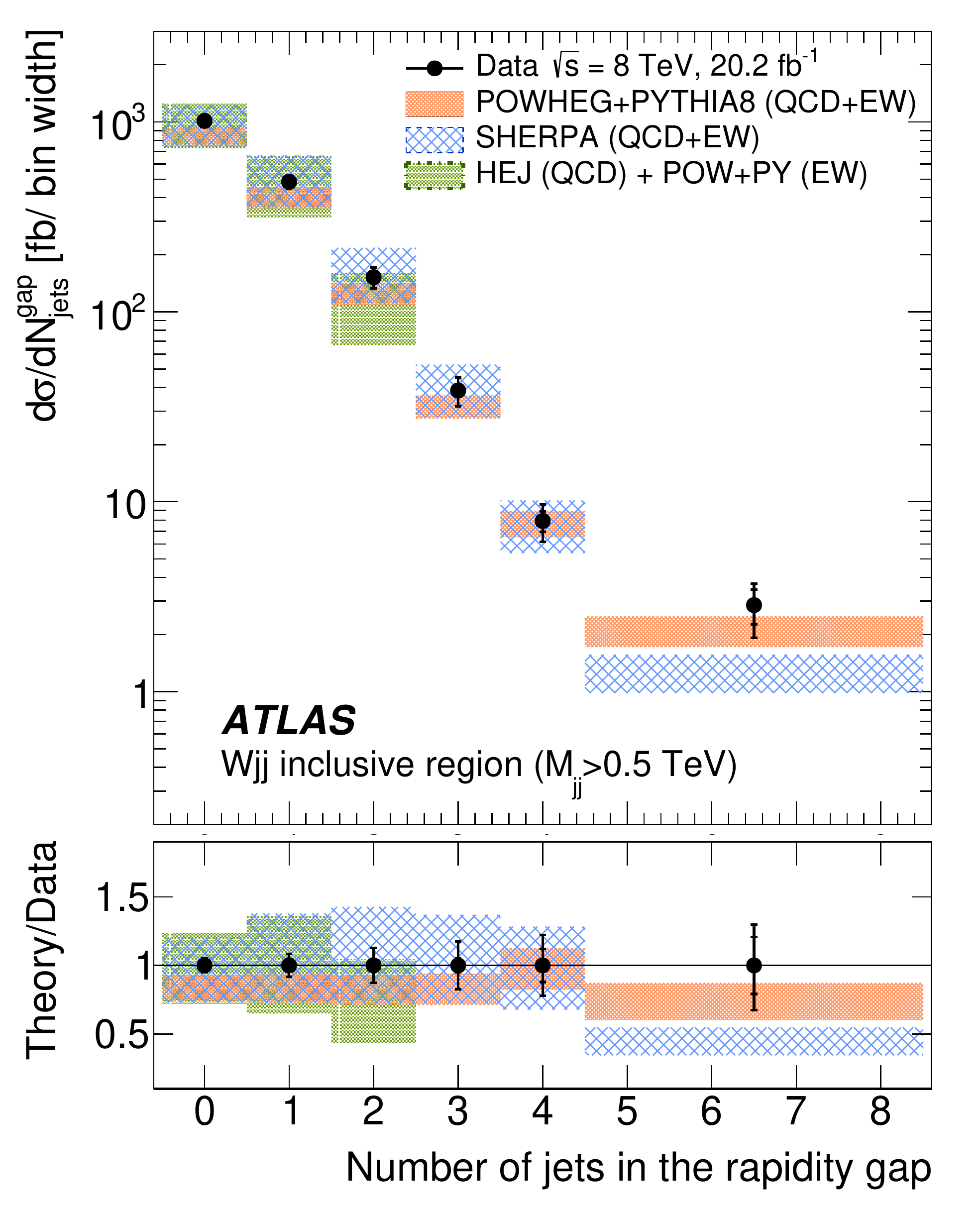}
\includegraphics[width=0.49\textwidth]{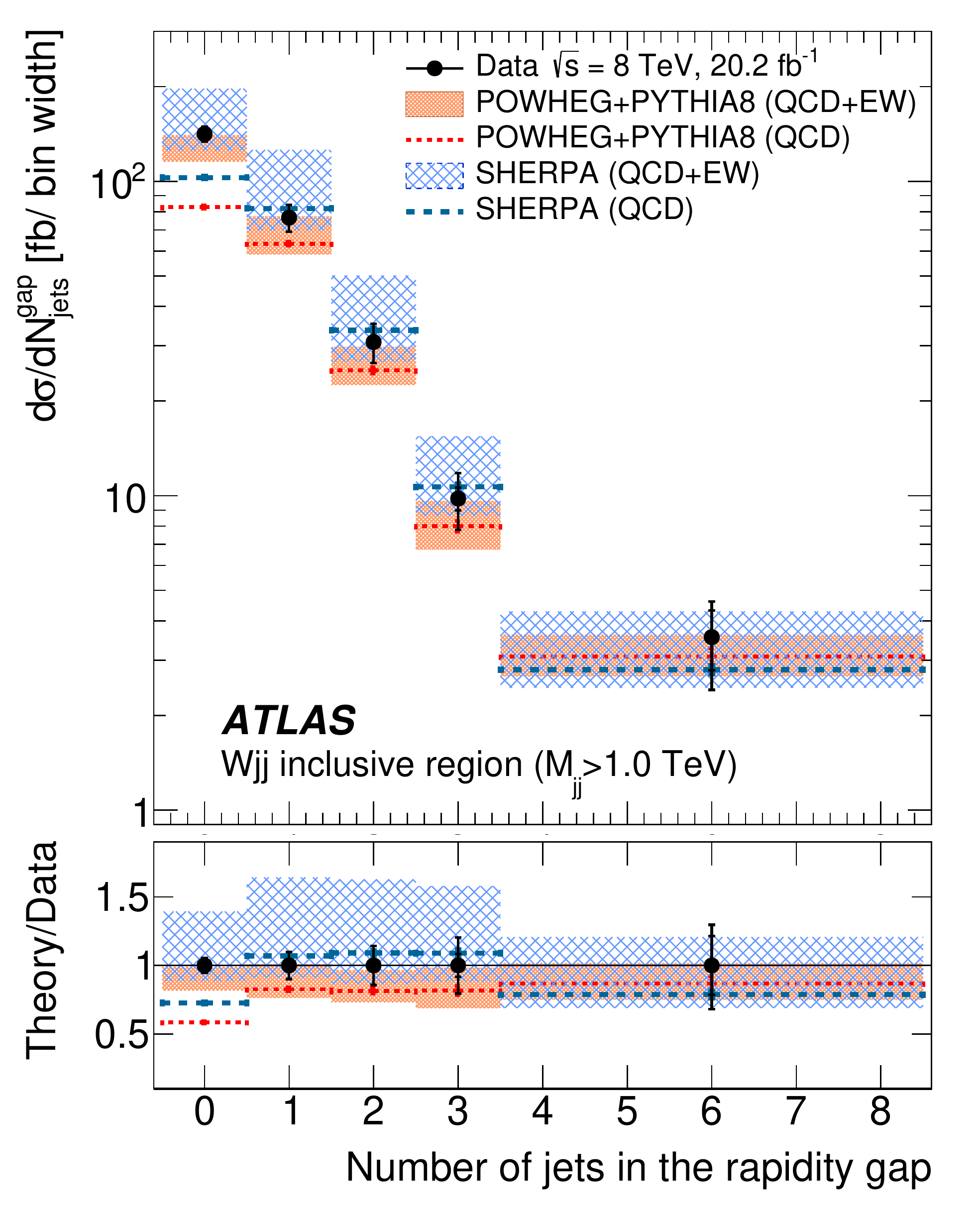}
\includegraphics[width=0.49\textwidth]{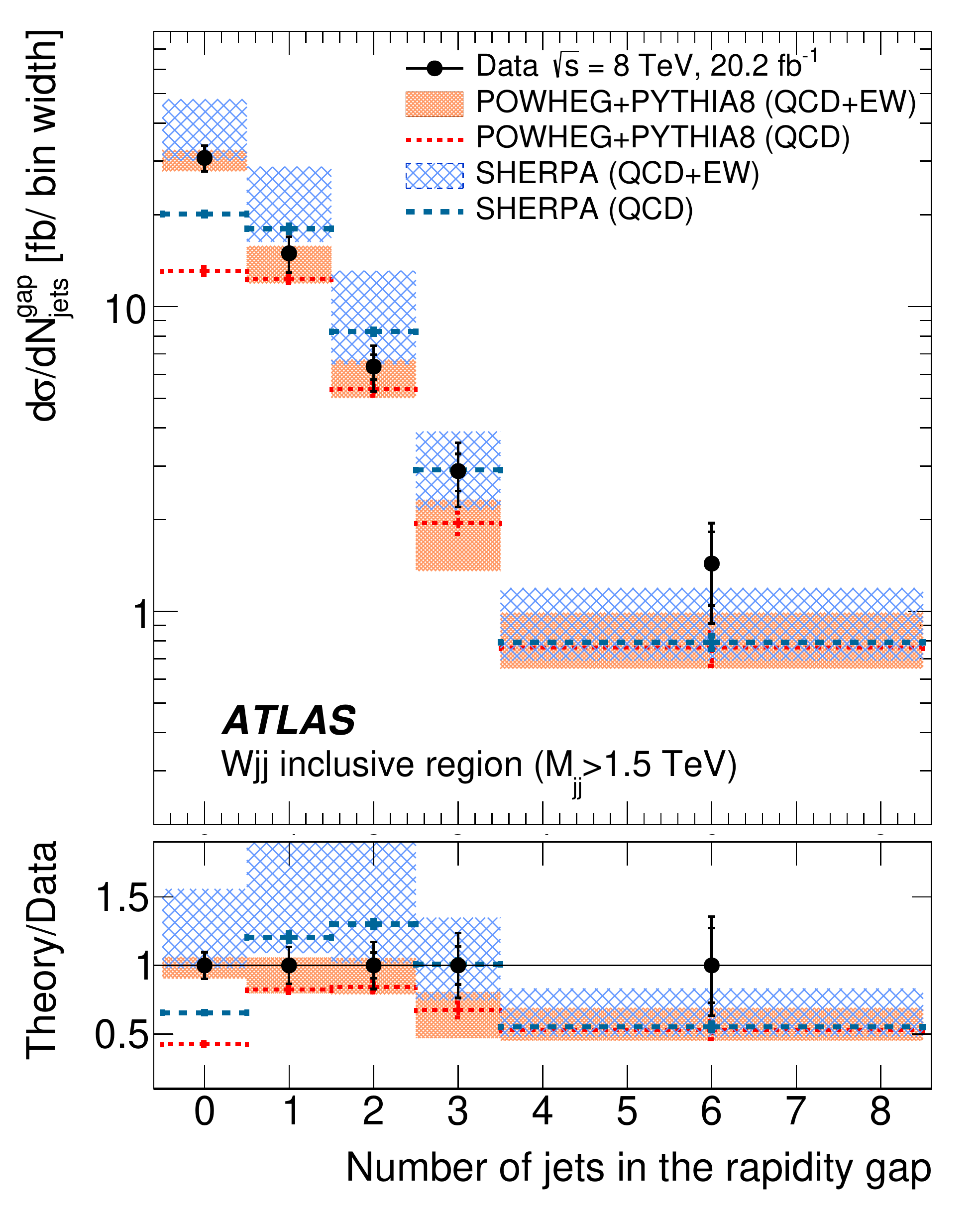}
\includegraphics[width=0.49\textwidth]{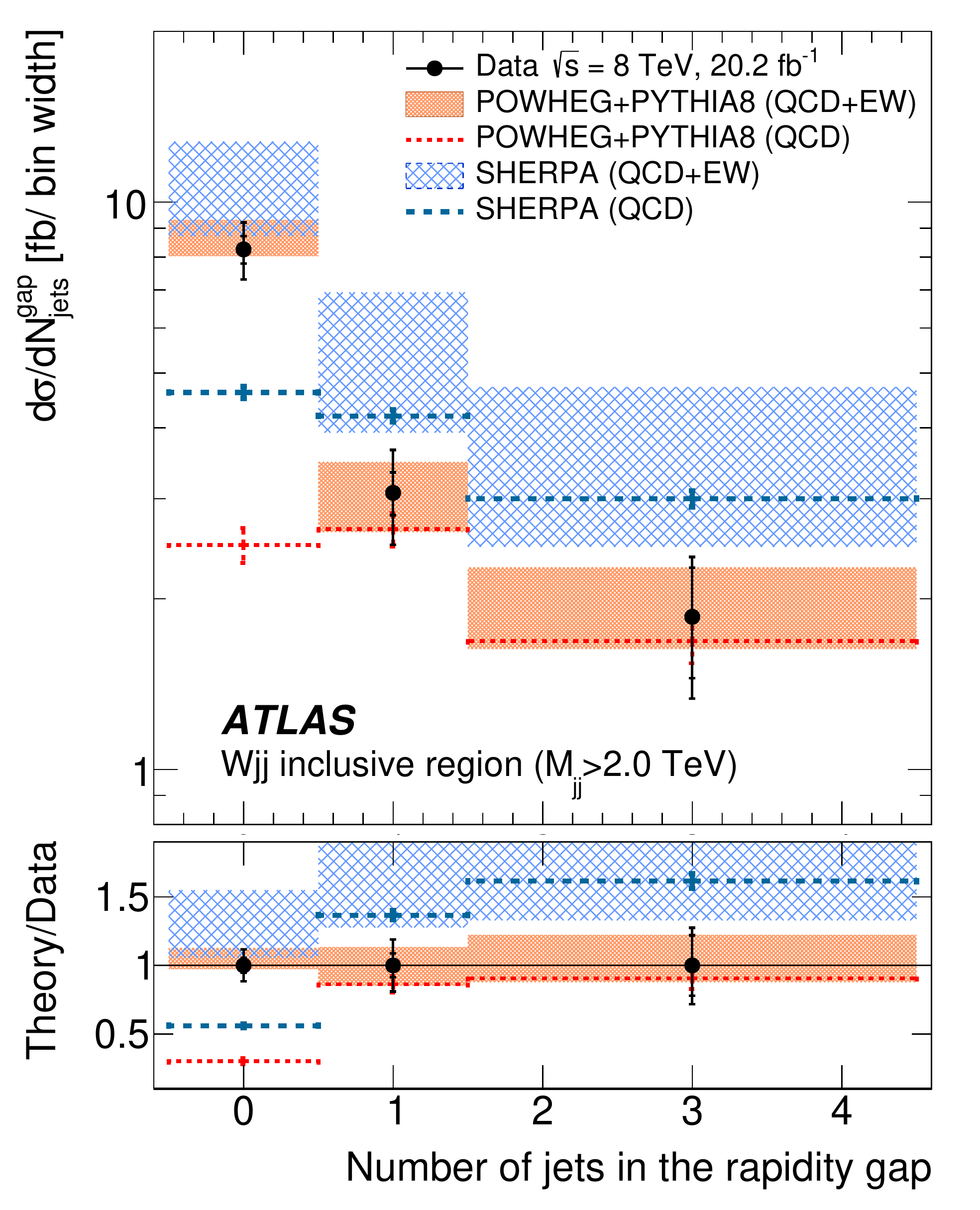}
\caption{Differential \wjets production cross sections as a function of the number of hard jets in the rapidity interval 
between the two leading jets in the inclusive fiducial region with four thresholds on the dijet invariant mass (0.5~\TeV, 
1.0~\TeV, 1.5~\TeV, and 2.0~\TeV).  Both statistical (inner bar) and total (outer bar) measurement uncertainties 
are shown, as well as ratios
of the theoretical predictions to the data (the bottom panel in each distribution). }
\label{unfolding:aux:AUX16}
\end{figure}
\begin{figure}[htbp]
\centering
\includegraphics[width=0.49\textwidth]{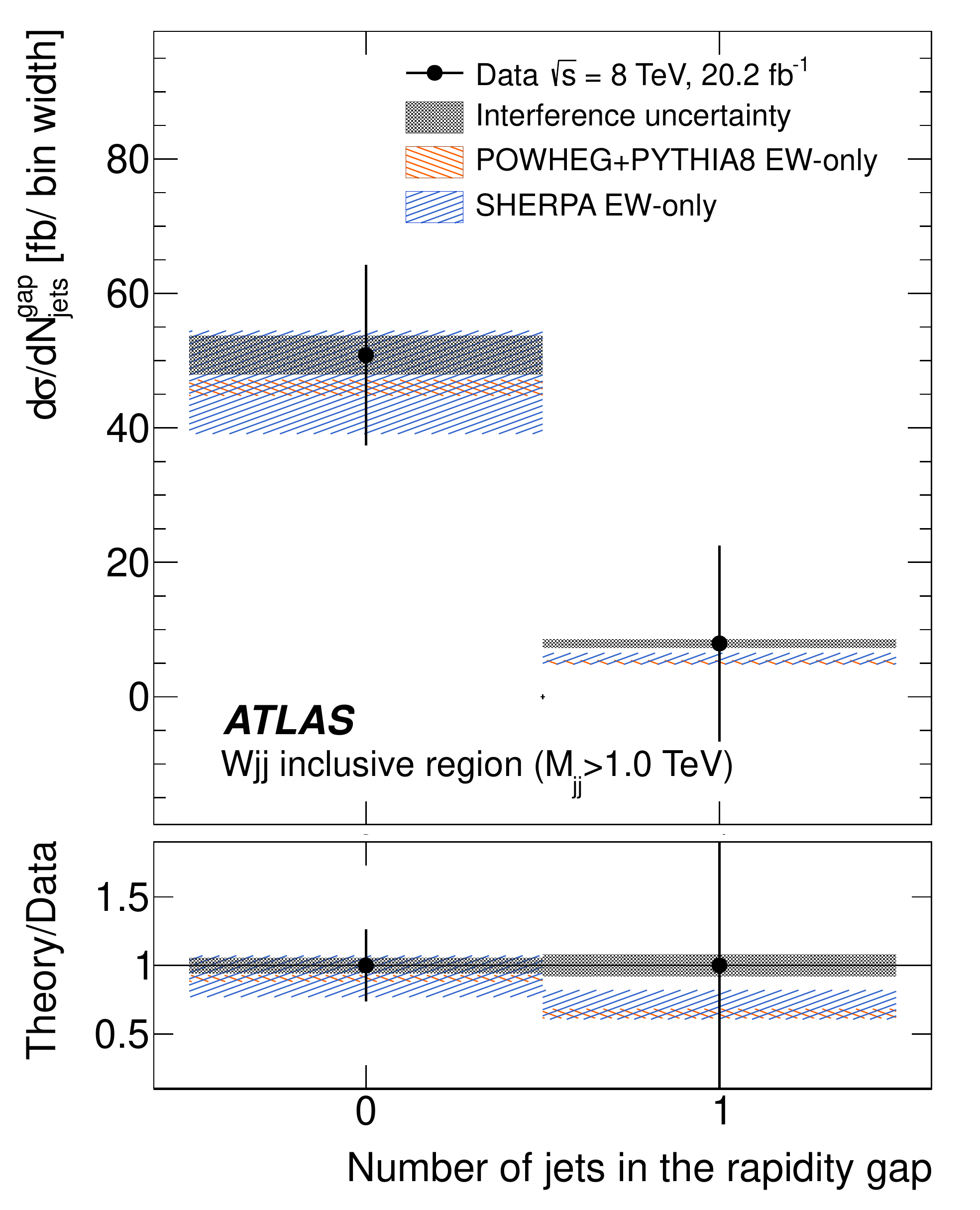}
\includegraphics[width=0.49\textwidth]{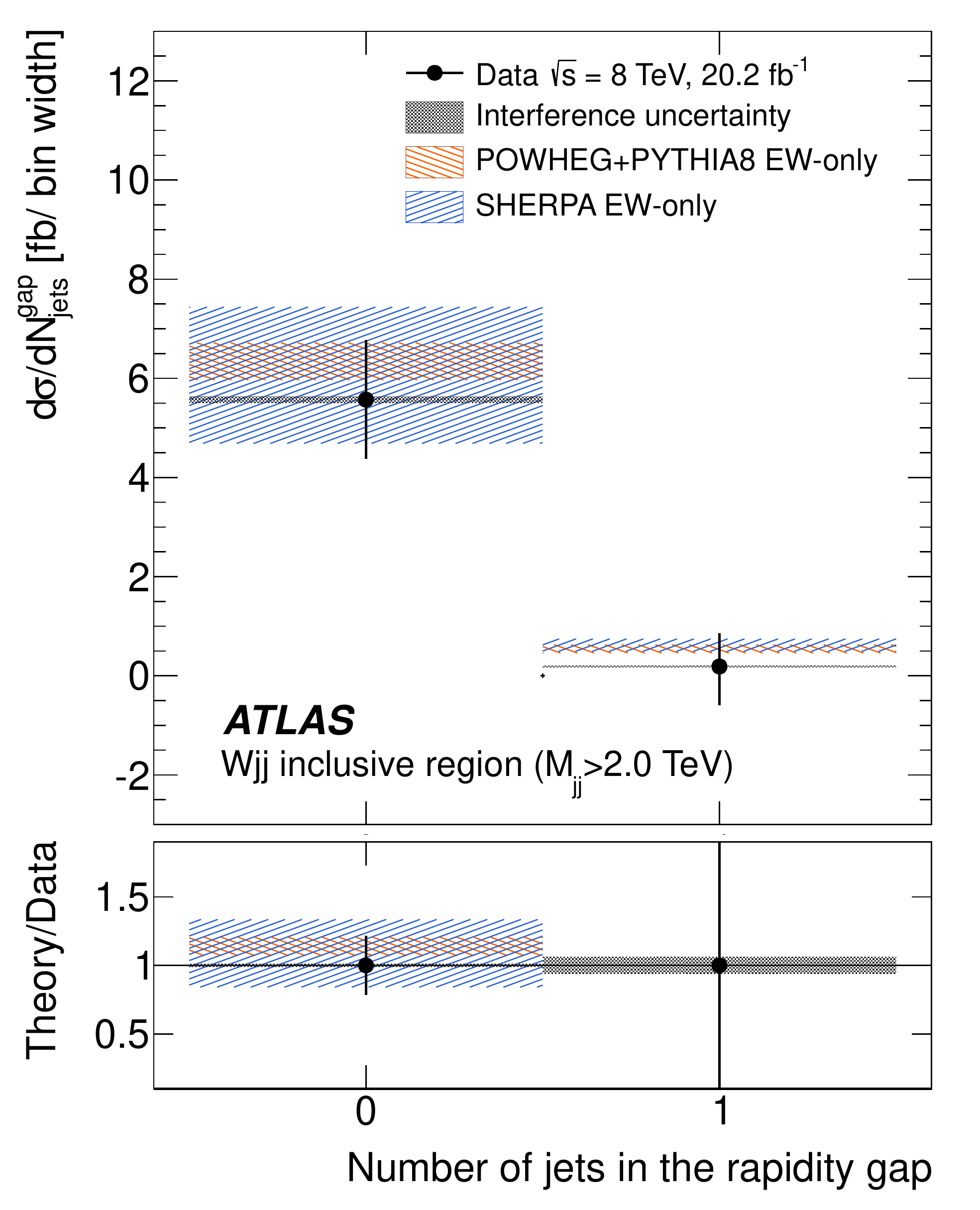}
\includegraphics[width=0.49\textwidth]{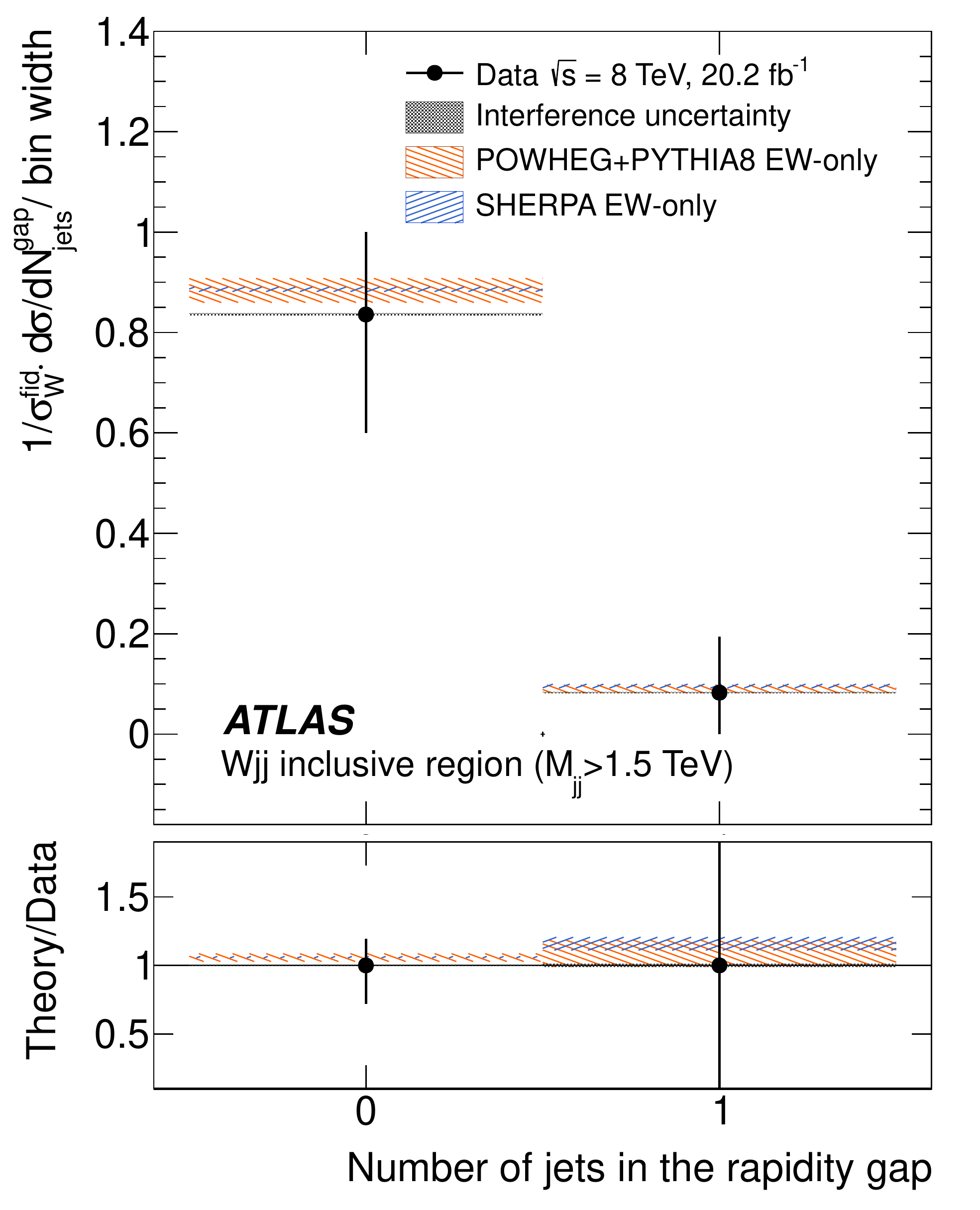}
\includegraphics[width=0.49\textwidth]{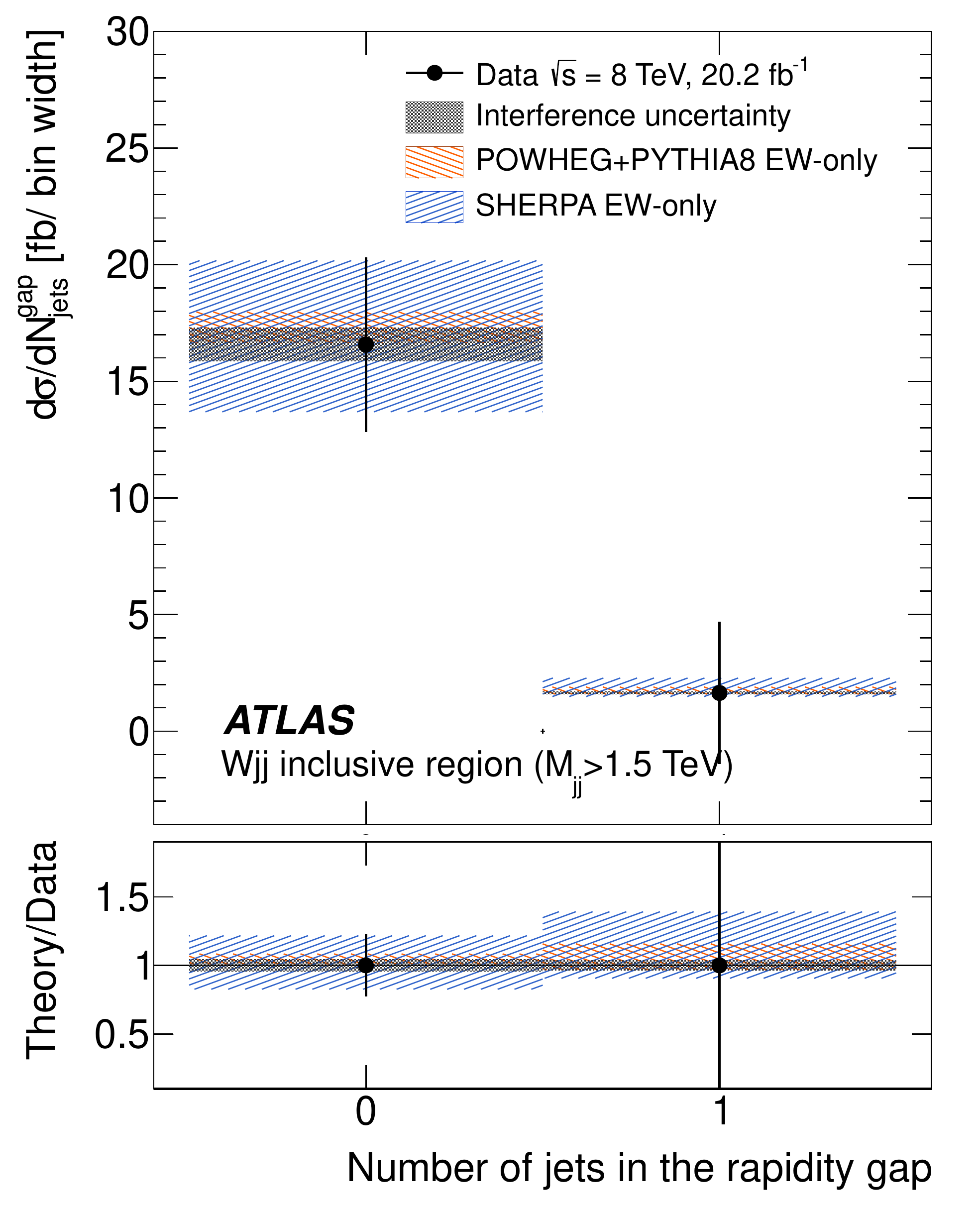}
\caption{Differential electroweak \wjets production cross sections as a function of the number of hard jets in the 
rapidity gap between the two leading jets in the inclusive fiducial region with $\mjj >1.0$~\TeV~(top left), 
1.5~\TeV~(top right and bottom left), and 2.0~\TeV (bottom right).  The region with $\mjj > 1.5$~\TeV, includes both 
absolute (top right) and normalized (bottom left) distributions.  Both statistical (inner bar) and total (outer bar) 
measurement uncertainties are shown, as well as ratios of the theoretical predictions to the data (the bottom panel 
in each distribution). }
\label{unfolding:aux:AUX3}
\end{figure}
\begin{figure}[htbp]
\centering
\includegraphics[width=0.49\textwidth]{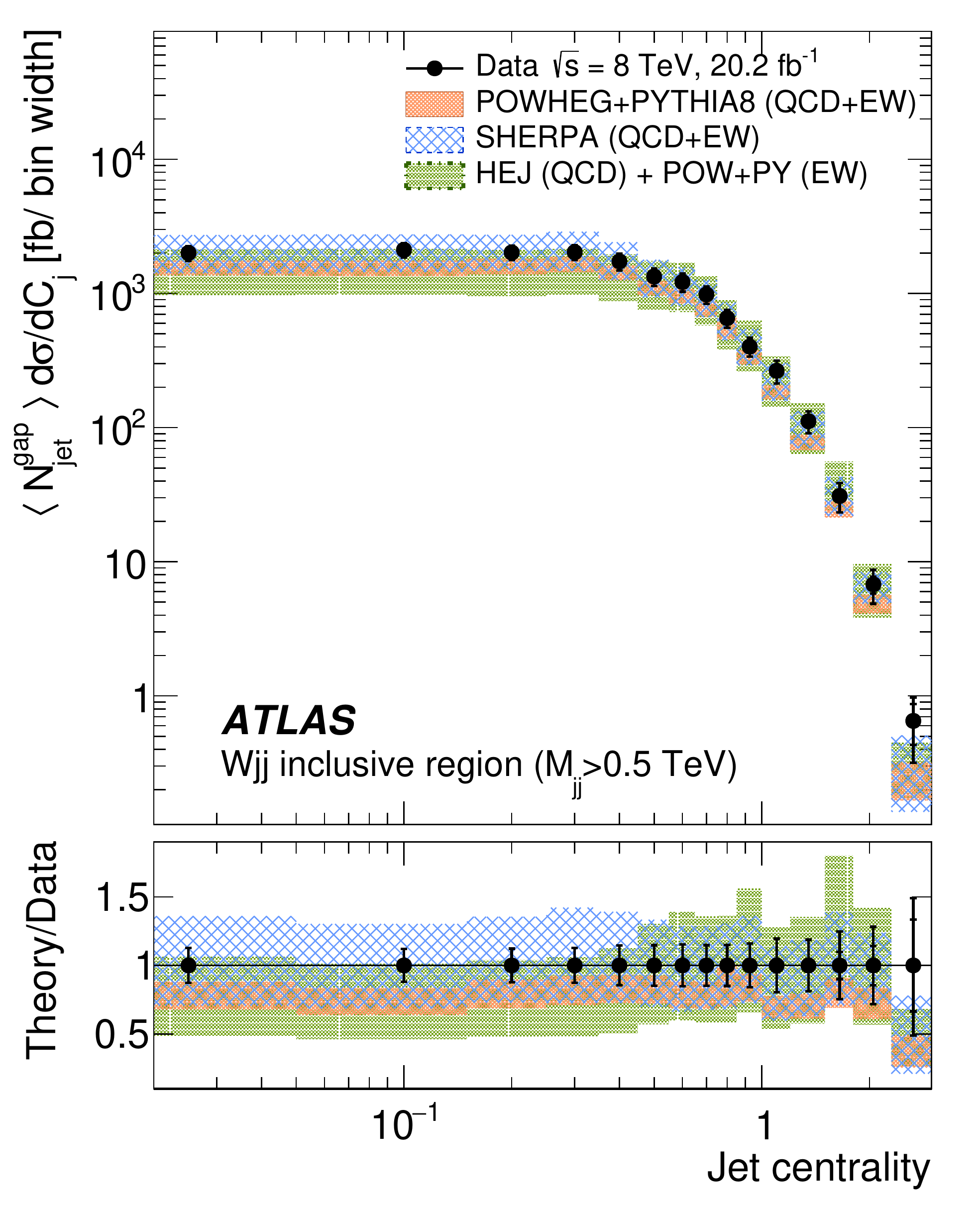}
\includegraphics[width=0.49\textwidth]{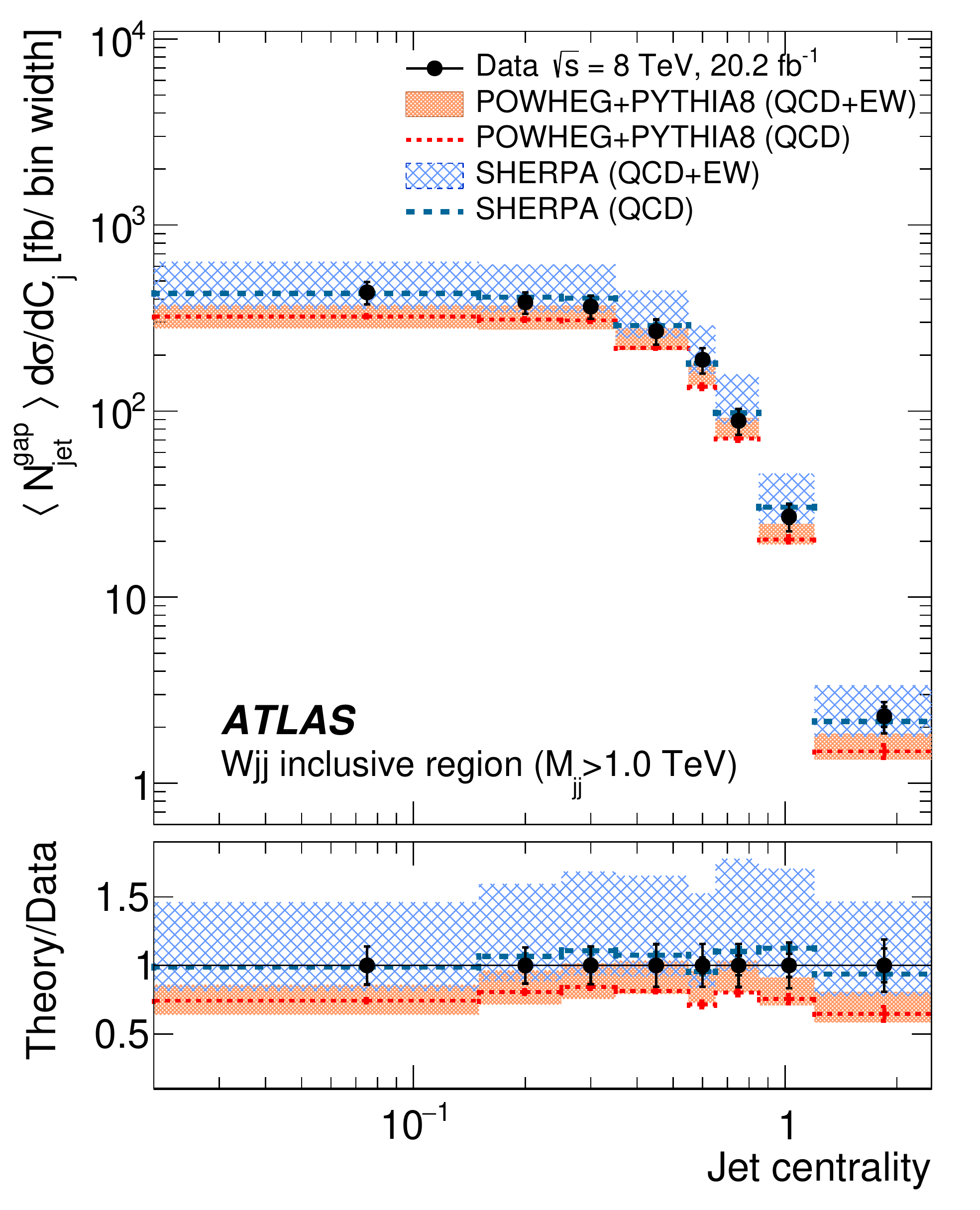}
\includegraphics[width=0.49\textwidth]{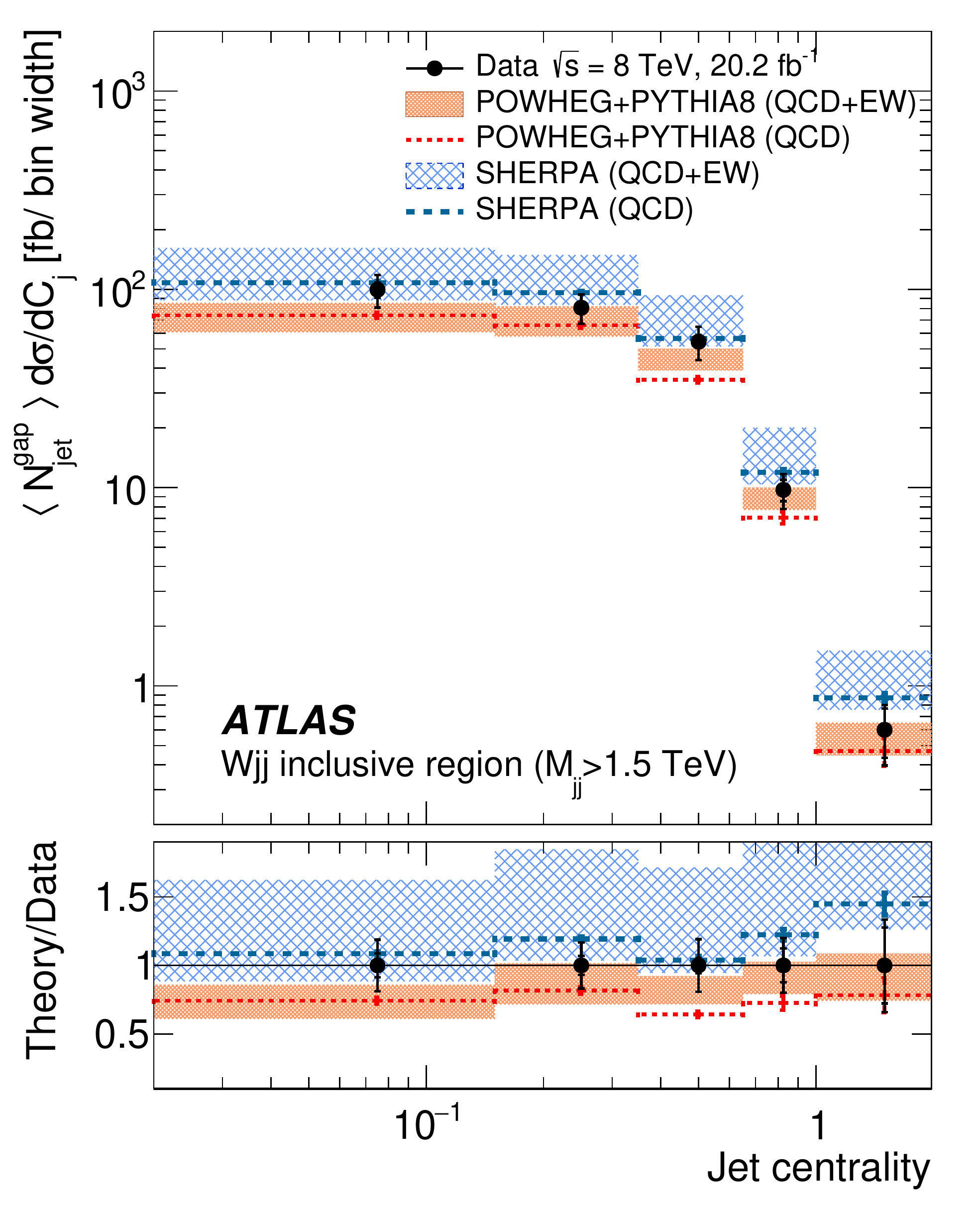}
\includegraphics[width=0.49\textwidth]{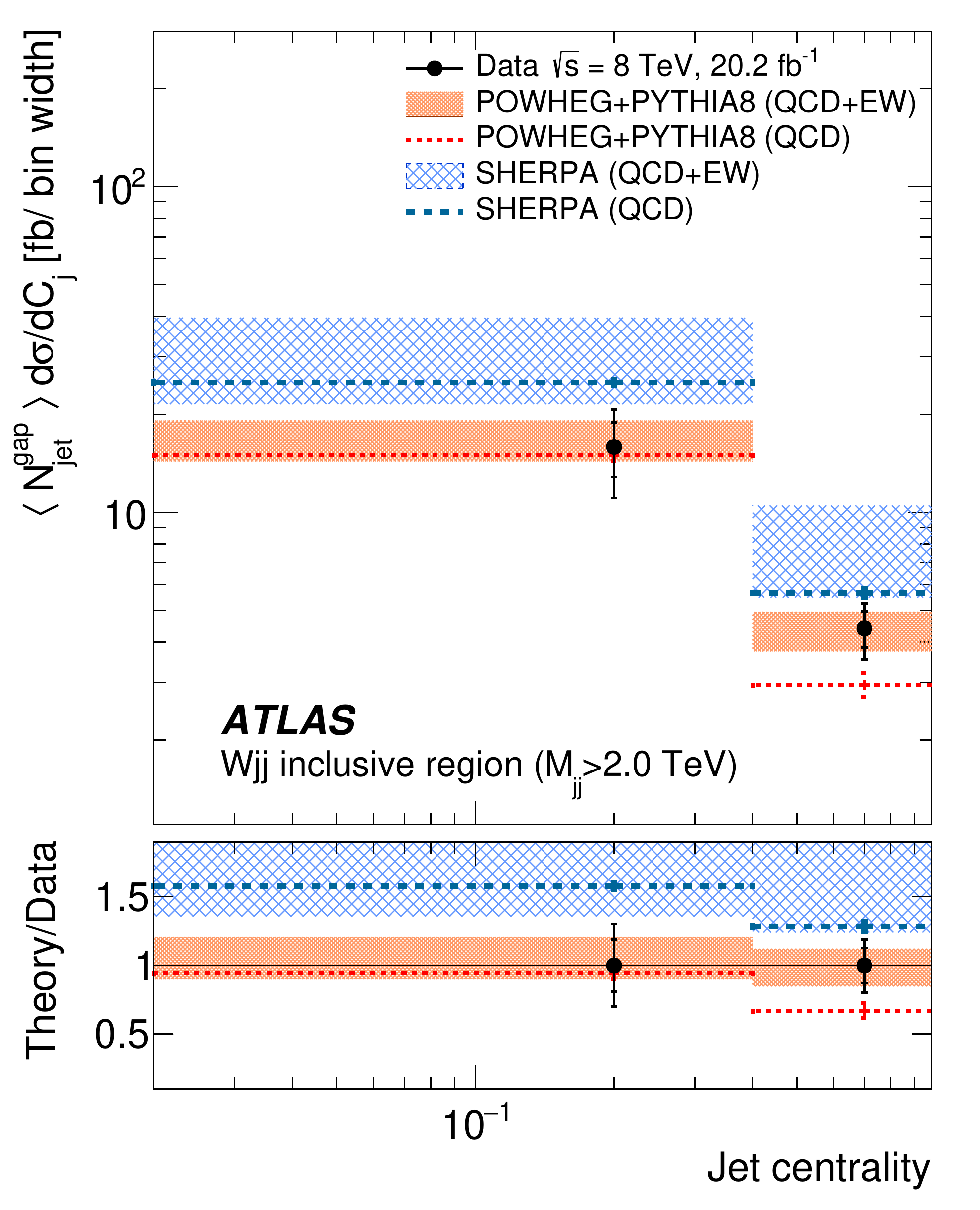}
\caption{Differential \wjets production cross sections as a function of jet centrality
in the inclusive fiducial region with four thresholds on the dijet invariant mass (0.5~\TeV, 1.0~\TeV, 1.5~\TeV, 
and 2.0~\TeV).  Both statistical (inner bar) and total (outer bar) measurement uncertainties are shown, as well as 
ratios of the theoretical predictions to the data (the bottom panel in each distribution). }
\label{unfolding:aux:AUX23}
\end{figure}
\begin{figure}[htbp]
\centering
\includegraphics[width=0.49\textwidth]{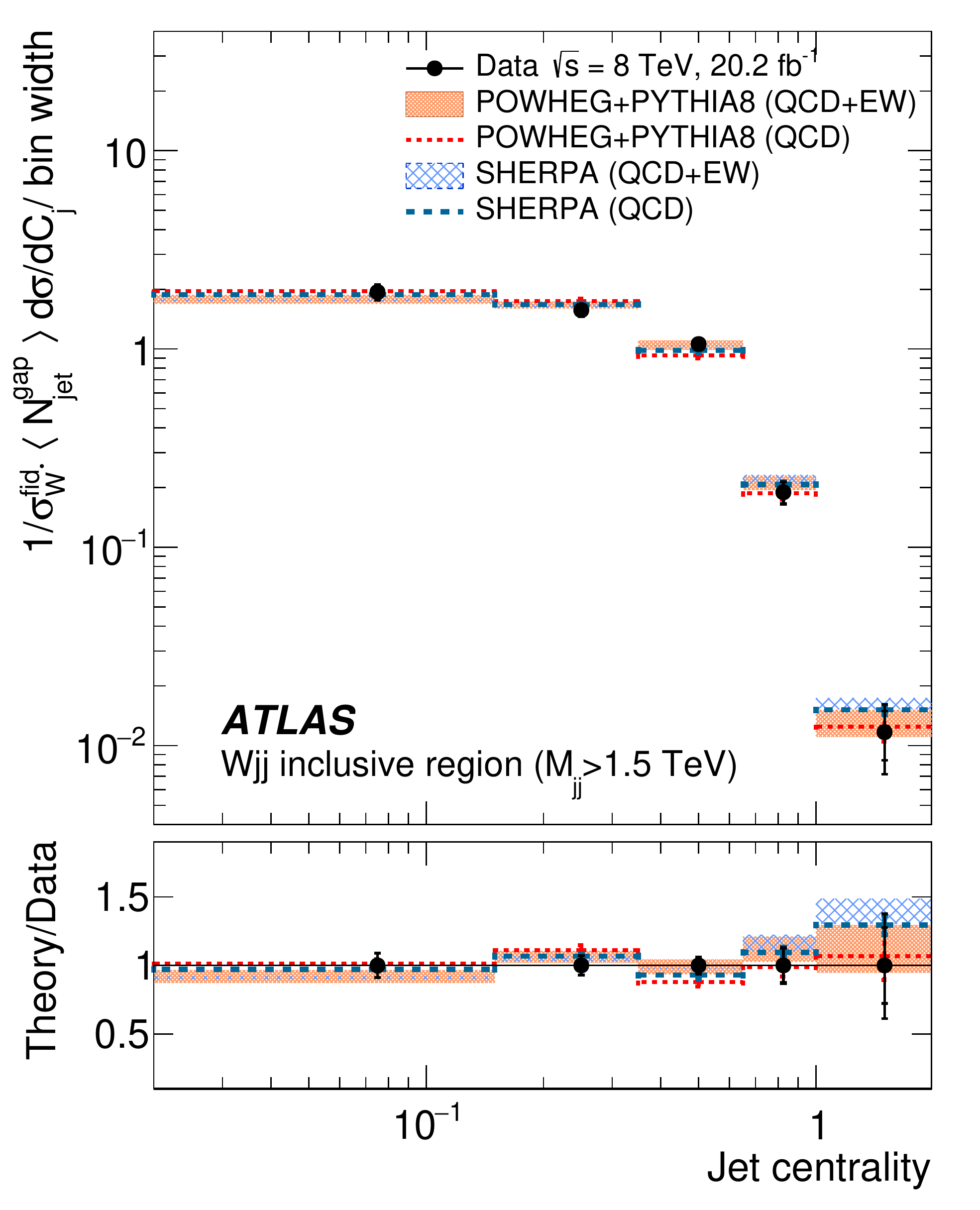}
\includegraphics[width=0.49\textwidth]{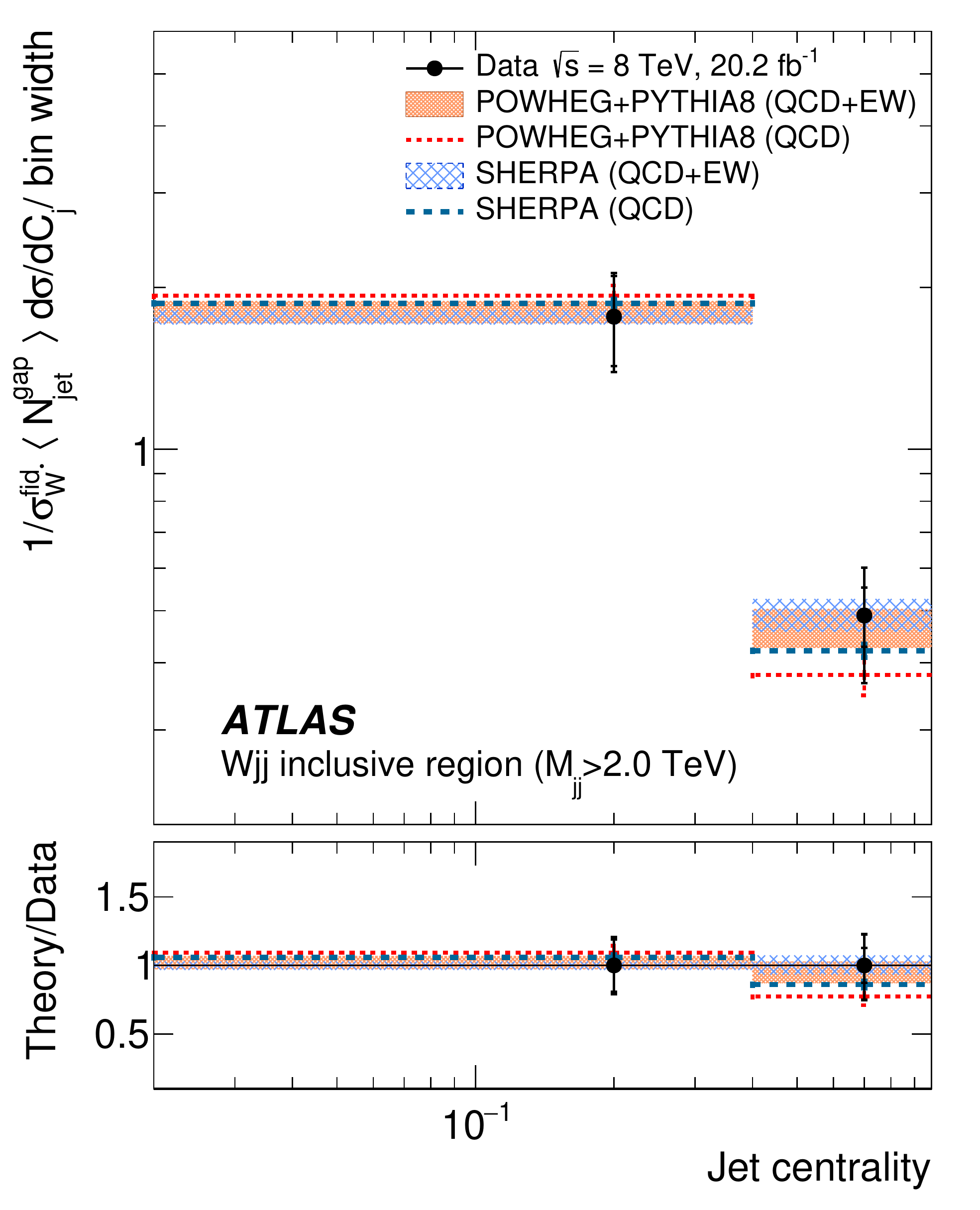}
\includegraphics[width=0.49\textwidth]{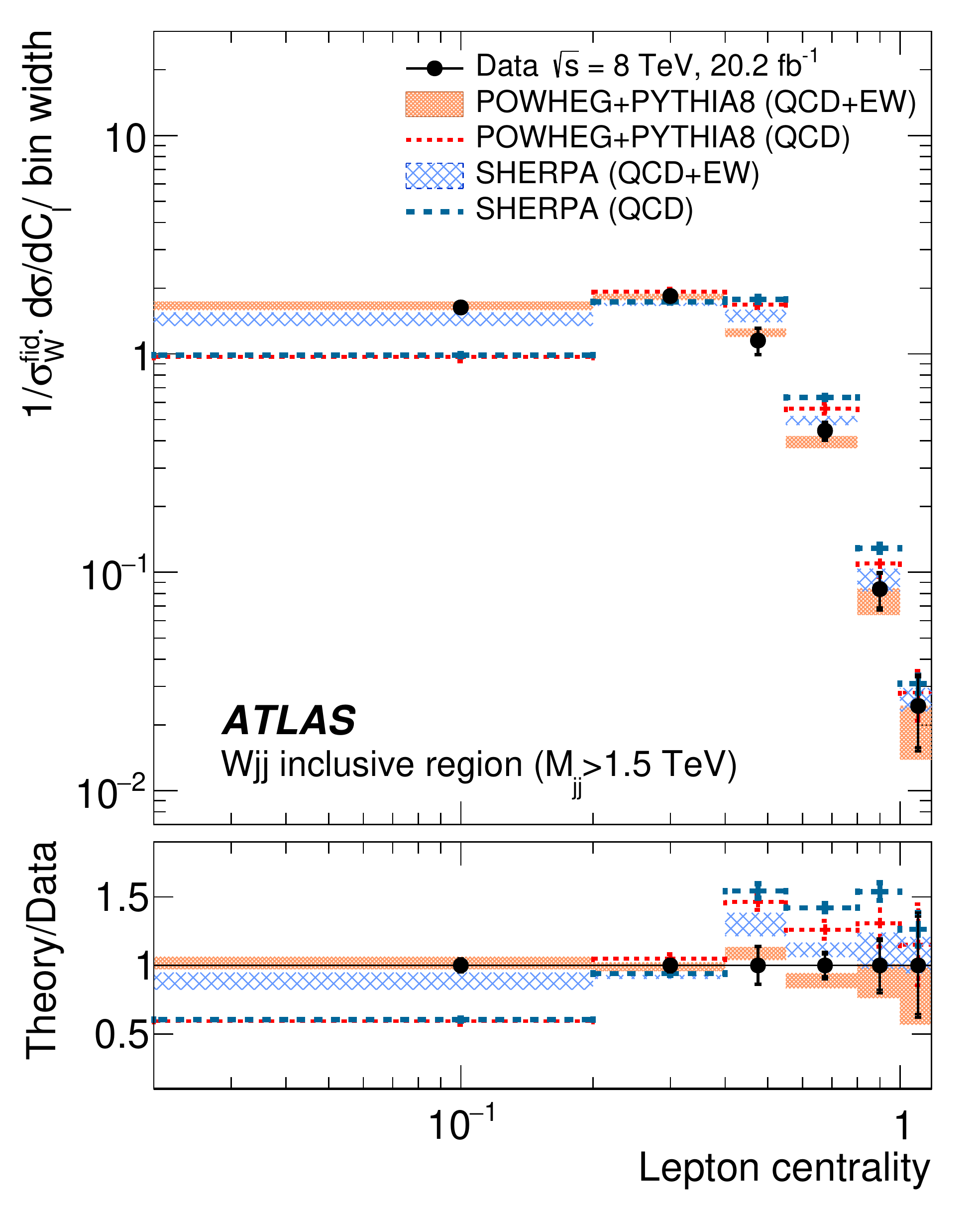}
\includegraphics[width=0.49\textwidth]{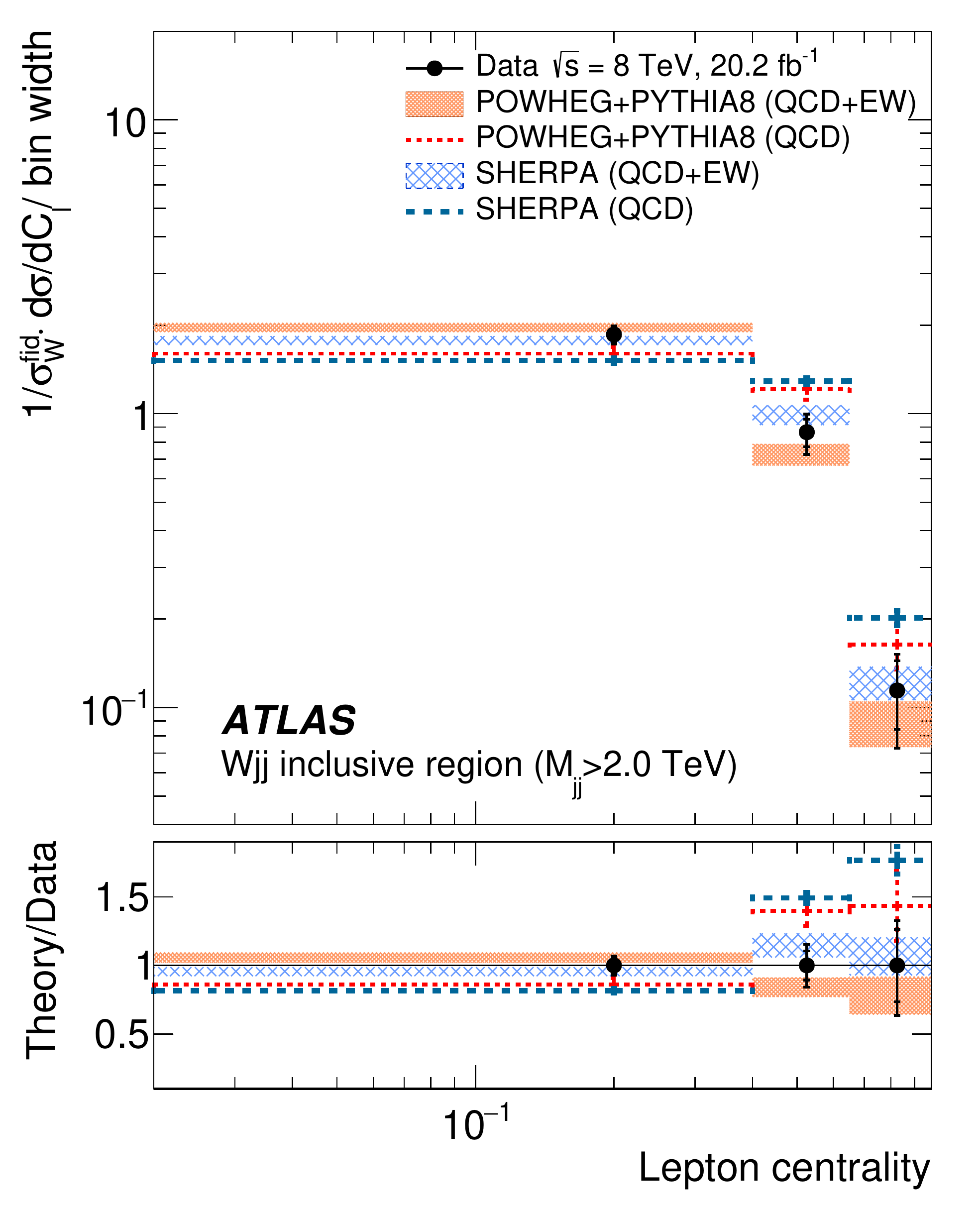}
\caption{Unfolded normalized differential \wjets production cross sections as a function of jet centrality (top) 
and lepton centrality (bottom) for the inclusive fiducial region with $\mjj > 1.5$~\TeV~(left) and 2.0~\TeV~(right).  
Both statistical (inner bar) and total (outer bar) measurement uncertainties are shown, as well as ratios
of the theoretical predictions to the data (the bottom panel in each distribution). }
\label{unfolding:aux:AUX14}
\end{figure}
\begin{figure}[htbp]
\centering
\includegraphics[width=0.35\textwidth]{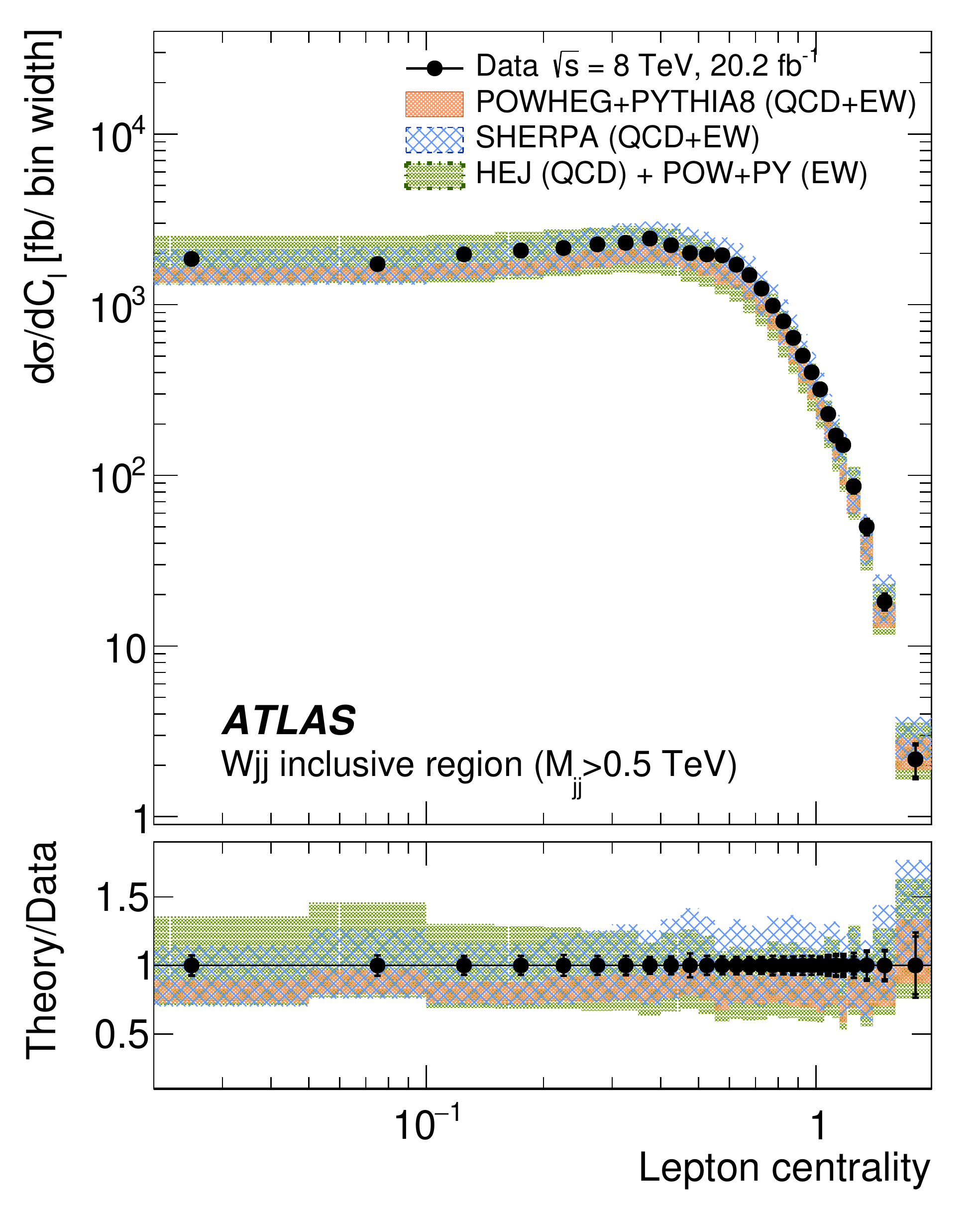}
\includegraphics[width=0.35\textwidth]{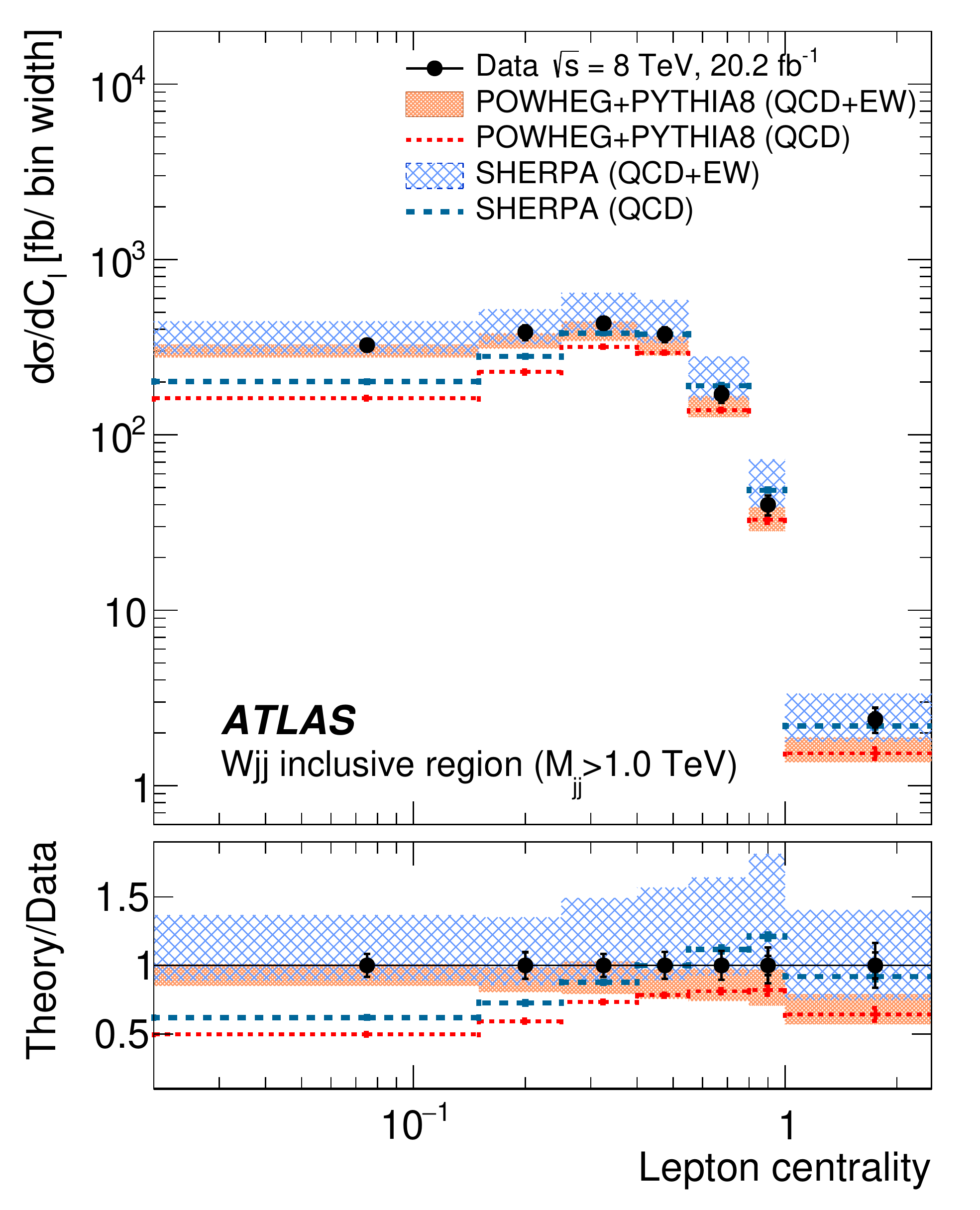}
\includegraphics[width=0.35\textwidth]{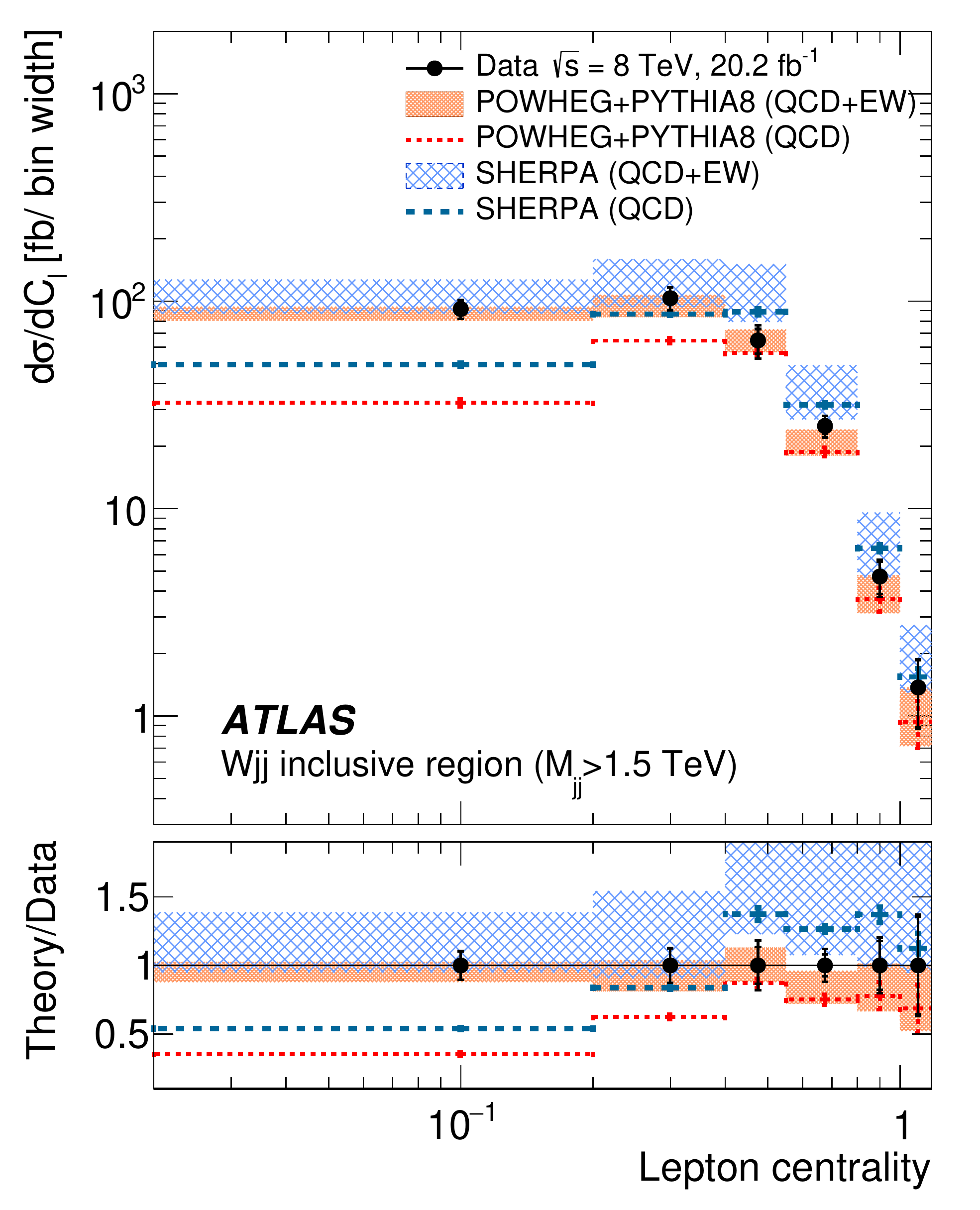}
\includegraphics[width=0.35\textwidth]{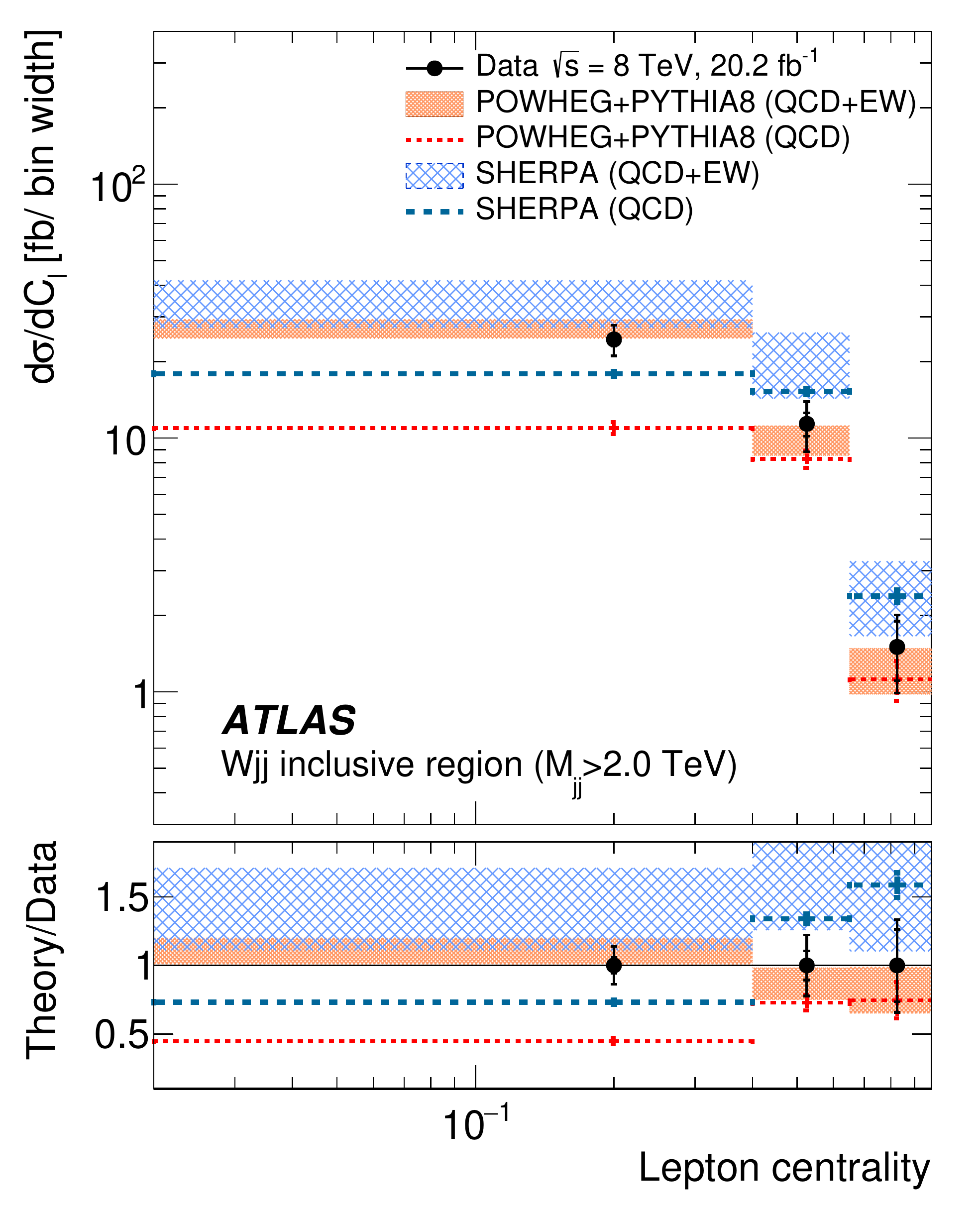}
\includegraphics[width=0.35\textwidth]{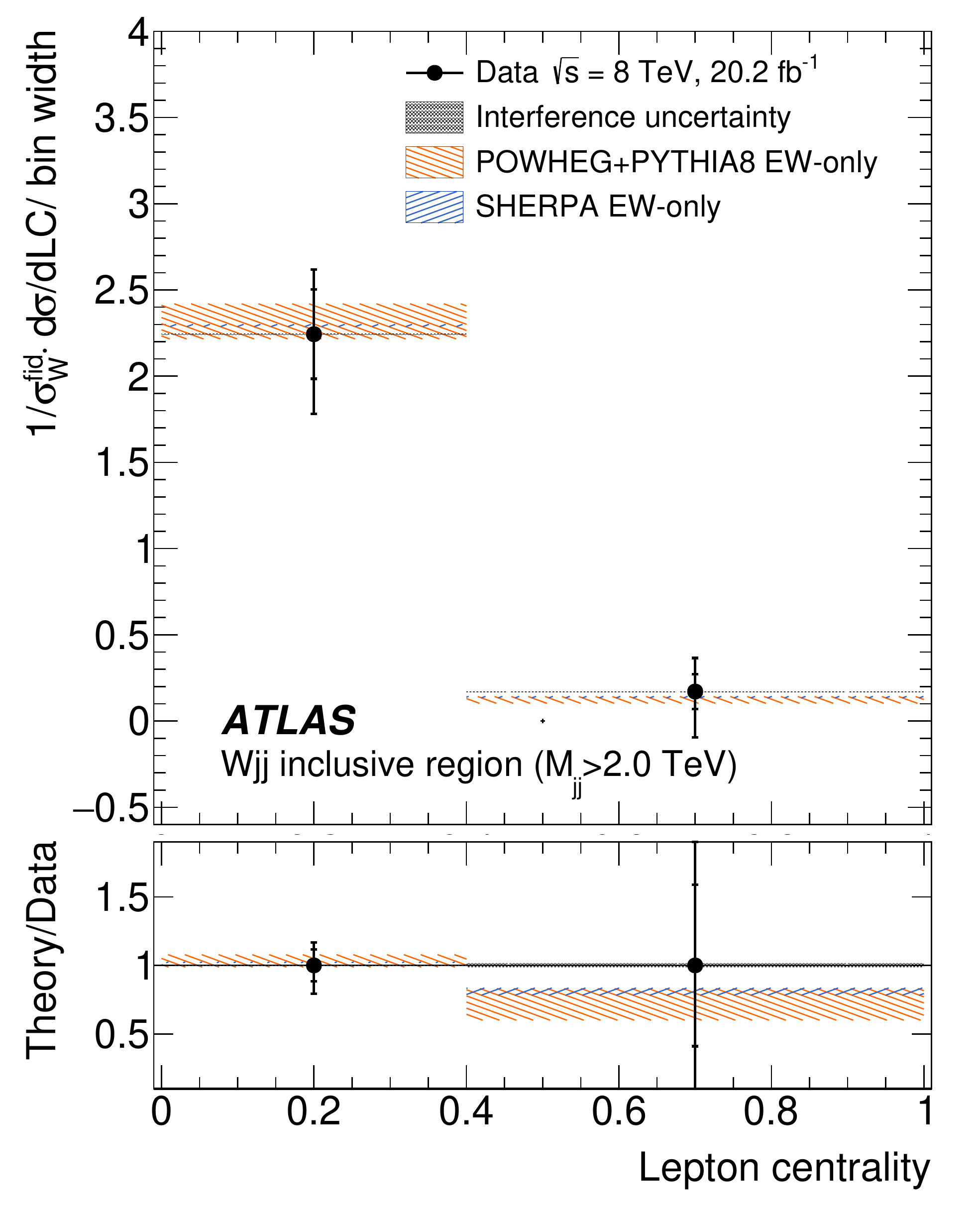}
\caption{Unfolded differential \wjets production cross sections as a function of lepton centrality
in the inclusive fiducial region with four thresholds on the dijet invariant mass (0.5~\TeV, 1.0~\TeV, 1.5~\TeV, 
and 2.0~\TeV).  The bottom plot shows the normalized distribution for $\mjj> 2.0$~\TeV.  Both statistical 
(inner bar) and total (outer bar) measurement uncertainties are shown, as well as ratios of the theoretical 
predictions to the data (the bottom panel in each distribution). }
\label{unfolding:combined_measurementLC1Dinclusiveabs}
\end{figure}
\begin{figure}[htbp]
\centering
\includegraphics[width=0.49\textwidth]{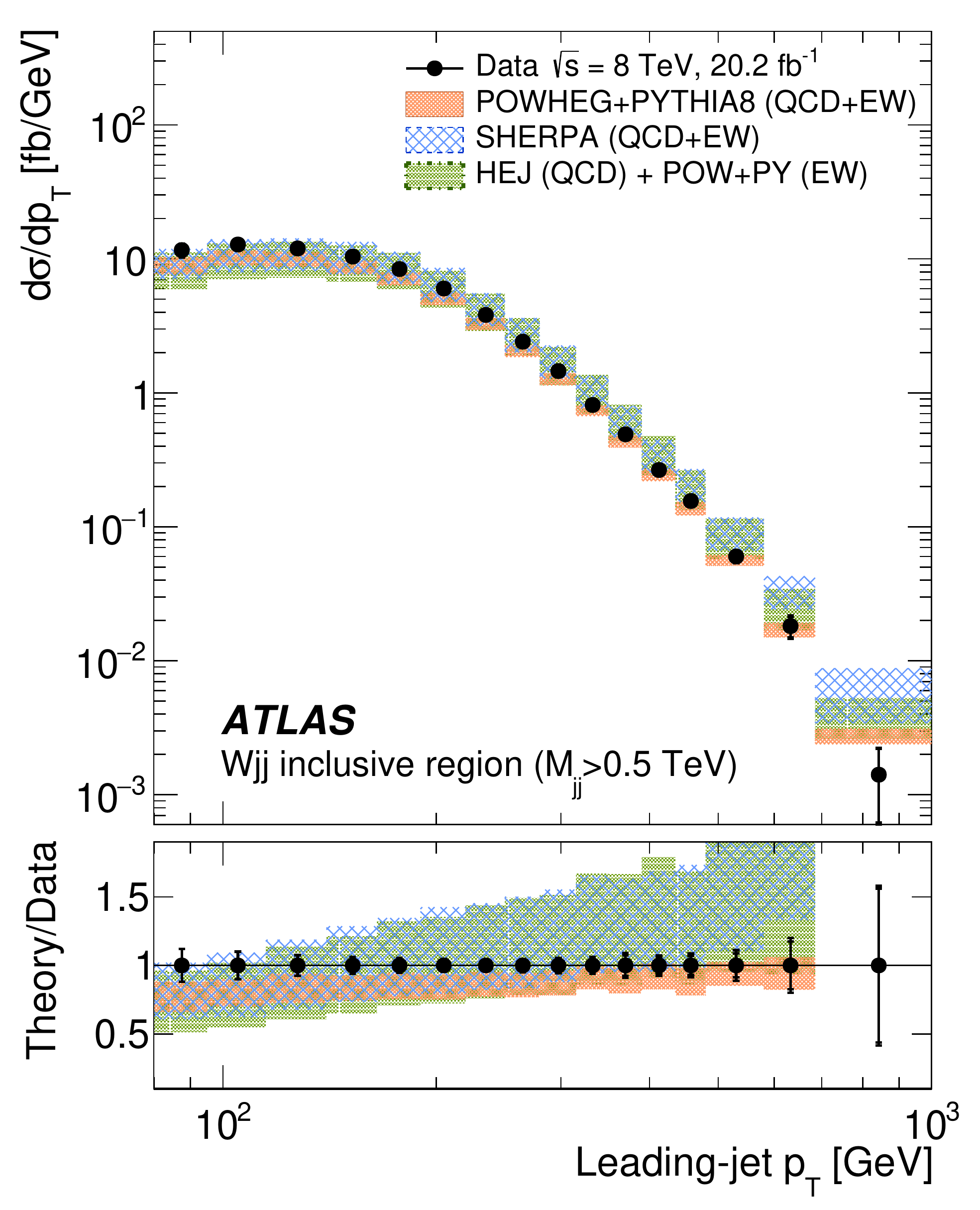}
\includegraphics[width=0.49\textwidth]{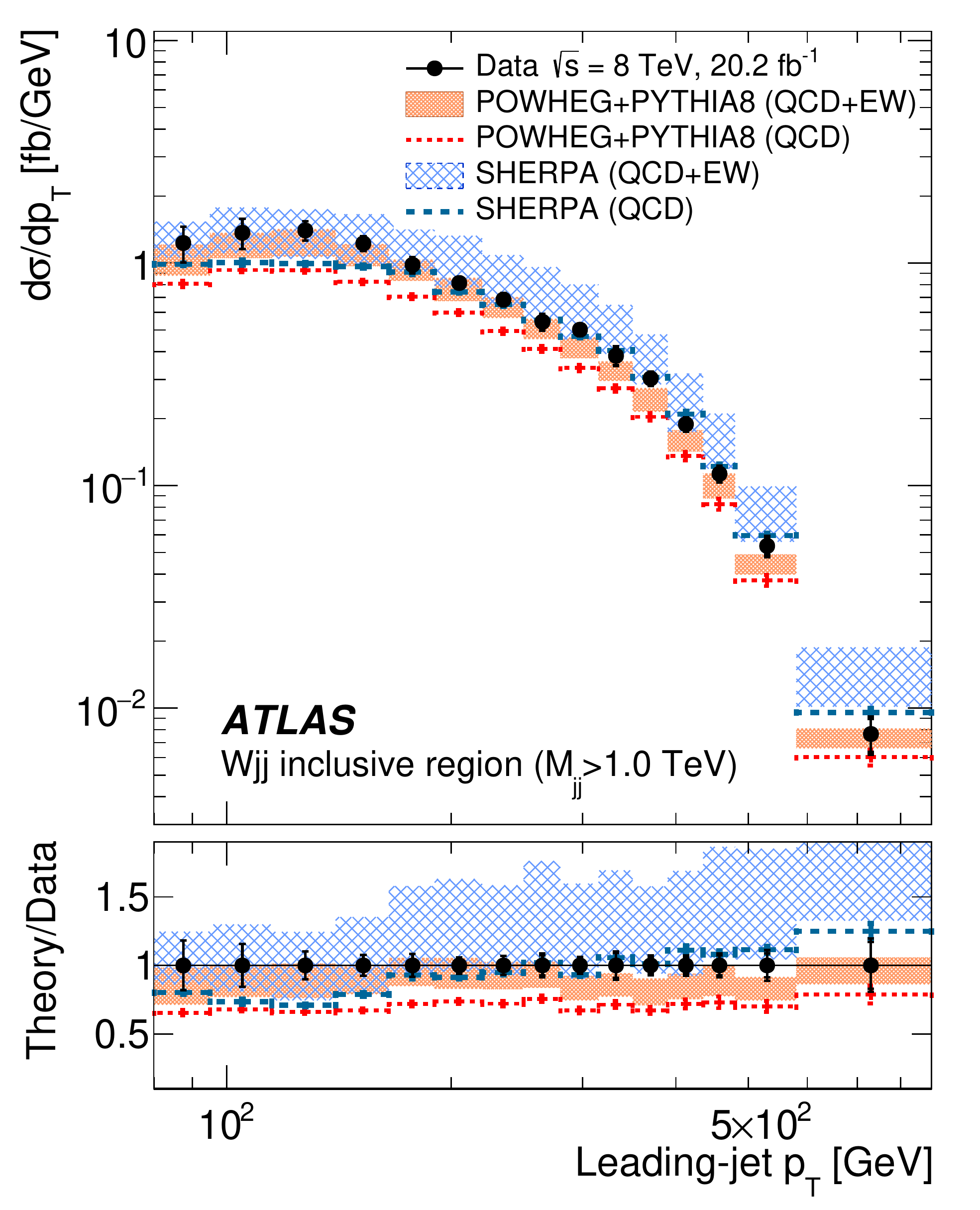}
\includegraphics[width=0.49\textwidth]{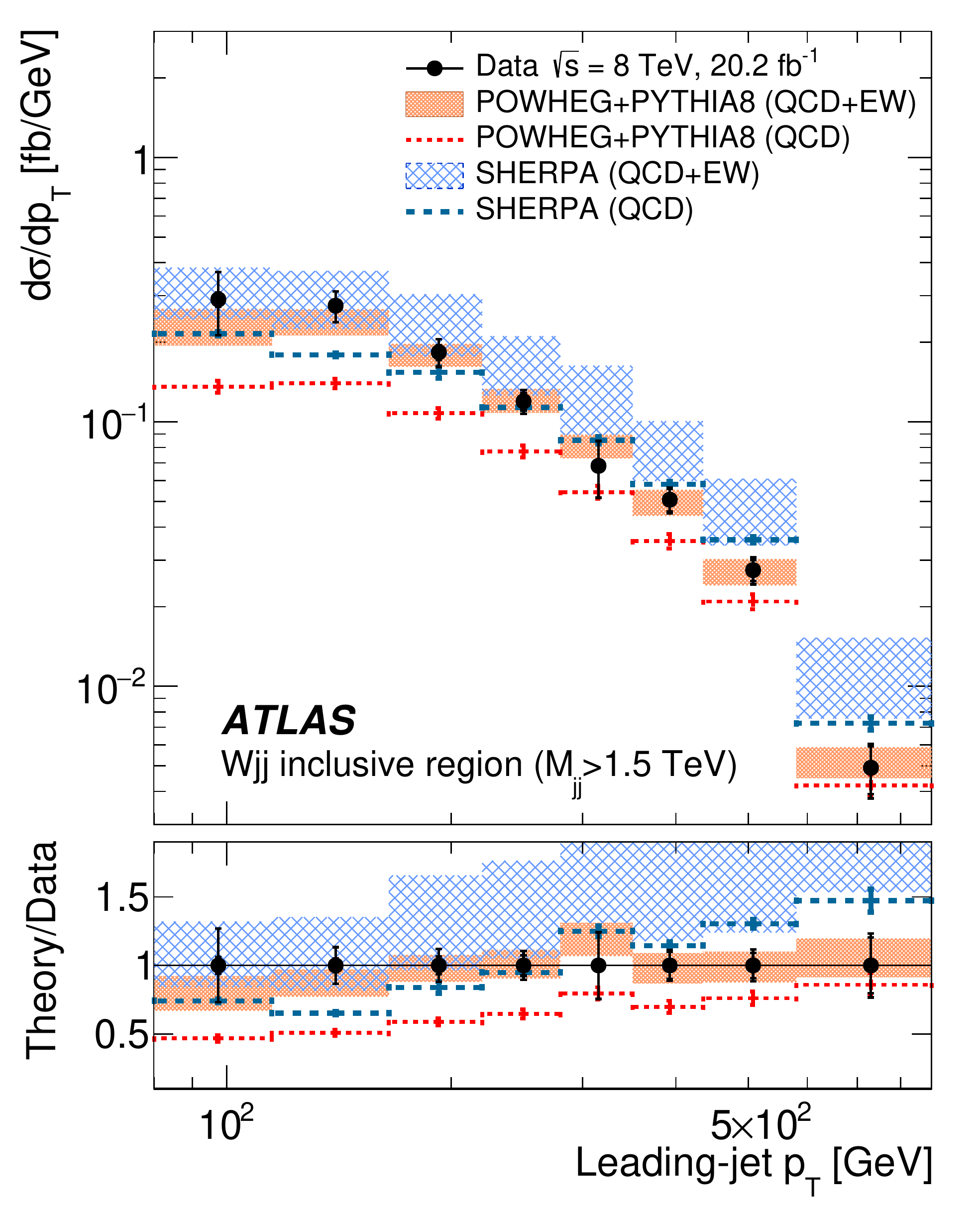}
\includegraphics[width=0.49\textwidth]{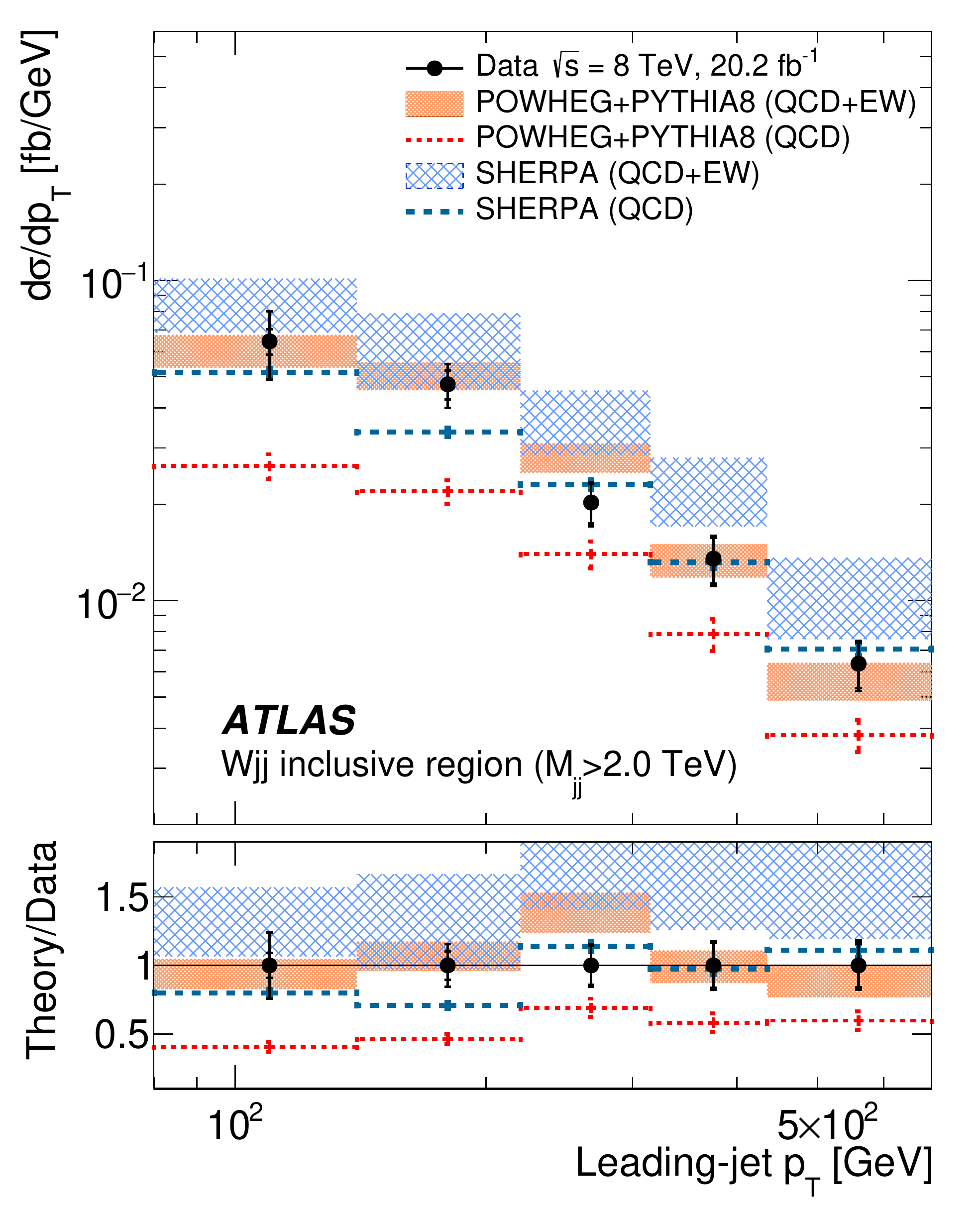}
\caption{Unfolded absolute differential \wjets production cross sections as a function of leading-jet $\pt$ for the 
inclusive fiducial region when the dijet invariant mass threshold is progressively raised in 500~\GeV~increments from 
0.5~\TeV~(top left) to 2.0~\TeV~(bottom right).  Both statistical (inner bar) and total (outer bar) measurement 
uncertainties are shown, as well as ratios of the theoretical predictions to the data (the bottom panel in each 
distribution).}
\label{unfolding:combined_measurementj1pt1Dinclusive}
\end{figure}
\begin{figure}[htbp]
\centering
\includegraphics[width=0.49\textwidth]{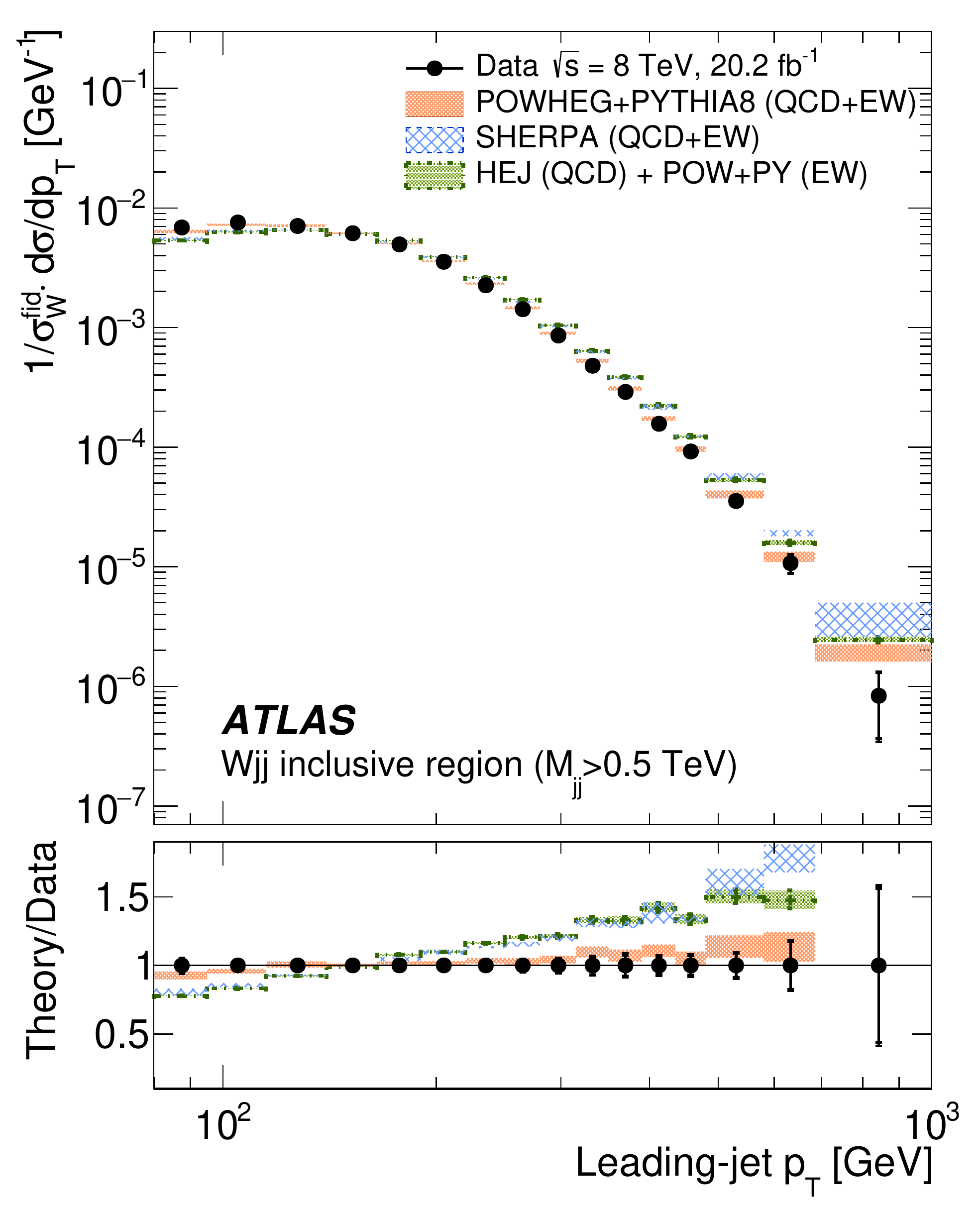}
\includegraphics[width=0.49\textwidth]{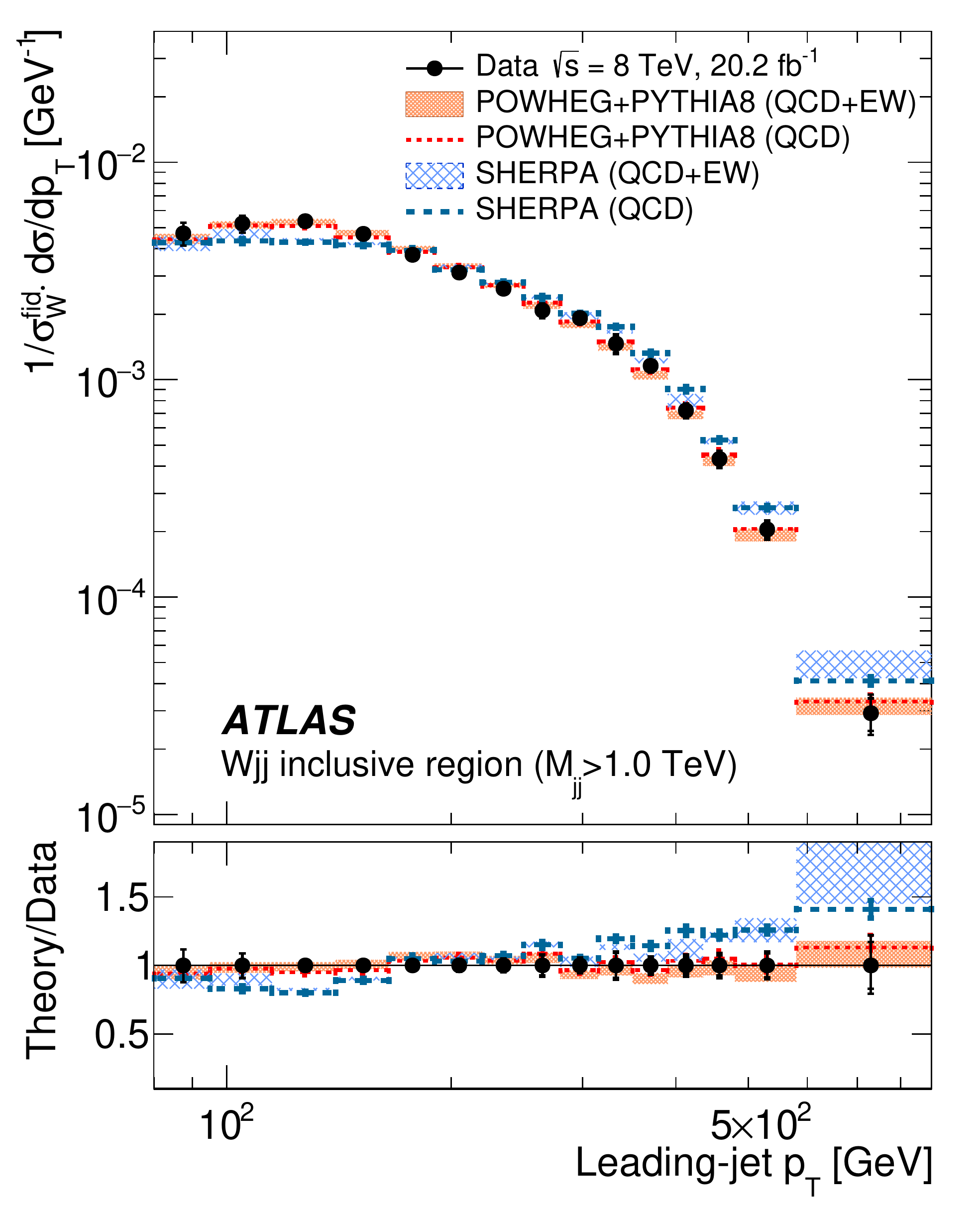}
\includegraphics[width=0.49\textwidth]{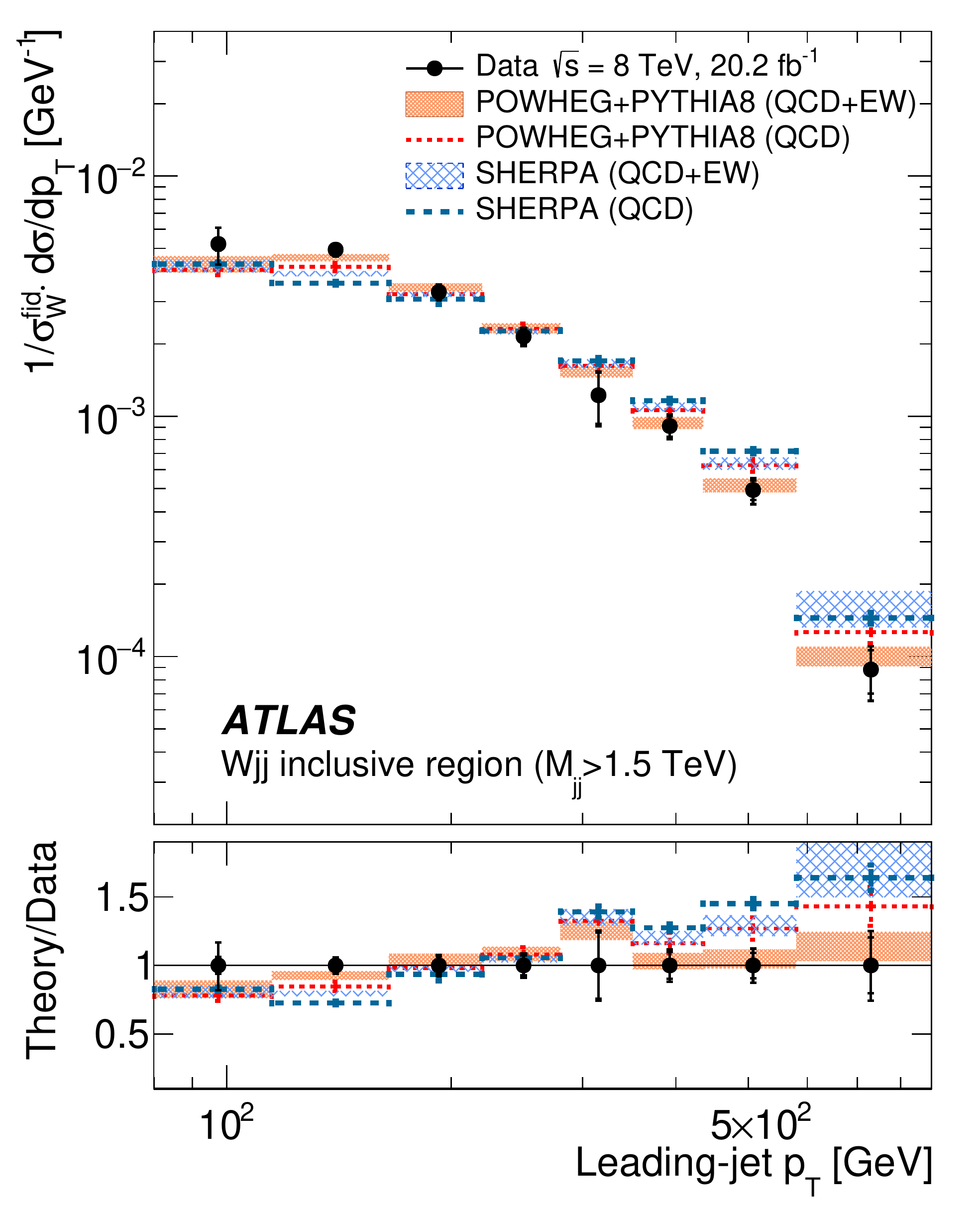}
\includegraphics[width=0.49\textwidth]{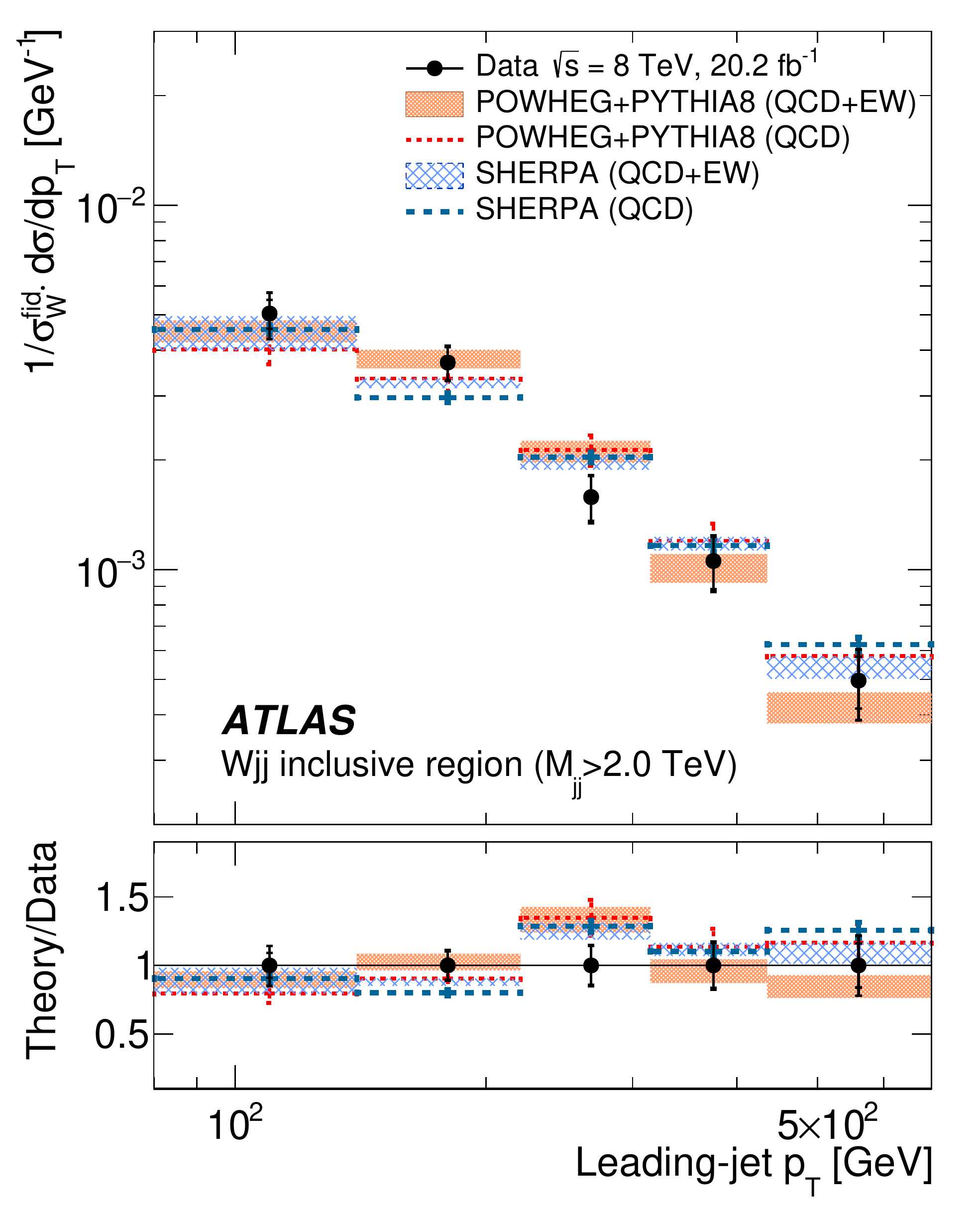}
\caption{Unfolded normalized differential \wjets production cross sections as a function of the leading-jet \pt
in the inclusive fiducial region with four thresholds on the dijet invariant mass (0.5~\TeV, 1.0~\TeV, 1.5~\TeV, and 
2.0~\TeV).  Both statistical (inner bar) and total (outer bar) measurement uncertainties are shown, as well as ratios
of the theoretical predictions to the data (the bottom panel in each distribution). }
\label{unfolding:aux:AUX7}
\end{figure}
\begin{figure}[htbp]
\centering
\includegraphics[width=0.35\textwidth]{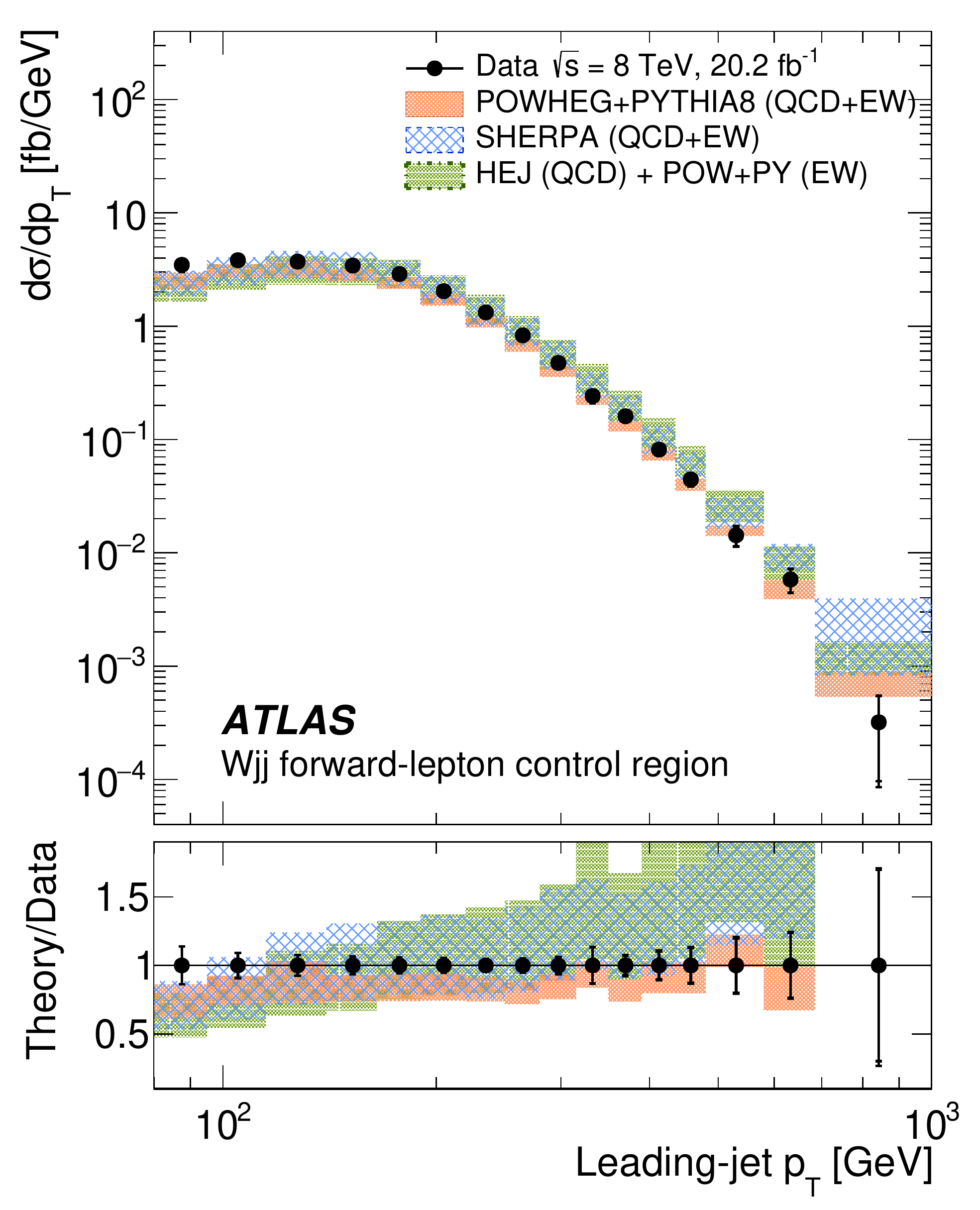}
\includegraphics[width=0.35\textwidth]{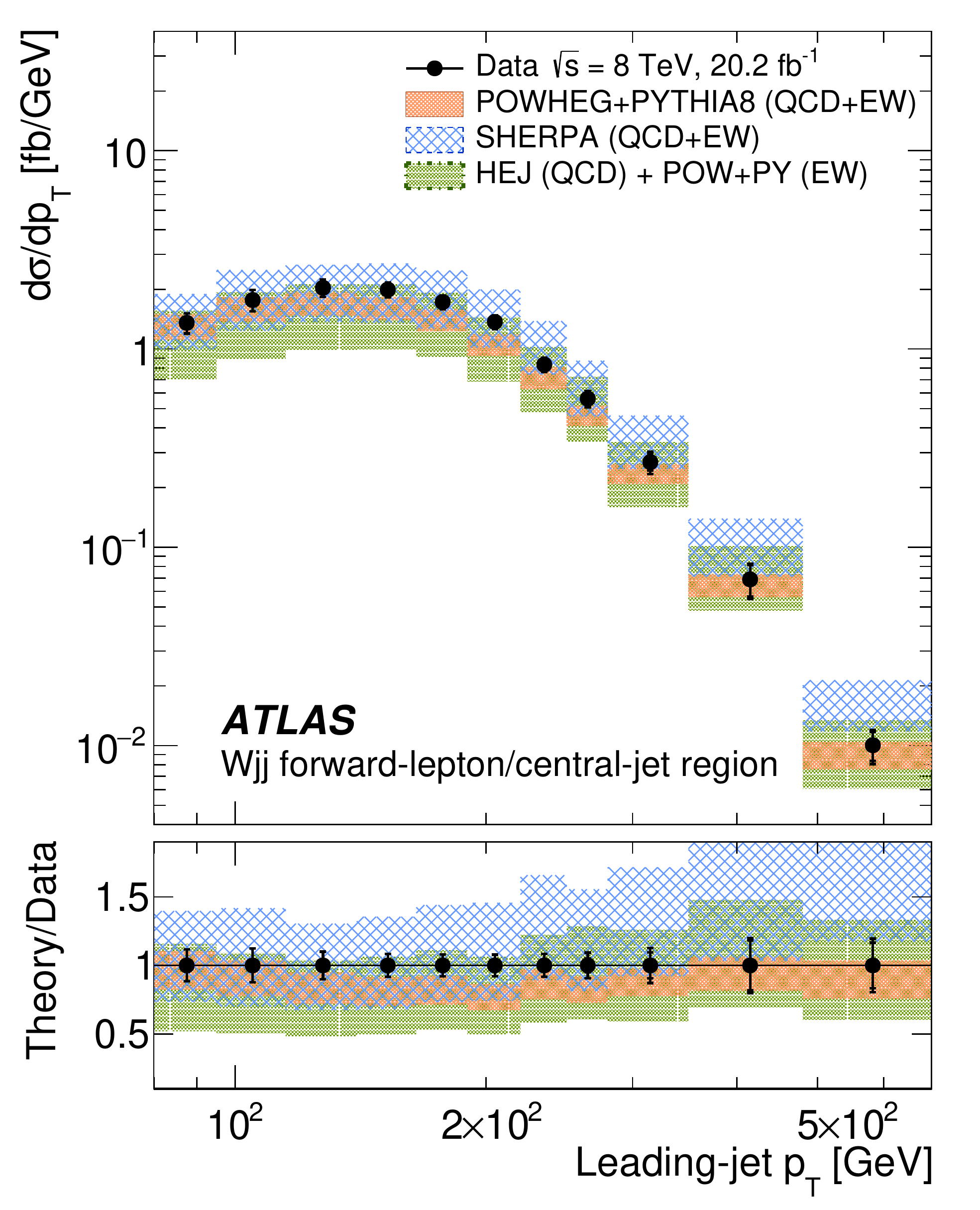}
\includegraphics[width=0.35\textwidth]{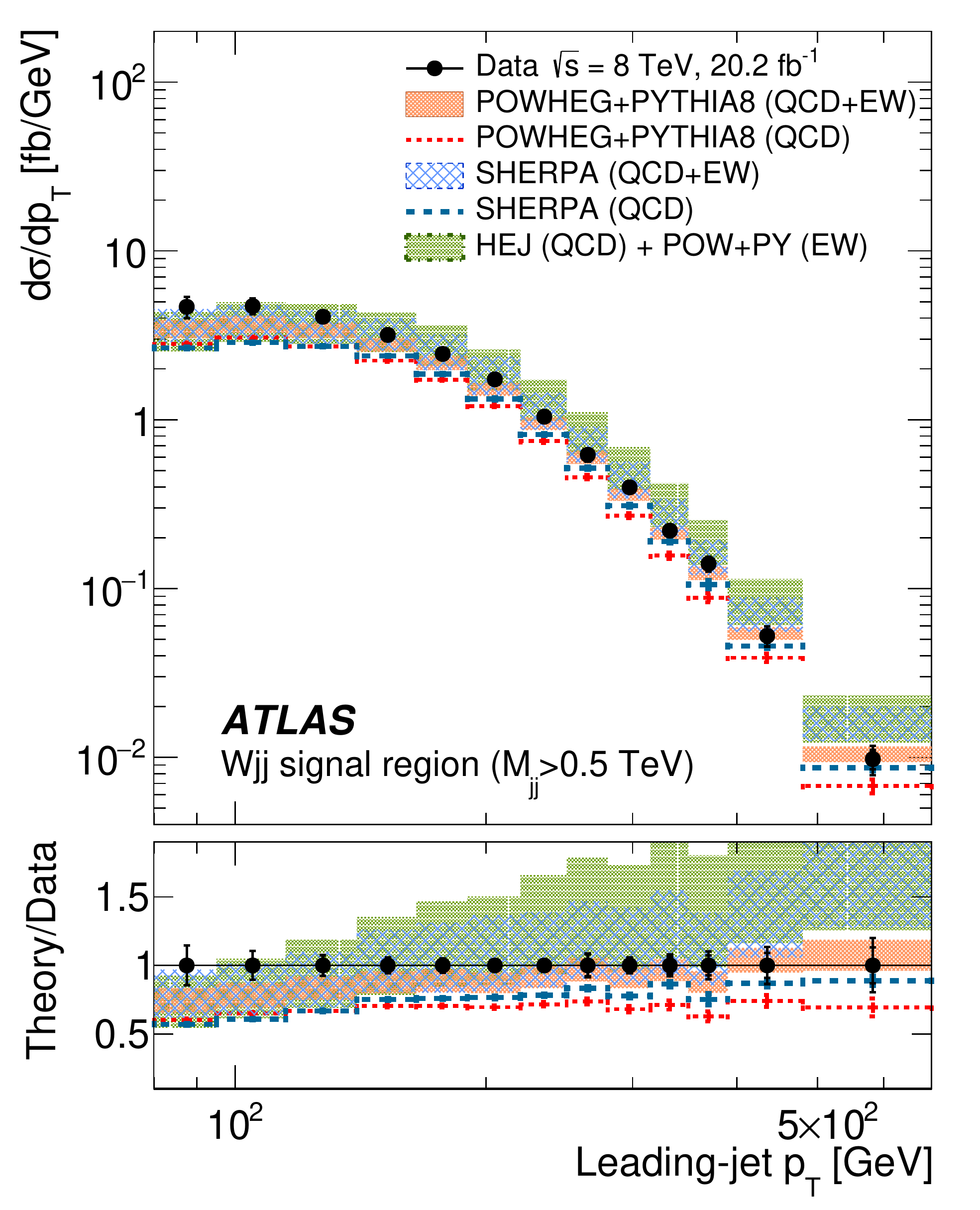}
\includegraphics[width=0.35\textwidth]{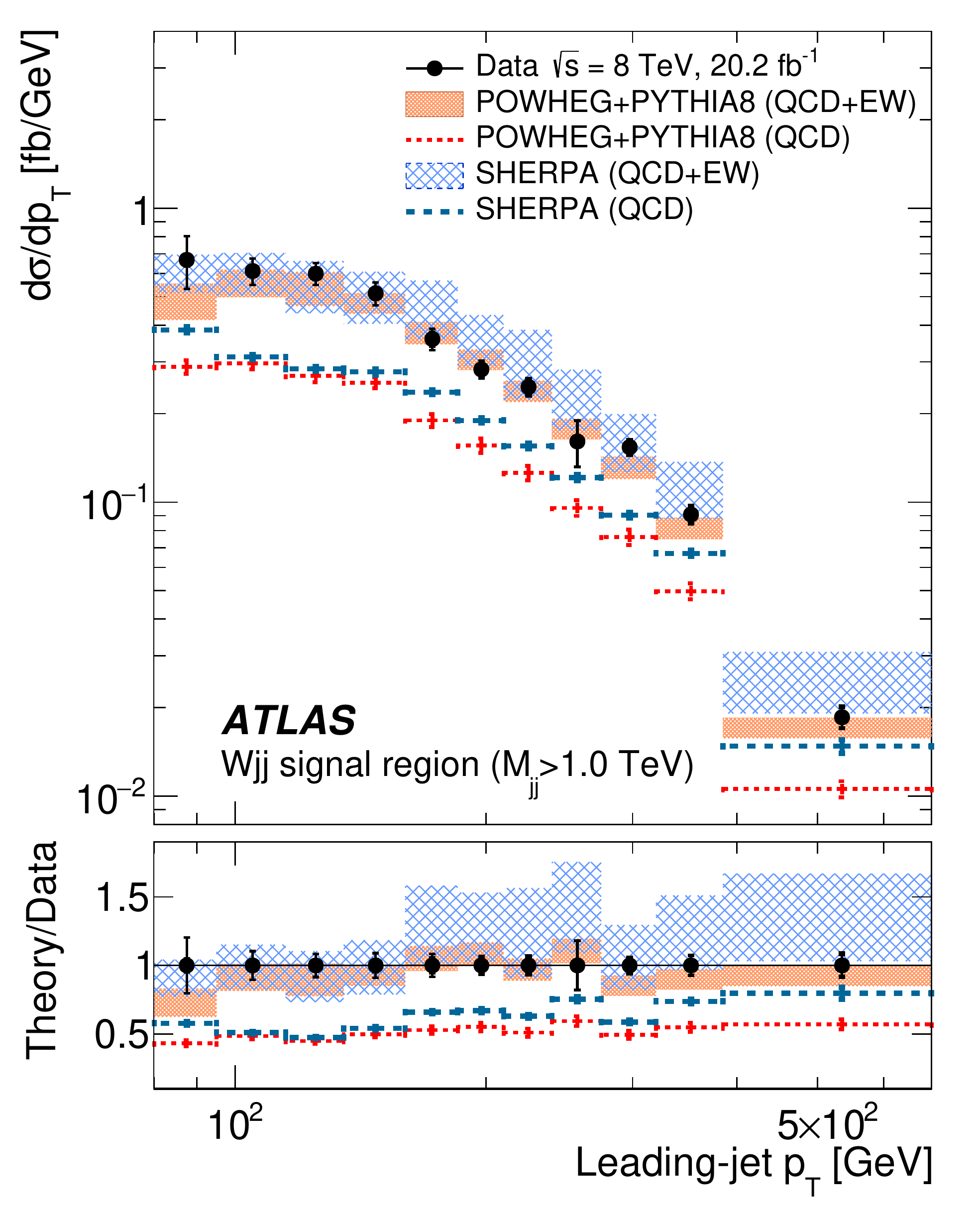}
\includegraphics[width=0.35\textwidth]{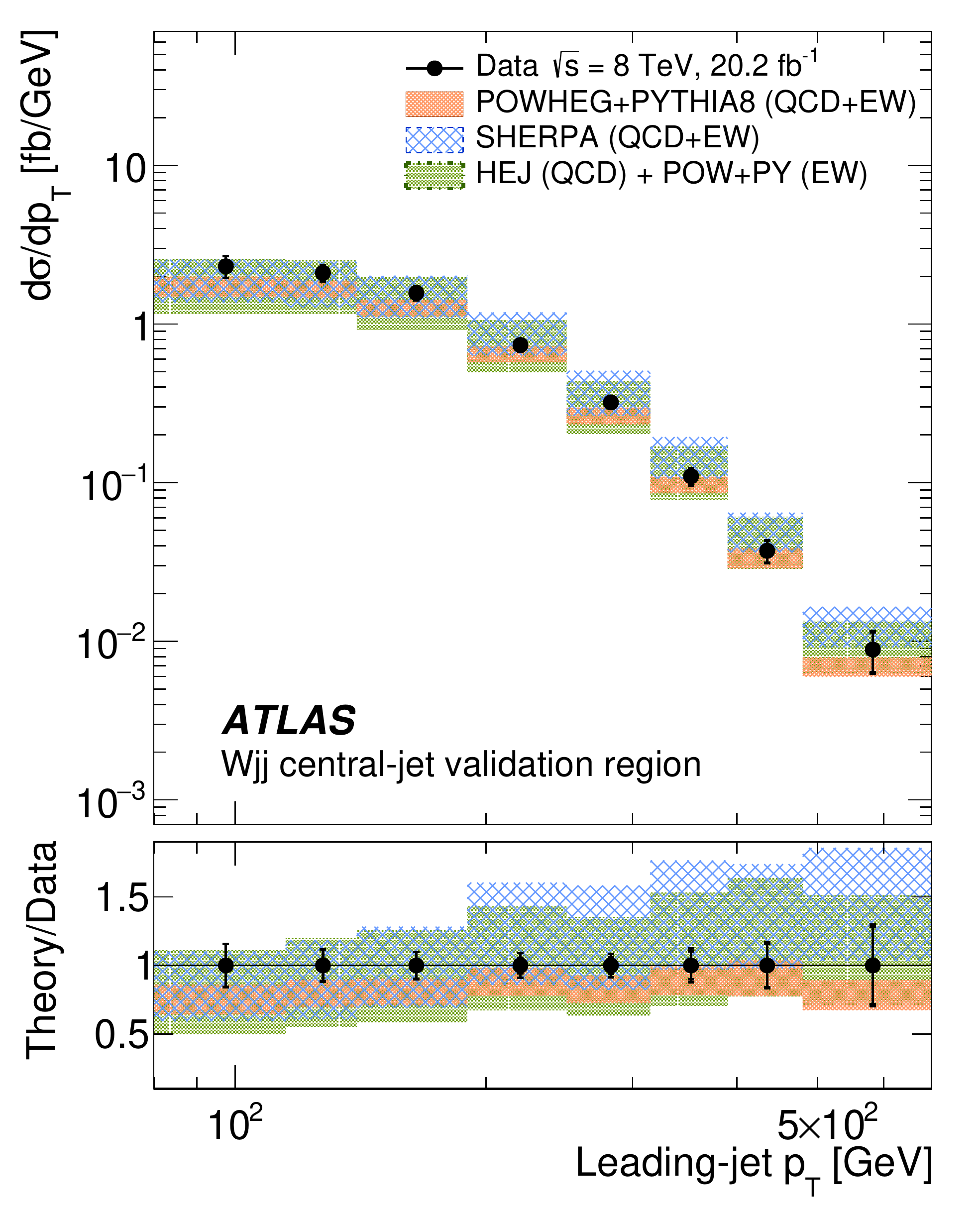}
\includegraphics[width=0.35\textwidth]{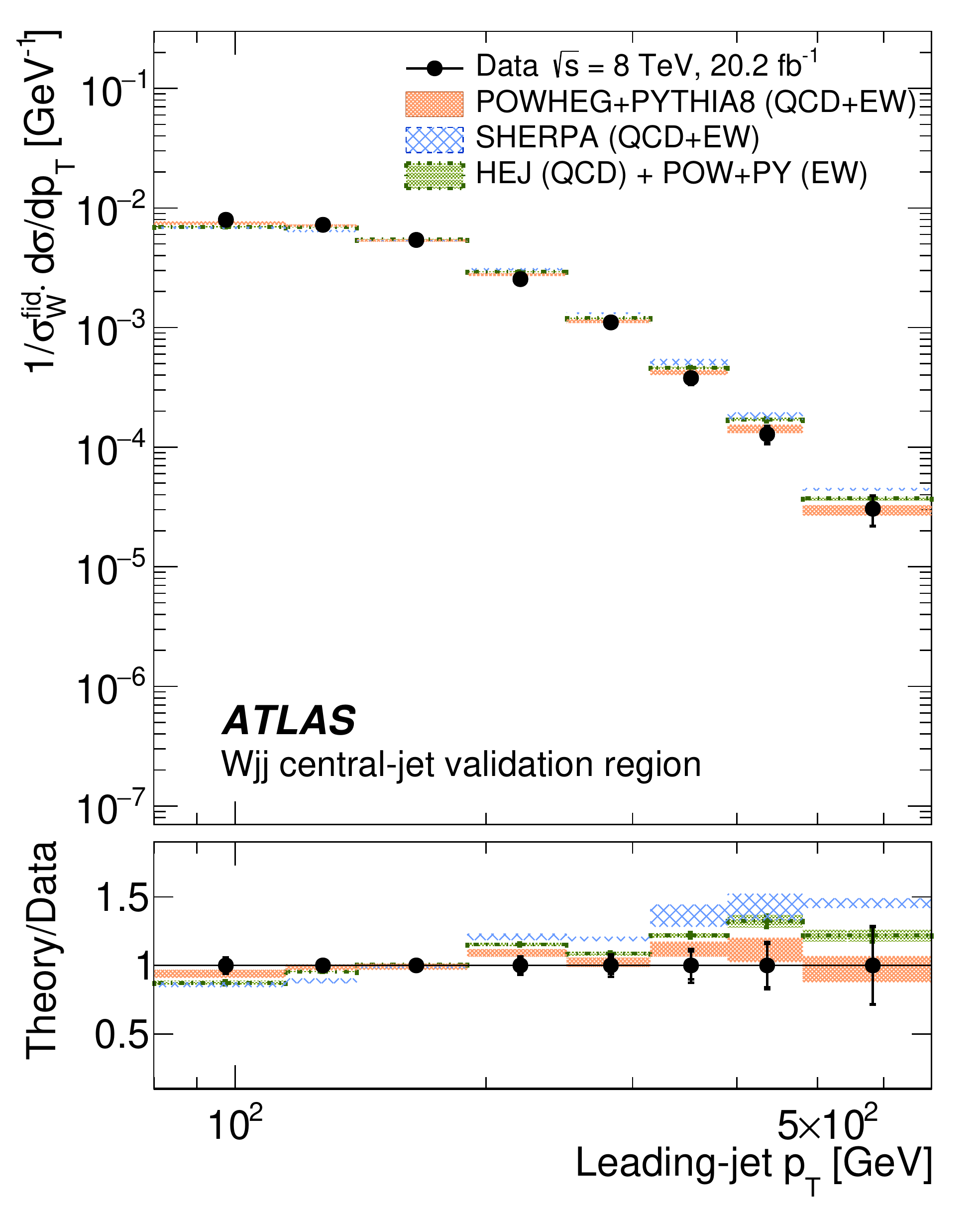}
\caption{Unfolded absolute differential \wjets production cross sections as a function of leading-jet $\pt$ for the
forward-lepton control region (top left), forward-lepton/central-jet fiducial region (top right), the signal 
regions with $\mjj>0.5$~TeV (middle left) and 1.0~\TeV (middle right), and the central-jet validation region 
(bottom).  The absolute (left) and normalized (right) distributions are shown in the central-jet region.  Both 
statistical (inner bar) and total (outer bar) measurement uncertainties are shown, as well as ratios of the 
theoretical predictions to the data (the bottom panel in each distribution). }
\label{unfolding:combined_measurementj1pt1DantiLC}
\end{figure}
\begin{figure}[htbp]
\centering
\includegraphics[width=0.49\textwidth]{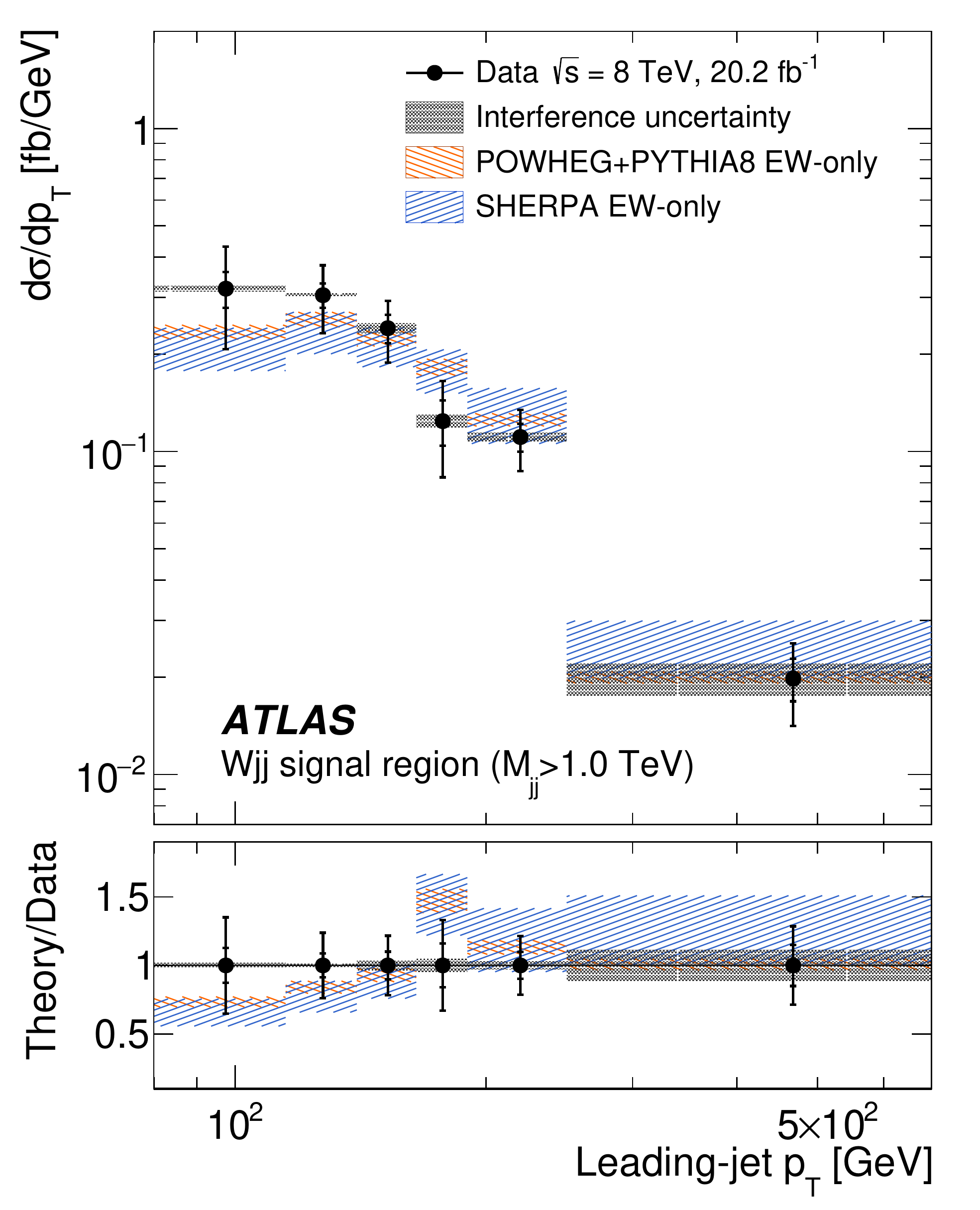}
\includegraphics[width=0.49\textwidth]{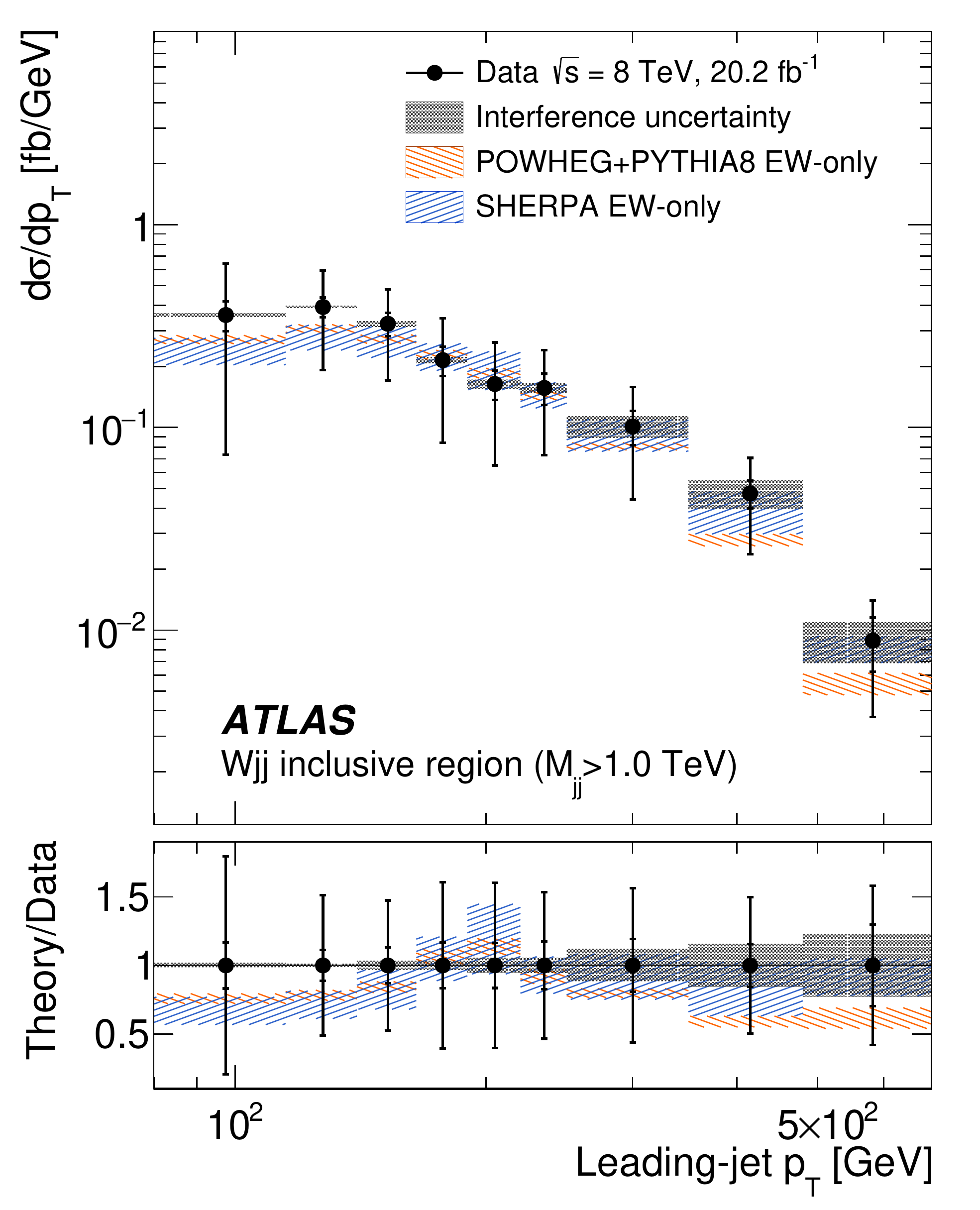}
\includegraphics[width=0.49\textwidth]{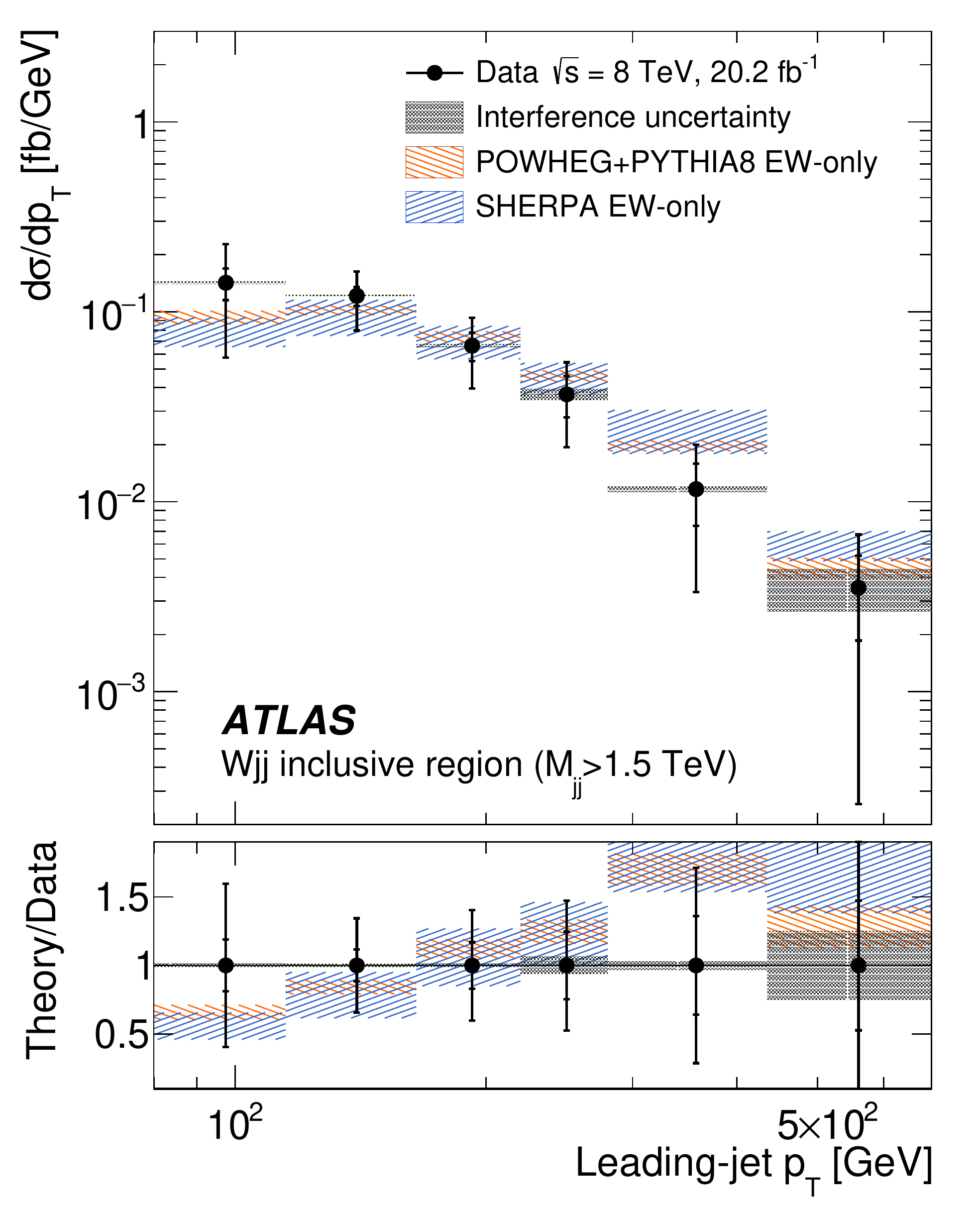}
\includegraphics[width=0.49\textwidth]{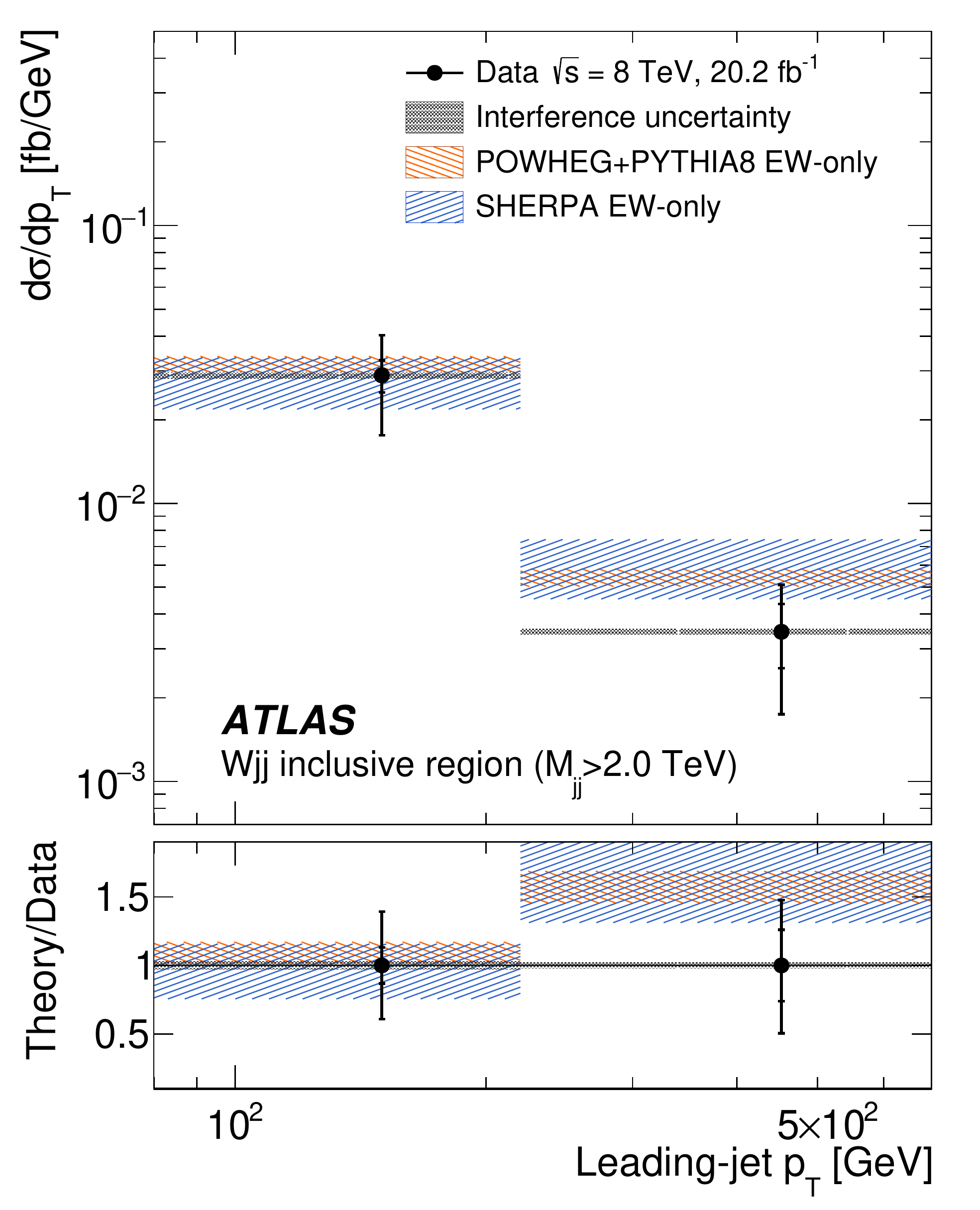}
\caption{Differential electroweak \wjets production cross sections as a function of the leading-jet $\pt$ in the 
high-mass signal region and the inclusive fiducial region with three thresholds on the dijet invariant mass (1.0~\TeV, 
1.5~\TeV, and 2.0~\TeV).  Both statistical (inner bar) and total (outer bar) measurement uncertainties are shown, as 
well as ratios of the theoretical predictions to the data (the bottom panel in each distribution). }
\label{unfolding:aux:AUX4}
\end{figure}
\begin{figure}[htbp]
\centering
\includegraphics[width=0.35\textwidth]{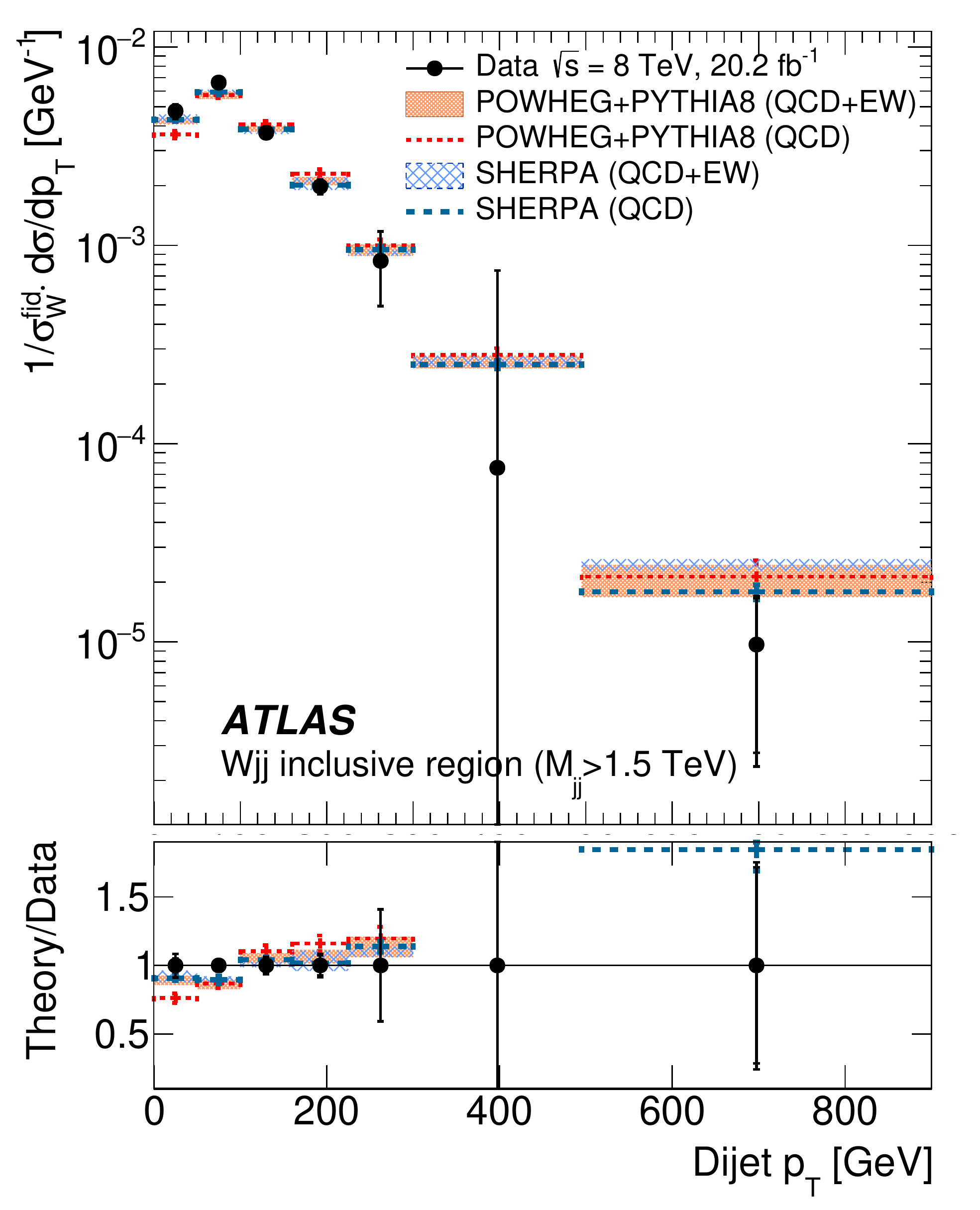}
\includegraphics[width=0.35\textwidth]{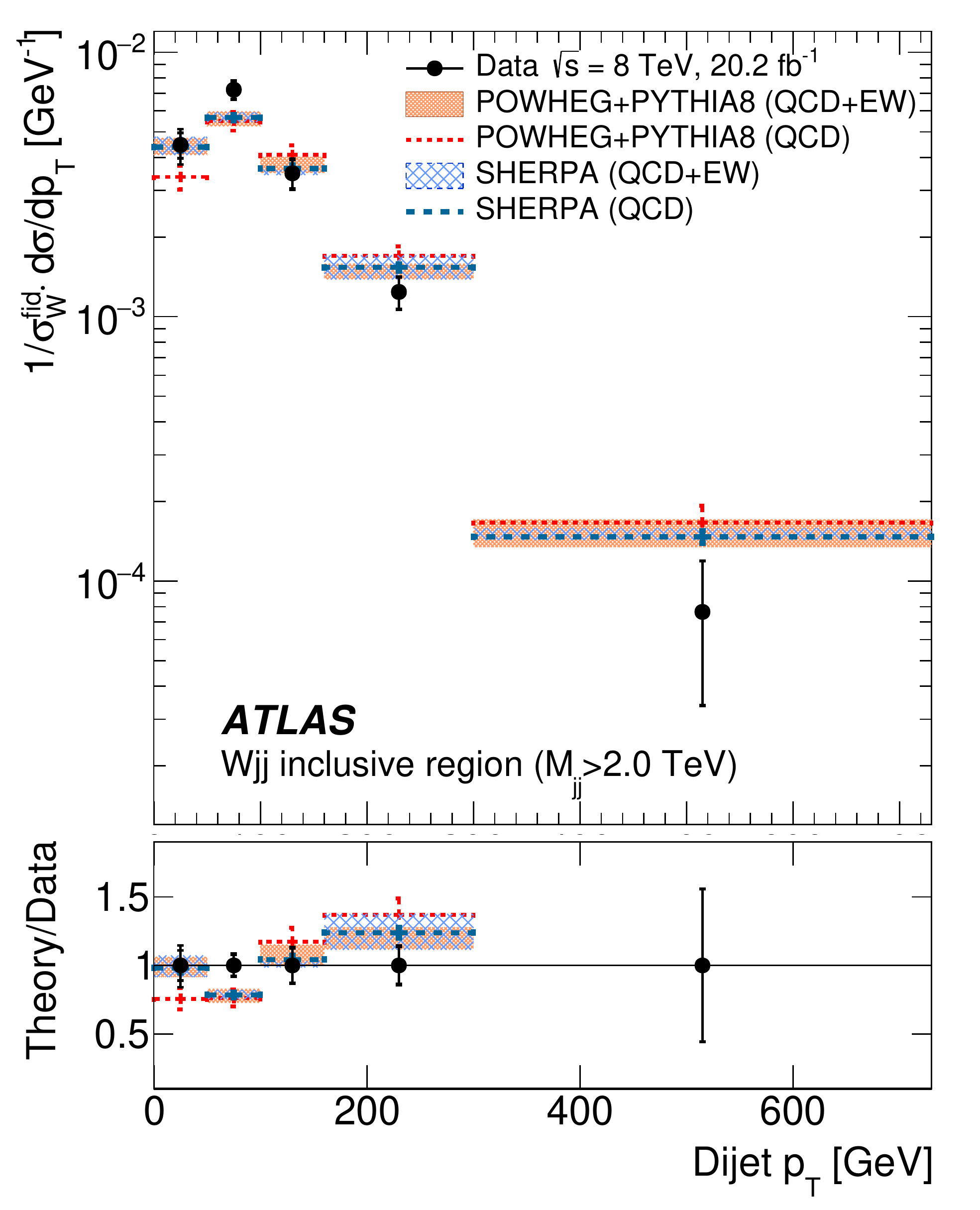}
\includegraphics[width=0.35\textwidth]{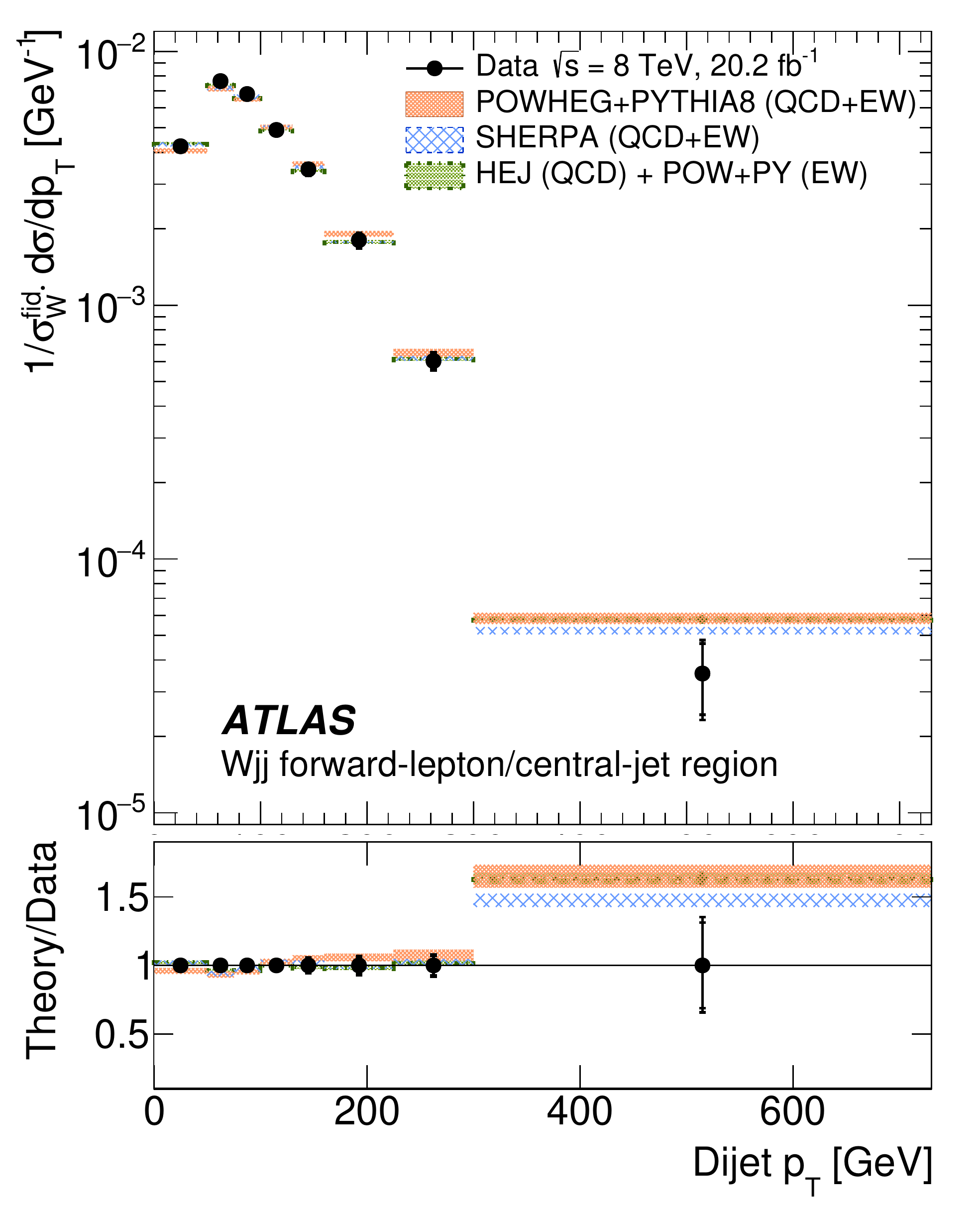}
\includegraphics[width=0.35\textwidth]{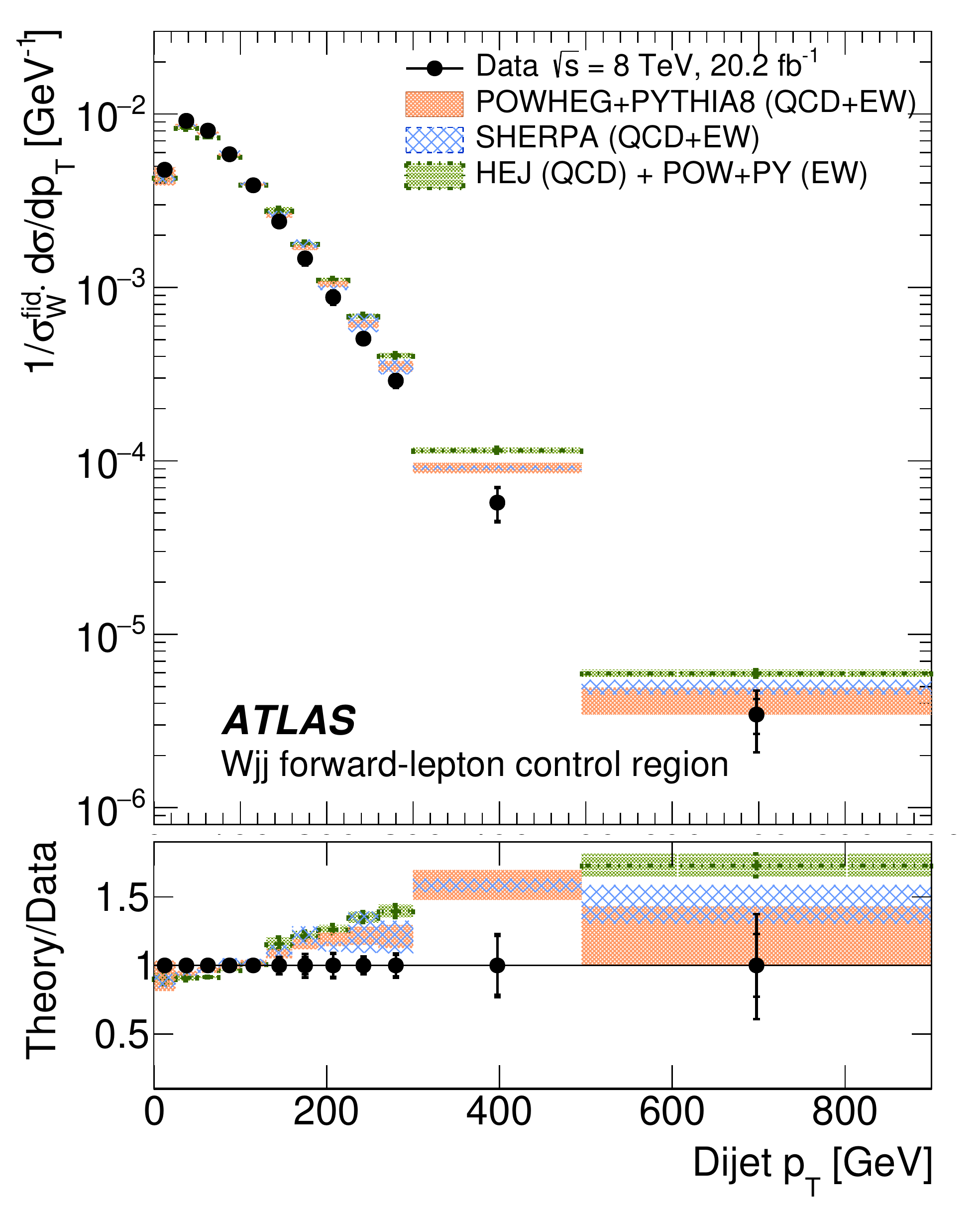}
\includegraphics[width=0.35\textwidth]{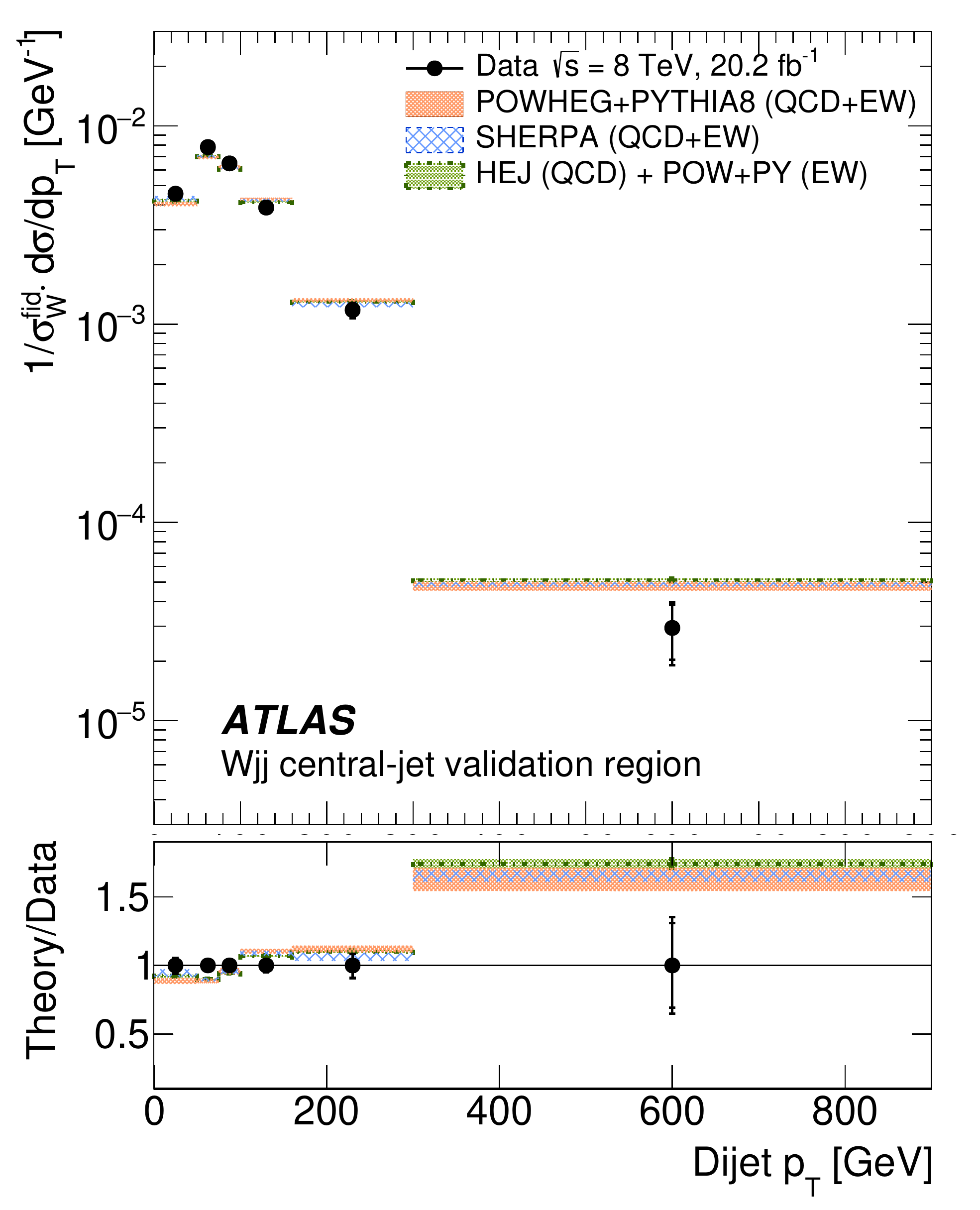}
\includegraphics[width=0.35\textwidth]{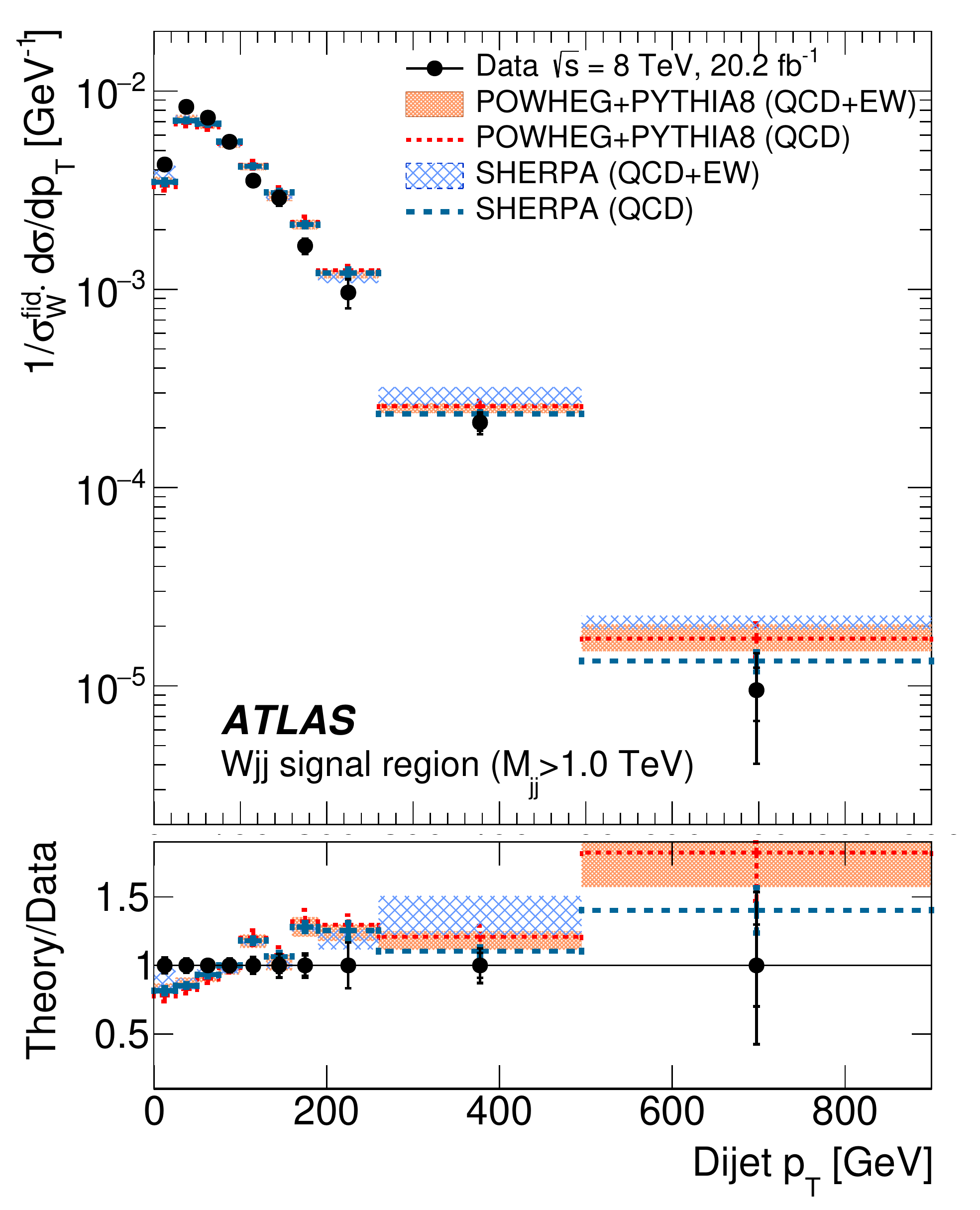}
\caption{Unfolded normalized differential \wjets production cross sections as a function of dijet $\pt$ in the 
inclusive (top), forward-lepton/central-jet (middle left), forward-lepton (middle right), central-jet (bottom left), 
and high-mass signal (bottom right) fiducial regions.  The inclusive regions show the distributions for \mjj thresholds 
of 1.5~\TeV~(left) and 2.0~\TeV~(right).  Both statistical (inner bar) and total (outer bar) measurement uncertainties 
are shown, as well as ratios of the theoretical predictions to the data (the bottom panel in each distribution).}
\label{unfolding:unfolding:combined_measurementDijetPtNorm}
\end{figure}
\begin{figure}[htbp]
\centering
\includegraphics[width=0.35\textwidth]{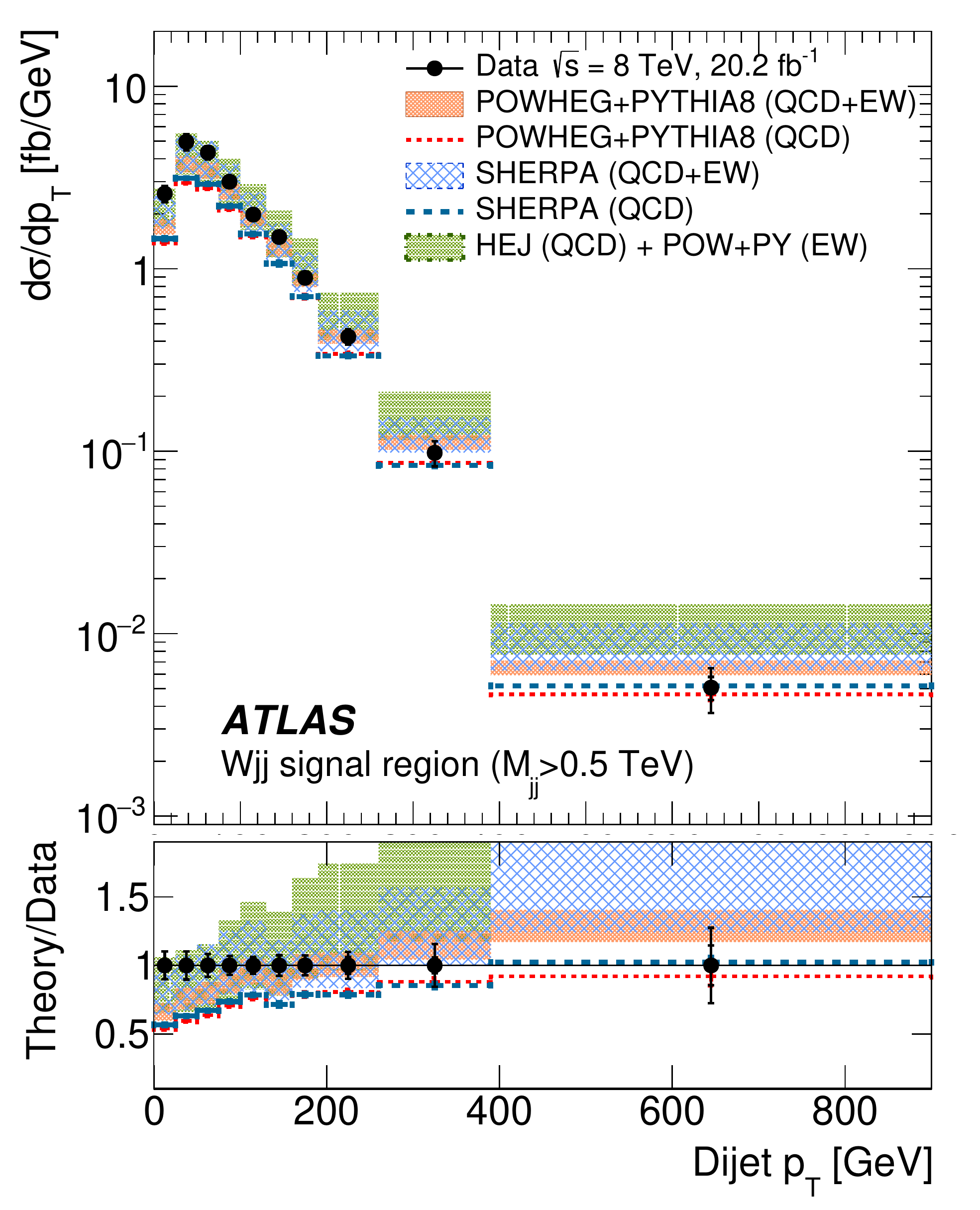}
\includegraphics[width=0.35\textwidth]{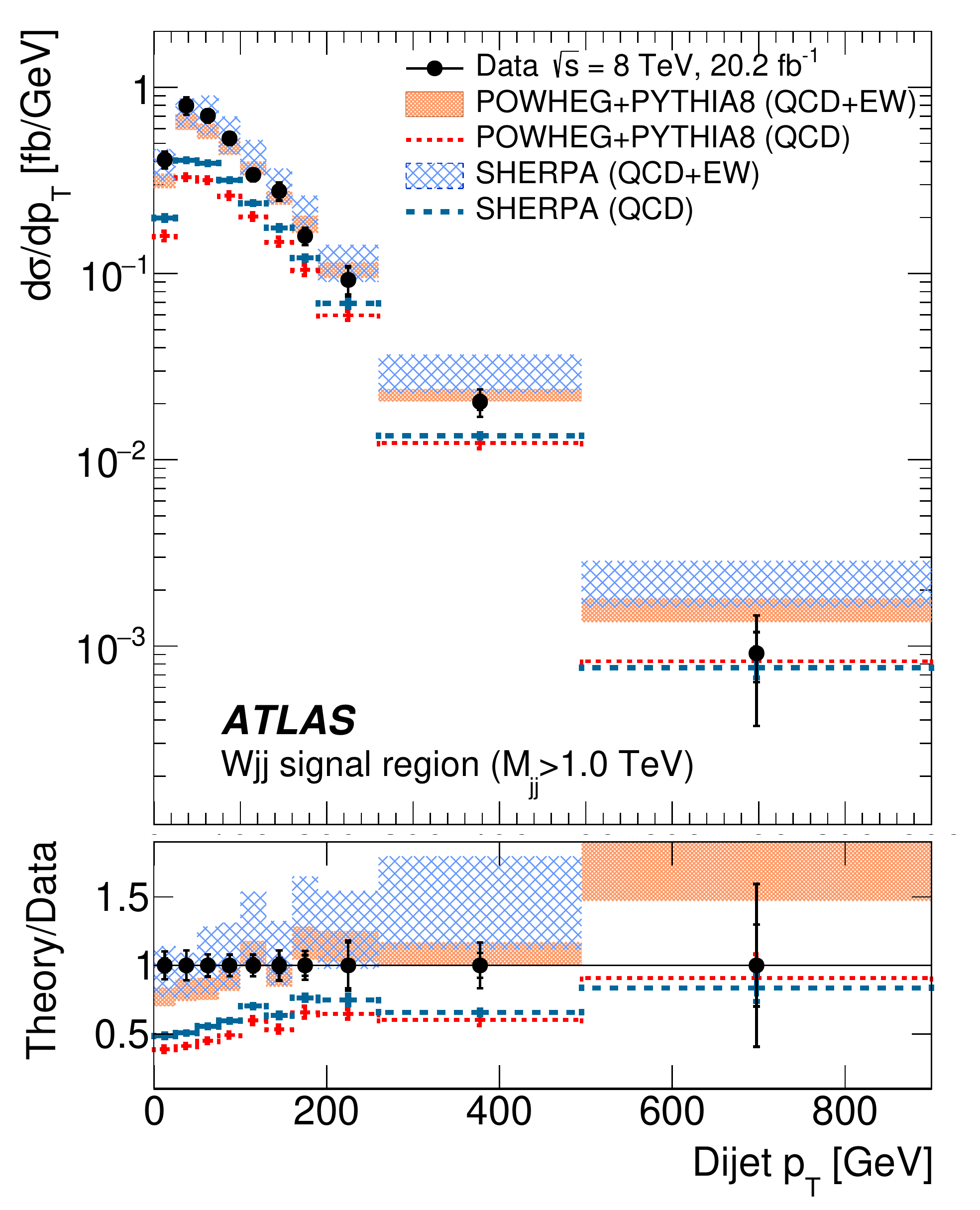}
\includegraphics[width=0.35\textwidth]{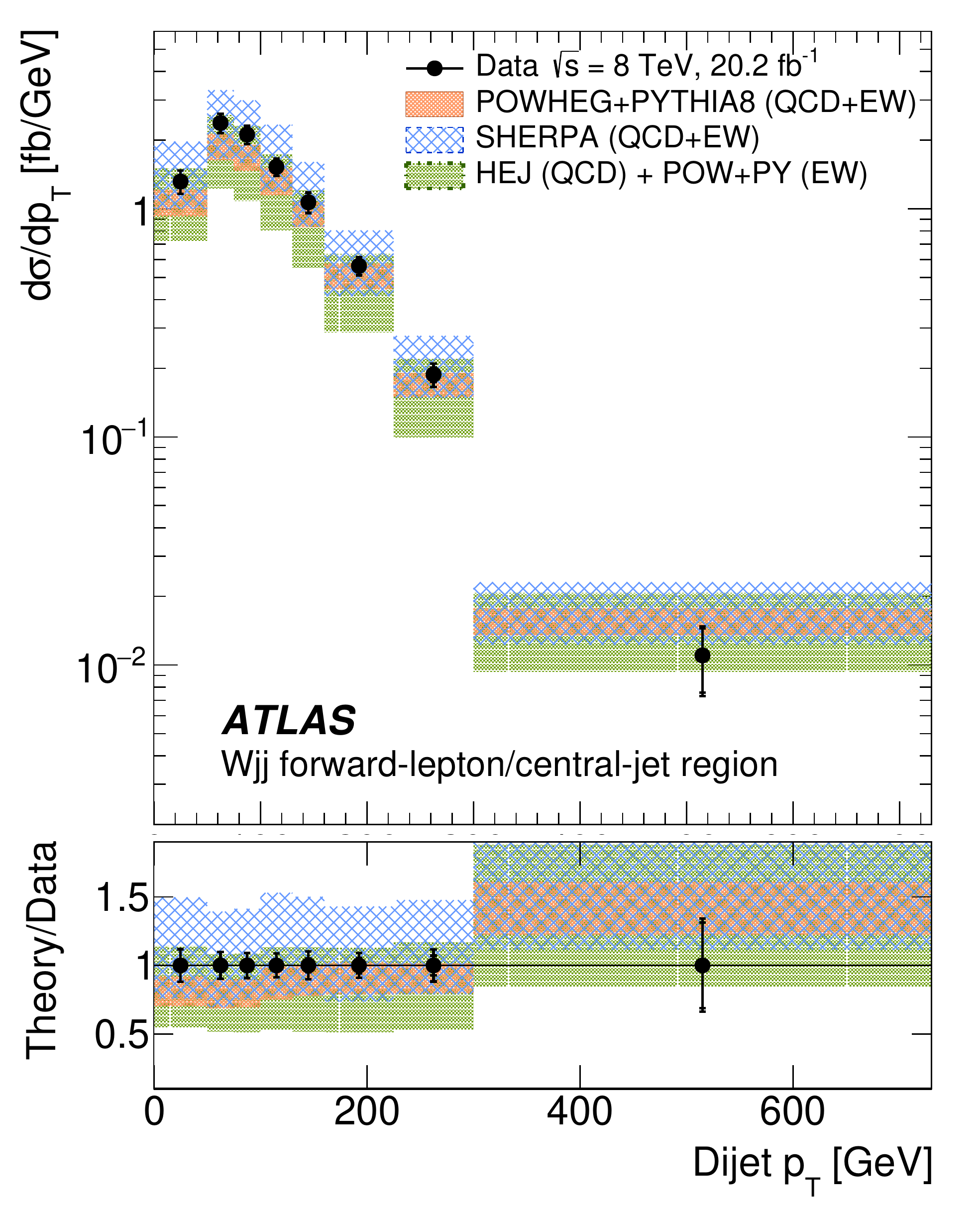}
\includegraphics[width=0.35\textwidth]{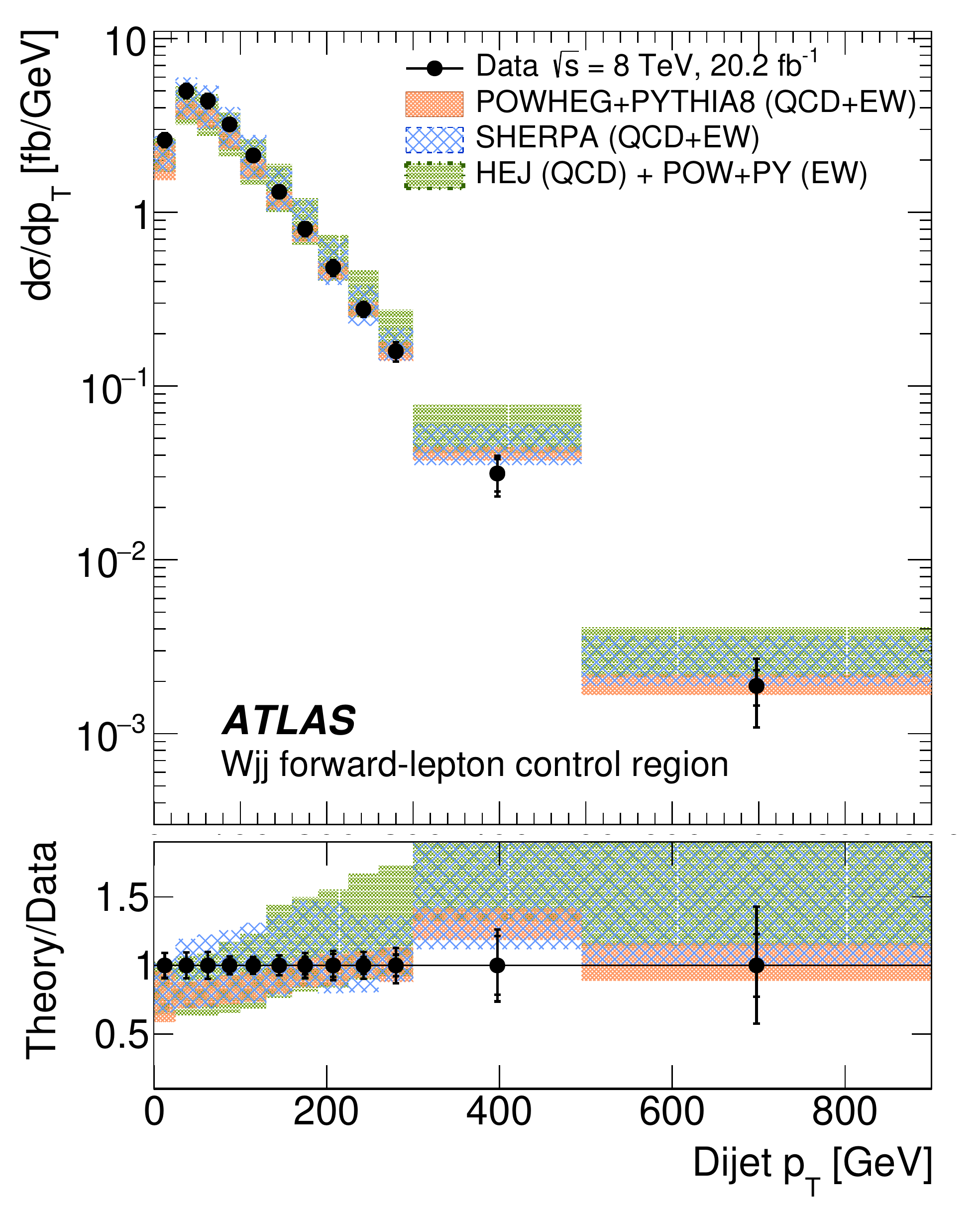}
\includegraphics[width=0.35\textwidth]{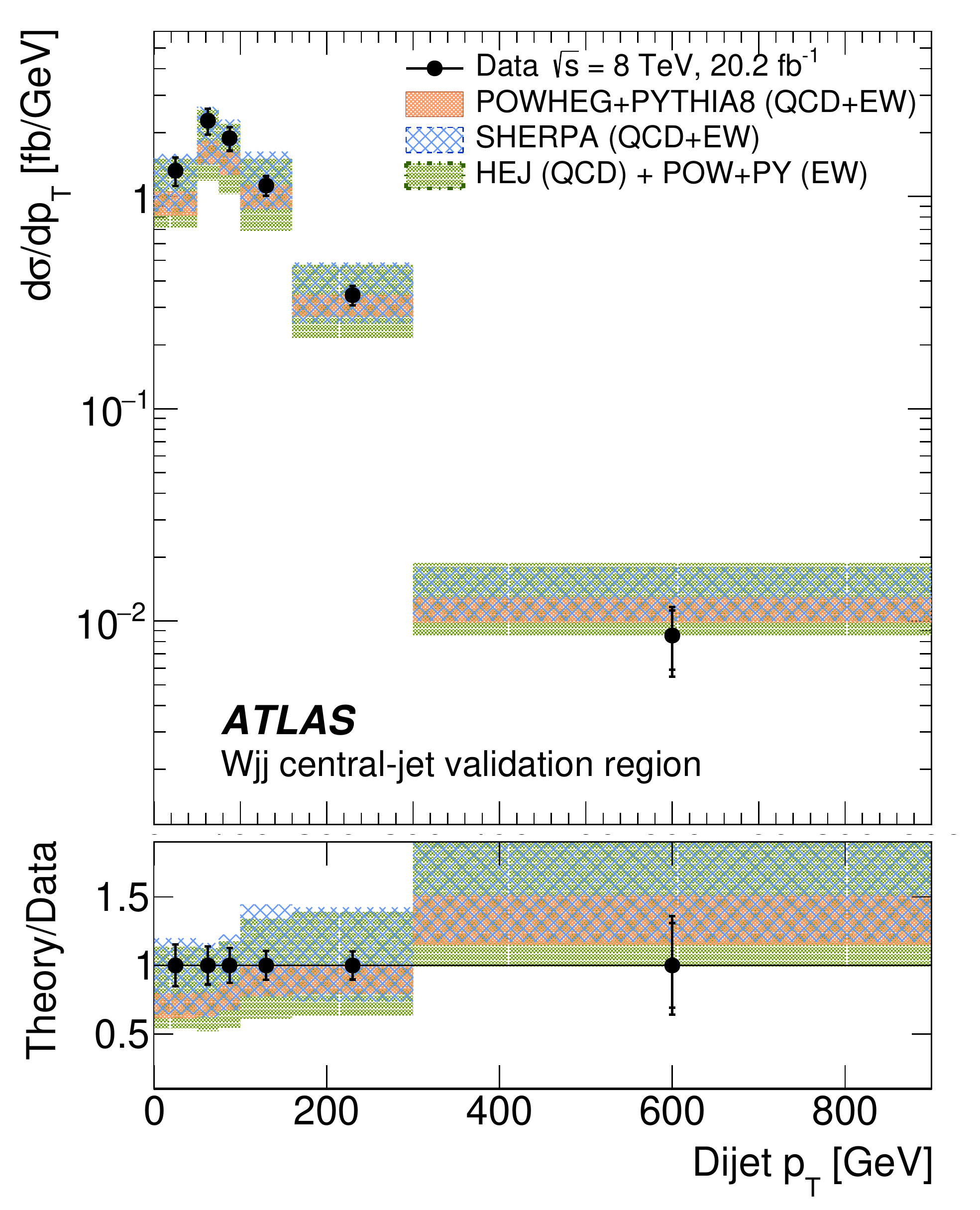}
\caption{Differential \wjets production cross sections as a function of dijet $\pt$ in the signal, high-mass 
signal, forward-lepton/central-jet, forward-lepton, and central-jet fiducial regions.  Both statistical
(inner bar) and total (outer bar) measurement uncertainties are shown, as well as ratios
of the theoretical predictions to the data (the bottom panel in each distribution). }
\label{unfolding:aux:AUX21}
\end{figure}
\begin{figure}[htbp]
\centering
\includegraphics[width=0.49\textwidth]{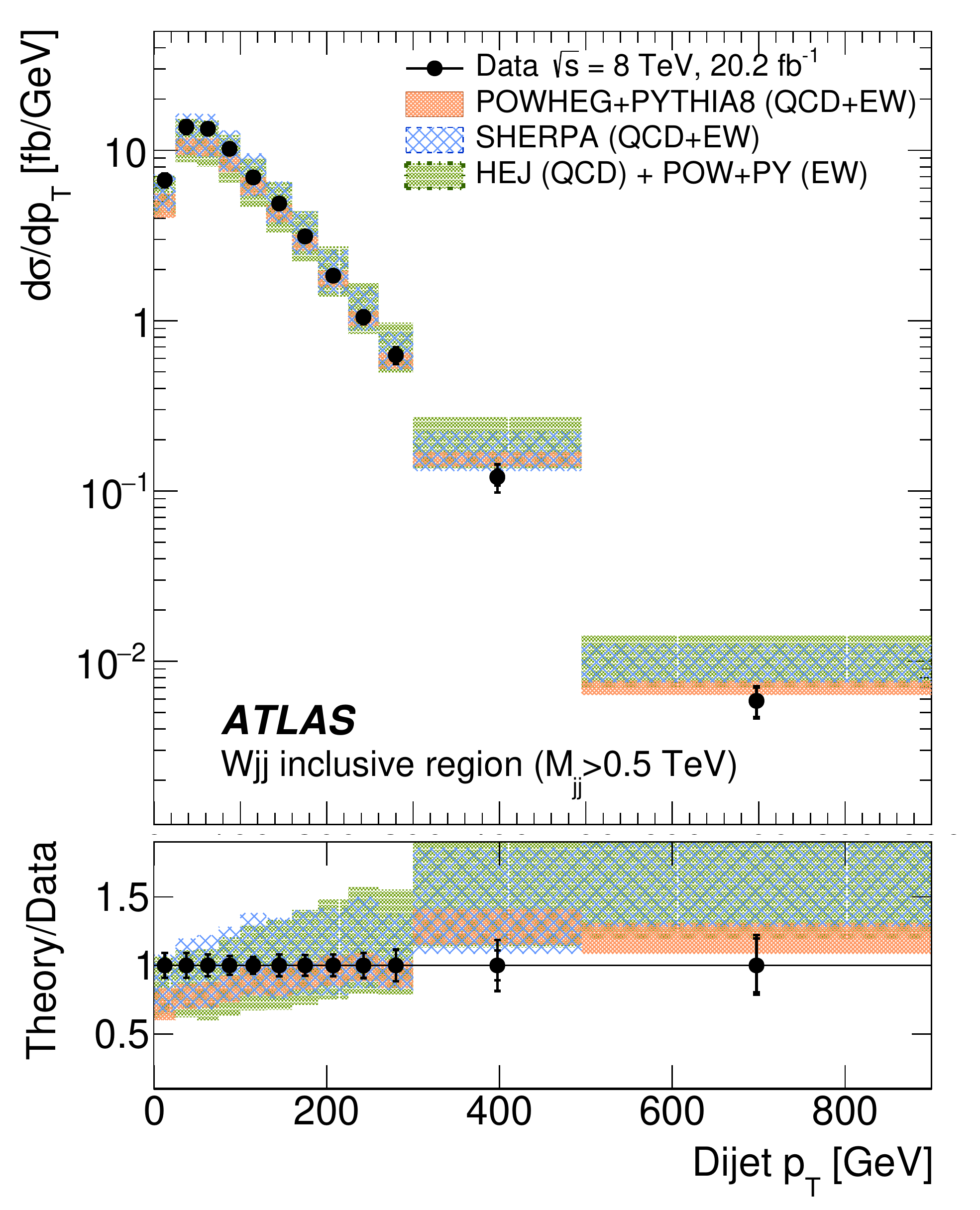}
\includegraphics[width=0.49\textwidth]{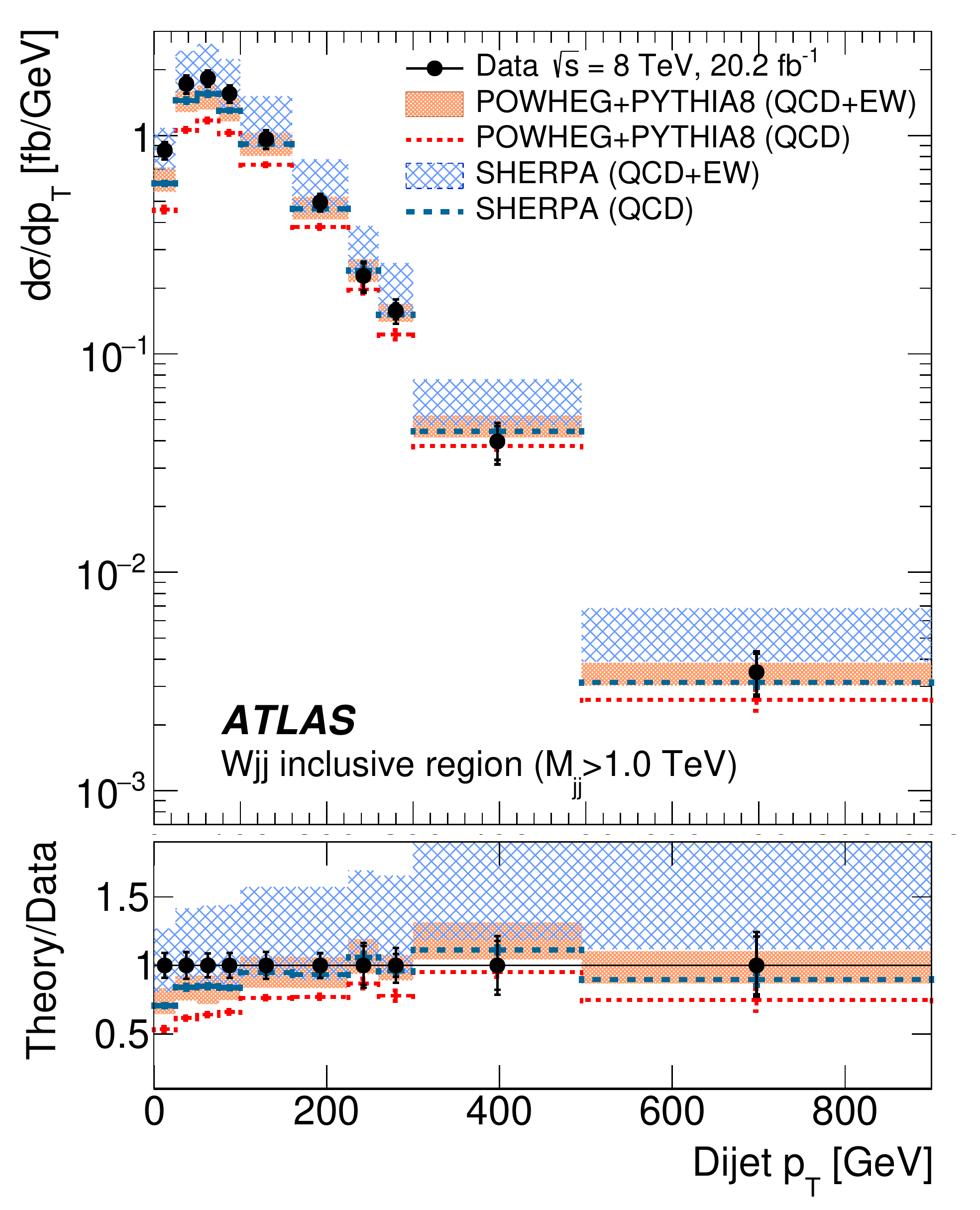}
\includegraphics[width=0.49\textwidth]{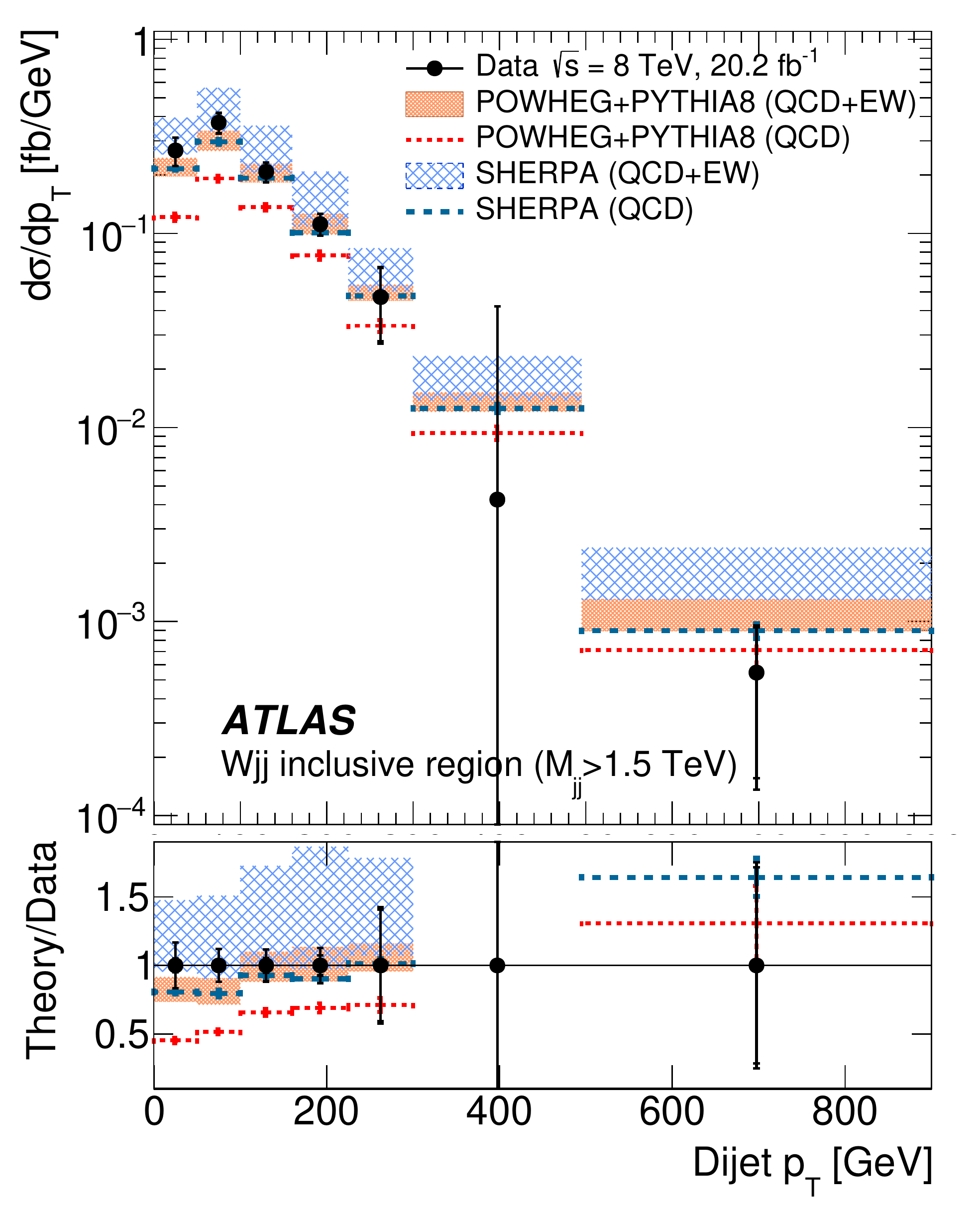}
\includegraphics[width=0.49\textwidth]{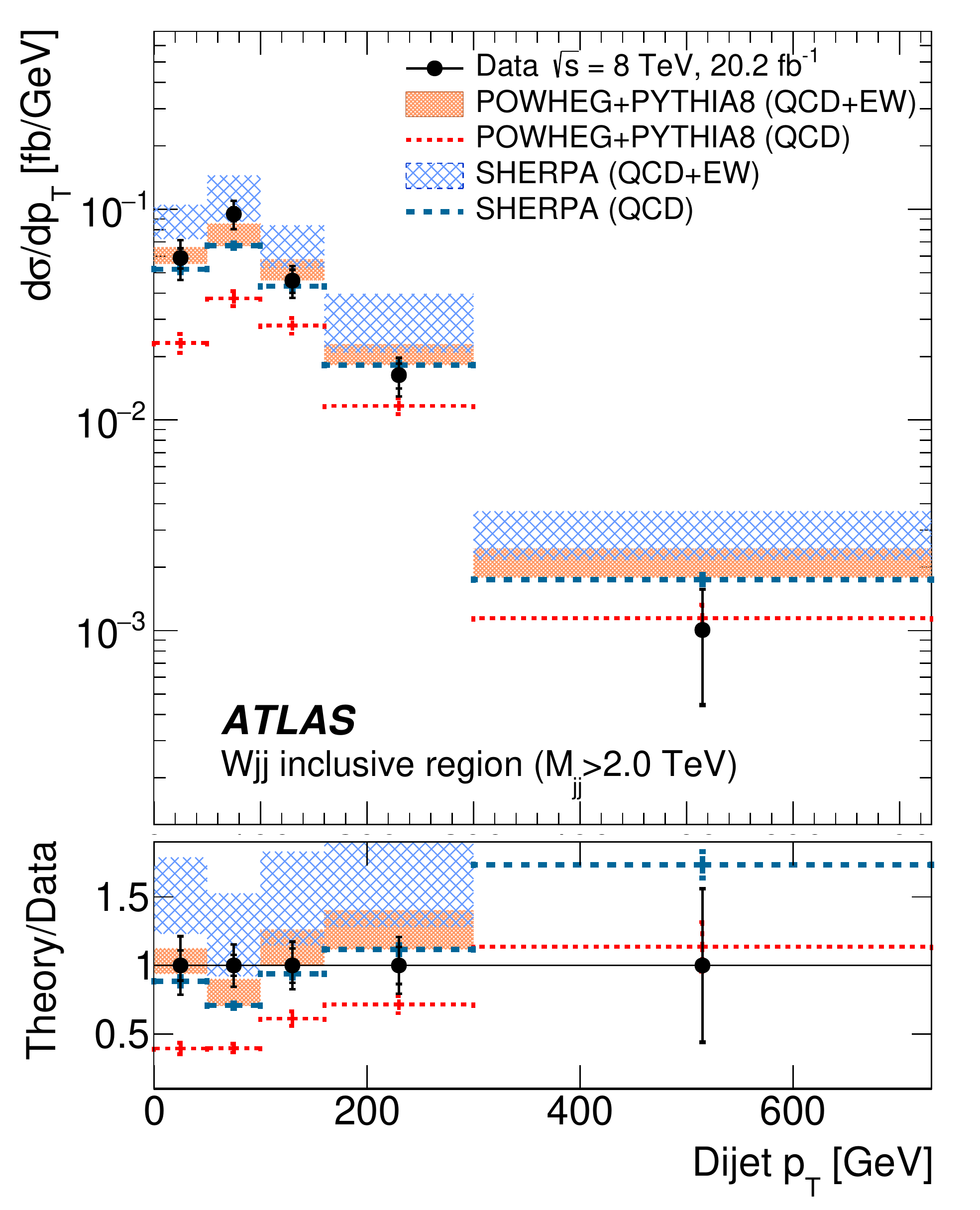}
\caption{Differential \wjets production cross sections as a function of dijet $\pt$ in the inclusive fiducial 
region with four thresholds on the dijet invariant mass (0.5~\TeV, 1.0~\TeV, 1.5~\TeV, and 2.0~\TeV).  Both 
statistical (inner bar) and total (outer bar) measurement uncertainties are shown, as well as ratios
of the theoretical predictions to the data (the bottom panel in each distribution).}
\label{unfolding:aux:AUX22}
\end{figure}
\begin{figure}[htbp]
\centering
\includegraphics[width=0.49\textwidth]{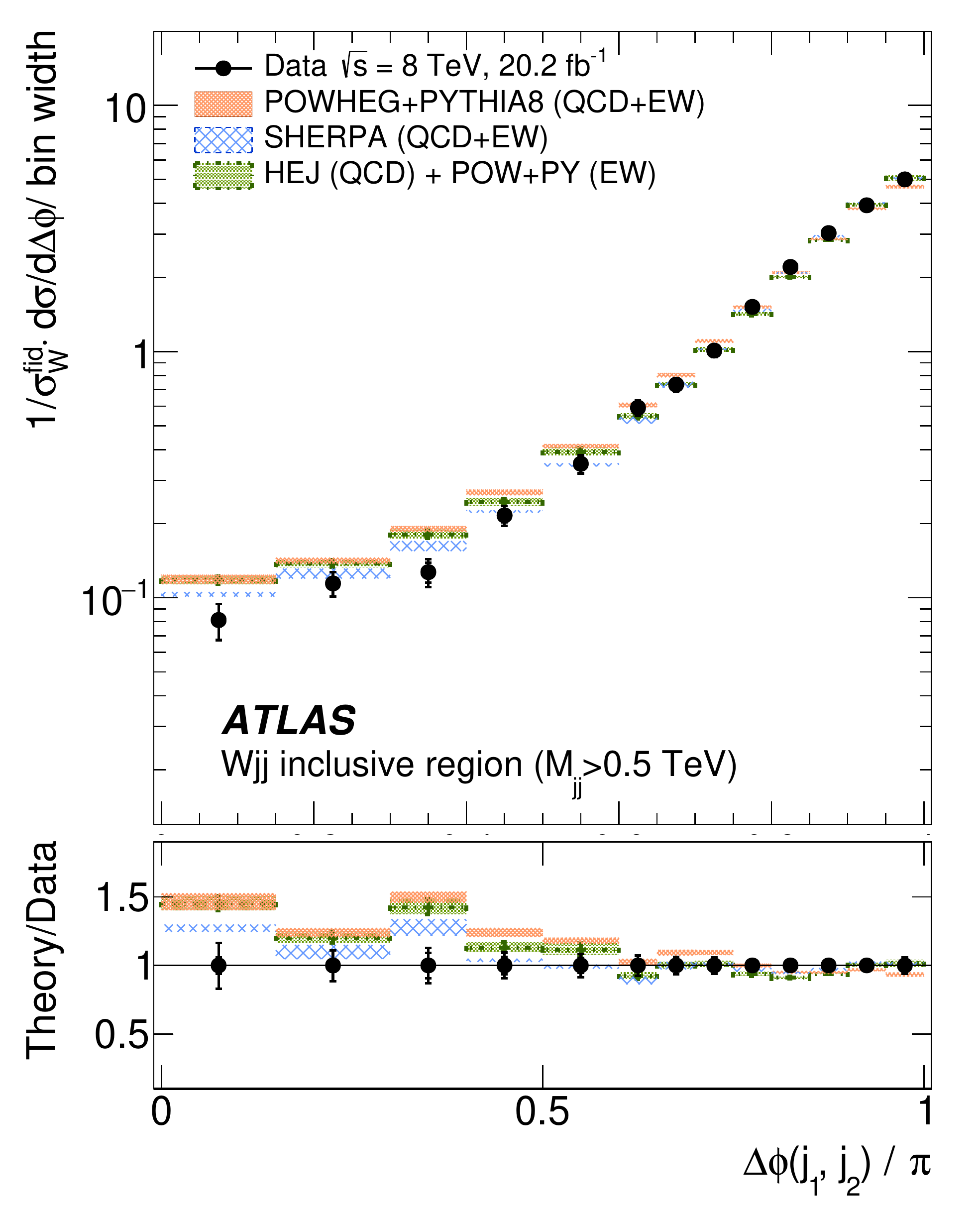}
\includegraphics[width=0.49\textwidth]{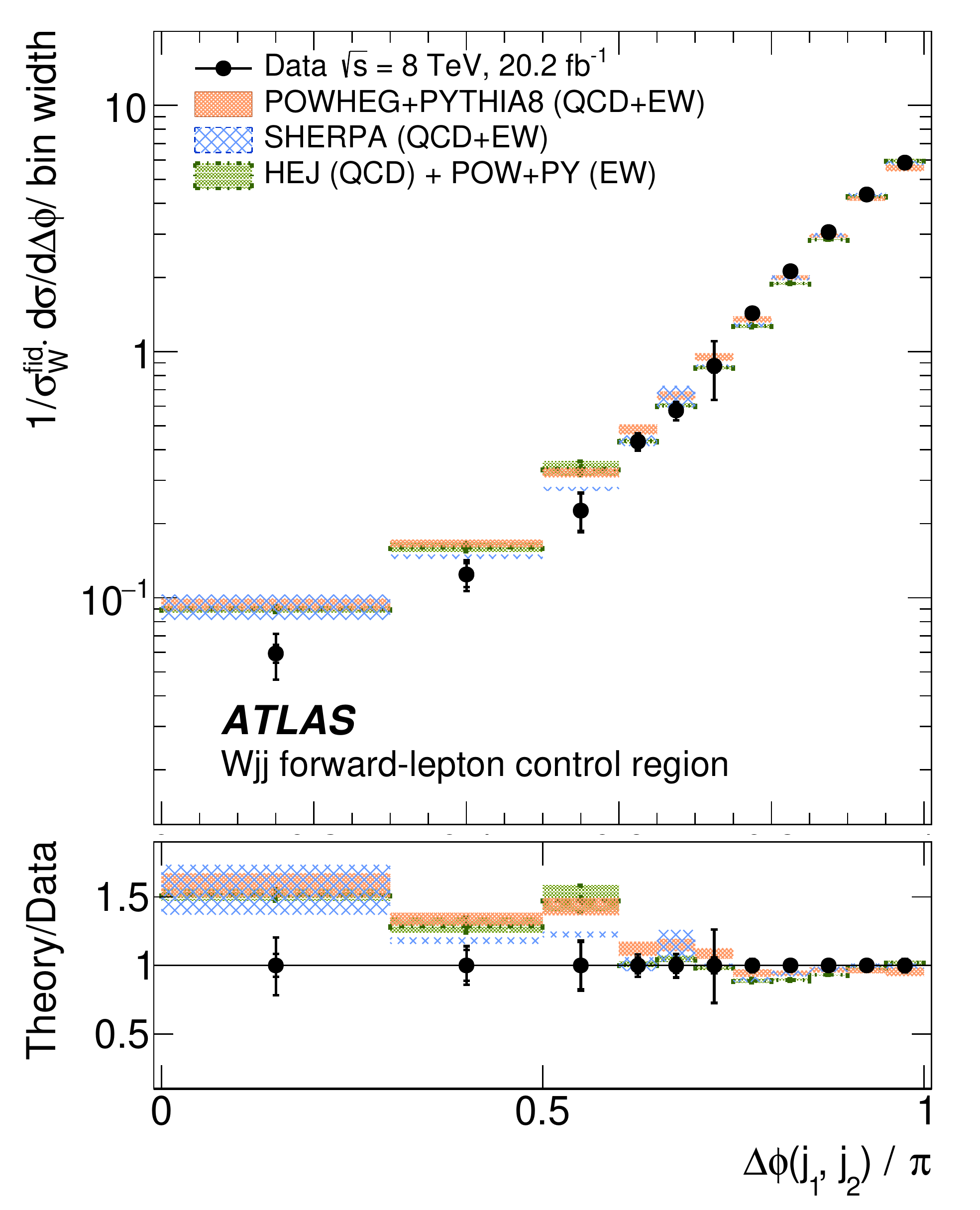}
\includegraphics[width=0.49\textwidth]{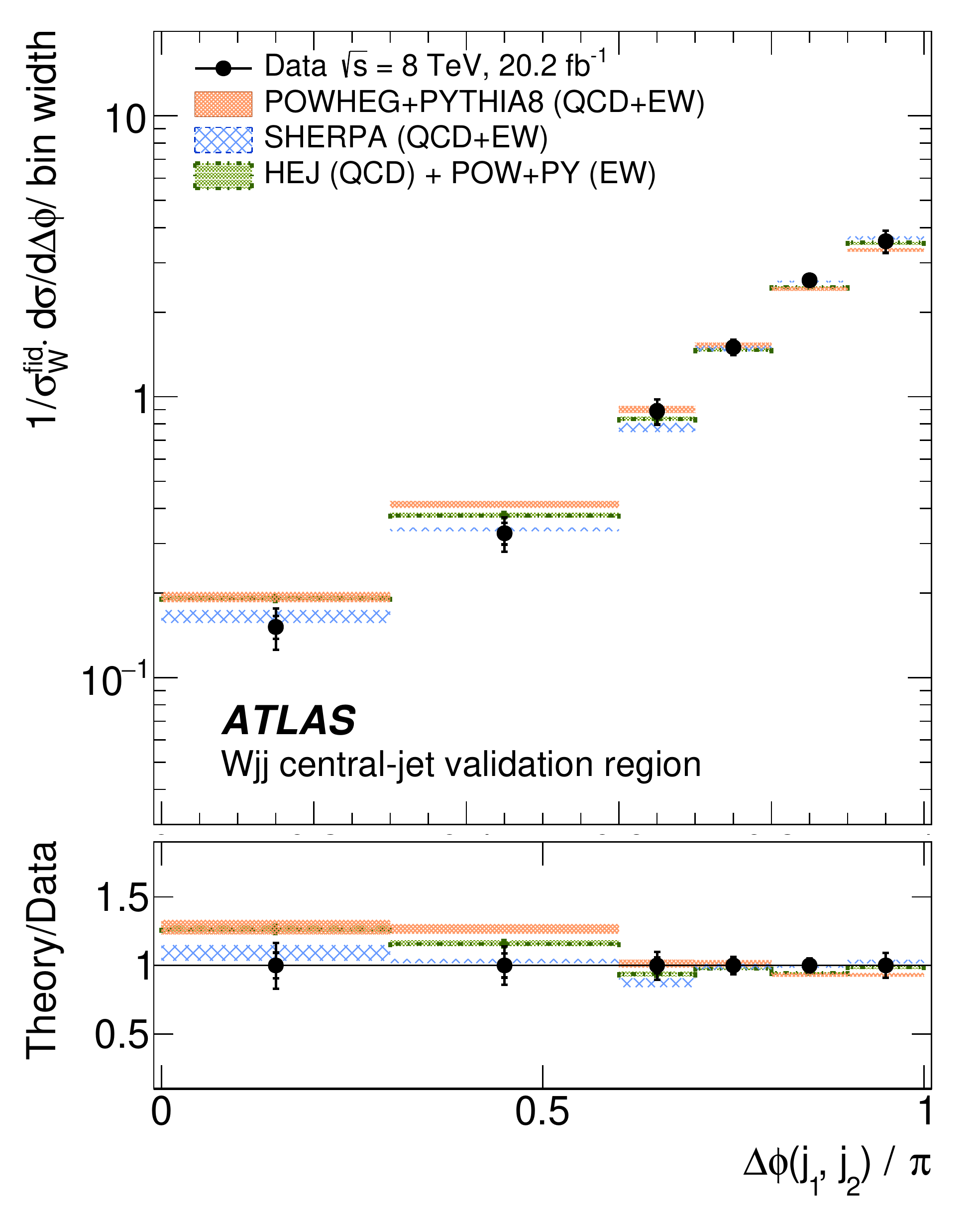}
\includegraphics[width=0.49\textwidth]{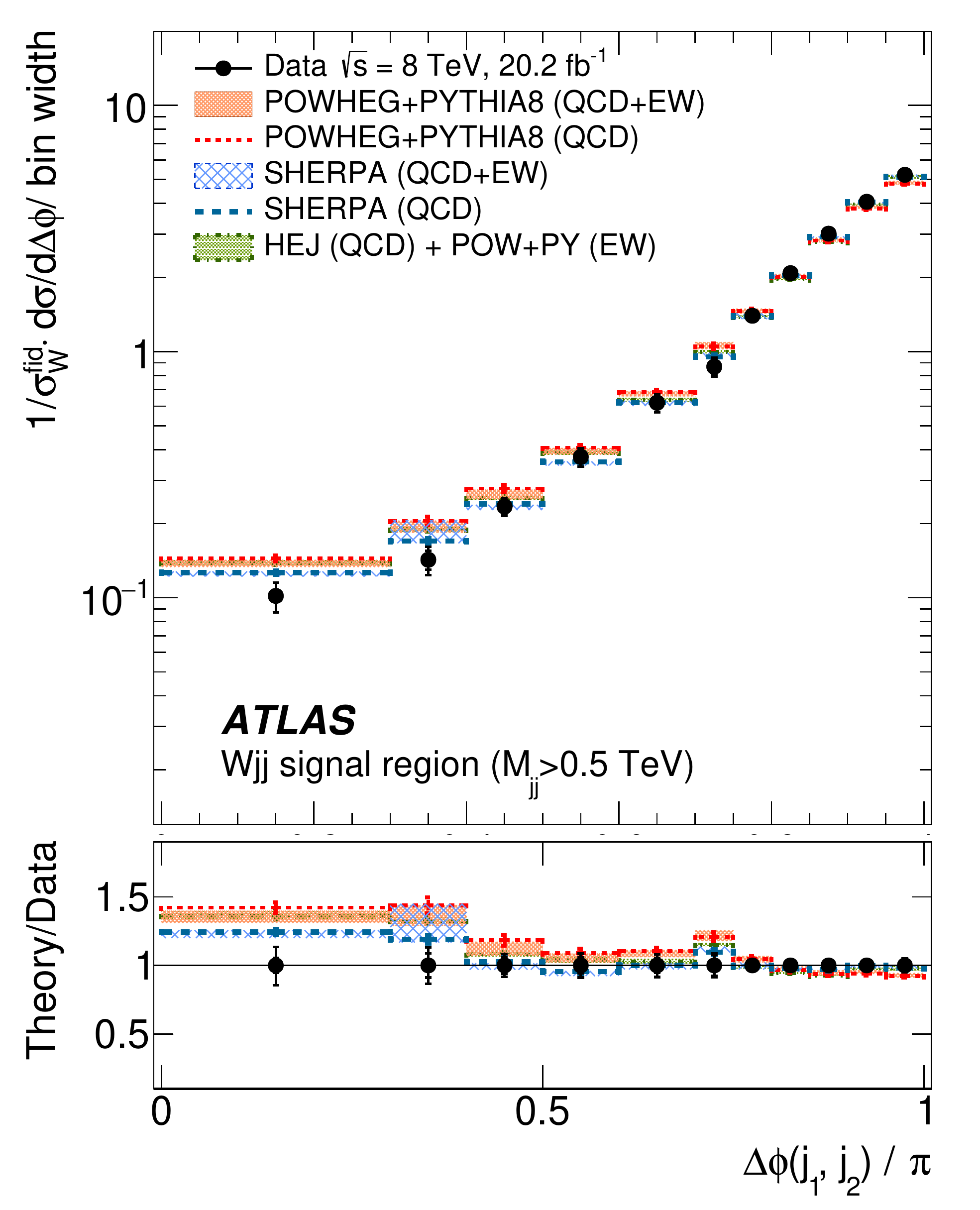}
\caption{Unfolded normalized differential production cross sections as a function of $\Delta\phi(j_1,j_2)$ for the 
inclusive, forward-lepton control, central-jet validation, and signal fiducial regions.  Both statistical
(inner bar) and total (outer bar) measurement uncertainties are shown, as well as ratios of the theoretical 
predictions to the data (the bottom panel in each distribution). }
\label{unfolding:combined_measurementdphi121DinclusiveNorm}
\end{figure}
\begin{figure}[htbp]
\centering
\includegraphics[width=0.49\textwidth]{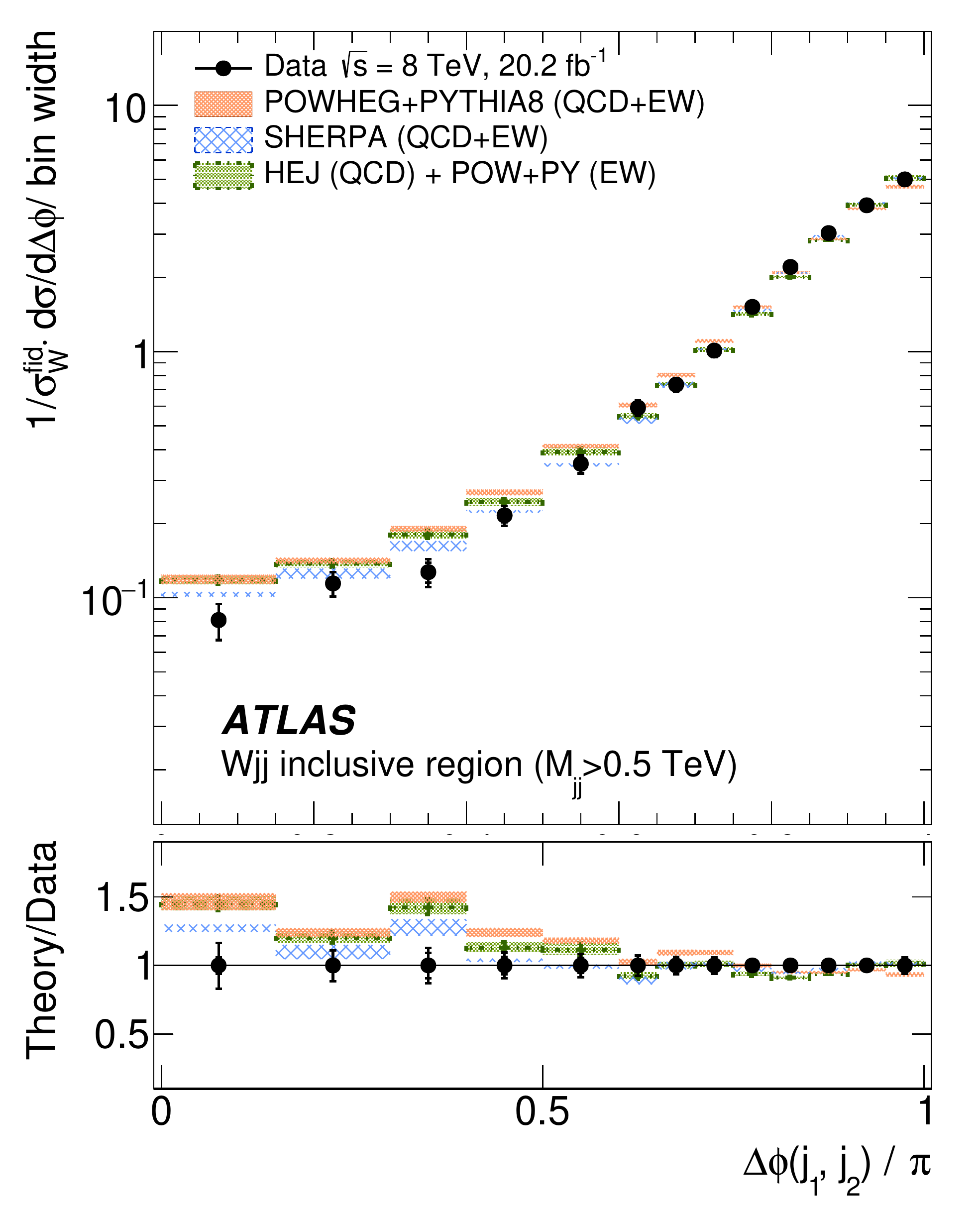}
\includegraphics[width=0.49\textwidth]{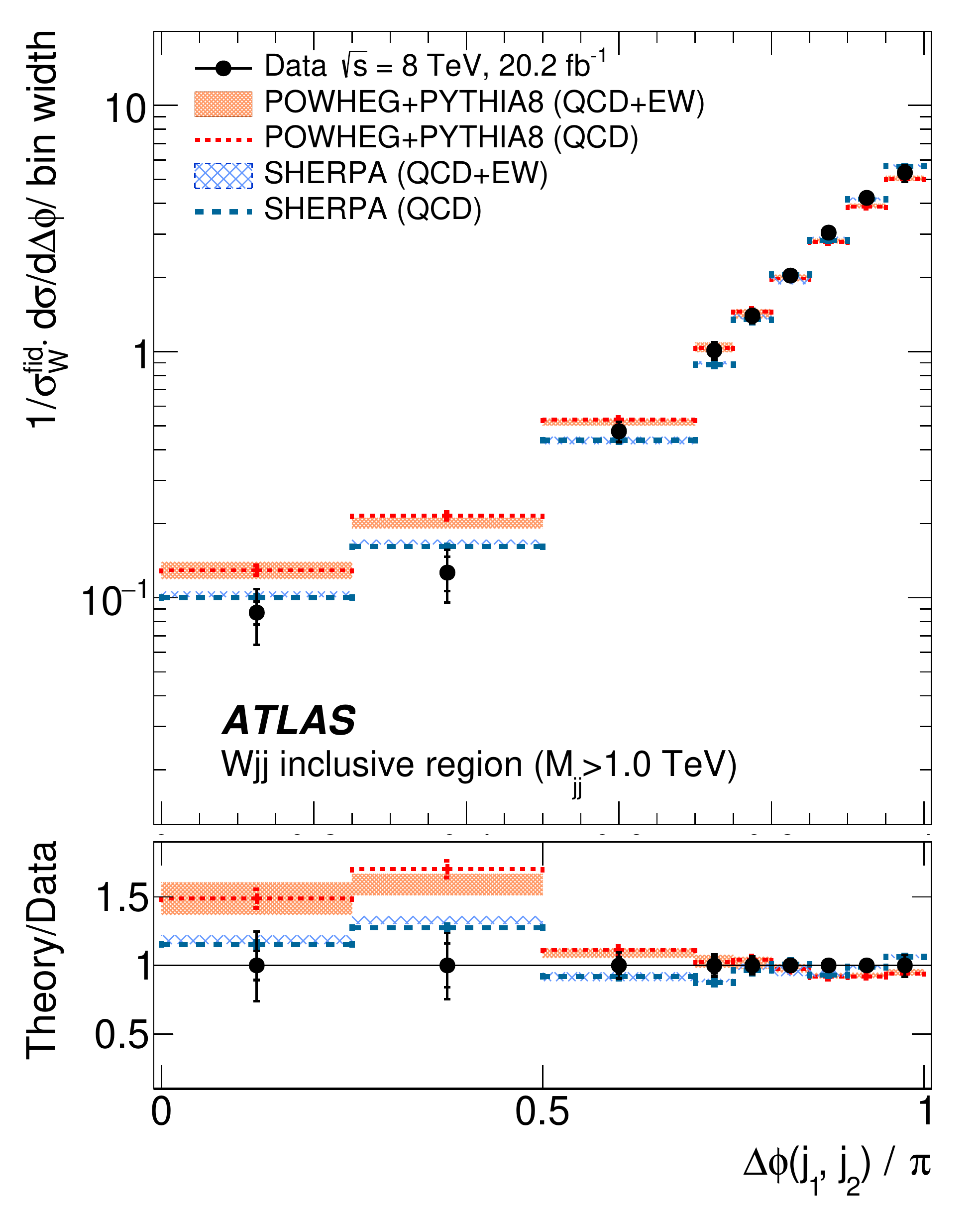}
\includegraphics[width=0.49\textwidth]{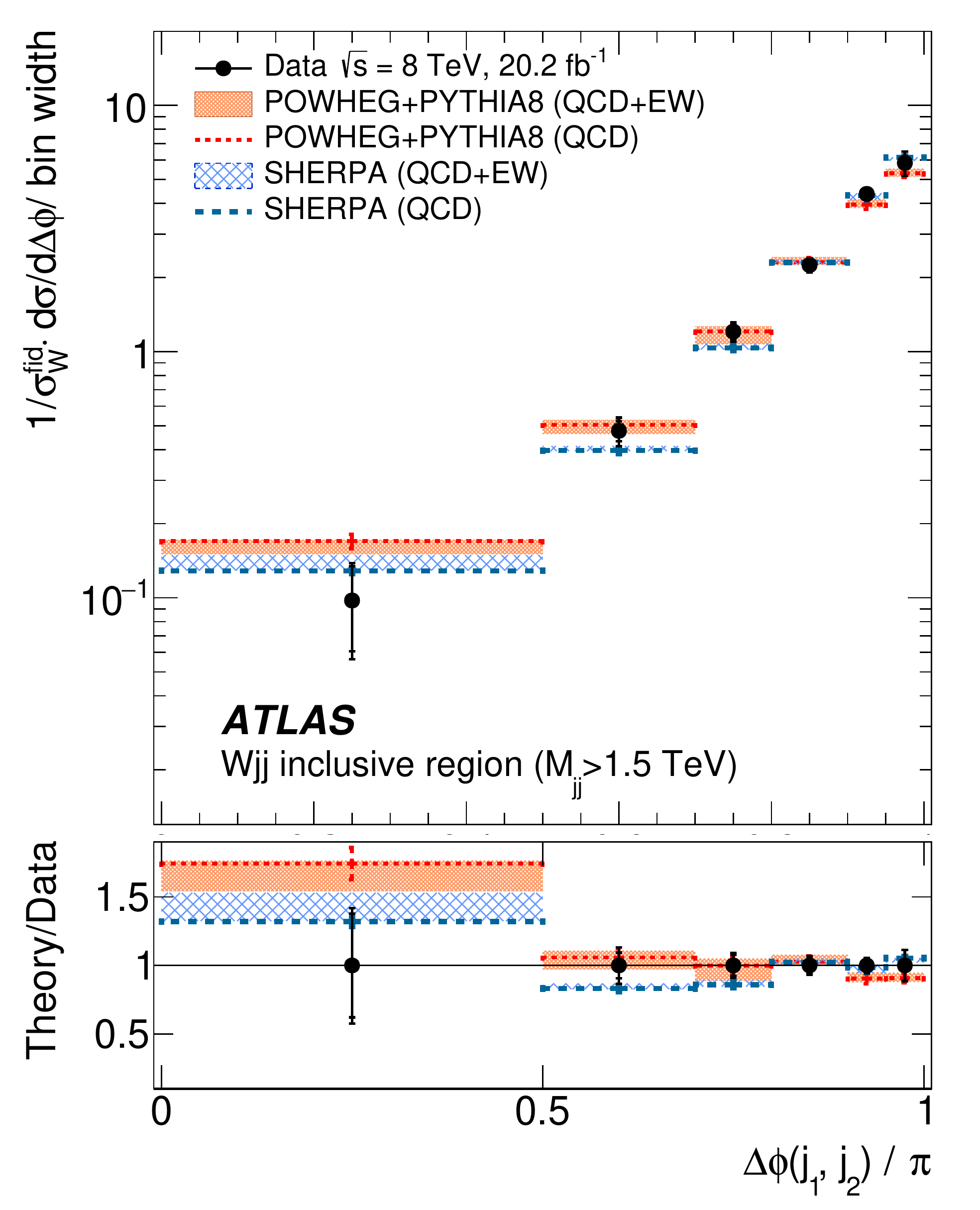}
\includegraphics[width=0.49\textwidth]{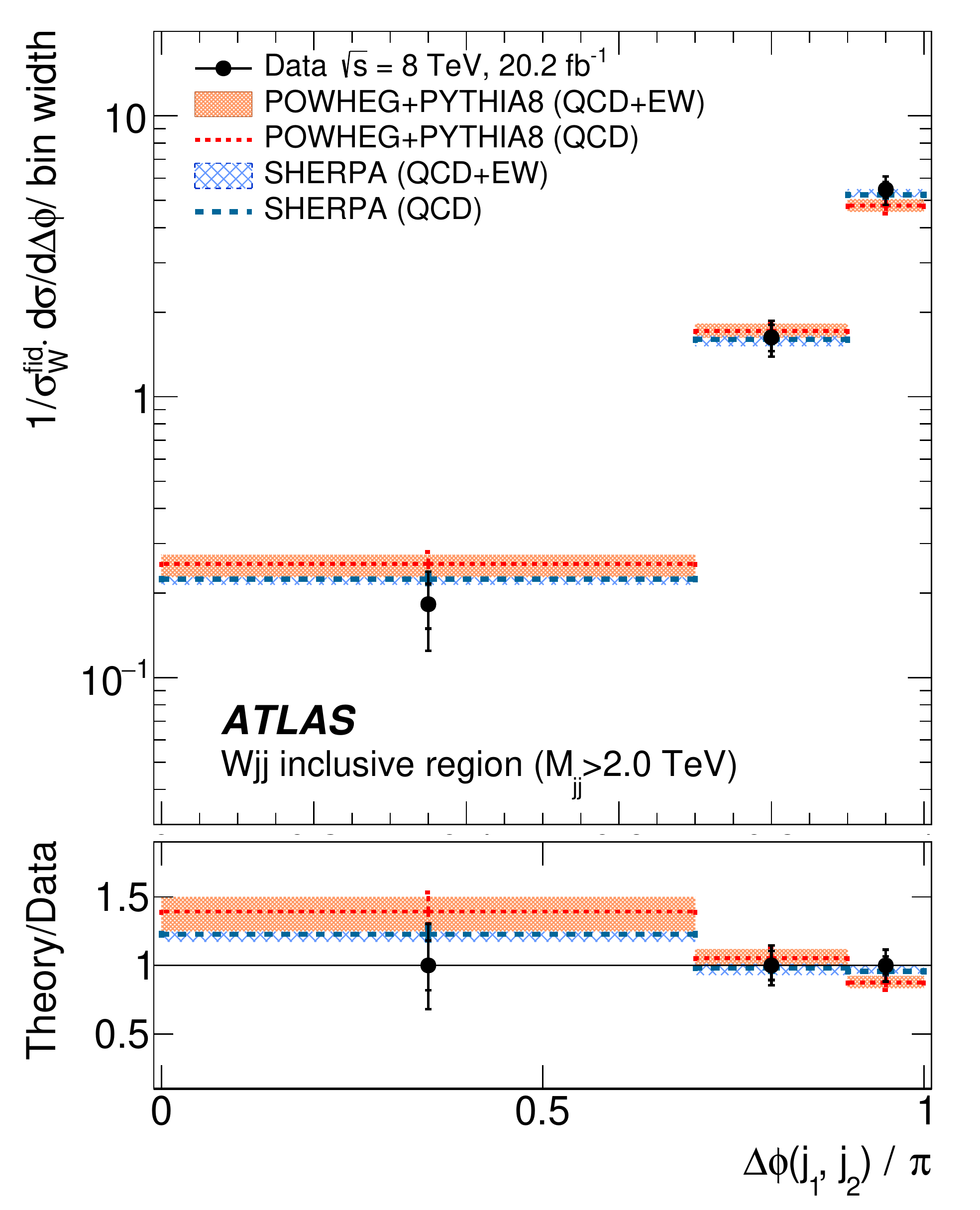}
\caption{Unfolded normalized differential \wjets production cross sections as a function of $\Delta\phi(j_1,j_2)$
in the inclusive fiducial region with four thresholds on the dijet invariant mass (0.5~\TeV, 1.0~\TeV, 1.5~\TeV, 
and 2.0~\TeV).  Both statistical (inner bar) and total (outer bar) measurement uncertainties are shown, as well as 
ratios of the theoretical predictions to the data (the bottom panel in each distribution).}
\label{unfolding:aux:AUX11}
\end{figure}
\begin{figure}[htbp]
\centering
\includegraphics[width=0.35\textwidth]{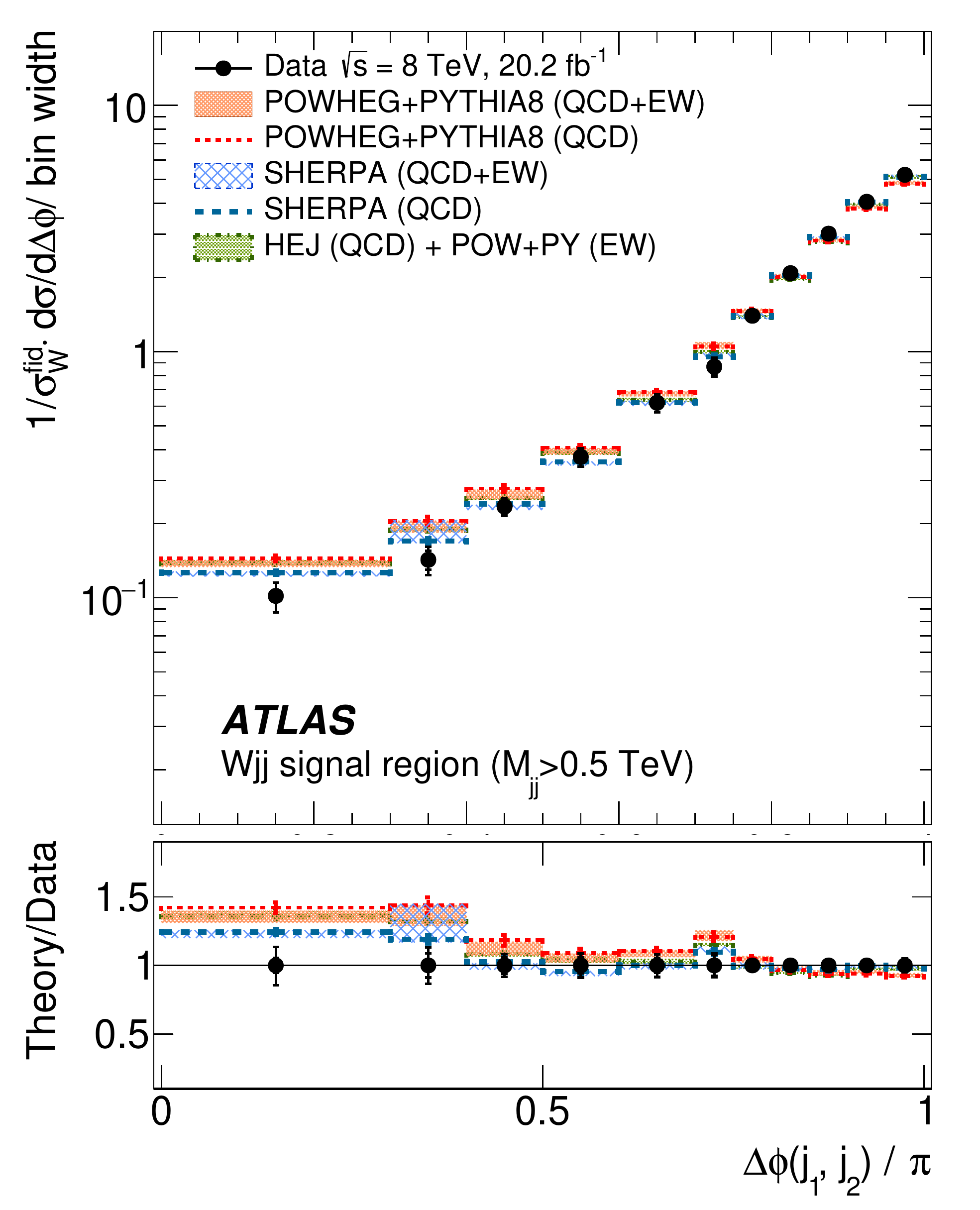}
\includegraphics[width=0.35\textwidth]{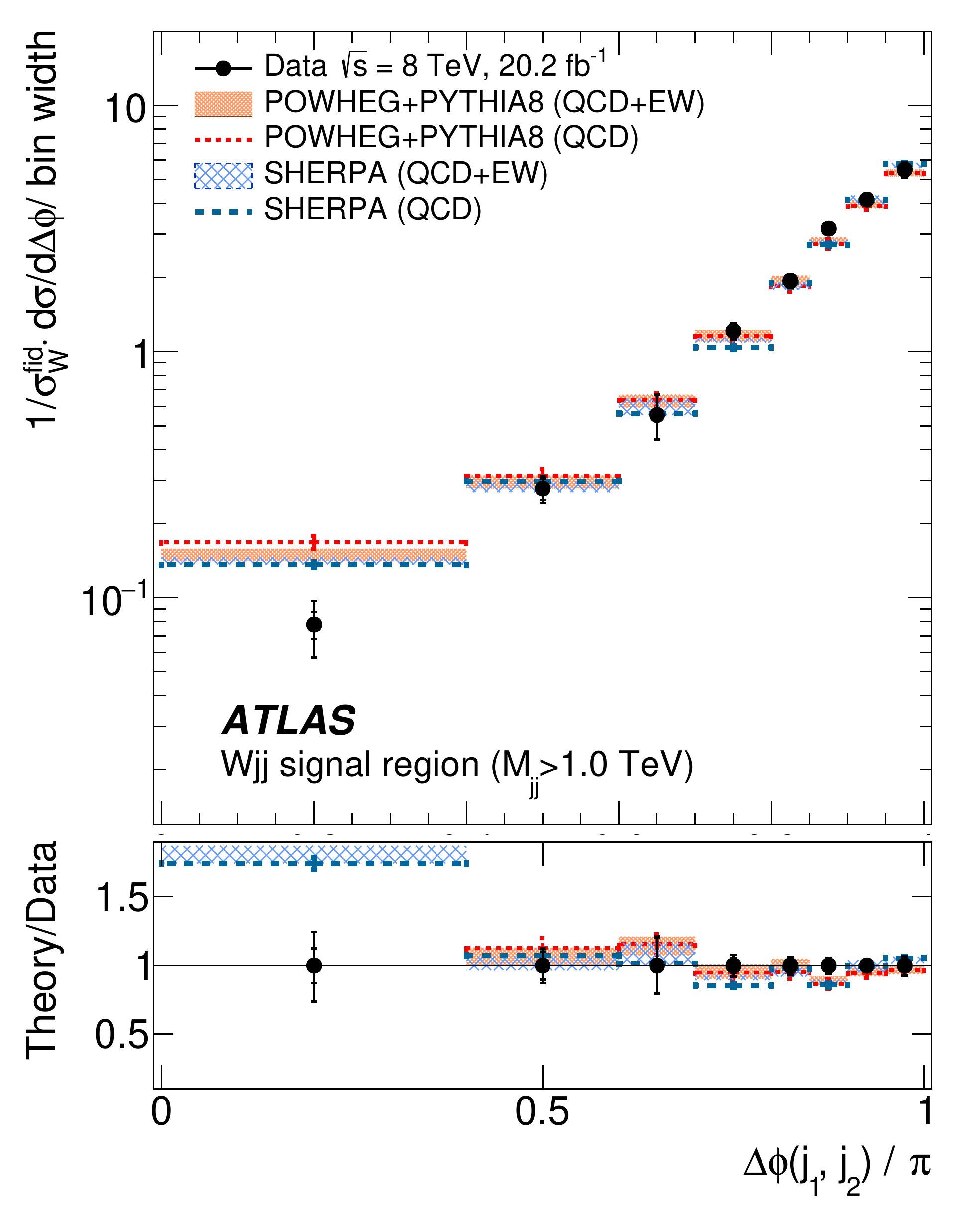}
\includegraphics[width=0.35\textwidth]{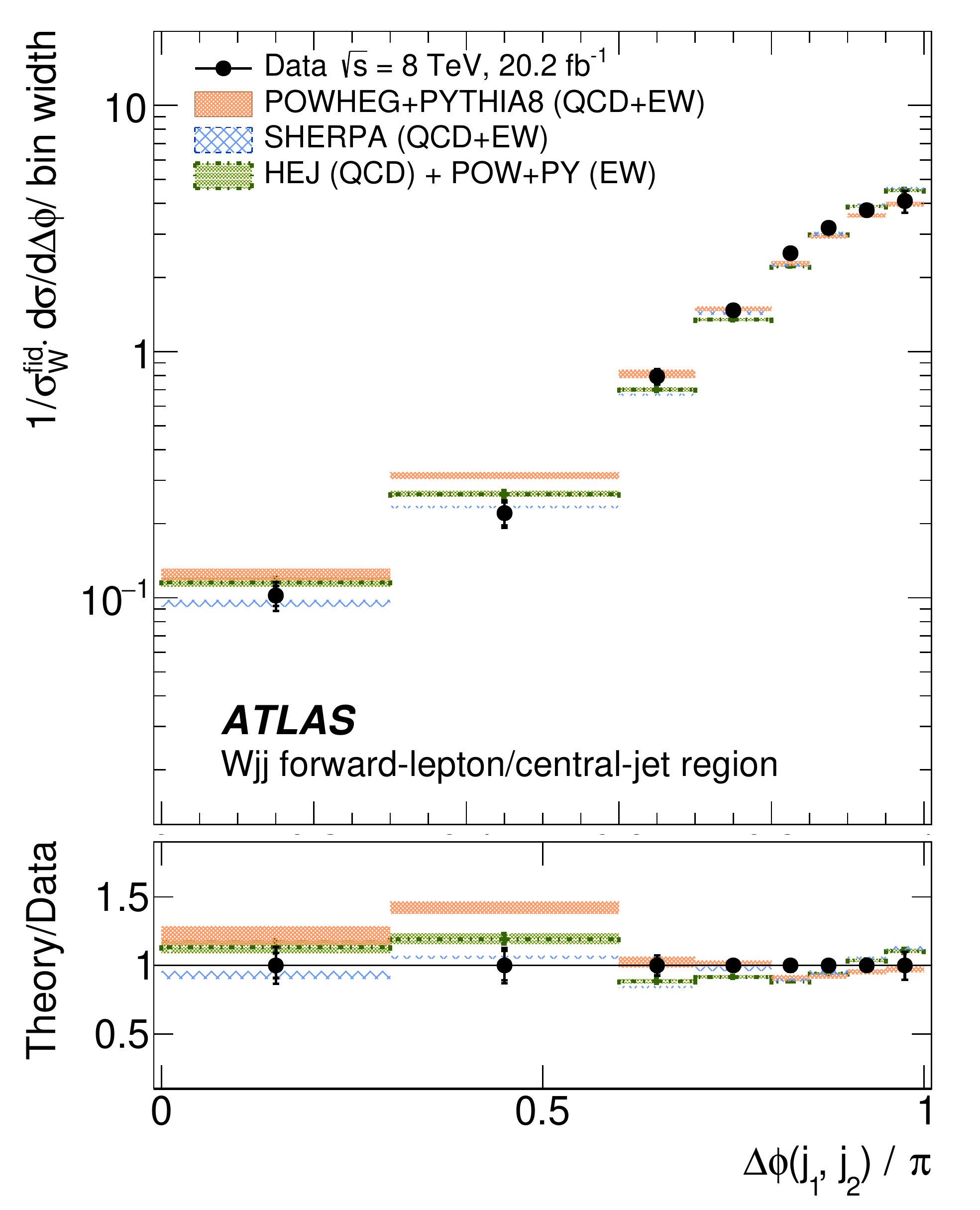}
\includegraphics[width=0.35\textwidth]{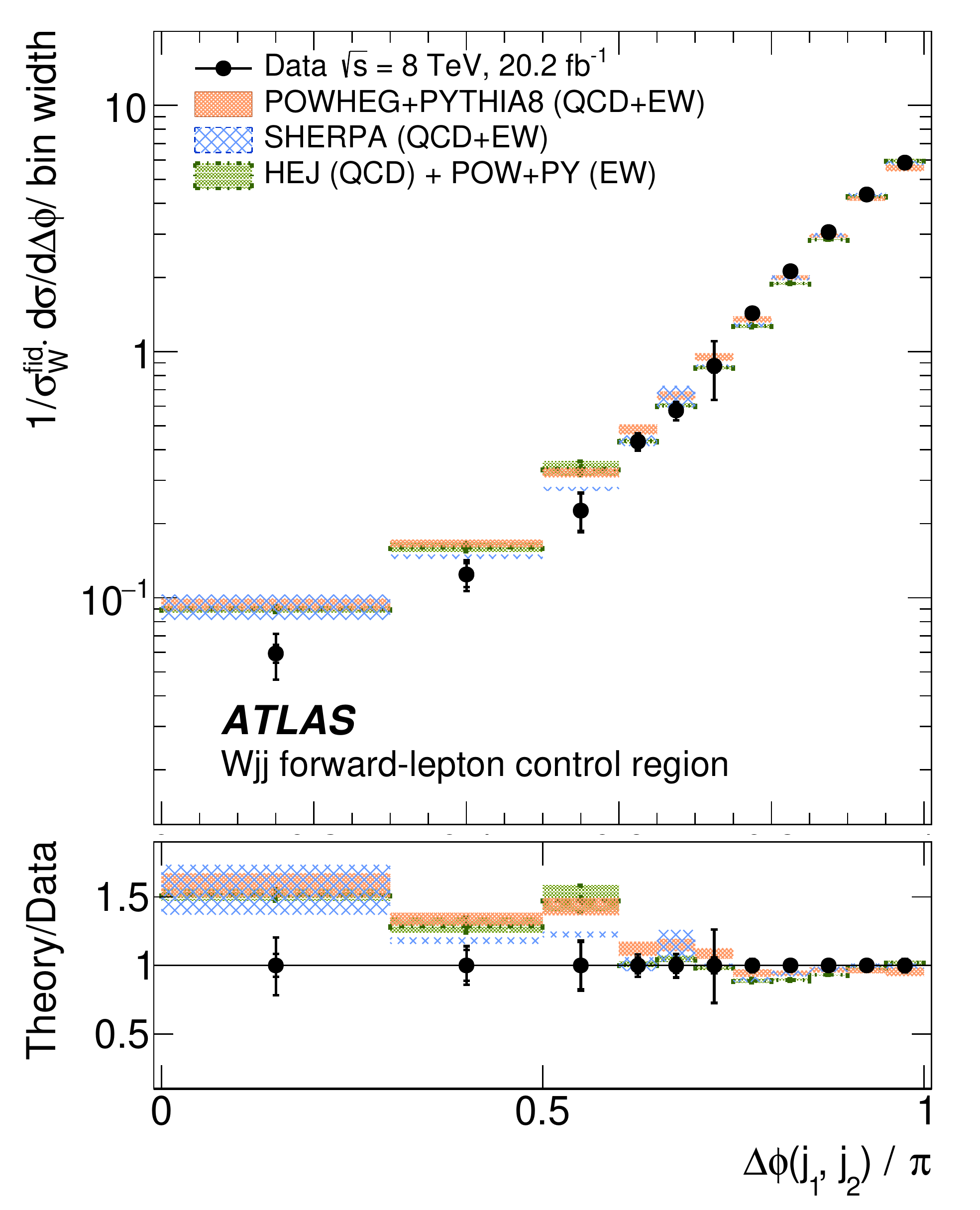}
\includegraphics[width=0.35\textwidth]{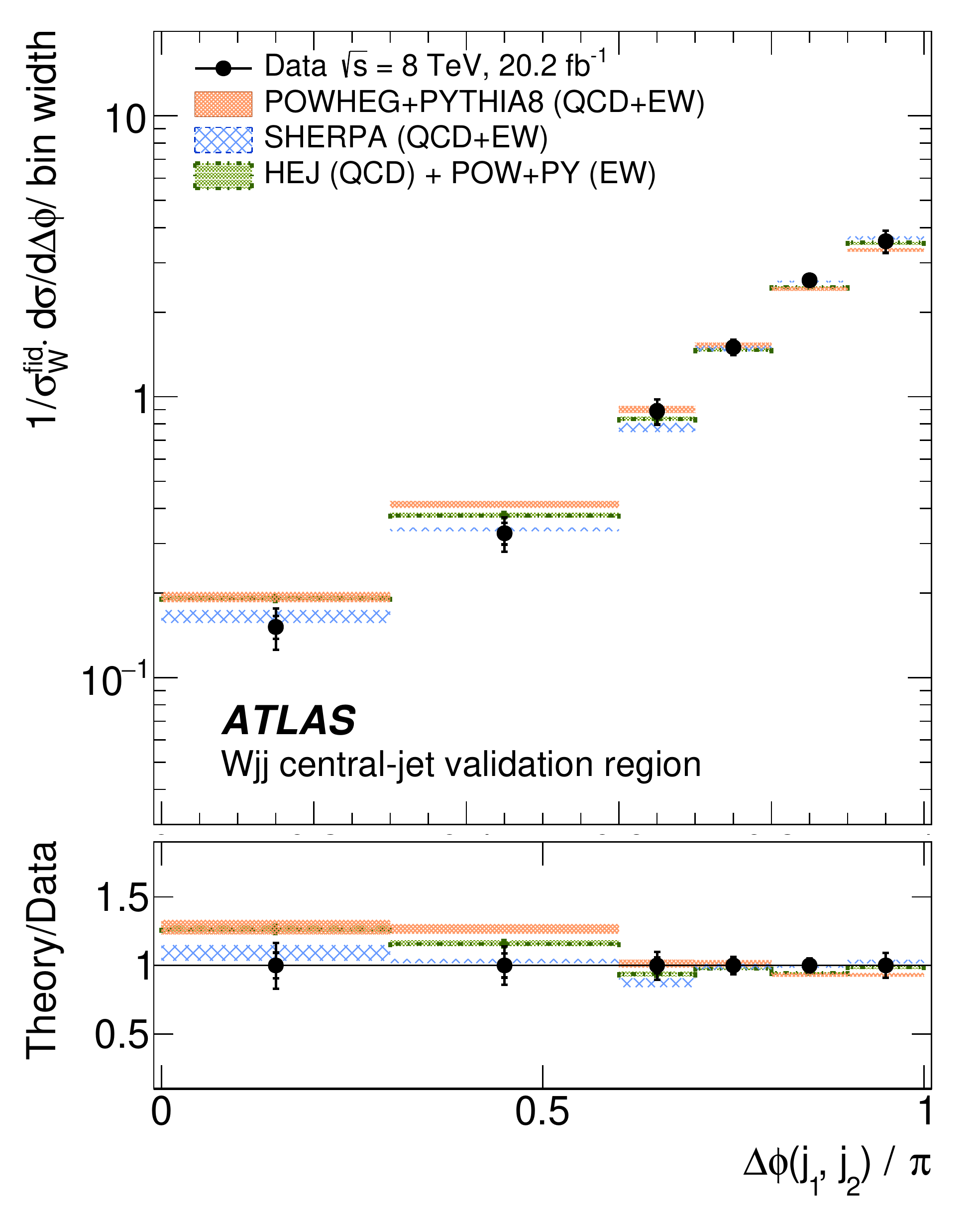}
\caption{Unfolded normalized differential \wjets production cross sections as a function of $\Delta\phi(j_1,j_2)$ in 
the signal, high-mass signal, forward-lepton/central-jet, forward-lepton, and central-jet fiducial regions.  Both 
statistical (inner bar) and total (outer bar) measurement uncertainties are shown, as well as ratios
of the theoretical predictions to the data (the bottom panel in each distribution). }
\label{unfolding:aux:AUX12}
\end{figure}
\begin{figure}[htbp]
\centering
\includegraphics[width=0.35\textwidth]{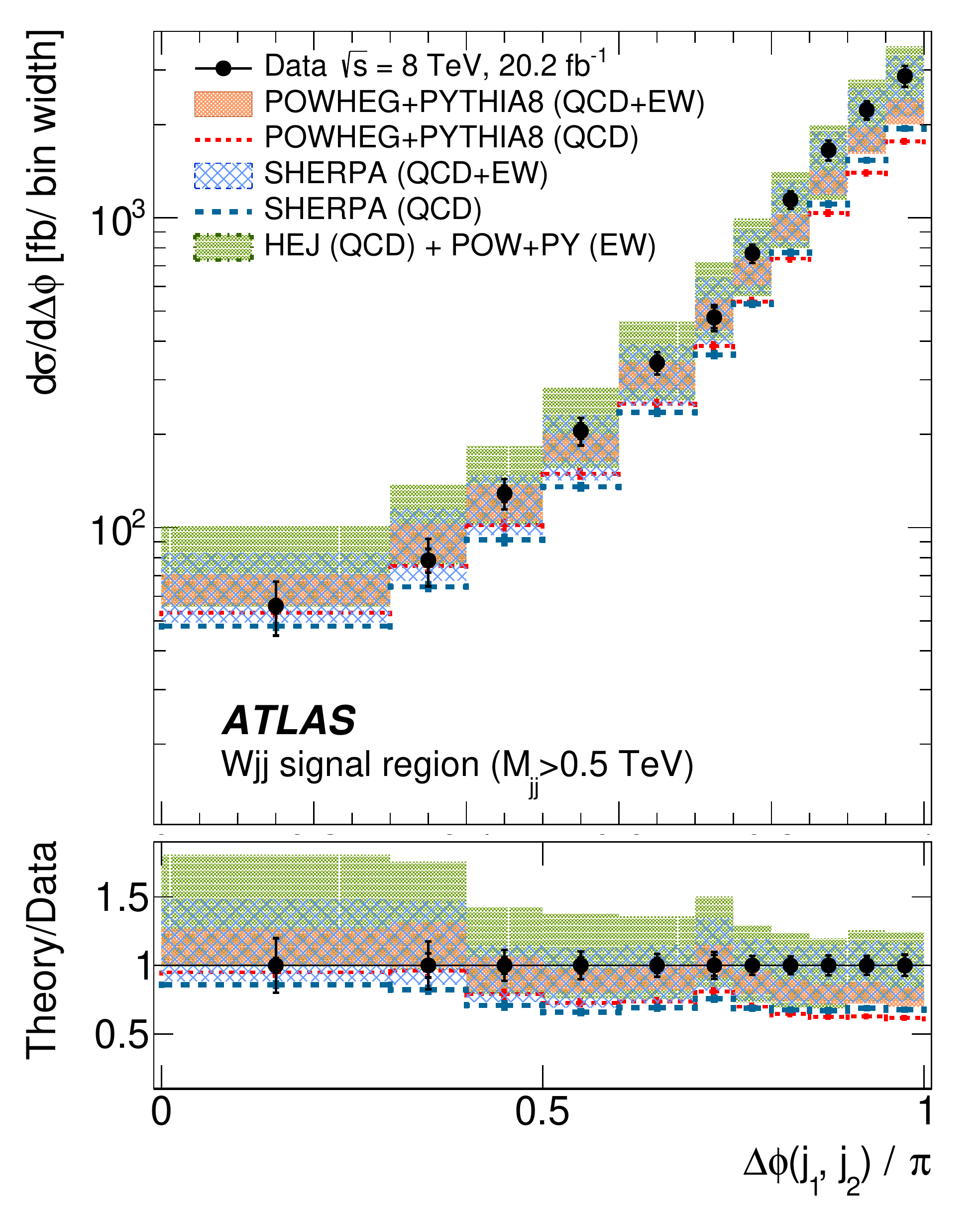}
\includegraphics[width=0.35\textwidth]{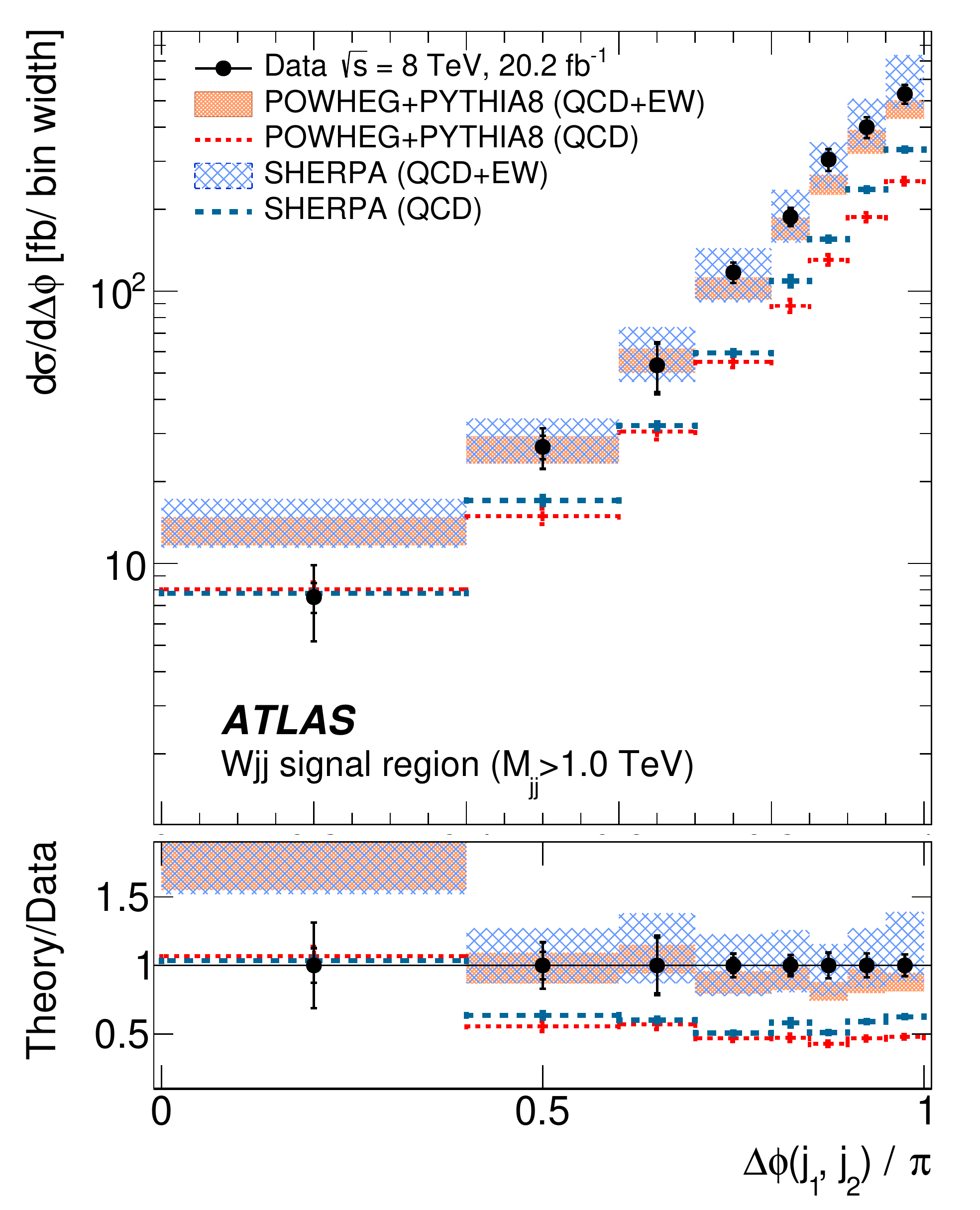}
\includegraphics[width=0.35\textwidth]{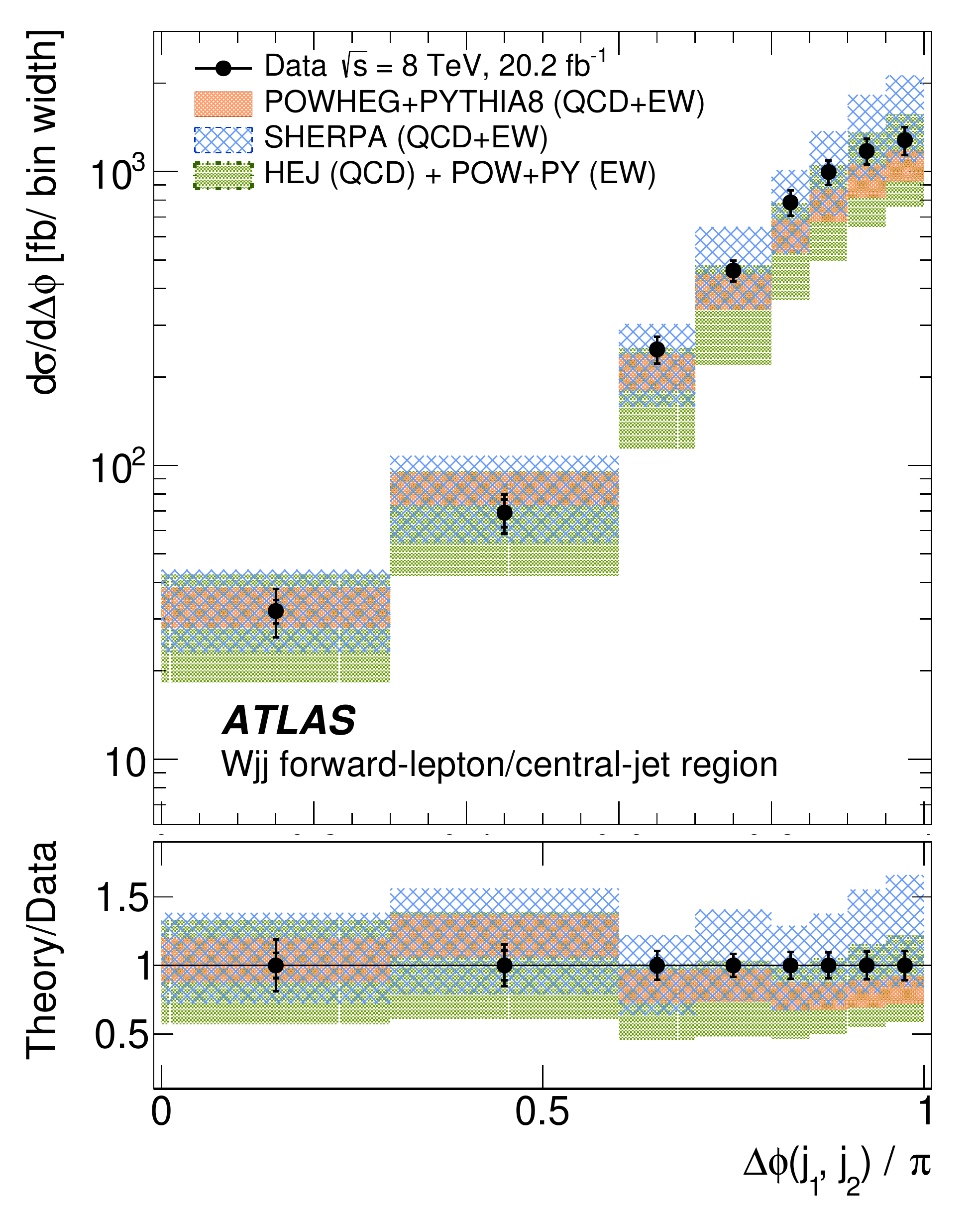}
\includegraphics[width=0.35\textwidth]{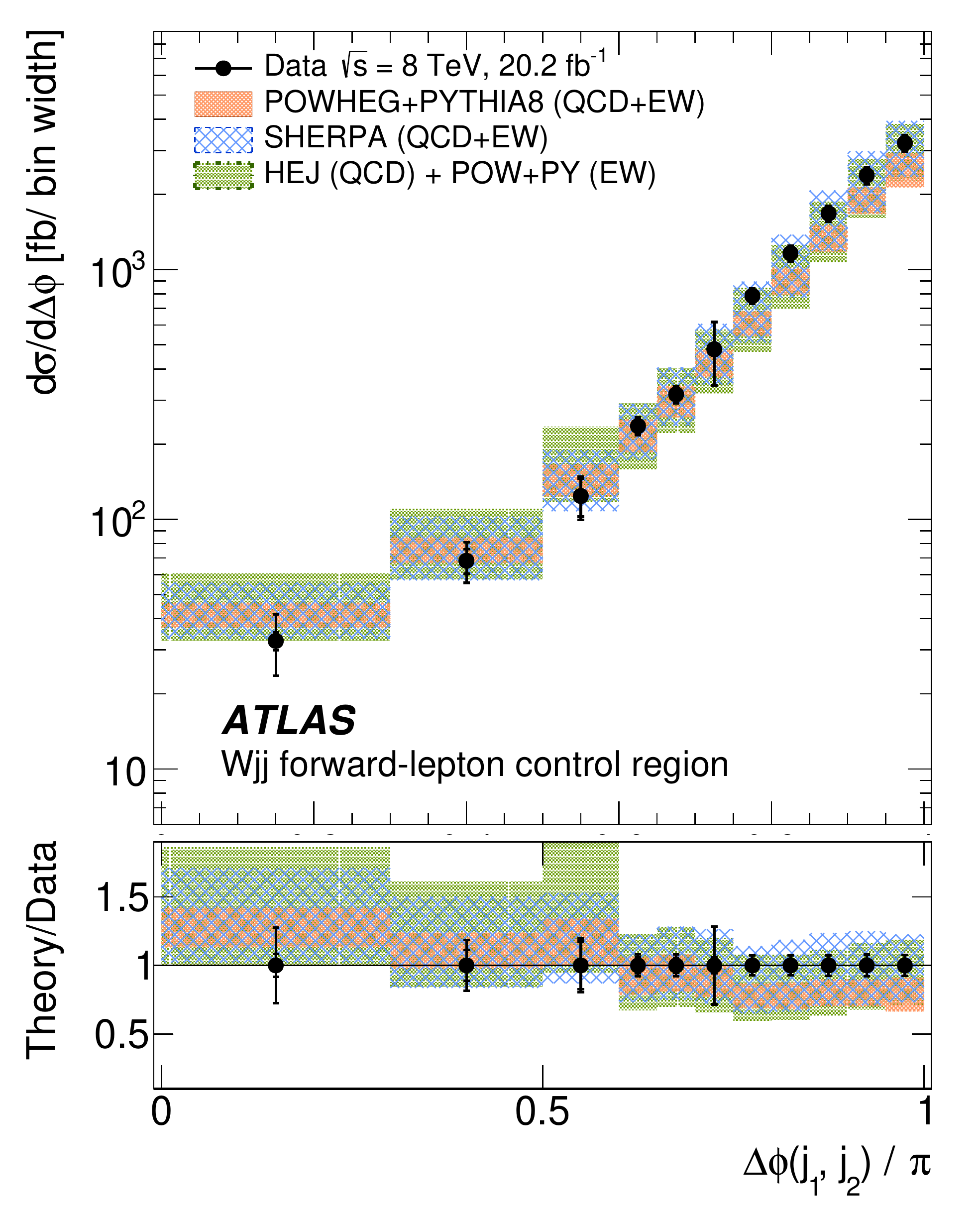}
\includegraphics[width=0.35\textwidth]{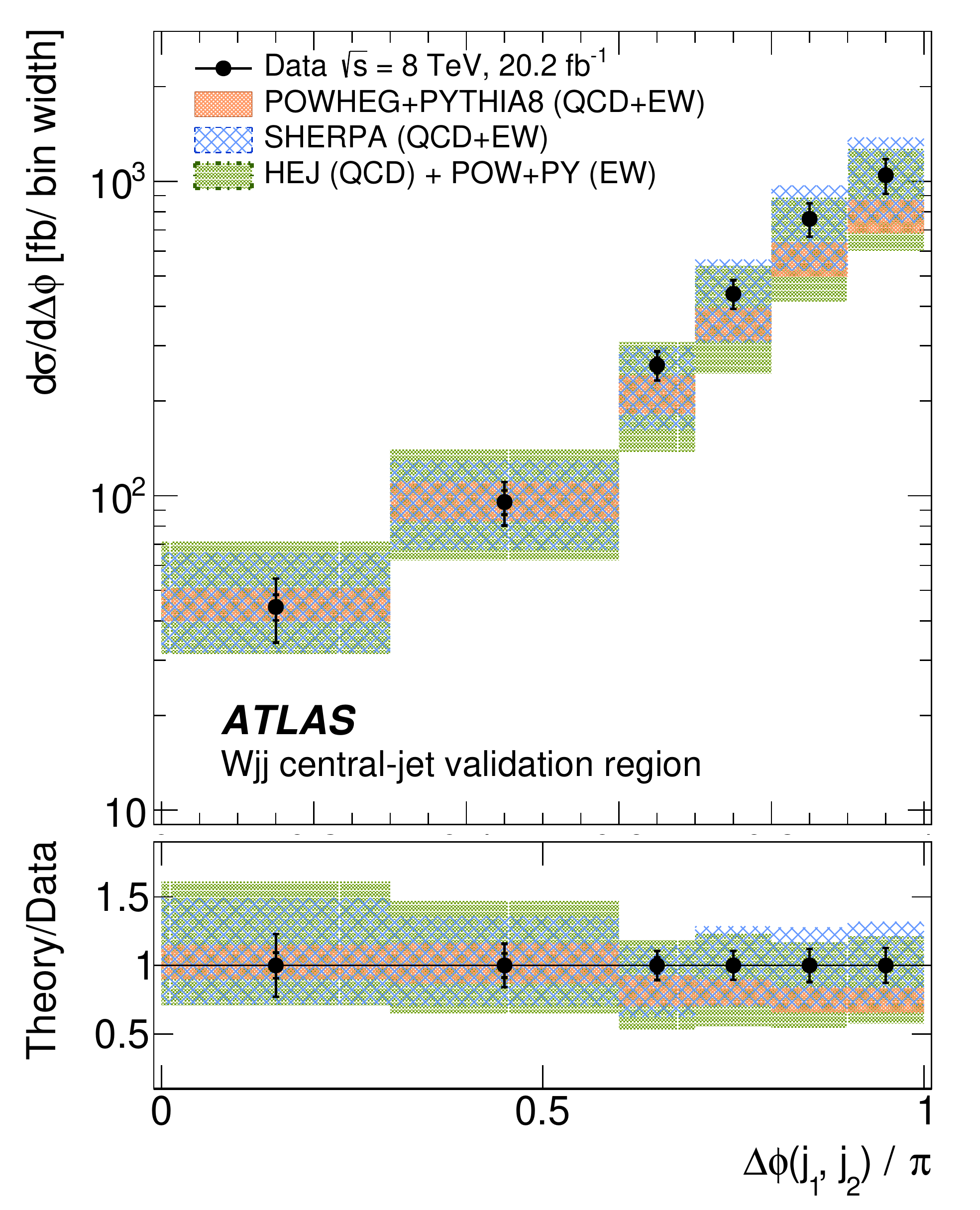}
\caption{Differential \wjets production cross sections as a function of $\Delta\phi(j_1,j_2)$ in the signal, 
high-mass signal, forward-lepton/central-jet, forward-lepton, and central-jet fiducial regions.  Both statistical
(inner bar) and total (outer bar) measurement uncertainties are shown, as well as ratios
of the theoretical predictions to the data (the bottom panel in each distribution). }
\label{unfolding:aux:AUX19}
\end{figure}
\begin{figure}[htbp]
\centering
\includegraphics[width=0.49\textwidth]{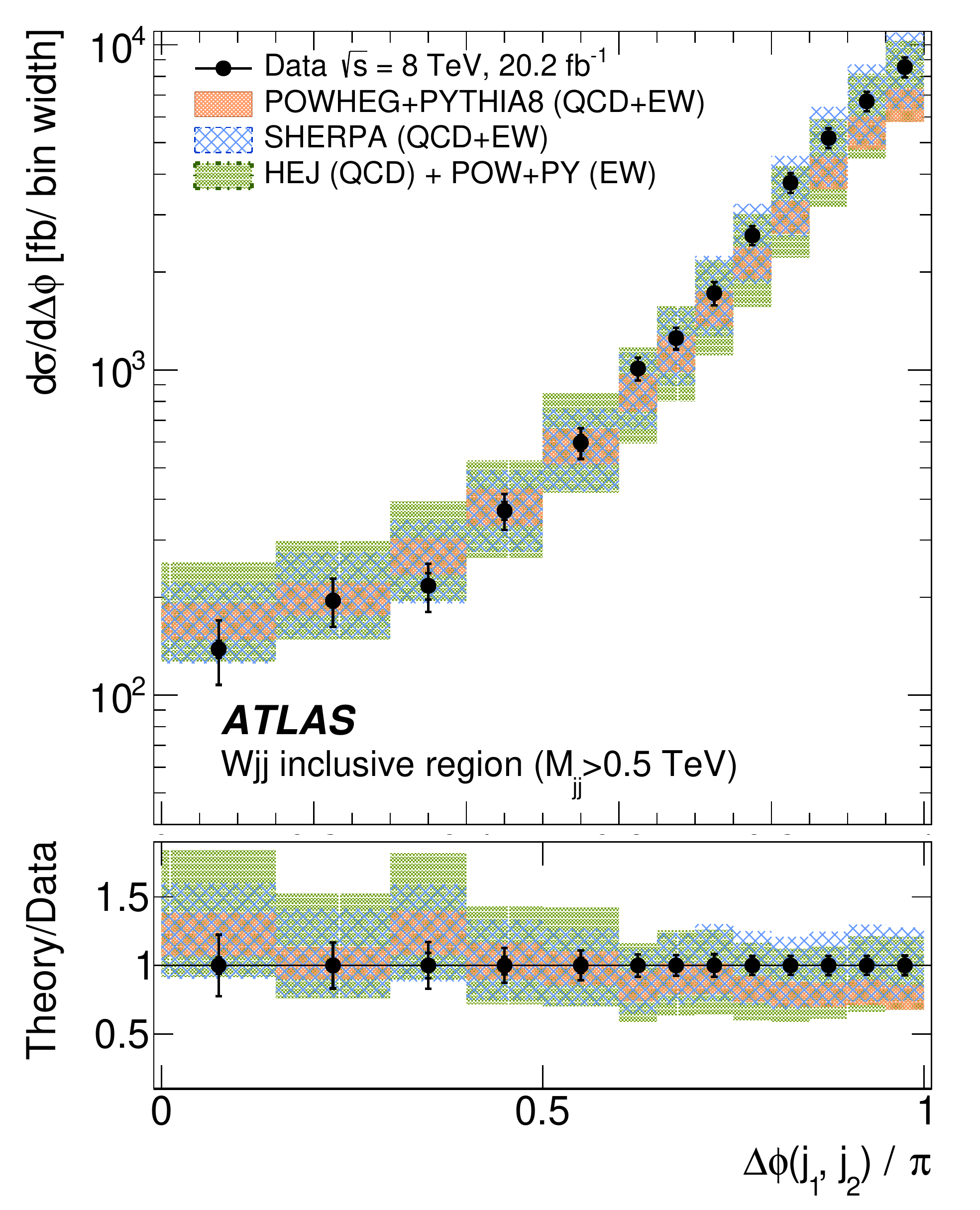}
\includegraphics[width=0.49\textwidth]{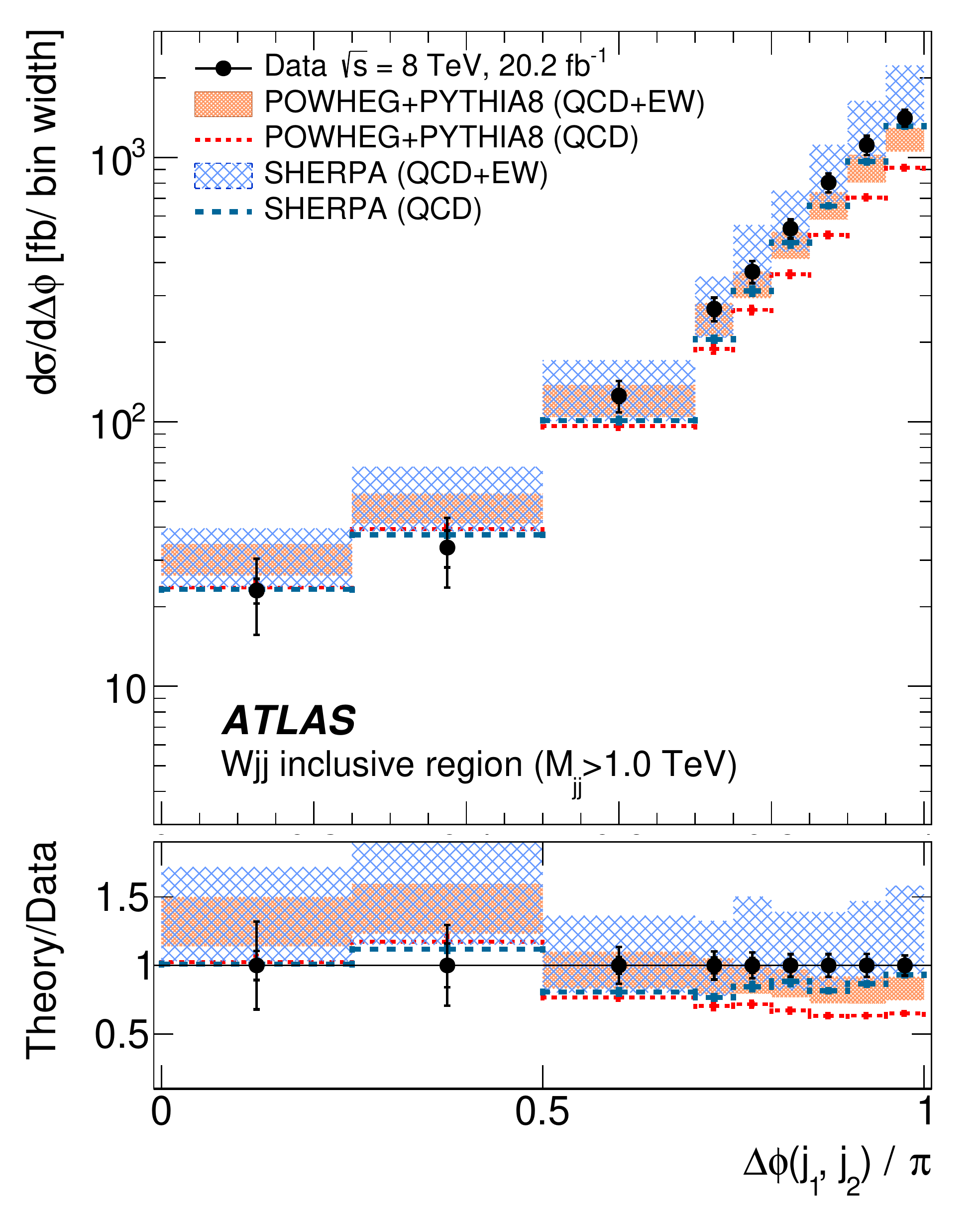}
\includegraphics[width=0.49\textwidth]{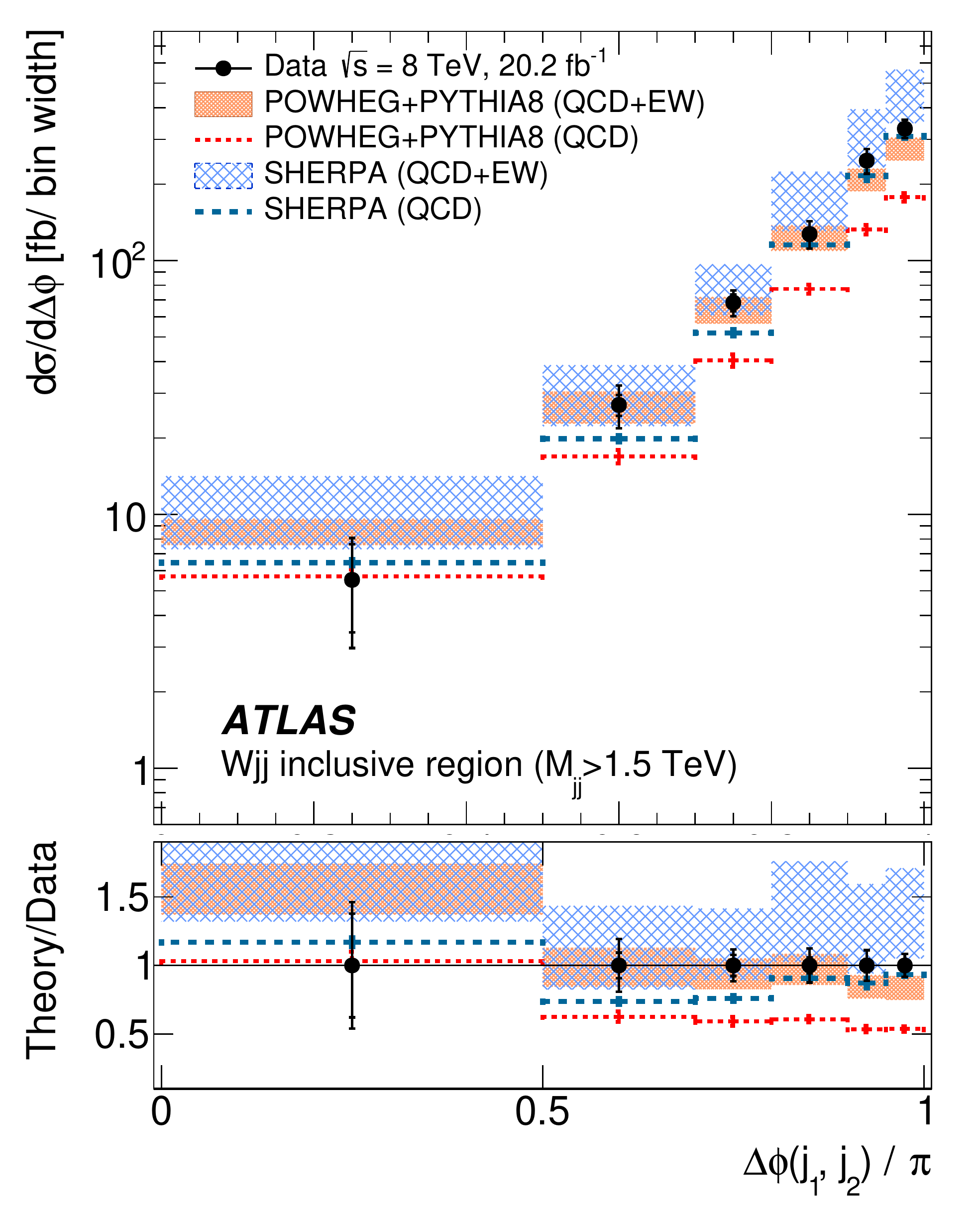}
\includegraphics[width=0.49\textwidth]{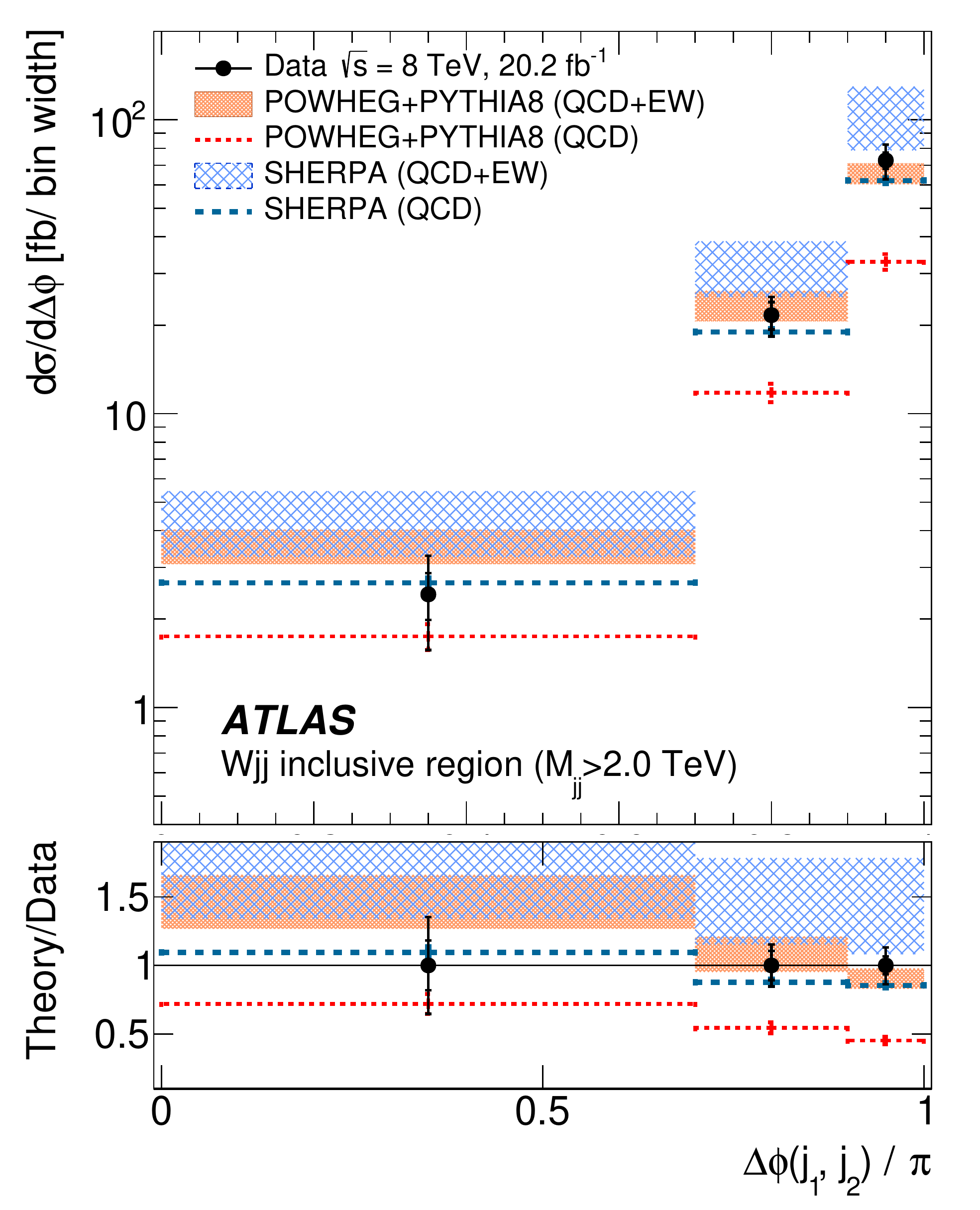}
\caption{Differential \wjets production cross sections as a function of $\Delta\phi(j_1,j_2)$
in the inclusive fiducial region with four thresholds on the dijet invariant mass (0.5~\TeV, 1.0~\TeV, 1.5~\TeV, 
and 2.0~\TeV).  Both statistical (inner bar) and total (outer bar) measurement uncertainties are shown, as well as ratios
of the theoretical predictions to the data (the bottom panel in each distribution). }
\label{unfolding:aux:AUX20}
\end{figure}
\begin{figure}[htbp]
\centering
\includegraphics[width=0.35\textwidth]{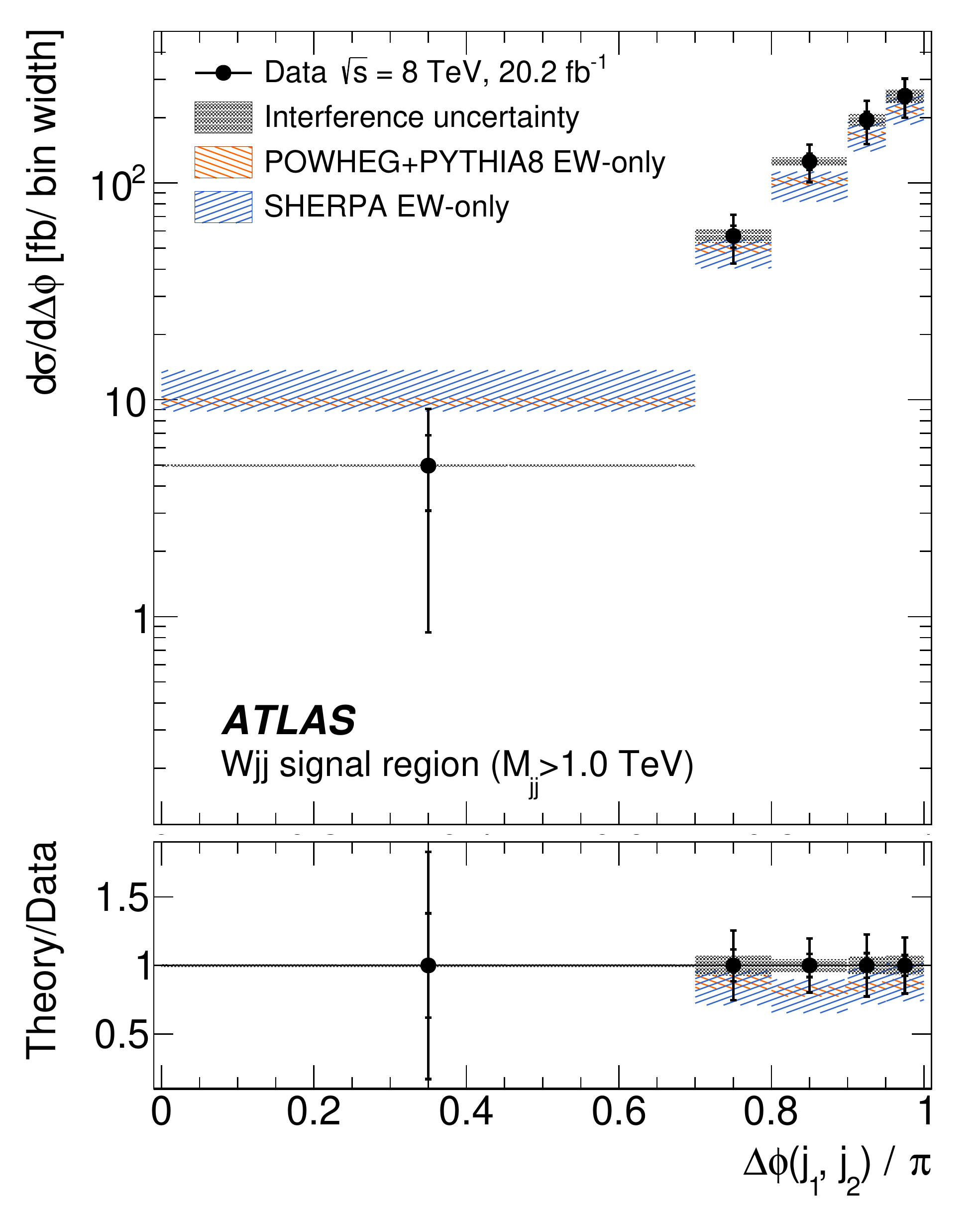}
\includegraphics[width=0.35\textwidth]{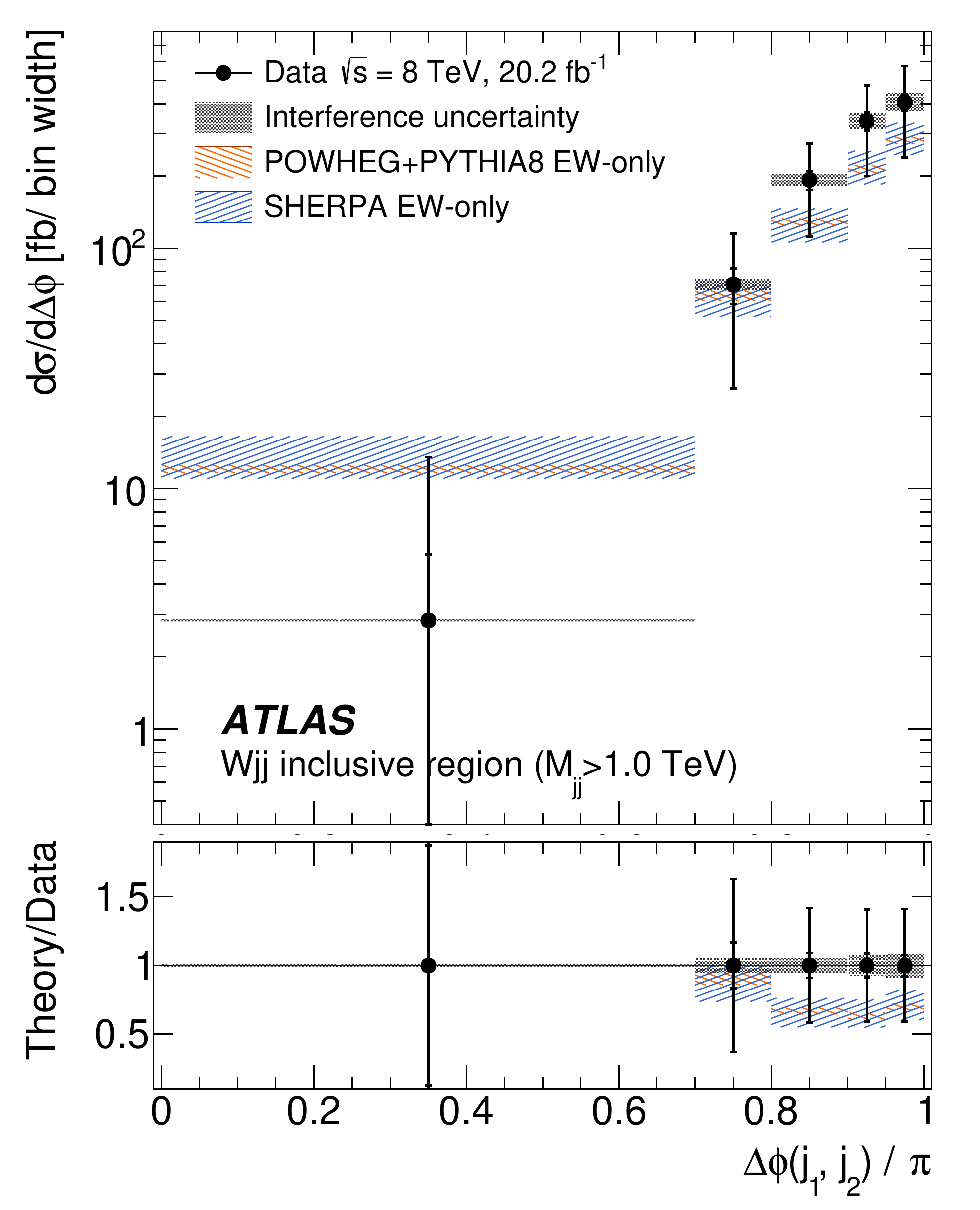}
\includegraphics[width=0.35\textwidth]{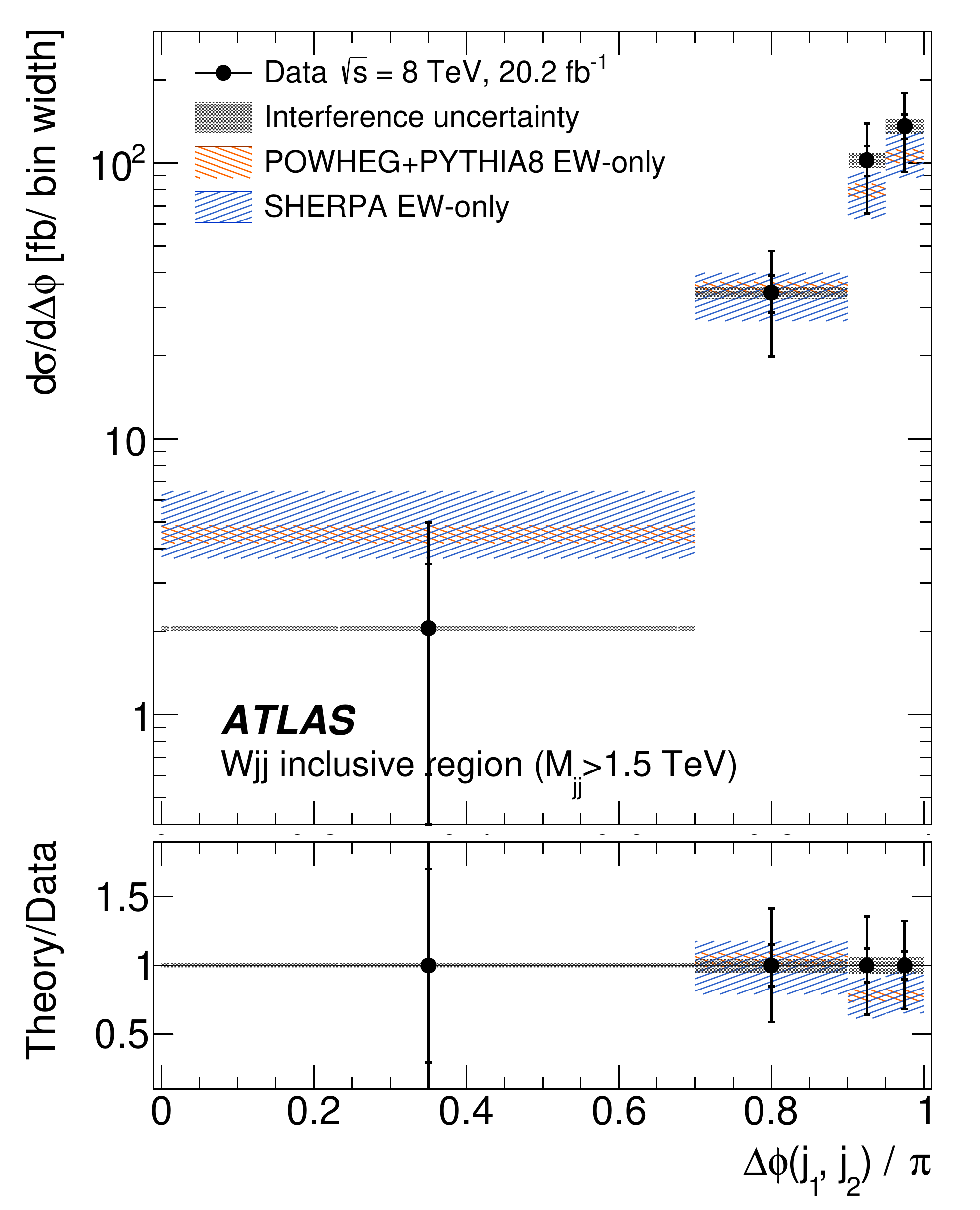}
\includegraphics[width=0.35\textwidth]{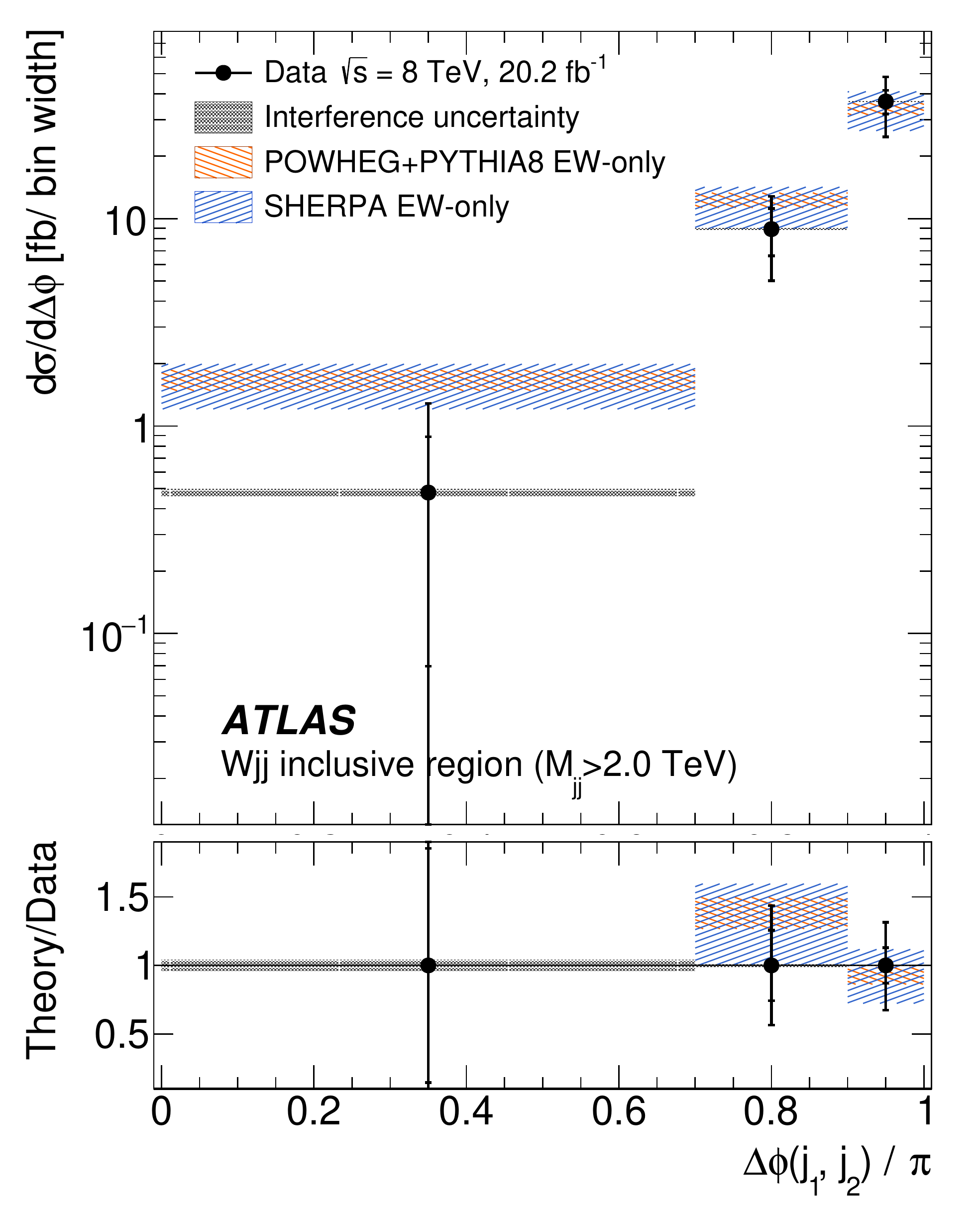}
\includegraphics[width=0.35\textwidth]{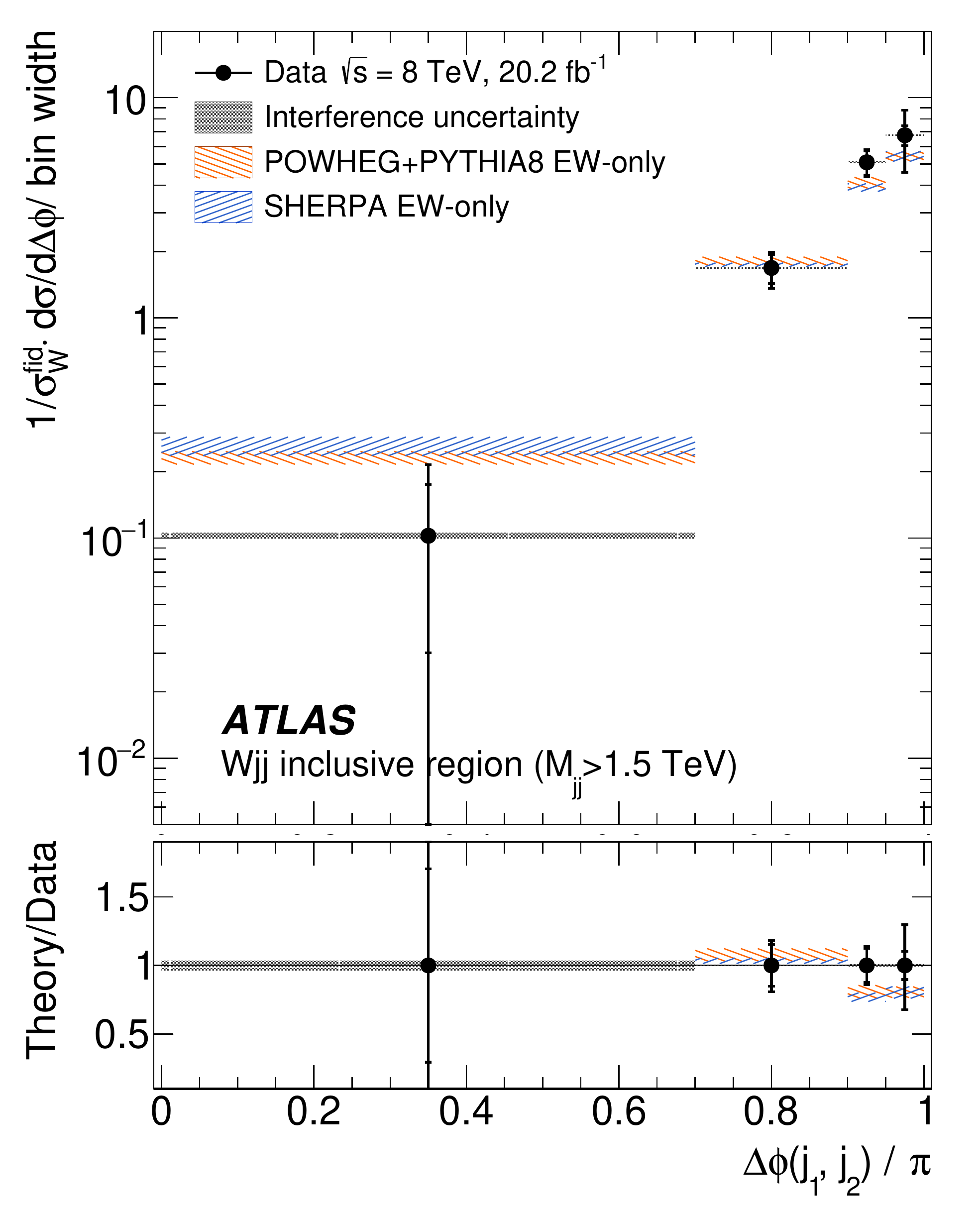}
\includegraphics[width=0.35\textwidth]{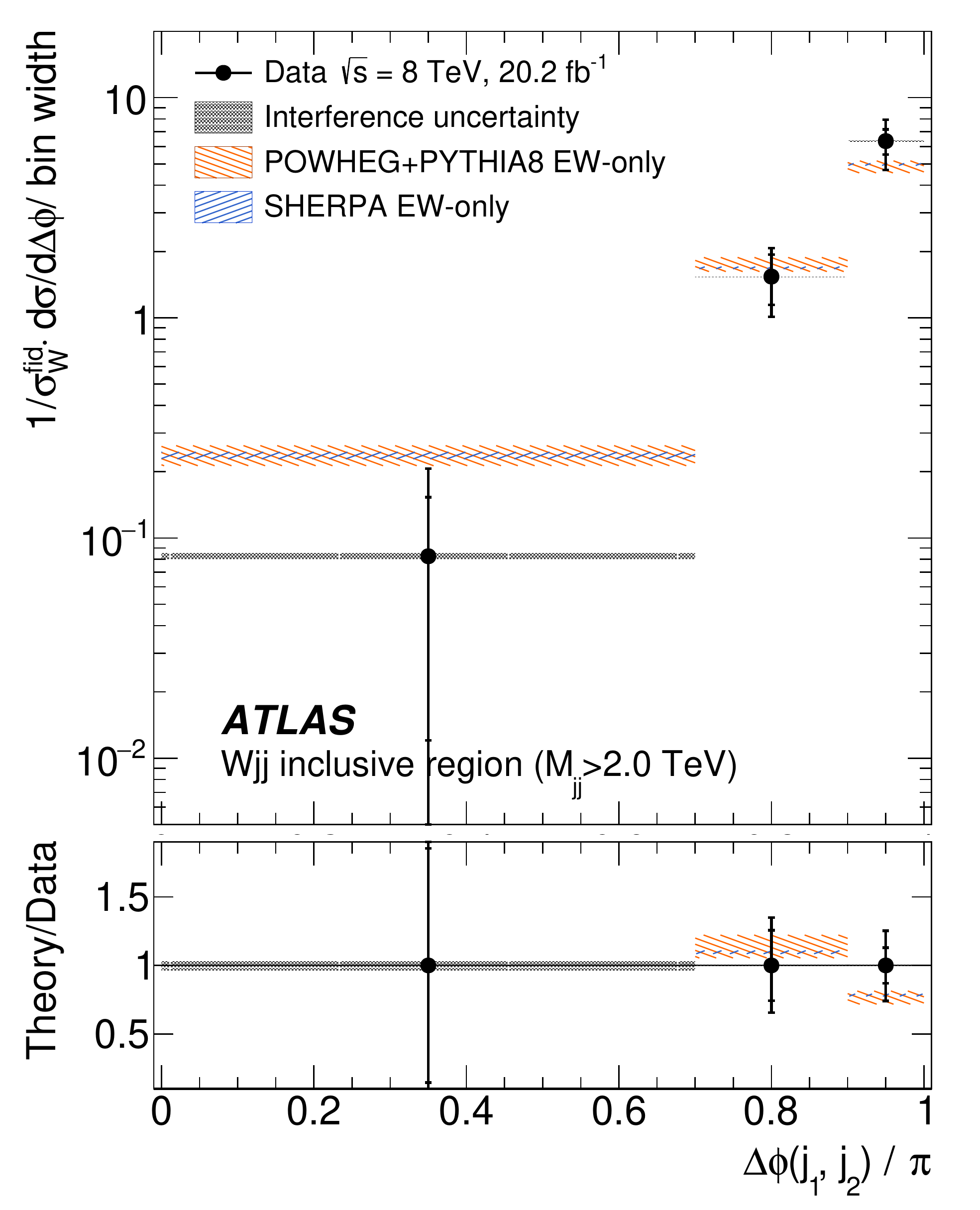}
\caption{Differential electroweak \wjets production cross sections as a function of $\Delta\phi(j_1,j_2)$ in 
the high-mass signal region and the inclusive fiducial region with three thresholds on the dijet invariant mass 
(1.0~\TeV, 1.5~\TeV, and 2.0~\TeV).  The bottom two distributions are normalized, the rest are absolute.  
Both statistical (inner bar) and total (outer bar) measurement uncertainties are shown, as well as ratios
of the theoretical predictions to the data (the bottom panel in each distribution). }
\label{unfolding:aux:AUX6}
\end{figure}

This section includes normalized and absolute differential QCD+EW and EW \wjets 
production cross-section measurements not directly discussed in the main text (Figs.~\ref{unfolding:combined_measurementdijetmass1Dinclusive}--\ref{unfolding:aux:AUX6}).  
The complete set of measured differential spectra is available in \textsc{hepdata}\,\cite{HEPDATA}.

\clearpage

%% file: atlas_authlist.tex
\begin{flushleft}
{\Large The ATLAS Collaboration}

\bigskip

M.~Aaboud$^\textrm{\scriptsize 137d}$,
G.~Aad$^\textrm{\scriptsize 88}$,
B.~Abbott$^\textrm{\scriptsize 115}$,
J.~Abdallah$^\textrm{\scriptsize 8}$,
O.~Abdinov$^\textrm{\scriptsize 12}$,
B.~Abeloos$^\textrm{\scriptsize 119}$,
S.H.~Abidi$^\textrm{\scriptsize 161}$,
O.S.~AbouZeid$^\textrm{\scriptsize 139}$,
N.L.~Abraham$^\textrm{\scriptsize 151}$,
H.~Abramowicz$^\textrm{\scriptsize 155}$
% ...
\bigskip
\\
% ...
$^{8}$ Department of Physics, The University of Texas at Arlington, Arlington TX, United States of America\\
% ...
$^{12}$ Institute of Physics, Azerbaijan Academy of Sciences, Baku, Azerbaijan\\
% ...
$^{88}$ CPPM, Aix-Marseille Universit{\'e} and CNRS/IN2P3, Marseille, France\\
% ...
$^{115}$ Homer L. Dodge Department of Physics and Astronomy, University of Oklahoma, Norman OK, United States of America\\
% ...
$^{119}$ LAL, Univ. Paris-Sud, CNRS/IN2P3, Universit{\'e} Paris-Saclay, Orsay, France\\
% ...
$^{137}$ $^{(a)}$ Facult{\'e} des Sciences Ain Chock, R{\'e}seau Universitaire de Physique des Hautes Energies - Universit{\'e} Hassan II, Casablanca; $^{(b)}$ Centre National de l'Energie des Sciences Techniques Nucleaires, Rabat; $^{(c)}$ Facult{\'e} des Sciences Semlalia, Universit{\'e} Cadi Ayyad, LPHEA-Marrakech; $^{(d)}$ Facult{\'e} des Sciences, Universit{\'e} Mohamed Premier and LPTPM, Oujda; $^{(e)}$ Facult{\'e} des sciences, Universit{\'e} Mohammed V, Rabat, Morocco\\
% ...
$^{139}$ Santa Cruz Institute for Particle Physics, University of California Santa Cruz, Santa Cruz CA, United States of America\\
% ...
$^{151}$ Department of Physics and Astronomy, University of Sussex, Brighton, United Kingdom\\
% ...
$^{155}$ Raymond and Beverly Sackler School of Physics and Astronomy, Tel Aviv University, Tel Aviv, Israel\\
% ...
$^{161}$ Department of Physics, University of Toronto, Toronto ON, Canada\\
% ...
$^{a}$ Also at Department of Physics, King's College London, London, United Kingdom\\
$^{b}$ Also at Institute of Physics, Azerbaijan Academy of Sciences, Baku, Azerbaijan\\
$^{c}$ Also at Novosibirsk State University, Novosibirsk, Russia\\
$^{d}$ Also at TRIUMF, Vancouver BC, Canada\\
$^{e}$ Also at Department of Physics {\&} Astronomy, University of Louisville, Louisville, KY, United States of America\\
% ...
\end{flushleft}
